%% file: Bertemes2018_draft4.tex
\documentclass[a4paper,fleqn,usenatbib]{mnras}

\usepackage{mathptmx}

\usepackage[T1]{fontenc}
\usepackage{ae,aecompl}
\usepackage{fixltx2e}

\usepackage{graphicx}	
\usepackage{amsmath}	
\usepackage{amssymb}	

\usepackage{xcolor}		
\usepackage{mathrsfs} 
\usepackage{breakcites} 
\usepackage{caption} 
\usepackage{multicol}
\usepackage{rotating} 
\usepackage{subfigure}
\usepackage[labelfont=bf]{caption}

\usepackage{letltxmacro} 
\usepackage{xparse}

\input{mycommands.tex}

\LetLtxMacro\oldcitep\citep 
\RenewDocumentCommand{\citep}{O{} O{} m}{\oldcitep[#1][#2]{#3}}
\NewDocumentCommand{\citex}{O{} O{} m}{\oldcitep{#3}}

\LetLtxMacro\oldcitet\citet 
\RenewDocumentCommand{\citet}{O{} O{} m}{\oldcitet[#1][#2]{#3}}

\title[CO- vs dust-based gas masses \& dynamics in Stripe82]{Cross-calibration of CO- vs dust-based gas masses \\ and assessment of the dynamical mass budget \\ in Herschel-SDSS Stripe82 galaxies} 

\author[Bertemes et al.]{
Caroline Bertemes,$^{1}$ \thanks{E-mail: c.bertemes@bath.ac.uk} 
Stijn Wuyts,$^{1}$
Dieter Lutz,$^{2}$
Natascha M. F\"orster Schreiber, $^{2}$
\newauthor
Reinhard Genzel, $^{2}$
Robert F. Minchin, $^{3}$ 
Carole G. Mundell, $^{1}$,
David Rosario, $^{4}$
\newauthor
Am\'elie Saintonge,  $^{5}$
Linda Tacconi $^{2}$
\\
$^{1}$ Department of Physics, University of Bath, Claverton Down, Bath, BA2 7AY, UK\\
$^{2}$ Max-Planck Institut f\"ur extraterrestrische Physik, 85741 Garching, Germany \\
 $^{3}$ Arecibo Observatory, HC3 Box 53995, Arecibo, PR 00612, USA \\
 $^{4}$ Department of Physics, University of Durham, South Road, Durham DH1 3LE, UK\\
 $^{5}$ University College London, Gower Street, London WC1E 6BT, UK
}

\date{Accepted XXX. Received YYY; in original form ZZZ}

\pubyear{2017}

\begin{document}
\label{firstpage}
\pagerange{\pageref{firstpage}--\pageref{lastpage}}
\maketitle

\begin{abstract}
We present a cross-calibration of CO- and dust-based molecular gas masses at $z \leqslant 0.2$. Our results are based on a survey with the IRAM 30-m telescope collecting CO(1-0) measurements of 78 massive ($\log \mstar / \Msun > 10$) galaxies with known gas-phase metallicities, and with IR photometric coverage from WISE (22 \micron ) and Herschel SPIRE ($250$, $350$, $500 \micron$). 
We find a tight relation ($\sim 0.17$ dex scatter) between the gas masses inferred from CO and dust continuum emission, with a minor systematic offset of $0.05$ dex.  The two methods can be brought into agreement by applying a metallicity-dependent adjustment factor ($\sim 0.13$ dex scatter). We illustrate that the observed offset is consistent with a scenario in which dust traces not only molecular gas, but also part of the \HI\  reservoir, residing in the \Hmol -dominated region of the galaxy. 
Observations of the CO(2-1) to CO(1-0) line ratio for two thirds of the sample indicate a narrow range in excitation properties, with a median ratio of luminosities $ \left\langle R_{21} \right\rangle \sim 0.64 $. 
Finally, we find dynamical mass constraints from spectral line profile fitting to agree well with the anticipated mass budget enclosed within an effective radius, once all mass components (stars, gas and dark matter) are accounted for.
\end{abstract}

\begin{keywords}
{galaxies: ISM -- galaxies: evolution -- galaxies: fundamental parameters -- galaxies: kinematics and dynamics -- radio lines: galaxies -- surveys}
\end{keywords}



\section{Introduction}
\label{sec:Intro}

The stellar build up in galaxies proceeds for the most part gradually, with about $90$\% of star formation occurring on the so-called Main Sequence \citep[MS; ]{Brinchmann2004, Noeske2007_MS, Peng2010_Qing}, a near-linear relation between the star formation rate (SFR) and stellar mass which separates regular galaxies from starbursting and passive galaxies. The relative ratio of starbursts to MS galaxies remains roughly unchanged out to $z \sim 2$ \citep{Rodighiero2011, Sargent2012}, the epoch at which the cosmic SFR density peaked \citep{Madau2014_SFH_rev}. The tightness and shape of the relation suggests that most of the stellar mass growth is governed by a self-regulated equilibrium of star formation, the gas inflows which fuel it and outflows \citep[e.g. ][]{Lilly2013}.
Addressing to which degree variations in SFR can be attributed to variations in the efficiency of star formation requires knowledge of their gas content.  The fuel for star formation is held in large molecular gas clouds, which cool down efficiently to $\sim 10 \ {\rm K}$ due to dust shielding, the formation of molecules and (mostly) CO emission \citep[see, e.g, ][ for a review]{Heyer2015}. 
Although $\Hmol$ is the most abundant molecule in these reservoirs, it cannot be observed directly because its first excited state, the $J = 1$ rotational state, is $175$ K above the ground state. At the low prevalent temperatures, almost no molecules occupy this state. Any assessment of the molecular gas content is thus confined to indirect methodologies.

One avenue is to study the optically thick CO rotational transition lines originating from the surface of molecular gas clouds in order to indirectly infer their mass integrated over a galactic region or the galaxy as a whole. This approach adopts a metallicity-dependent conversion factor $\alpha_{\rm CO}$ \citep{Genzel2012, Bolatto2013}. {Historically, CO-based studies were first conducted predominantly} for very luminous infrared outliers including star-bursts \citep[e.g.][]{Radford1991, Solomon1997}. Over the past two decades however, studies in the Local Universe have aimed to target a more complete census of galaxies, including regular galaxies \citep{KenneyYoung1988, Saintonge2011, Saintonge2017}, normal, non-interacting disks \citep{Braine1993}, early type galaxies \citep{Combes2007, Young2011}, AGN hosts \citep{Garcia-Burillo2003} and isolated galaxies \citep{Lisenfeld2011}.
Also at higher redshifts, the advent of state-of-the-art facilities with increased sensitivity has enabled studies probing down to normal galaxies, responsible for the bulk of cosmic star formation  \citep{Genzel2015, Tacconi2010, Daddi2010, Geach2011, Magdis2012a, Magnelli2012, Bauermeister2013}.

In recent years, an alternative approach based on studying dust as a gas tracer has emerged, since molecules form primarily on the surface of larger dust grains acting as a coolant and as a protection against dissociating radiation. Wide area and deep pencil-beam observations with the Herschel Space Telescope together with advanced interferometric observations over broad bandwidths with NOEMA and ALMA have accumulated high-quality infrared continuum spectral energy distributions (SEDs) and/or monochromatic flux measurements in the Rayleigh-Jeans tail of the dust emission for large samples of galaxies. From those, dust masses can be derived and combined with metallicity-dependent dust-to-gas scaling relations \citep[e.g., ][]{Leroy2011,Sandstrom2013} to yield a measure of the gas reservoir. While intrinsically lacking some of the other merits of spectral line surveys (use as kinematic tracer, probe of detailed molecular gas conditions, ability to map at higher spatial resolution by exploiting the dynamic range in velocities) for the sake of measuring bulk gas reservoirs, dust-based approaches have been appealing due to the reduced telescope time they typically require.

It is encouraging that the global picture of rapidly declining molecular gas fractions over the past 10 gigayears is recovered by studies employing CO and dust methods alike \citep{Daddi2010, Genzel2010, Genzel2015, Riechers2010, Tacconi2010, Tacconi2013, Tacconi2017, Geach2011, Magdis2012a, Magnelli2012}. Results based on the two tracers can further be brought into quantitative agreement via the application of zero-point corrections \citep{Genzel2015, Tacconi2017}. However, one should bear in mind that for most of the above evolutionary studies, statements on the agreement between methods refer largely to the ensemble of galaxies, and in the case of IR SED analyses frequently rely on stacking. 
Samples of galaxies for which measurements in both tracers can be compared on an individual object basis remain sparse, and where available resulting gas masses can differ by up to $\sim 0.5$ dex depending on the exact prescription \citep{Decarli2016}.  The situation, especially at higher look-back times, becomes more dire if requiring constraints on the gas-phase metallicity extracted directly from optical line ratios rather than inferred from a mass-metallicity relation, which itself features substantial scatter.

Direct cross-calibrations between CO- and dust-based gas masses exploiting large statistical samples with known metallicities therefore remain indispensable, even at lower redshifts.  Examples of such recent comparisons between CO and far infrared luminosities of nearby galaxies, or of the gas masses based thereupon, include \citet[][20 submm-selected galaxies]{Bourne2013}, \citet[][12 nearby galaxies most of which are UltraLuminous InfraRed Galaxies ULIRGs]{Scoville2014}, and \citet{Groves2015} who study the galaxy-integrated and resolved radial trends for 36 KINGFISH-HERACLES-THINGS galaxies spanning a wider dynamic range. Here, we present an IRAM 30-m CO survey, carried out in the region of the Herschel Stripe82 Survey \citep[HerS,][]{Viero2014_HerS} and complemented with Arecibo observations, targeting 78 massive ($\log \mstar  \gtrsim 10$) star-forming galaxies.  Augmented with 14 COLD GASS \citep{Saintonge2011} galaxies with matching sample definition and data requirements, our sample covers a larger dynamic range in luminosity than \citet{Bourne2013} and \citet{Scoville2014}, and sensitively increases number statistics for this type of analysis within our considered mass range.  Specifically, the dust method we explore is based on a WISE + SPIRE sampling of the IR SED with the aim of making our results directly applicable to the thousands of galaxies in overlap between these two legacy data sets.

The goal of this paper is to establish at $z \sim 0$ a statistical cross-calibration between CO-and dust-based gas masses based on the same set of galaxies with known metallicities and relying on direct observations rather than stacking. 
Once such a cross-calibration is established, it can be applied to Herschel legacy data sets. Further, when carefully accounting for the cosmic evolution of the atomic-to-molecular gas ratio \citep{ObreschkowRawlings2009}, the cross-calibration could in principle function as a zero point for high-redshift studies with NOEMA and ALMA. 

The paper is organised as follows: Section \ref{sec:SampleObs} describes our sample and observations.  In Section  \ref{sec:Methods}, we outline the methods used to determine gas and dynamical masses.  We present our results on the cross-calibration of gas mass measurements, the excitation properties and dynamical mass budget of galaxies in our sample in Section \ref{sec:Results}, and proceed to a summary in Section \ref{sec:Summary}. Throughout this paper, we assume a standard cosmology with $H_0 = 70 \ \kms \ Mpc^{-1}$, $\Omega_m = 0.3$ and  $\Omega_{\Lambda} = 0.7$. We further adopt a \citet{Chabrier2003_IMF_rev} IMF.


\section{Sample and Observations}
\label{sec:SampleObs}

Here we present our sample, the selection criteria tied to it, and the direct observables obtained from CO and HI observations and the ancillary IR and optical data.

\subsection{Sample Selection}
\label{sec:SampleSel}
\begin{figure}
\centering
\includegraphics[width=0.5\textwidth]{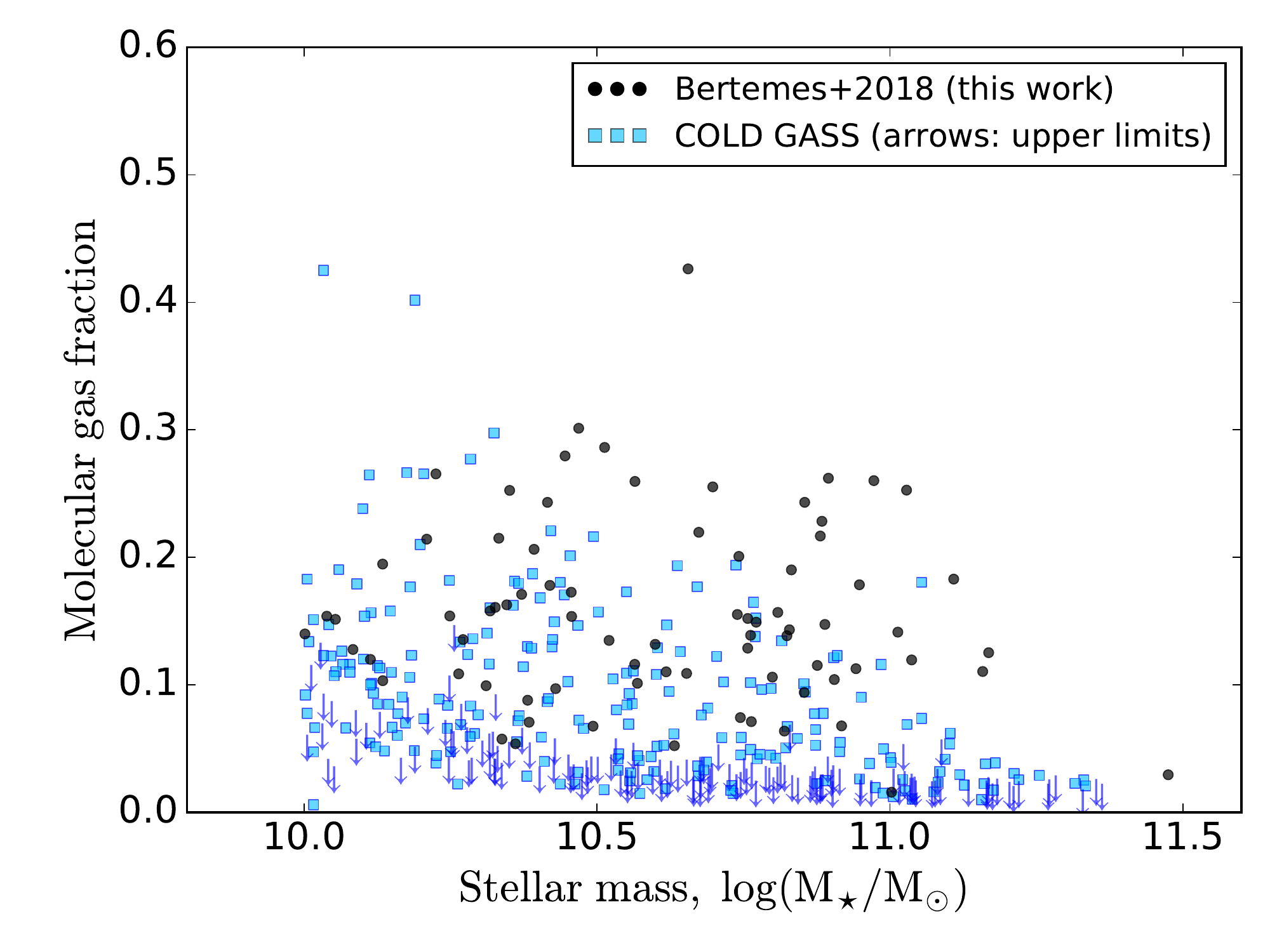}
\caption{Our Stripe82 CO sample in the $f_{\rm gas}$ vs stellar mass $M_{\star}$ plane (black dots).  COLD GASS measurements (blue squares) and upper limits (blue arrows) are shown for reference.  For both surveys the molecular gas fractions are derived consistently using the same CO methodology.  Our sample covers a broad parameter space with the exception of passive galaxies.
}
\label{fig:Sample_fgas_vs_Mstar}
\end{figure}

Our sample is chosen with the goal of cross-calibrating CO- and dust-based gas masses on, first of all, the same set of galaxies, and second, only galaxies with known metallicities { based on strong optical nebular lines}.

To ensure that the dust method can be applied to our sample in addition to the CO method, we rely on IR data from the Herschel Stripe82 Survey \citep[HerS,][]{Viero2014_HerS} and the Wide-field Infrared Survey Explorer \citep[WISE,][]{Wright2010_WISE} All-Sky Release.
The HerS imaging survey mapped an area of $79 \ {\rm deg}^2$ within the SDSS/Stripe82 field. Observations were taken in the three IR photometric bands (250, 350, and 500 \micron) of the Spectral and Photometric Imaging Receiver (SPIRE) instrument. The focus on Stripe82 guarantees access to a multitude of multi-wavelength data from other surveys.
The WISE mid-IR All-Sky survey covered the entire sky in four imaging bands centered around 3.4, 4.6, 12, and 22 \micron. 
For our purpose, we require WISE 22 \micron\ photometry combined with HerS  detections in at least one of the SPIRE 250, 350 and 500 \micron\ bands ($30 \ {\rm mJy}$, $3 \sigma$ at $250$ \micron).  { In our final sample, all 78 galaxies are $3 \sigma$-detected at $22$, $250$ and $350$ \micron , and 58 of them have a $3 \sigma$ detection at $500$ \micron . That being said, flux measurements of lower significance are still included in the fit, with appropriate weights. For the WISE $22 \micron$ band, we adopt the profile-fitting (PRO) photometry given that none of our sources are extended at $22 \micron$ (i.e., all have reduced $\chi ^2 < 3$ from the profile-fitting routine). }

\begin{figure}
\centering
\hspace*{-0.7cm}
\includegraphics[width=0.55\textwidth]{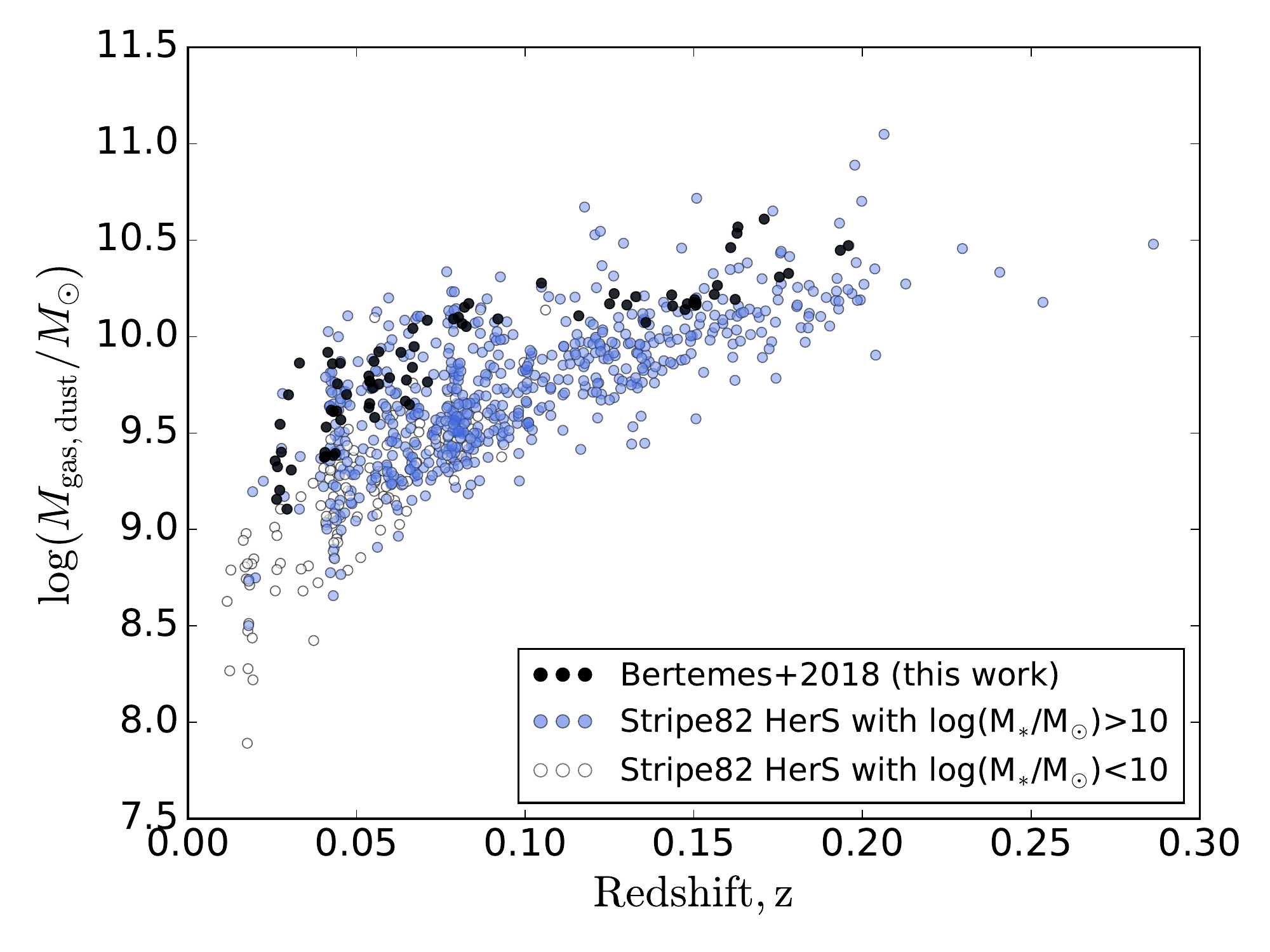}
\caption{Dust-based molecular gas masses as a function of redshift. The black dots represent our sample; the blue dots show the underlying
massive galaxy population with ${\rm S/N} \geqslant 3$ detections in the WISE 22um band and at least one SPIRE band. For reference, galaxies with stellar mass below $10^{10} \Msun$ satisfying otherwise identical criteria are marked with open circles.}
\label{fig:Mgasdust_vs_z}
\end{figure}

\begin{figure}
\centering
\hspace*{-0.7cm}
\includegraphics[width=0.55\textwidth]{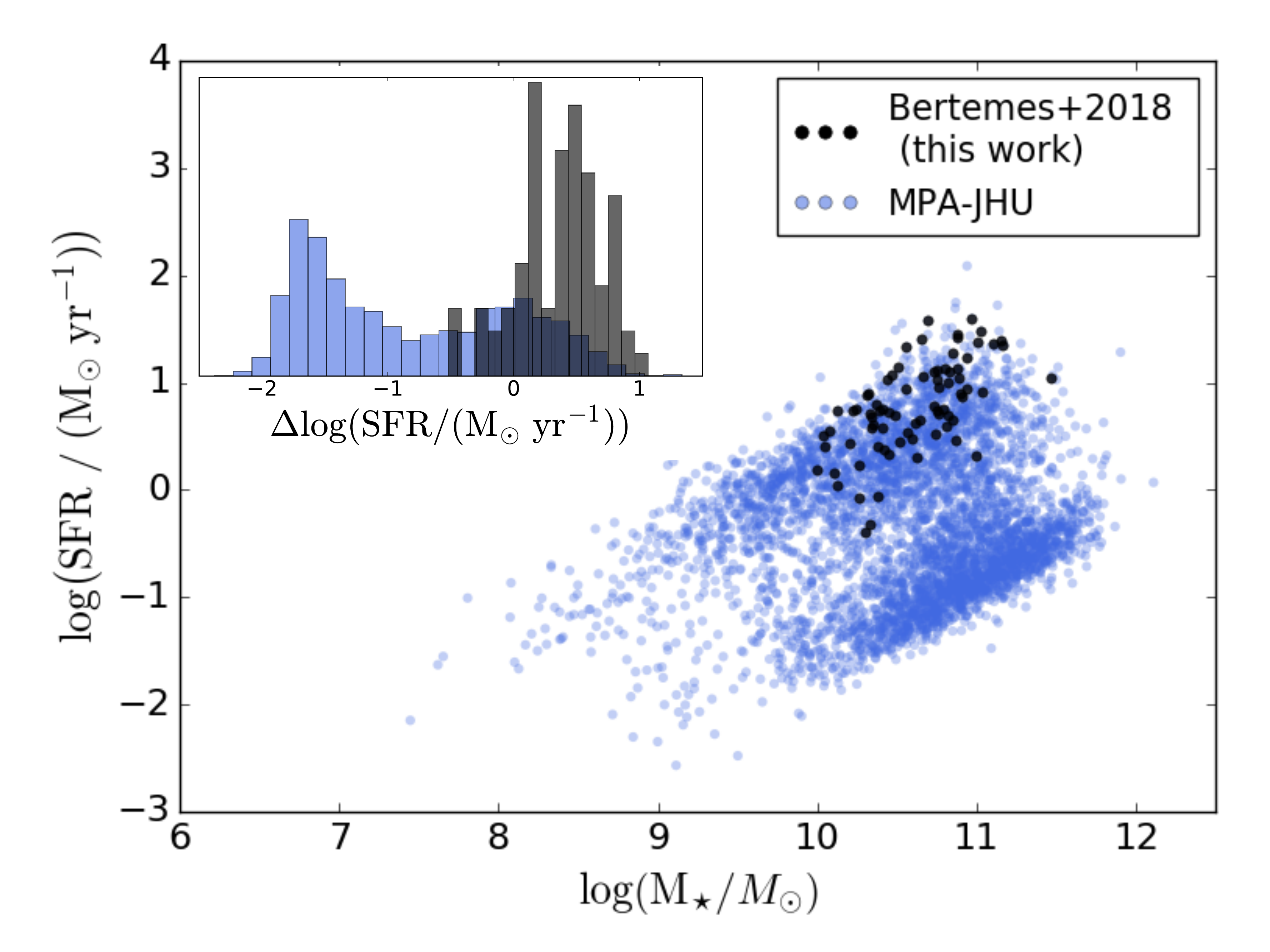}
\caption{SFR as a function of stellar mass, $M_{\star}$, for our sample (black dots) and for the underlying MPA-JHU population (blue dots). Our galaxies overlap with the star-forming part of the MPA-JHU sample. Specifically, the upper left corner displays a histogram contrasting the offset from the MS, $\Delta {\rm \log(SFR)} = {\rm \log \ SFR - \log \ SFR_{MS} }$, for our sample and the MPA-JHU population with $\log \ \mstar / \Msun >10 $. While our sample is preferentially drawing from more actively star-forming galaxies, it spans the full range of the MS population, within $[-1.2 \sigma, +2.4 \sigma]$ from the midline of the MS distribution.}
\label{fig:SFR_vs_Mstar}
\end{figure}

In order to have access to reliable gas-phase metallicity measurements based on strong optical lines, and to prevent potential non-stellar contributions to the SFR diagnostics we employ, we require our targets to lie on the star-forming branch of the \citet{BPT1981} diagram\footnote{See Appendix \ref{app:AGN_contr} for an assessment of potential biases induced when applying our dust-based gas mass methodology to galaxies that do feature contributions from an active galactic nucleus to their infrared SED.}.
From the 431 possible objects fulfilling this and the IR-based conditions outlined above, we choose a subset of 78 galaxies at $0.025 < z < 0.2$ with stellar masses $M_{\star} > 10^{10} \Msun$. In our selection, we aim to maximise the range of galactic properties. As a result, our sample spans an order of magnitude in $M_{\star}$, nearly two orders of magnitude in SFR, and gas mass fractions $f_{\rm gas}$ typically ranging from $\sim 2 \%$ to $\sim 30 \%$ according to our CO-based estimates (originally chosen based on the dust-based gas masses). Figure \ref{fig:Sample_fgas_vs_Mstar} summarises the span in $f_{\rm gas}$ and $M_{\star}$ in more detail compared to the Data Release 3  COLD GASS galaxies from \citet{Saintonge2011}. We do not target passive galaxies. 
At the lowest redshift, we target gas-poor galaxies while at higher redshifts, our galaxies were selected to fall into the gas-rich tail, according to the dust-based prediction (see Figure \ref{fig:Mgasdust_vs_z}). This approach serves to build a sample of considerable dynamic range (comparable to the range in gas masses spanned by all galaxies with $\log \mstar / \Msun > 10 $) in the most efficient manner: targeting the abundant lower mass and thus intrinsically fainter population at short distances, while selecting  the more rare luminous systems from a larger cosmic volume. In more detail, the position of our galaxies in the ${\rm SFR}-\mstar$ plane, as well as the distribution of their MS offset $\Delta {\rm \log(SFR)} = {\rm \log \ SFR - \log \ SFR_{MS} }$, are summarised in Figure \ref{fig:SFR_vs_Mstar}, which contrasts them to the underlying population of SDSS galaxies from the MPA-JHU database (Section \ref{sec:GalProps}). Rather than adopting a prescription for the MS from the literature (which can vary depending on the adopted SFR indicator and details of sample selection), we determine a MS with constant sSFR based on the underlying MPA-JHU population with $\log \ \mstar /  \Msun> 10$. This effectively corresponds to a linear MS with zero-point $ \log {\rm sSFR} = \log {\rm SFR} - \log \mstar = -10.19 $.  
It is visible that, compared to a random sample, our selection is preferentially drawing from more (molecular) gas-rich and therefore more star-forming galaxies. However, all of our galaxies are contained within $\Delta {\rm \log (SFR)} \sim [-1.2 \sigma, +2.4 \sigma]$. We note that our sample includes 20 LIRGs ($L_{\rm IR}>10^{11}L_{\odot}$), and no ULIRGs ($L_{\rm IR}>10^{12}L_{\odot}$).
Postage stamps for our 78 galaxies (based on 3 SDSS bands, $gri$) can be found in Appendix \ref{app:postage_stamps}.

In additon to the newly observed Stripe82 sources, we also include all 14 star-forming galaxies (including 2 LIRGs and no ULIRGs) from the COLD GASS survey for which dust-based gas masses could be computed (i.e. detected in the WISE 22um band and at least one of the three SPIRE bands \citep{Pilbratt2010_Herschel, Griffin2010_HerS} from the SPIRE Point Source Catalogue (SPSC\footnote{\url{https://www.cosmos.esa.int/web/herschel/spire-point-source-catalogue}}). {As it turns out, all of them have $3 \sigma$ detections in all four IR bands used for our analysis}. Henceforth, we will simply refer as ``our sample'' to the combination of our 78 CO observations and the 14 additional COLD GASS sources.

\subsection{CO observations and data reduction}
\label{sec:CO_DataRed}

Our CO observations were taken with the IRAM 30-m telescope using the Eight Mixer Receiver (EMIR) with the Wideband Line Multiple Autocorrelator (WILMA) backend. For all targets, the redshifted CO(1-0) line could be observed in the upper sideband of the 3 mm E090 band ($89-117 \rm{GHz}$). We additionally observed the redshifted CO(2-1) line whenever it was covered by the same frequency setup for the 1.3 mm E230 band. This was the case for 56 out of our 78 galaxies. Exposure times range from $6-108$ minutes with median value ${\rm <t_{exp}>=36}$ minutes, adjusted on an object-to-object basis until a significant ($> 5 \sigma$) detection was obtained.

The data reduction is performed using the \textsc{CLASS} software\footnote{\url{ http://www.iram.fr/IRAMFR/GILDAS/}} {following procedures outlined by \citet{Saintonge2011}}. Each scan is corrected for platforming and baseline-subtracted. 
Individual scans are combined into a single spectrum for each galaxy, with velocity bins corresponding to $\sim 22 \ \kms$. 
The signal is converted from instrumental units to units of \Jykms\ using the 2013 wavelength-dependent values of the IRAM 30-m telescope efficiencies.\footnote{\url{http://www.iram.es/IRAMES/mainWiki/Iram30mEfficiencies}} Within an appropriately defined window, the signal is summed up to yield the CO line flux. {Flux errors are computed as follows: Next to directly propagating the RMS error per bin to the line flux (leading to a mean observational error of $\sim 10$\% in our CO observations), we also include a flux calibration error of 10\%, as well as the error on the aperture correction (obtained by perturbing the correction model described below), leading to a median total error of 16\% on our CO flux measurements.} We present a gallery of the resulting line profiles in Appendix \ref{app:spectra}. All fluxes, errors and other direct observables are listed in Table \ref{tab:RawData}.

We follow the procedure outlined by  \citet{Lisenfeld2011} to compute aperture corrections \citep[see also][]{Stark2013, Rowlands2015}.  Briefly, a model galaxy is assumed, and the emission before and after convolution with a Gaussian beam is compared. 
The correction factor thus corresponds to the ratio between the total extrapolated CO intensity and the observed inner part: $f_{\rm corr} = I_{\rm CO, total} / I_{\rm CO, observed}$.  
We assume that the CO emission follows an exponential disk distribution with scale length equal to $0.2 * R_{25}$, where $R_{25}$ is the $25 \ {\rm mag \, arcsec^{-2}}$ isophotal radius in the $g$ band. {We tested that our conclusions are not altered significantly when deriving the scale length from the half-light radius taken from \citet{Simard2011} instead. The correction can be expressed as follows:}
\begin{align}
	I_{\rm CO, \ observed} / I_{\rm CO, \ total }= 4 \int_{0} ^{\infty}  \int_{0} ^{\infty} \exp \left(-\frac{ \sqrt{x^2+y^2} }{h_{\rm CO}}\right) \nonumber \\
	\cdot \exp \left(- \ln 2 \left[ \left( \frac{2x}{\rm HPBW} \right)^2 + \left( \frac{2y \cos i}{\rm HPBW} \right)^2 \right] \right) \ dx \ dy
\end{align}
where $h_{\rm CO}$ is the scale length of the CO emission, $i$ is the galaxy's inclination (see Section \ref{sec:lpfit}) and ${\rm HPBW}$ denotes the half-power beam width. The latter is well fit by ${\rm HPBW} = 2460  \cdot (\nu_{\rm rest} / {\rm GHz})^{-1} (1+z) $, where $\nu_{\rm rest}$ denotes the rest frequency of the given line.\footnote{\url{http://www.iram.es/IRAMES/telescope/telescopeSummary/telescope_summary.html}} 
{For the median redshift of our sample ($<z> \sim 0.065$) this corresponds to HPBW $\sim 22 ''$ for the CO(1-0) line and $\sim 11''$ for CO(2-1), translating to physical sizes of $25.0$ and $12.5 \ \rm kpc$, respectively.}
Aperture effects are therefore always more pronounced for CO(2-1) than for CO(1-0), with correction factors for the latter limited to $\lesssim 40$\% and with a median value of $\sim 9.6$\%. 

Our aperture-corrected CO(1-0) line fluxes range from $\sim 0.9$ to $\sim 67.5 \, \Jykms$, with a median of $\sim 7.6 \, \Jykms$. {Signal-to-noise ratios reach from $\sim 5$ to $\sim 25$, with $<S/N> \, \sim 10$.} 
Hereafter, all analysis and line fluxes quoted in tables, text or figures include the aperture correction. For completeness, we include the aperture correction factors that were applied as separate entries in Table \ref{tab:RawData}.

\subsection{ {HI observations and mass estimates}}
\label{sec:HI}

\begin{figure}
\centering
\includegraphics[width=0.5\textwidth]{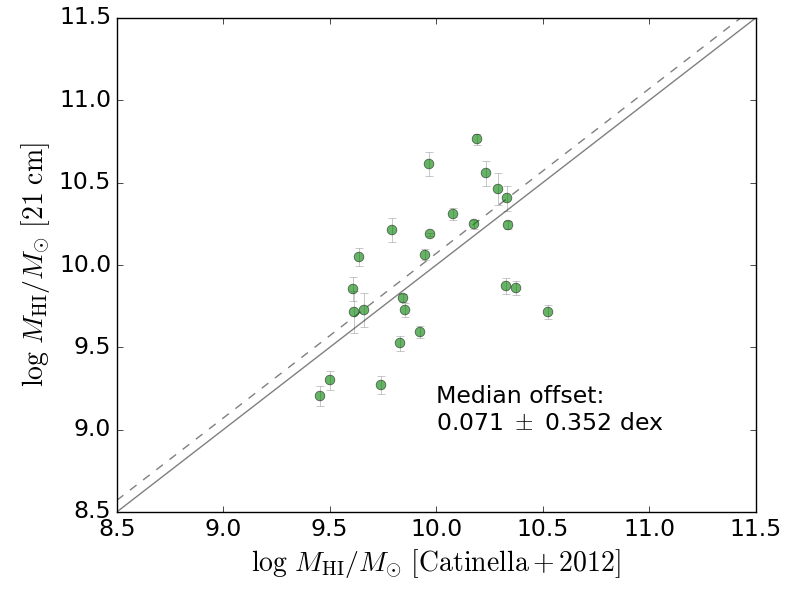}
\caption{Comparison between \HI\ masses obtained by 21 cm HI line observations for a subset of our galaxies, and \HI\ mass estimates predicted by a scaling relation taken from \citet{Catinella2012}. In the absence of an \HI\ measurement, we rely on this scaling relation whenever discussing the atomic-to-molecular gas ratio, or \HI\ masses directly. The 1-to-1 line (black solid) and the median offset (black dashed) are shown for reference. {The error bars represent RMS noise in the flux measurements and do not contain uncertainties in the conversion of flux to inferred \HI\ mass.} }
\label{fig:HI_obs_vs_Cat}
\end{figure}

Complementing the IRAM 30-m Stripe82 CO program, we carried out 21 cm HI line observations for a subset of our sample with the Arecibo 305-meter telescope. We observed and detected 24 sources, yielding $S/N$ levels ranging from {$\sim 3$ to $\sim 50$ with median $<S/N> \sim 9$. The median ratio between the peak signal and RMS noise per channel corresponded to $5.5$, with values ranging from $\sim 3$ to $\sim 30$.} Exposure times ranged from $\sim 3$ to $\sim 30$ minutes, and data reduction followed standard procedures {similar to those outlined in, e.g., \citet{Minchin2010}. In brief, spectra were polarisation-combined, scaled to $\rm Jy$, baseline-subtracted with a linear baseline and boxcar smoothed to yield velocity bins of widths ranging from $\sim 50 \ \kms$ to $\sim 130 \ \kms$.} For reference, a gallery of our 21 cm spectra recorded is presented in Appendix \ref{app:spectra}. 

We proceed to derive direct constraints on the HI mass for the 24 sources with HI measurements (out of 92 galaxies) as follows \citep[see, e.g. ][]{Catinella2012}
\begin{align}
	 \left( M_{\HI} / \Msun \right)  = 2.356 \cdot 10^5 (1+z)^{-1}  \left(  \frac{D_{\rm L}}{\rm Mpc}  \right) ^2    \left( \frac{S_{\rm \HI}}{\Jykms}  \right)  
\end{align}
{For 37 of the remaining sources that have a ${\rm NUV-r}$  color measurement available,} whenever discussing the atomic-to-molecular gas ratio ($M_{\HI} / M_\Hmol $) or \HI\ masses, we adopt the empirical scaling relation based on the GALEX Arecibo SDSS Survey (GASS) DR2 presented in \citet{Catinella2012}, {for which they find a scatter of $\sim 0.29$ dex:}
\begin{align}
	 \log \left( M_{\HI} / \mstar \right)  = -0.338 \log \mu_{\star} - 0.235 \left( {\rm NUV-r} \right) + 2.908
\end{align}
Here $\mu_{\star}$ is the stellar surface density: $\mu_{\star} = \mstar \ (2 \pi R_{50}^2)^{-1}$. We corrected the ${\rm NUV-r}$ colour \citep[from the GALEX database;][]{Bianchi2011_GALEX} for Galactic extinction, using $A_{\rm NUV-r} = 1.9807 \ A_r$, where the extinction in the r-band $A_r$ was obtained from the SDSS database.  A k-correction was applied to take into account the small redshift difference between galaxies in our sample and those studied by \citet{Catinella2012}. 
{For the other 31 sources without ${\rm NUV-r}$  color measurement,} 
we instead conducted a best fit using the tabulated binned median values in \citet{Catinella2012}, yielding: 
\begin{align}
	 \log \left( M_{\HI} / \mstar \right)  = -1.12  \log \mu_{\star} + 8.76 
\end{align}
{The scatter in this relation amounts to $\sim 0.36$ dex for the total, unbinned sample.} {Throughout the rest of the paper, errors on $M_{\HI}$ correspond to the RMS noise whenever a direct observation was made, and are set to the scatter in the respective scaling relation otherwise.}

Figure \ref{fig:HI_obs_vs_Cat} contrasts the HI masses obtained by our Arecibo survey and the values that the scaling relation would have yielded. Our results confirm the validity of the HI scaling relation based on GASS. 

\subsection{Ancillary data and galaxy properties}
\label{sec:GalProps}

The location of our sample in Stripe82 allows for access to a broad range of ancillary multi-wavelength data. 
Other than the WISE + SPIRE IR photometry, essential optical diagnostics to compute derived galaxy properties are taken from SDSS DR7 \citep{Abazajian2009_DR7} or catalogues based thereupon.  Specifically, we use MPA-JHU stellar masses \citep{Kauffmann03_MSFR_GALS, Salim2007} and SFRs \citep{Brinchmann2004}, {and uncertainties tabulated by these authors}. 
{Further, following \citet{Genzel2015} and \citet{Tacconi2017}, we take the strong optical line ratios $\oiii / \Hbeta$ and $\nii / \Halpha$ from SDSS  and in Table \ref{tab:DerivedData} list the gas-phase metallicities derived according to the \citet{PP04_metal} O3N2 calibration:}
\begin{align}
	&Z = 12 + \log (O/H)   \nonumber \\
	= \, &8.73 - 0.32 \cdot  \log ( \left( \OIII / \Hbeta \right)  /   \left( \NII / \Halpha \right)   )   
\end{align} 

Where relevant, structural properties based on the rest-optical emission are taken from \citet{Simard2011} who fitted Sersic models to the two-dimensional surface brightness profiles. The inclination $i$, which enters the calculation of aperture corrections and dynamical masses, is computed from the tabulated ellipticity measurements assuming a { disk thickness $h = 1/10$ for local galaxies \citep[e.g. ][]{Hall2012}. The latter is defined as the ratio of scale height to scale length \citep[see, e.g. ][]{Wuyts2016}: }
\begin{align}
	\cos (i)^2 = ((1-\epsilon)^2-h^2) / (1-h^2) 
\end{align}


\section{Methods}
\label{sec:Methods}

We now reprocess the direct observables presented in Section \ref{sec:SampleObs} to obtain the physically more meaningful quantities of gas and dynamical mass. We first discuss how we translate our CO (1-0) fluxes into a galaxy-integrated molecular gas content and then proceed to describe our alternative approach starting from the IR-based dust masses. Finally, we outline the procedure of line profile fitting to determine the dynamical (i.e. total) mass enclosed within $1 \ R_e$.

\subsection{CO-based molecular gas masses}
\label{sec:CO_method}

Studies of the Milky Way and Local Group galaxies have shown that the integrated CO(1-0) line luminosity $L_{\rm CO}^{\prime}$ is closely related to the number of molecular clouds, which in turn is related to the virial mass of the cloud system \citep[e.g.][]{Solomon1987,  Bolatto2013}. The total molecular gas mass of an ensemble of (near)-virialised clouds can therefore be expressed as:
\begin{align}
M_{\rm gas, \ CO}  = \alpha _{\rm CO} \cdot L_{\rm CO}^{\prime}
\end{align}

To calculate $L_{\rm CO}^{\prime}$ from our aperture-corrected CO fluxes $S_{\rm CO}$, we follow the procedure from \citet{Solomon1997}:
\begin{align}
\frac{L_{\rm CO}^{\prime}} { \left( K \ \kms \ {\rm pc}^2 \right)}      & \overset{\mathrm{def}}{=} \int_{\rm source} \int_{\rm line} T_R(v) \  dv \ dA  \nonumber \\
	= 3.25 \cdot &10^7  \left( \frac{S_{\rm CO}}{\Jykms}  \right)     \left(  \frac{\nu_{\rm obs}}{\rm Hz}  \right)  ^{-2}    \left(  \frac{D_{\rm L}}{\rm Mpc}  \right) ^2     \left( 1+z  \right)^{-3}    
\end{align}
In the definition above,  $T_R$ denotes the wavelength-dependent surface brightness temperature in the Rayleigh-Jeans regime. $\nu_{\rm obs} = \nu_{\rm em} (1+z)^{-1}$ is the observed frequency, and $D_L$ denotes the luminosity distance.
For Milky Way-like systems, the conversion factor $\alpha _{\rm CO}$ can be assumed to be constant: $\alpha _{\rm CO, \ MW} = 3.2 \ \Msun / (  K \ \kms \ {\rm pc}^2)  $. In general, however, the conversion is metallicity-dependent \citep[][hereafter G12 and B13, respectively]{Leroy2011, Genzel2012, Bolatto2013}. We follow the approach used in, e.g., \citet{Genzel2015}, \citet{Tacconi2017}, and account for metallicity by applying the geometric mean of the B13 and G12 correction factors $\chi_{\rm G12, \ B13} (Z)$, where: 
\begin{align}
	\chi_{\rm G12} (Z) &= 10^{-1.27 \cdot (12+\log(O/H)-8.67)}  \\
	\chi_{\rm B13} (Z) &= 0.67 \cdot \exp \left( 0.36 \cdot 10^{-(12+\log(O/H)-8.67)} \right) 
\end{align}
\label{equ:G12_B13}
{We tested that similar conclusions are reached when adopting the prescription for $\alpha_{\rm CO}$ presented in 
\citet{Accurso2017}, which in addition to a metallicity-dependence features a secondary dependence on the offset from the main sequence.} 
Finally, we apply a $36$\% mass correction for helium, such that our final gas masses are computed as:
\begin{align}
\lMgasCO =\log \left( 3.2 \  L_{\rm CO}^{\prime} \right)  + \frac{1}{2} \log \left(  \chi_{\rm B13} \cdot \chi_{\rm G12}  \right)  +  \log \left( 1.36 \right) \label{equ:MgasCO}
\end{align}
{We note that in the error calculation on \MgasCO , we consciously choose to adopt the fractional errors from the flux measurements (with median value $16$\%; see section \ref{sec:CO_DataRed}). The errors thus reflect observational uncertainties, rather than uncertainties in the conversion factor.}

\subsection{Dust-based gas masses}
\label{sec:dust_method}

In this Section, we describe how we infer a galaxy-integrated dust mass from the IR data described in Section \ref{sec:SampleSel}, and how we proceed to derive an estimate of the gas mass based thereupon.

\begin{figure}
\centering
\includegraphics[width=0.5\textwidth]{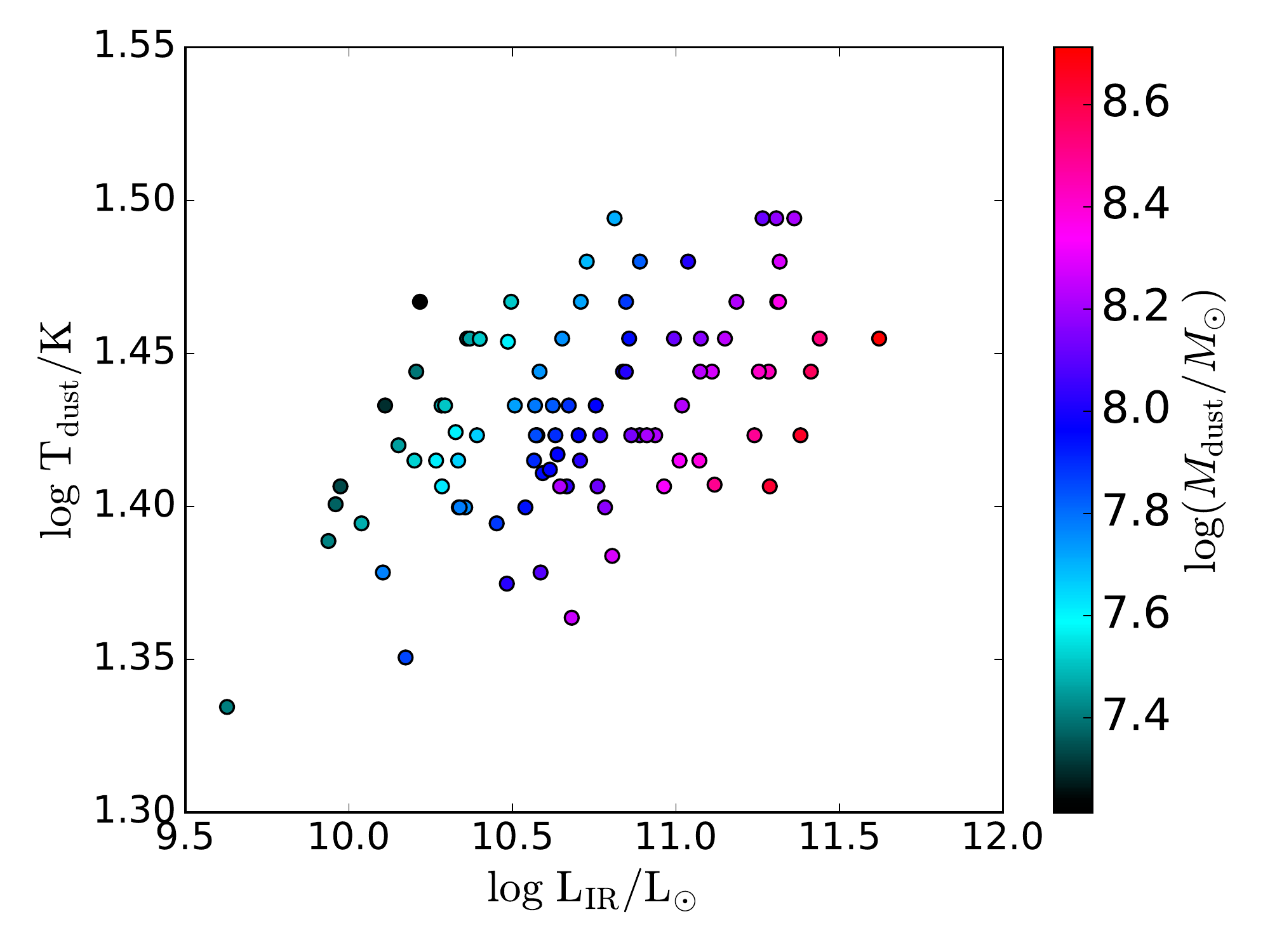}
\caption{Dust mass \Mdust\ (shown in colour) as a function of the dust temperature, $T_{\rm dust}$, and the total IR luminosity, $L_{\rm IR}$. }
\label{fig:LIR_Tdust_Mdust}
\end{figure}

To this date, there are a number of different dust-based approaches in use, which can, broadly speaking, be classified into three types:  One route is based on probing the FIR spectrum at a single point in the long-wavelength Rayleigh-Jeans tail, where the dependence on dust temperature is minor, such that the dust mass scales nearly linearly with the monochromatic IR luminosity \citep[e.g. ][]{Scoville2014, Groves2015}. Second, one can probe the FIR SED with several filters and fit physically motivated dust emission models \citep[e.g.]{DraineLi2007}. 
Finally, one may use a simplification of the second approach, where the fitting is done based on modified blackbody emission \citep[however, see the discussion in][for a critical assessment of the latter]{Berta2016}. Our method fits to WISE+SPIRE SEDs and applies a conversion to gas masses from \citet{Genzel2015}, which itself is derived from FIR SED fitting of \citet{DraineLi2007} models.

For each object, we fit the empirical \citet[][hereafter DH02]{DaleHelou2002} templates to the WISE + SPIRE SED.  The dust temperature $T_{\rm dust}$ corresponding to the best-fitting template is taken from \citet[][Table A.1]{Magnelli2014}, who fitted single modified blackbody functions with emissivity index $\beta=1.5$ to the DH02 template set.  
As a second parameter, the best-fitting amplitude yields a measure of the total infrared luminosity or equivalently SFR \citep[see][]{Kennicutt1998}.  {Uncertainties on our fit are derived using a Monte Carlo technique.} We follow \citet{Genzel2015} in computing the dust mass as: 
\begin{align}
	\left( \frac{M_{\rm dust}}{\Msun } \right) = 1.2 \cdot 10^{15}   \left( \frac{\rm SFR}{\mpyr}  \right)     \left(  \frac{T_{\rm dust}}{K} \right)^{-5.5} \label{eq:Mdust_formula}
\end{align}
Figure \ref{fig:LIR_Tdust_Mdust} displays the range in $L_{\rm IR}$ and $T_{\rm dust}$ for the galaxies in our sample, and the dust mass estimates based thereupon. 

We then translate \Mdust\ into a gas mass estimate using the metallicity-dependent conversion based on sightline observations through five Local Group galaxies by \citet{Leroy2011}: 
\begin{align}
	\log \left( M_{\rm gas, \, dust} \right) = \lMdust + 2 - 0.85 * ( (12+\log({\rm O/H})) - 8.67)  \label{equ:dust_conversion}
\end{align}
We remind the reader that the spatially resolved study conducted by \citet{Leroy2011} by definition established a conversion between dust mass and total gas mass along the line of sight within CO-emitting regions, i.e. \Hmol\ + \HI . 
In contrast, the CO method is used to probe the purely molecular gas phase (i.e. \Hmol\ only), which holds the immediate fuel for star formation. We will return to this point in Section \ref{sec:Comparison} when contrasting both methods. 
The dust-based gas masses obtained from equation \ref{equ:dust_conversion} are displayed as a function of redshift in Figure \ref{fig:Mgasdust_vs_z}, which highlights our sample compared to the underlying distribution of Stripe82 HerS galaxies. 
{We note that our adopted errors on $M_{\rm gas, \, dust}$ (which typically correspond to  $\sim 6$\%) are propagated directly from the above-mentioned statistical uncertainties in the IR SED fit. Similarly to the CO-based approach, our error calculation thus accounts for observational uncertainties, and does not include any systematic uncertainty in the dust-to-gas conversion.}


\subsection{Line profile fits and enclosed mass components}
\label{sec:lpfit}

Encoded in their width and spectral shape, the CO line profiles contain information about the dynamical mass enclosed within the CO-emitting region. To each of our spectra, we apply a fit based on gas emission from an inclined rotating disk. In Section \ref{sec:Mdyn}, we will then use the best-fit models as a means to inspect the total mass budget within the effective radius.

At each radius $r$, we calculate the relative brightness of the light tracer (i.e. CO(1-0)) based on the assumption that molecular gas resides in an exponential disk with half-light radius $R_e$ taken from \citet{Simard2011}. The emission from each position within the galaxy will contribute to the line profile at a specific velocity, which depends on the radius, the inclination and the angle with respect to the major axis of the disk. 
The general line profile shape is thus modelled as the superposition of the signals emitted over all radial velocities $v_{\rm rad}(r) = v(r) \sin (i) \cos (\phi)$ convoluted with the beam width, where $v(r) = (GM(<r)/r)^{1/2}$ denotes the circular velocity at a given radius, $i$ the inclination and $\phi$ the azimuthal angle. {We calculate the circular velocity based on the model radial mass distribution $M_{\rm tot}(<r) = M_{\rm baryon} (<r) + M_{\rm DM} (<r) $.} 
The baryonic mass profile is taken to follow the S\'ersic model fit to the optical image by Simard et al. (2011). Since the stellar content dominates, other components which may not follow a S\'ersic profile can be neglected. For the dark matter distribution, we adopt a NFW halo with virial radius tied to redshift and input halo mass following standard cosmology \citep[e.g.][]{Mo1998} and concentration set by redshift and halo mass according to \citet{Dutton2014}.
We fit the CO(1-0) line profile with the above model, leaving three parameters free: mass of the S\'ersic component, mass of the NFW component, and amplitude of the emission (i.e. CO(1-0) line flux). As initial guesses to be perturbed, we take $\mstar + \MgasCO + M_{\HI}$ as the mass of the S\'ersic component, a halo mass calculated according to the $\mstar - M_{\rm halo}$ relation from \citet{Moster2013}, and an arbitrary value for the flux. We then choose the model yielding the lowest value for 
\begin{align}
\chi^2 = \sum \left( \frac{F_{\rm obs, \ j}-F_{\rm model, \ j}}{{\rm RMS_{obs}}} \right) ^{2}  
\end{align}
where the summation is over spectral bins and $F_{\rm obs, \ j}$ and $F_{\rm model, \ j}$ are the modelled and observed fluxes in bin $j$, and ${\rm RMS_{obs}}$ is the mean RMS error per bin determined outside of the line of our observed spectra. {Errors on our dynamical mass estimates are derived similarly to those on \Mgasdust : by running 100 Monte Carlo realisations and using that $68$\% of the resulting dynamical masses fall within $\pm 1\sigma$, which leads to a median error of $\sim 27$\%. }

{We note that our galaxy-integrated line profiles do not generally leave us in a position to robustly constrain the mass of each of the components separately due to strong degeneracies, but in some cases the extra flexibility in setting the shape of the mass distribution improves the quality of fit.  For our purpose, we are only extracting the total dynamical mass (i.e., the sum of both components), which is most robustly constrained within one effective radius.}


\section{Results}
\label{sec:Results}

We now turn to analysing the excitation properties of our sample in Section \ref{sec:Excitation}, and then proceed to compare and cross-calibrate our CO- and dust-based gas masses in Section \ref{sec:Comparison}. In Section \ref{sec:Mdyn}, as a sanity check, we further evaluate the overall mass budget of our galaxies by comparing dynamical mass constraints to the sum of enclosed masses from individual components (stars, gas, dark matter).

\subsection{Excitation properties}
\label{sec:Excitation}
\begin{figure}
\centering
\includegraphics[width=0.5\textwidth]{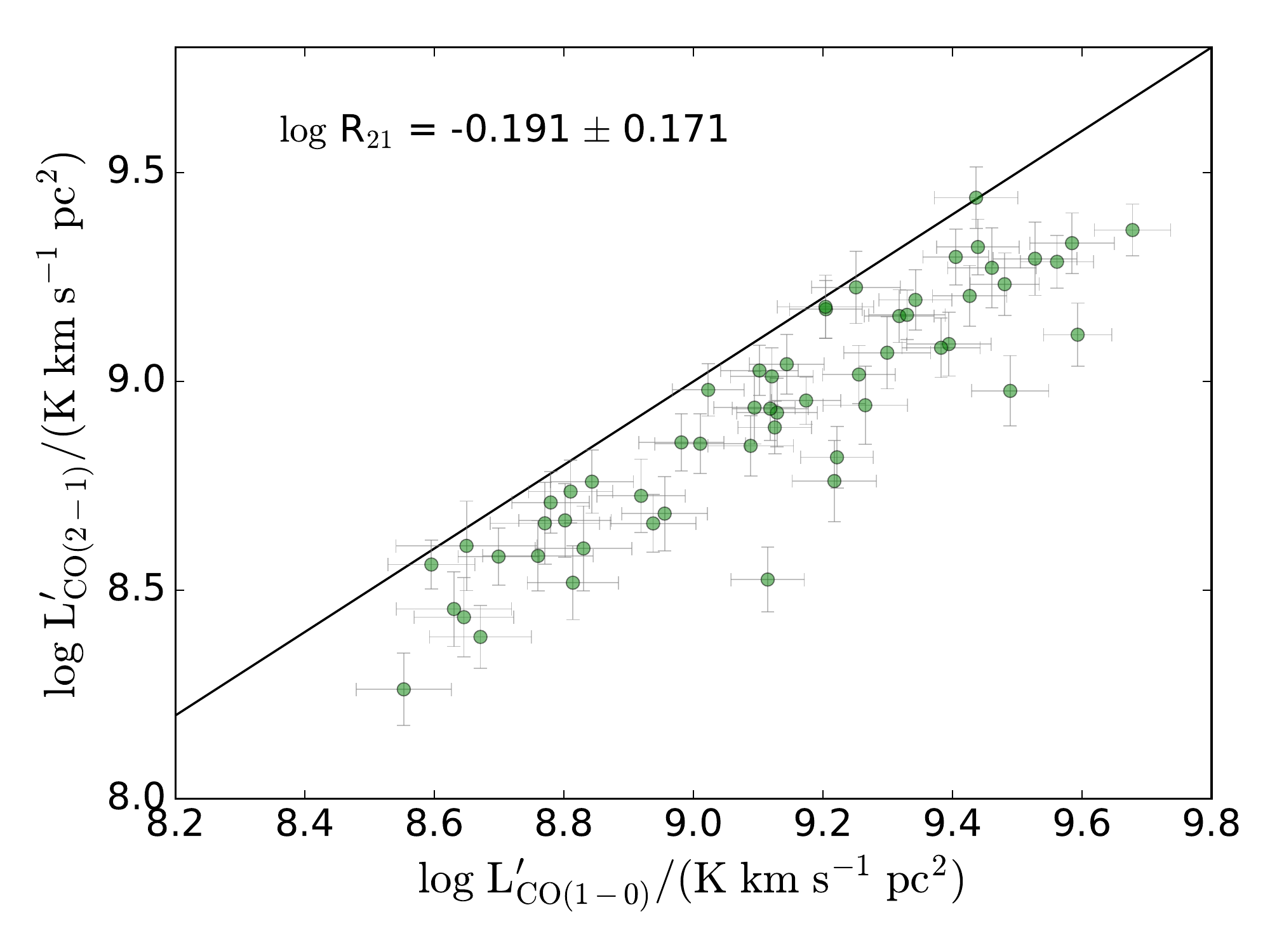}
\caption{A tight relation is observed between the integrated luminosity of the CO(2-1) line vs the CO(1-0) line, with median excitation $<R_{21}> \sim 0.64$ and scatter $\sim 0.17$ dex. }
\label{fig:LCO21_vs_10}
\end{figure}

Figure \ref{fig:LCO21_vs_10} compares the aperture-corrected CO(2-1) and CO(1-0) line luminosities of the 56 galaxies for which CO(2-1) observations could be taken in the same frequency setup (effectively all objects with $z<0.14$). We find a tight relation with  $\sim 0.17$ dex scatter, suggesting that only a very limited range in excitation properties (at least in the low-$J$ regime) is present within our sample. No significant trends of $R_{21} = L_{\rm CO(2-1)} / L_{\rm CO(1-0)} $  with intrinsic galaxy properties (sSFR, mass, metallicity, size) or angular size are found, confirming the uniform excitation properties of the galaxies in our sample and the lack of observationally induced variations in the line ratio once aperture corrections are accounted for. Our median value for the line ratio, $ \left\langle R_{21} \right\rangle \sim 0.64 $, is broadly consistent with literature values ranging from $0.5-0.8$ in the case of spatially resolved {and galaxy-integrated} star-forming disks \citep[e.g.][]{Leroy2013, Rosolowsky2015, Saintonge2017}.  

Finally, from simultaneous fitting of the line profiles of CO(1-0) and CO(2-1), leaving the extent of CO(2-1) as an extra free parameter, we infer comparable sizes for the two CO tracers, with the CO(2-1) size formally being $~10$\% smaller ($\pm 0.16$ dex).

\subsection{Comparing CO- and dust-based gas masses}
\label{sec:Comparison}

\begin{figure}
\centering
\includegraphics[width=0.5\textwidth]{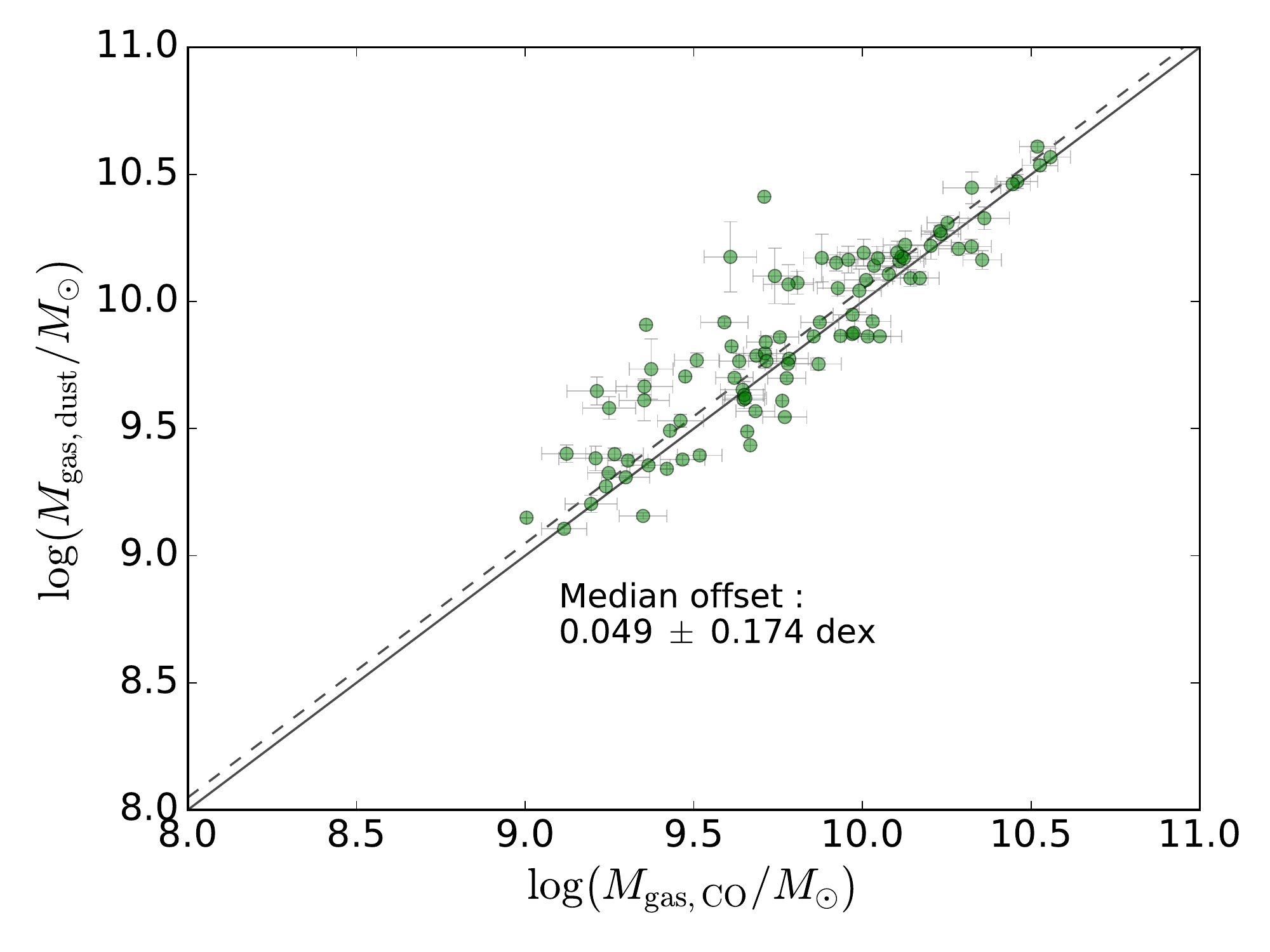}
\caption{Comparison between dust-based gas masses and CO-based gas masses. Cold gas masses from the two methods correlate strongly, albeit with a modest offset towards higher values for the dust-based inference. The 1-to-1 line (black solid) and the median offset (black dashed) are shown for reference. {The errorbars on $\MgasCO$ include RMS noise, a flux calibration error of $10\%$, and the statistical uncertainty on the aperture correction (see Section \ref{sec:CO_DataRed} ), while the statistical errorbars on $\Mgasdust$ were derived by perturbing the IR SED and re-running our SED fitting procedure (see Section \ref{sec:dust_method}).} }
\label{fig:Comparison}
\end{figure}

We now turn to the comparison between inferences of the cold gas reservoirs based on CO(1-0) and the dust continuum emission.  Figure \ref{fig:Comparison} clearly shows an encouragingly tight relation, with standard deviation  $\sim 0.17$ dex.  In detail, the dust-based gas masses show a very minor systematic offset from their CO-based counterparts by $\sim 0.05$ dex, {with the significance of the offset corresponding to just $2.7 \ \sigma$ for the ensemble (where $\sigma =  0.174 / \sqrt{92}$ for 92 targets).} 
As explained in Section \ref{sec:dust_method}, one might naively expect a finite offset between the two methods, given our use of CO as a tracer of purely molecular gas, while the \citet{Leroy2011} dust-to-gas conversion included measures of the HI column density as well.

\begin{figure}
\centering
\includegraphics[width=0.5\textwidth]{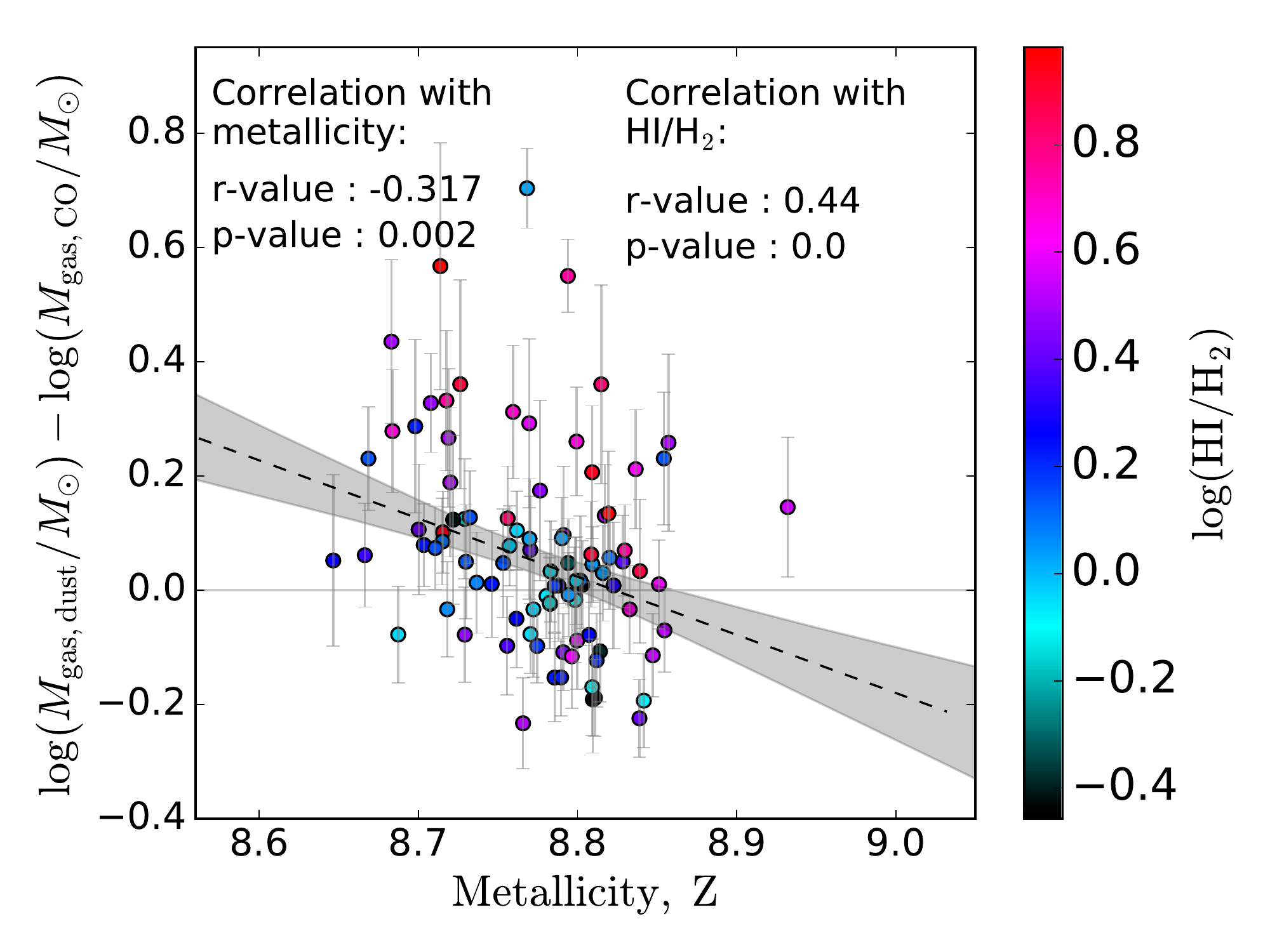}
\caption{The offset between our dust-based and CO-based gas masses, $\lMgasdust - \lMgasCO$, reduces with increasing metallicity. A secondary dependence on atomic-to-molecular gas ratio (colour coding) is notable, such that \HI -rich systems have relatively higher dust-based gas masses compared to the CO reference.}
\label{fig:Offset_dependencies}
\end{figure} 

We explore the dependencies of the residuals in Figure \ref{fig:Offset_dependencies}, focusing on metallicity and atomic-to-molecular gas ratio. Our findings suggest that the residuals reduce with metallicity, albeit the scatter is considerable.
We note that the x- and y-axes in this plot are not independent, since both our \MgasCO\ and \Mgasdust\ calculations depend on metallicity (see Sections \ref{sec:CO_method} and \ref{sec:dust_method}). Each data point is colour-coded by a measure of the galaxy-integrated atomic-to-molecular gas ratio $M_{\HI} / M_{\Hmol} $ (as introduced in Section \ref{sec:HI}), which exhibits a strong relation to the residuals. We note, however, that any such secondary dependence on $M_{\HI} / M_{\Hmol} $ is impractical to implement in a recipe to recalibrate the dust-based gas mass, by lack of CO coverage over the large-area far-IR continuum surveys for which such a recipe can be of use. When repeating this exercise for different galaxy properties, we report no significant dependencies on sSFR, galaxy mass or size. 

\begin{figure}
\centering
\includegraphics[width=0.5\textwidth]{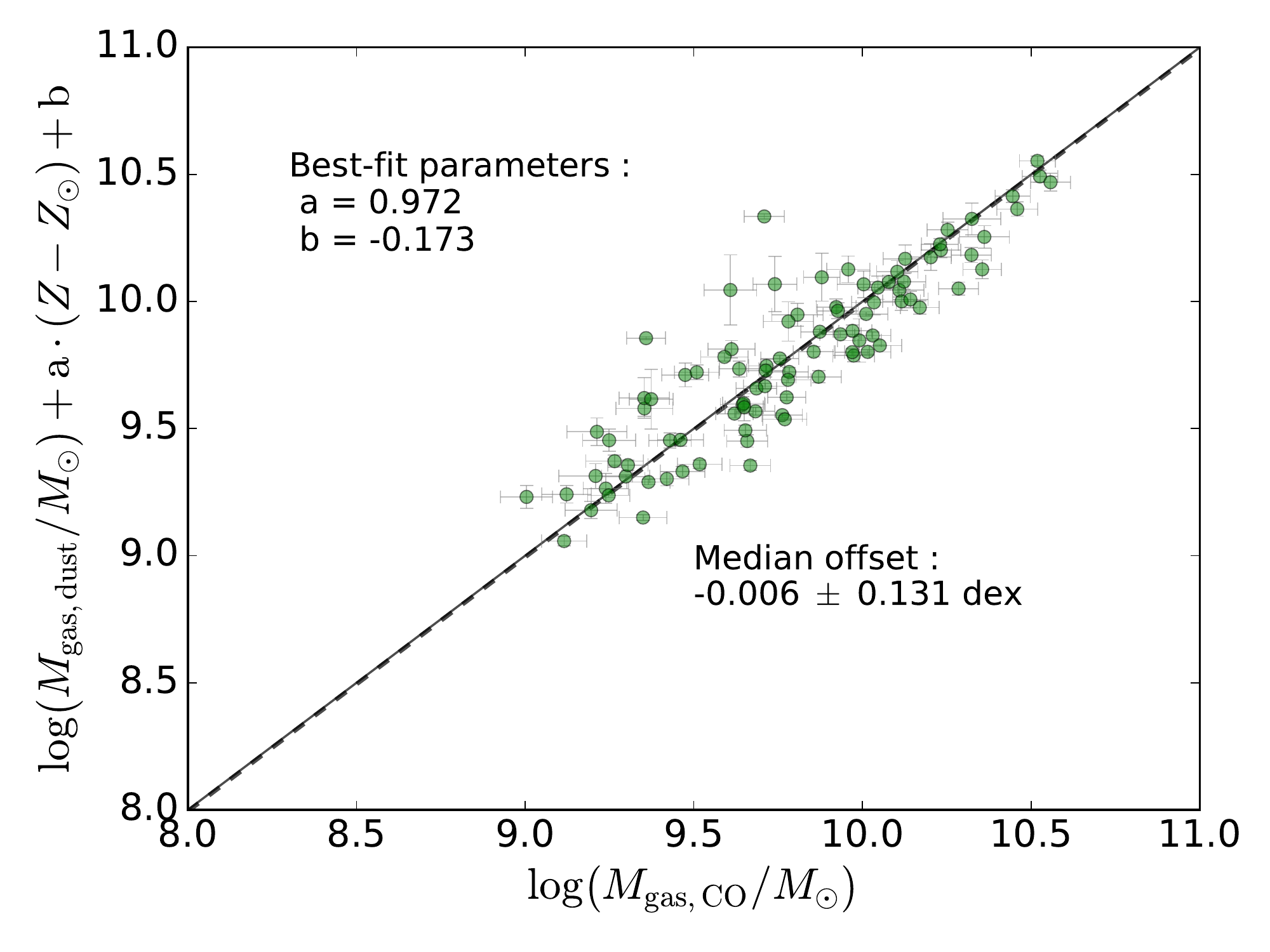}
\caption{Same as Figure \ref{fig:Comparison}, but using ``corrected" dust-based gas masses. The latter were obtained from cross-calibrating the CO-and dust-based gas masses shown in figure \ref{fig:Comparison} adopting a linear dependence on metallicity. The 1-to-1 line (black solid) and the median offset (black dashed) are shown for reference. {Error bars on the y axis represent the statistical error on \Mgasdust\ and do not contain systematic uncertainties in the metallicity calibration.}}
\label{fig:Recipe}
\end{figure}

We aim to establish a way to predict purely molecular gas masses from dust. To this end, we apply a linear fit depending on metallicity to our dust-based gas masses to improve consistency with the CO-based estimates. We thus minimise the expression: $\left[ \lMgasdust + a \cdot (Z - \Zsun) + b \right] - \lMgasCO$. Figure \ref{fig:Recipe} shows our thus adjusted dust-based gas masses as a function of \lMgasCO . Logically, the data points lie around the 1-to-1 line without remaining systematics, and inclusion of the metallicity-dependent term allowed for a slight reduction in scatter, {from $\sim 0.17$ dex to $\sim 0.13$ dex}. Our best-fit parameter values are $a \sim 0.97$ and $b \sim -0.17$. Combining these with Equation \ref{equ:dust_conversion} (i.e the \citet{Leroy2011} conversion from dust to total gas mass) yields:

\begin{align}
\label{equ:dust_to_molgas}
	\log \left( M_{\rm mol \ gas, \, dust} \right)	 &= \lMdust + 1.83 \\ \nonumber
	&+ 0.12 * ( (12+\log({\rm O/H})) - 8.67)
\end{align}

Thus, the metallicity dependence of the \textit{molecular} gas-to-dust ratio is greatly reduced compared to the metallicity dependence of the {\it total} gas-to-dust ratio \citep[see also, e.g.,][]{Cortese2016}. 

However, one should keep in mind that Eq \ref{equ:dust_to_molgas} follows our default approach of adopting the geometric mean of prescriptions by G12 and B13 for the metallicity dependence of the $\alpha_{\rm CO}$ conversion factor (see Eq. \ref{equ:MgasCO}). If instead we were to adopt $\chi_{\rm G12}$ ($\chi_{\rm B13}$) the $0.12$ coefficient in front of the metallicity term would become $-0.37$ ($0.61$). We hence conclude that no significant conclusions can be drawn on deviations from a constant {\it molecular} gas-to-dust ratio across the metallicities sampled, and a metallicity dependence as steep as prescribed by \citet{Leroy2011} is ruled out when considering just gas in the molecular phase. 

\begin{figure}
\centering
\setlength{\fboxsep}{0pt}%
\setlength{\fboxrule}{0.5pt}%
\fbox{
\includegraphics[width=0.4\textwidth]{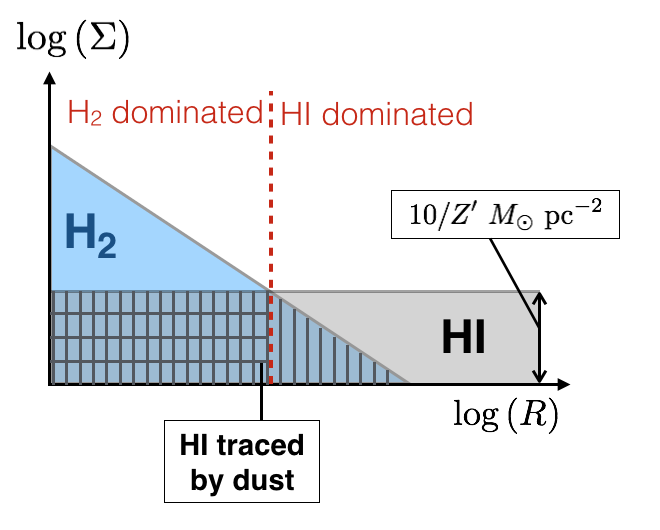}
}
\caption{Toy model sketching the atomic and molecular gas mass profiles within a galaxy. We assume an exponential disk distribution for \Hmol\ and flat surface mass density for HI (at least over the radial range containing appreciable amounts of molecular gas). We use this model to derive by how much our dust-based gas masses would be affected assuming that the dust traces not only \Hmol , but also part or all of the \HI . 
}
\label{fig:Toy_Model}
\end{figure}

\begin{figure}
\centering
\includegraphics[width=0.48\textwidth]{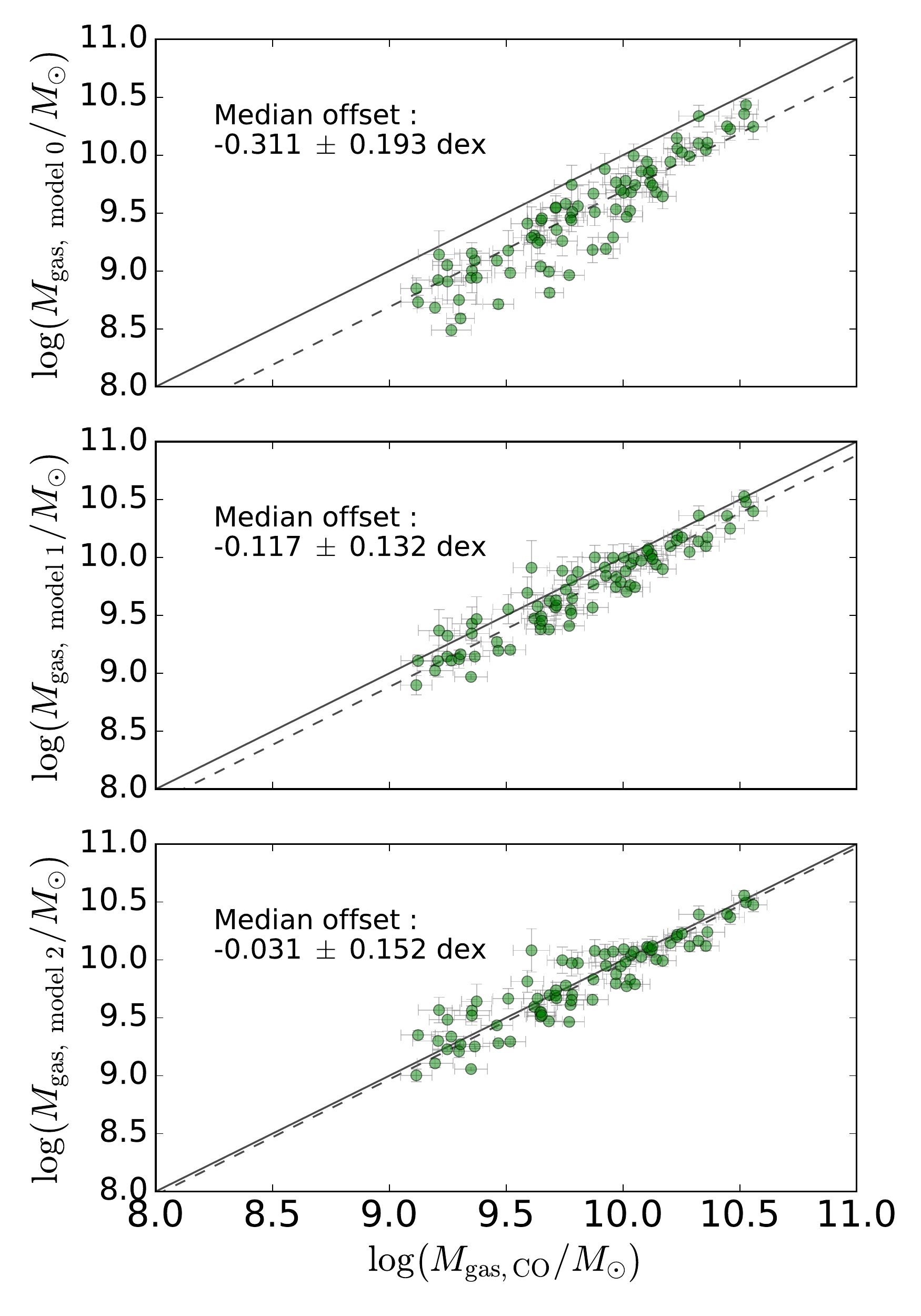}
\caption{Same as figure \ref{fig:Comparison}, but with our toy model adjustment for HI applied to the dust-based gas masses. Here \Mgasdust\ is adjusted and contrasted to \MgasCO\ according to three scenarios. Model 0 is based on the assumption that dust traces all cold gas (\HI\ + \Hmol). Model 1 assumes that at any radius at most as much \HI\ as \Hmol\ is traced by dust, and model 2 assumes that only \HI\ in the \Hmol -dominated region is traced by dust. While model 0 is clearly overshooting the adjustment, model 2 (in which dust traces HI gas in the molecular-dominated regime) yields the best agreement. The 1-to-1 line (black solid) and the median offset (black dashed) are shown for reference.} 
\label{fig:Toy_Model_applied}
\end{figure}

As motivated earlier, it would be plausible to expect a finite excess in \Mgasdust\ compared to the CO reference indicating that dust not only traces gas in the molecular phase, but also (or at least in part) the atomic gas phase. 
We consider a toy model assuming that \HI\ resides in a disk of constant surface mass density, at least out to those galactic radii traced by molecular gas. {We adopt a value of $10/Z' \ {\rm \Msun \ pc^{-2}} $ for the surface mass density motivated by observations of nearby galaxies \citep{Bigiel2008, Wong2013} and in broad agreement with theoretical work based on radiative transfer calculations of the \HI -to-\Hmol\ transition threshold \citep{Sternberg2014}
}, where $Z' = Z / Z_{\odot}$ {with solar metallicity $Z_{\odot} = 8.67$ \citep{Asplund2004_Zsun}}. We further assume that the molecular gas distribution follows an exponential profile with scale length adopted from the {\it g}-band image, in agreement with the observed exponential decline in CO brightness in the most nearby galaxies \citep[see, e.g. ][]{Leroy2009, Schruba2011}. The resultant mass profile is sketched in Figure \ref{fig:Toy_Model}, showing surface density $\Sigma$ as a function of radius $R$: the central region is dominated by the molecular phase, whereas \HI\ prevails at larger radii. We then downscale \Mgasdust\ (computed as outlined in Section \ref{sec:dust_method}) according to three different scenarios. First, following the hypothesis that dust traces the total cold gas mass (\HI\ + \Hmol ), {i.e. $M_{\rm gas, \ mol} = \Mgasdust \cdot (M_{\Hmol} / (M_{\HI}+M_{\Hmol})) $,} we find that the offset between dust- and CO-based mass estimates would change from $+ 0.05$ to $-0.34$ dex {(model 0 in Figure \ref{fig:Toy_Model_applied})}. 
Thus assuming that all atomic gas is associated with dust, including that residing at large galactocentric radii, clearly overshoots the downscaling. Therefore, unless there are conspiring systematic offsets in our adopted CO-to-\Hmol\ conversion factor $\alpha_{\rm CO}$ and dust-to-gas ratio each at the level of $\sim 50$\%, our analysis implies that dust does not trace the entire cold gas disk (\HI\ + \Hmol ). 
This interpretation is consistent with studies comparing the scale lengths of the 21cm and IR emission in nearby galaxies \citep{Thomas2004} and in the Virgo cluster \citep{Cortese2010_truncdust}, which showed evidence for truncated dust  (relative to \HI ). The idea that dust traces only part of \HI\ in a mixed phase yields a better agreement than the first scenario. {If all \HI\ in overlap with \Hmol\ is traced (model 1; vertically shaded region in Figure \ref{fig:Toy_Model}), the offset amounts to $-0.12$ dex ($0.13$ dex scatter).} If only the \Hmol -dominant phase is traced (model 2; horizontally shaded region in Figure \ref{fig:Toy_Model}), the offset changes to $- 0.031$ dex ($0.15$ dex scatter), which is consistent with our observations. {However we caution that, given that the initial offset of $0.05$ dex is minor enough ($2.7 \ \sigma$) to be a consequence of uncertainties in the measurement, our results do not allow us to discriminate between model 2 and the possibility that dust does not trace any \HI\ at all. The conclusion that not all \HI\ is traced by dust on the other hand is statistically robust. }

\subsection{Dynamical mass constraints}
\label{sec:Mdyn}

In addition to integrated CO line flux information, our IRAM survey also yields dynamical constraints on the mass budget through the width and profile shape of the galaxy-integrated CO line measurements. Here, we take advantage of these additional constraints to investigate the total mass budget enclosed within the effective radius. As discussed in Section \ref{sec:lpfit}, we fit the observed CO(1-0) line profile with inclined rotating disk models accounting for the finite beam width. We report the resultant dynamical masses in Table \ref{tab:DerivedData}. We note that consistent dynamical masses are obtained when fitting the line profiles of CO(2-1), where observed ($\Delta \ \log M_{\rm dyn} = 0.03 \pm 0.13$).

For a cross-check, we calculate an  independent estimate of the enclosed mass components based on the following mass profiles: 
For the stellar and molecular mass distribution, we assume a S\'ersic profile as fit by \citet{Simard2011}. 
For the atomic gas we adopt a uniform distribution with a threshold surface density of $12/(Z - Z_{\odot}) \Msun {\rm pc}^{-2}$ motivated by radiative transfer calculations \citep{Sternberg2014}, above which gas is expected to convert from the atomic to the molecular phase. Such a saturation effect has indeed been observed in spatially resolved studies of nearby galaxies \citep{Bigiel2008, Schruba2011}. We infer the total dark matter mass using the $\mstar - M_{\rm halo}$ relation derived through abundance matching \citep{Moster2013}, and adopt the halo concentration from \citep{Dutton2014}. By distributing the total masses $M_{\star}$, $M_{\HI}$, $M_{\Hmol}$ and $M_{\rm halo}$ according to these profiles, we calculate the mass enclosed within one effective radius ($R_e$) for each mass component. 

{We adopt the errors on $M_{\star}$ provided by MPA-JHU as noted in Section \ref{sec:GalProps}, while errors on $M_{\HI}$ and $M_{\Hmol}$ remain the same as throughout (see Section \ref{sec:HI}  and \ref{sec:CO_method}). Since our adopted halo masses are derived using the $\mstar - M_{\rm halo}$ relation, their uncertainties follow from a combination of the intrinsic scatter of $ \ 0.15$ dex in the \citet{Moster2013} relation, and the uncertainty on $M_{\star}$ as input variable. We account for that by drawing, for each galaxy, 100 stellar masses from a Gaussian centered around $\log M_{\star}$ with standard deviation $( ( d \log \ M_{\star} )^2 +0.15^2 )^{1/2}$, then computing the corresponding 100 halo masses and using the central 68th percentile to arrive at the final $\pm 1\sigma$ error on $M_{\rm halo}$.}

\begin{figure*}
\centering
\includegraphics[width=0.49\textwidth]{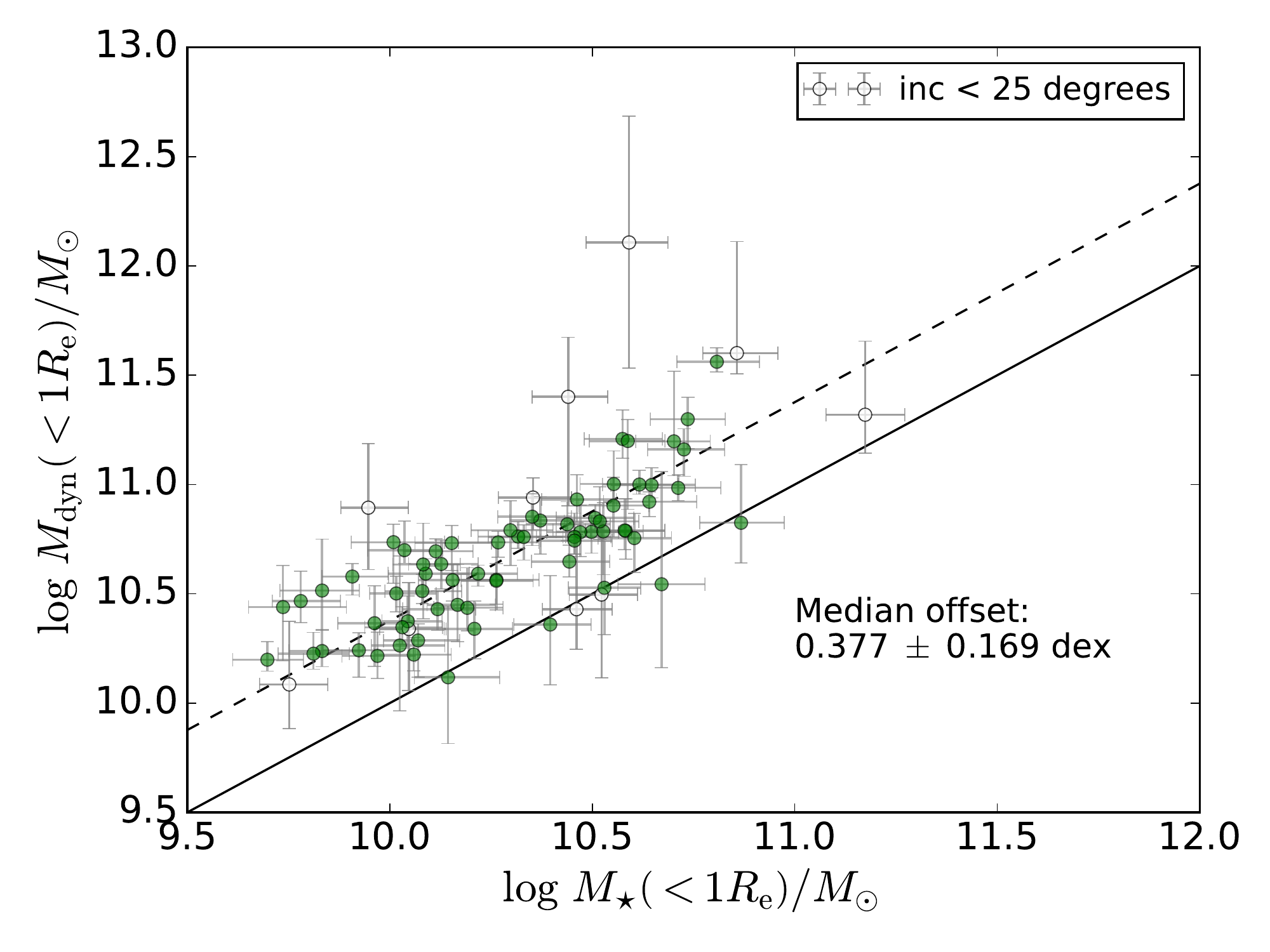}
\includegraphics[width=0.49\textwidth]{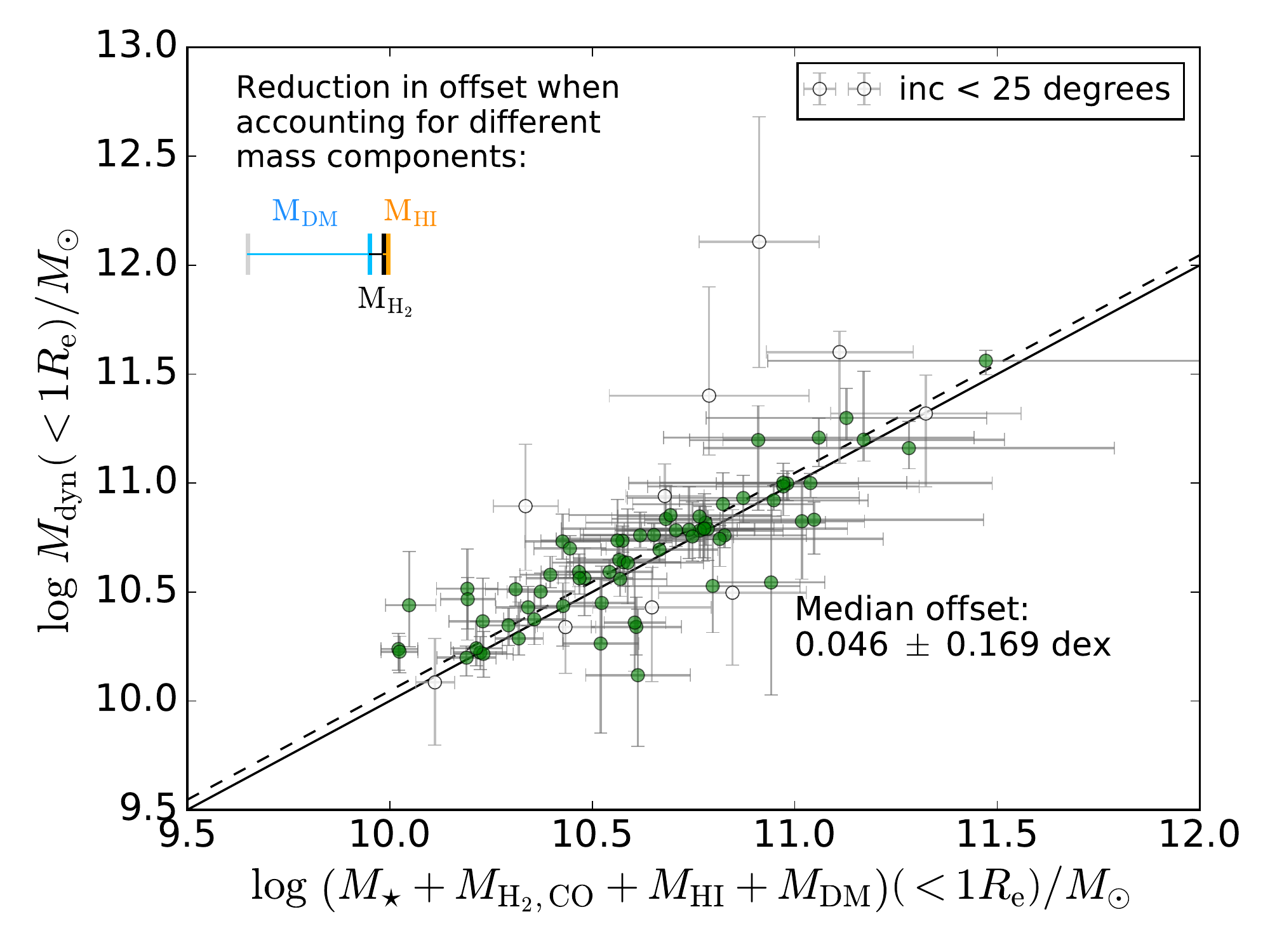}
\caption{Dynamical mass within $1 \ R_e$ versus different mass components enclosed within the same radius.  Objects with inclination < 25\deg are marked with open circles and excluded from the statistics due to significant uncertainties in inclination correction.  {\it Left:} Dynamical vs stellar mass.  Stellar mass alone typically accounts for less than half of the mass within $1 \ R_e$. \textit{Right}: Dynamical mass plotted against the sum of enclosed stellar mass, molecular gas mass, atomic gas mass and DM halo mass. Markers in the top left of the diagram indicate by how much the offset is reduced when adding each mass component. The offset is finally reduced to $\sim 0.05$ dex, with scatter $\sim 0.17$ dex. Our dynamical mass constraints thus agree well with the sum of enclosed mass components. {The 1-to-1 line (black solid) and the median offset (black dashed) are shown for reference. }}
\label{fig:Mdyn}
\end{figure*}

Figure \ref{fig:Mdyn} (left-hand panel) contrasts the dynamical mass estimates to the stellar masses of the respective galaxies. A sizeable systematic offset is evident, of $\sim 0.38 \pm 0.21 \rm \ dex$, thus leaving significant room for other mass components than stars. Low-inclination systems are shown for completeness (open circles), but are discarded from the statistics quoted given their results are dominated by uncertainties in the inclination correction. Next, we add our best estimate of all mass components that should, to various degrees, contribute to the mass budget within the effective radius. Accounting for the sum of all enclosed masses yields an excellent agreement to the dynamically inferred mass budget (right-hand panel of Figure \ref{fig:Mdyn}), with the offset dropping to $\sim 0.05 \pm 0.17 \rm \ dex$. This increases our confidence in our ability to quantify the (sum of) stellar masses and dark matter contributions in the inner disk regions. We note that this exercise should not be considered as a means to tighten constraints on the gas mass content (let alone identify subtle variations or re-scaling in conversion factors). This is because baryonic gas fractions remain below $\sim 35$\%  for nearly all objects in our sample, and gas-to-total (i.e. including dark matter) mass fractions are even smaller. 
On a sidenote, the prospects for kinematics to usefully constrain the (gas) mass budget at higher redshifts are more optimistic, given their gas-rich nature \citep[e.g.][and references therein]{Tacconi2017} and reduced (inner) dark matter fractions \citep[e.g.]{Wuyts2016, Genzel2017, Lang2017, Uebler2017}. For example, \citet{Tadaki2017} present an application to dynamically constrain $\alpha_{\rm CO}$ in the core region of massive z$\sim$2 SFGs.


\section{Summary and conclusion}
\label{sec:Summary}

In this paper, we presented an IRAM survey targeting the CO(1-0) emission line of 78 galaxies with known metallicities and with existing archival IR measurements from WISE and the HerS survey. We additionally included 14 COLD GASS galaxies with WISE and SPIRE coverage.
For 58 galaxies, we further obtained CO(2-1) measurements, thus enabling an analysis of our targets' excitation properties. 
We calculated CO- and dust-based gas masses to establish a cross-calibration between both methods for nearby galaxies. Our CO-based gas mass fractions $\MgasCO / (\MgasCO + \mstar )$ typically vary between $\sim 2-30$ percent, with median value $14.5$\%. Finally, we modelled the CO(1-0) line profiles to derive a dynamical mass constraint within one effective radius and compared the latter to the sum of enclosed mass components.

Our main results are:
\begin{list}{}{}
\item $\bullet$ In the comparison of CO(2-1) to CO(1-0) luminosities, we find a median ratio $<R_{21}>$ corresponding to $\sim 0.64$ with a small scatter of $\sim 0.17$ dex. Our results are consistent with $0.5< R_{21} <0.8$, the range of values reported in studies of the Local Universe \citep{Leroy2013, Rosolowsky2015, Saintonge2017}. We report no significant trends with sSFR, mass, metallicity, galaxy size, angular size, thus suggesting a very uniform distribution of excitation properties. 
\item $\bullet$ We report a tight relation {between CO- and dust-based gas masses} ($0.17$ dex scatter), with the dust-based gas masses showing a minor offset from their CO counterparts by about $0.05$ dex. We find that the offset decreases both with a decreasing measure of the atomic-to-molecular mass ratio and with increasing metallicity, although with considerable scatter.
We adjust our dust-based gas masses to match the CO-based reference by introducing a zero point and metallicity-dependent term (equation \ref{equ:dust_to_molgas}), effectively corresponding to a re-calibration of the dust-to-(molecular) gas ratio. We further showed that our offset is consistent with the hypothesis that dust traces not only \Hmol , but also part of the atomic hydrogen in overlap with the \Hmol -dominant molecular gas disk, but not the entire galaxy-integrated \HI\ reservoir. This result reflects our expectations, given that the gas-to-dust conversion we used \citep{Leroy2011} was established to yield the total gas mass (\HI\ + \Hmol) along the line of sight in \Hmol -dominated regimes. {However, we caution that the observed offset (at the $2.7 \sigma$ level) may also be consistent with being a consequence of observational uncertainties. }
\item $\bullet$ When contrasting the dynamical mass within one effective radius to the sum of all enclosed mass components, we find an excellent agreement ($\sim 0.05 \pm 0.17$ dex offset). 
{Given the overall modest total gas mass fractions ($\lesssim 35$\% of baryons, $\lesssim 25$\% of total except for one gas-rich outlier), this result adds little to our empirical constraints on gas mass conversion factors from CO and/or dust.  However, it serves as an encouraging demonstration that the assumptions going into the derivation of stellar mass (e.g. IMF) and inner dark matter components (dependent on overall halo mass as well as DM profile) are sound.}
\end{list}


\section*{Acknowledgements}

{We thank the anonymous referee for constructive and useful suggestions.} This work is based on observations carried out with the IRAM 30-m telescope, and the Arecibo 305-m telescope. IRAM is supported by INSU/CNRS (France), MPG (Germany) and IGN (Spain). The Arecibo Observatory is operated by SRI International under a cooperative agreement with the National Science Foundation (AST-1100968), and in alliance with Ana G. M\'endez-Universidad Metropolitana, and the Universities Space Research Association. This publication makes use of data products from the Wide-field Infrared Survey Explorer, which is a joint project of the University of California, Los Angeles, and the Jet Propulsion Laboratory/California Institute of Technology, funded by the National Aeronautics and Space Administration. This work also used observations of {\it Herschel}, an ESA space observatory with science instruments provided by European-led Principal Investigator consortia and with important participation from NASA.

\newpage

\bibliographystyle{mnras}
\bibliography{bibliography} 


\clearpage

\newpage

\appendix

\section{AGN contribution}
\label{app:AGN_contr}

{
While our sample was selected to consist of purely star-forming galaxies, we aim to quantify to what extent the calibration proposed herein can be applied to samples containing SFG+AGN composite galaxies. To this end, we simulate a grid of model AGN SEDs and add their photometry in the $22$, $250$, $350$ and $500 \ \micron$ bands with appropriate scaling to the observed WISE+SPIRE photometry of our star-forming sample. 
We apply our SED fitting procedure to these mock composite observations, and calculate dust-based gas masses for the ``contaminated" sources.

Our AGN IR model SEDs are taken from \citet{Kirkpatrick2015}, who compiled a library of templates based on stacking \textit{Spitzer} mid-IR
spectroscopy and multi-wavelength imaging data of luminous infrared galaxies. Specifically, to explore different AGN spectral shapes, we use the four different AGN templates in their comprehensive library (``AGN1" to ``AGN4" according to different redshift and IR luminosity cuts), plus the purest AGN template from their mid-IR library (``MIR1.0"). For a detailed discussion about the technique, underlying assumptions, possible selection biases and limitations of the models, we refer the reader to \citet{Kirkpatrick2015}. To get a rough estimate of the bolometric luminosity $L_\mathrm{bol, \ AGN}$ of the AGN model IR SEDs, we apply a bolometric correction factor of $10.09$ at $22 \ \micron$ as proposed by \citet{Runnoe2012}.  For each galaxy in our sample, we scale the mock photometric fluxes such that the ratio of the AGN bolometric luminosity and the integrated IR luminosity of our star-forming SED, $f_{AGN} = \log (L_\mathrm{bol, \ AGN}/L_\mathrm{IR, \ SF})$, ranges from $-1$ to $1$, thus accounting for various degrees of AGN contribution. We caution that the \citet{Runnoe2012} conversion was established for luminous AGN with $45.1 < \log L_\mathrm{bol, \ AGN} < 47.0$, while our parameter space explores bolometric luminosities ranging from $10^{43.3}$ to $10^{46.2} {\rm \ erg \ s^{-1}}$. We also note that our sample is situated at redshifts $<0.2$ while the \citet{Kirkpatrick2015} templates were established for $z>0.3$. As a result, our treatment of the AGN contribution is likely simplistic, and exploring how AGN properties might differ for our parameter space is beyond the scope of this work. Here we simply aim to provide a rough estimate of the general scale of AGN contribution effects.

The expected median over-/underprediction in dust-based gas masses, $\Delta \log \left( M_{\mathrm{gas,dust}} \right) $, is shown in Figure \ref{fig:MockAGN} as a function of the strength of AGN contribution. At $f_{AGN}>1$, the gas mass is overpredicted with the discrepancy increasing with the strength of the AGN. The trends result from the two dependencies stated in Eq. \ref{eq:Mdust_formula}: the dust mass (and therefore the predicted gas mass) increases linearly with $ L_\mathrm{IR, \ SF}$, while the dependence on dust temperature scales with a power of $-5.5$. The AGN contribution increases the total IR luminosity, while at the same time adding predominantly to the flux at $22 \micron$,  thus compensating by making the dust appear hotter. Our findings suggest that dust-based gas masses will be overpredicted by a factor $2$ ($1.5$) if the AGN luminosity $ L_\mathrm{bol, \ AGN}$ exceeds $\sim 5.4 L_\mathrm{IR, \ SF}$ ($3.3  L_\mathrm{IR, \ SF}$).
}

\begin{figure}
\centering
\hspace*{-0.3cm}
\includegraphics[width=0.5\textwidth]{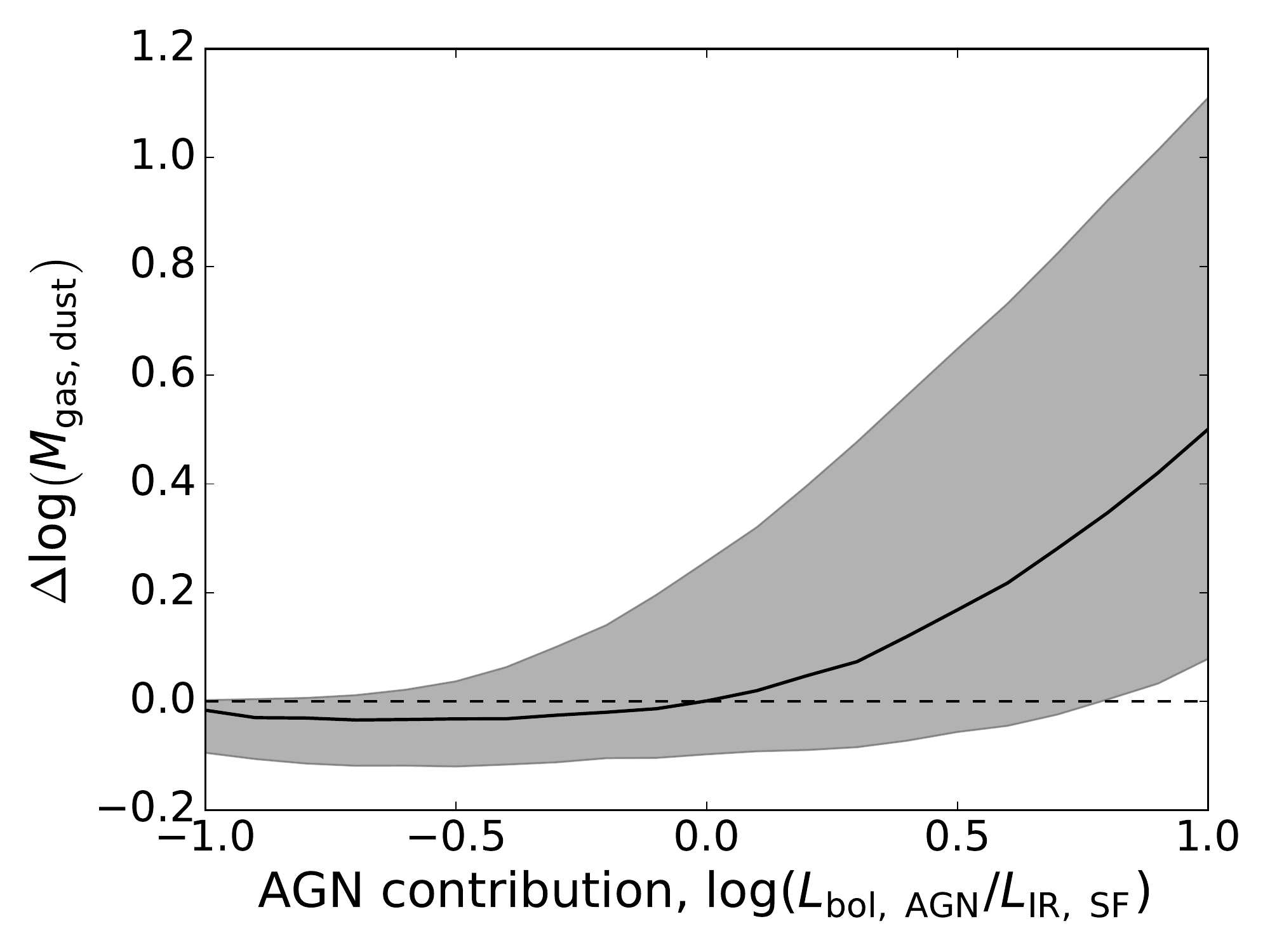}
\caption{{Expected variation in the prediction of \Mgasdust\ when fitting the SED of a model composite galaxy with AGN contribution, as opposed to a purely star-forming galaxy. The median trend (black solid line) is calculated ranging over all galaxies and for each of them, over five templates accounting for different AGN spectral shapes. The grey region displays the $68$ percentile range.}}
\label{fig:MockAGN}
\end{figure}

\newpage
\section{Catalog of observables and derived quantities}

In Tables \ref{tab:RawData} and \ref{tab:DerivedData}, we list the values of our direct observations and derived quantities for the Stripe82 sample.

\begin{table*} 
\captionof{table}{Directly observable quantities for our 78 Stripe82 galaxies: aperture-corrected CO(1-0) flux $\rm F_{CO(1-0)}$ and, where applicable, CO(2-1) flux $\rm F_{CO(2-1)}$, corresponding aperture correction factors ($f_{\rm corr, \ CO(1-0)}$, $f_{\rm corr, \ CO(2-1)}$), the luminosity ratio $R_{21} = F_{\rm CO(2-1)} / F_{\rm CO(1-0)} $, and, where applicable, \HI\ flux $\rm F_{\HI}$.}
{

\begin{tabular}{l c c c c c c c c c c}
\hline
ID & RA & DEC & $z$ & $\rm F_{CO(1-0)}$ & $f_{\rm corr, \ CO(1-0)}$ & $\rm F_{CO(2-1)}$ & $f_{\rm corr, \ CO(2-1)}$ &  $R_{21}$ &   $\rm F_{\HI }$ &  \\
 &  [hh:mm:ss.s] & [dd:mm:ss.s] & & $[{\rm Jy \ km \ s^{-1}}]$ & & $[{\rm Jy \ km \ s^{-1}}]$ & & & $[{\rm Jy \ km \ s^{-1}}]$ &  \\
\hline \\
216 & 01:46:02.5 & -00:17:36.3 & 0.055 & 18.1 $\pm$ 0.7 & 1.06 & 56.5 $\pm$ 1.6 & 1.62 & 3.1 $\pm$ 0.1 & / &  \\ 
586 & 02:18:02.1 & -00:33:34.7 & 0.144 & 3.6 $\pm$ 0.5 & 1.18 & / & / & / & / &  \\ 
621 & 02:12:19.8 & 01:04:41.4 & 0.151 & 6.4 $\pm$ 0.6 & 1.29 & / & / & / & / &  \\ 
742 & 01:15:12.7 & -00:11:53.6 & 0.196 & 4.2 $\pm$ 0.5 & 1.02 & / & / & / & / &  \\ 
745 & 01:44:55.5 & -00:40:27.9 & 0.15 & 2.9 $\pm$ 0.3 & 1.32 & / & / & / & / &  \\ 
210 & 01:04:44.7 & -00:53:41.2 & 0.066 & 2.1 $\pm$ 0.4 & 1.01 & 5.7 $\pm$ 1.1 & 1.19 & 2.7 $\pm$ 0.7 & / &  \\ 
335 & 01:21:19.4 & -00:46:50.2 & 0.054 & 9.2 $\pm$ 1.0 & 1.02 & 20.9 $\pm$ 1.3 & 1.51 & 2.3 $\pm$ 0.3 & / &  \\ 
749 & 01:44:50.7 & 00:09:21.9 & 0.148 & 3.5 $\pm$ 0.4 & 1.02 & / & / & / & / &  \\ 
310 & 01:14:56.3 & 00:07:50.5 & 0.041 & 27.9 $\pm$ 1.5 & 1.02 & 79.4 $\pm$ 2.6 & 1.68 & 2.8 $\pm$ 0.2 & / &  \\ 
576 & 01:48:21.4 & 00:46:57.9 & 0.143 & 6.3 $\pm$ 0.7 & 1.09 & / & / & / & / &  \\ 
315 & 01:14:25.7 & 00:32:09.9 & 0.041 & 10.7 $\pm$ 0.7 & 1.02 & 27.6 $\pm$ 1.5 & 1.99 & 2.6 $\pm$ 0.2 & 1.002 $\pm$ 0.113 &  \\ 
765 & 02:02:16.1 & 00:04:25.8 & 0.163 & 7.7 $\pm$ 0.6 & 1.13 & / & / & / & / &  \\ 
393 & 01:59:51.0 & 00:56:16.5 & 0.06 & 7.6 $\pm$ 0.8 & 1.04 & 25.7 $\pm$ 1.7 & 1.23 & 3.4 $\pm$ 0.4 & 2.581 $\pm$ 0.135 &  \\ 
62 & 02:18:59.6 & 00:19:48.0 & 0.03 & 41.1 $\pm$ 2.7 & 1.14 & 65.0 $\pm$ 3.3 & 1.59 & 1.6 $\pm$ 0.1 & 1.367 $\pm$ 0.131 &  \\ 
633 & 02:09:57.9 & -00:08:52.1 & 0.147 & 2.7 $\pm$ 0.4 & 1.06 & / & / & / & / &  \\ 
498 & 01:17:43.1 & 00:24:24.4 & 0.133 & 5.7 $\pm$ 0.6 & 1.04 & 11.0 $\pm$ 1.1 & 1.14 & 1.9 $\pm$ 0.3 & / &  \\ 
110 & 01:39:31.8 & -00:01:52.7 & 0.057 & 20.4 $\pm$ 0.8 & 1.02 & 46.2 $\pm$ 2.6 & 1.61 & 2.3 $\pm$ 0.2 & / &  \\ 
66 & 01:30:21.3 & -00:57:45.0 & 0.029 & 10.0 $\pm$ 1.3 & 1.17 & 37.1 $\pm$ 1.9 & 1.33 & 3.7 $\pm$ 0.5 & / &  \\ 
67 & 01:23:17.0 & -00:54:21.7 & 0.026 & 22.8 $\pm$ 2.2 & 1.3 & 75.4 $\pm$ 4.1 & 1.82 & 3.3 $\pm$ 0.4 & 0.681 $\pm$ 0.096 &  \\ 
495 & 01:11:42.2 & 00:46:04.7 & 0.125 & 4.1 $\pm$ 0.4 & 1.04 & 5.1 $\pm$ 0.9 & 1.11 & 1.2 $\pm$ 0.3 & / &  \\ 
72 & 01:19:34.4 & 00:39:13.2 & 0.054 & 9.9 $\pm$ 1.1 & 1.13 & 31.0 $\pm$ 1.5 & 1.47 & 3.1 $\pm$ 0.4 & / &  \\ 
657 & 02:02:21.6 & -00:54:17.5 & 0.162 & 2.9 $\pm$ 0.3 & 1.27 & / & / & / & / &  \\ 
770 & 02:05:10.5 & 00:33:46.1 & 0.156 & 4.0 $\pm$ 0.4 & 1.26 & / & / & / & / &  \\ 
793 & 01:57:01.5 & -00:16:44.5 & 0.045 & 14.8 $\pm$ 1.3 & 1.33 & 46.6 $\pm$ 2.3 & 1.49 & 3.2 $\pm$ 0.3 & 1.785 $\pm$ 0.326 &  \\ 
22 & 01:15:40.6 & 01:11:52.7 & 0.044 & 15.1 $\pm$ 1.2 & 1.18 & 37.7 $\pm$ 1.9 & 1.8 & 2.5 $\pm$ 0.2 & 2.364 $\pm$ 0.219 &  \\ 
345 & 01:56:09.0 & -00:16:15.8 & 0.047 & 10.4 $\pm$ 0.8 & 1.01 & 37.5 $\pm$ 1.3 & 1.41 & 3.6 $\pm$ 0.3 & 0.73 $\pm$ 0.134 &  \\ 
628 & 01:37:52.1 & 00:24:53.1 & 0.15 & 2.4 $\pm$ 0.4 & 1.01 & / & / & / & / &  \\ 
47 & 02:17:17.0 & -00:25:18.7 & 0.041 & 11.3 $\pm$ 1.4 & 1.04 & 23.9 $\pm$ 1.7 & 1.37 & 2.1 $\pm$ 0.3 & 1.525 $\pm$ 0.21 &  \\ 
579 & 02:05:13.6 & 00:24:02.9 & 0.136 & 2.0 $\pm$ 0.3 & 1.22 & / & / & / & / &  \\ 
4 & 01:21:20.4 & -00:36:51.5 & 0.054 & 9.8 $\pm$ 0.8 & 1.18 & 10.1 $\pm$ 1.1 & 1.34 & 1.0 $\pm$ 0.1 & / &  \\ 
525 & 02:11:16.7 & -00:50:02.4 & 0.083 & 6.4 $\pm$ 0.5 & 1.06 & 17.7 $\pm$ 1.0 & 1.29 & 2.8 $\pm$ 0.3 & 0.927 $\pm$ 0.236 &  \\ 
527 & 00:59:10.0 & -00:20:55.1 & 0.079 & 13.5 $\pm$ 0.8 & 1.02 & 17.9 $\pm$ 1.4 & 1.38 & 1.3 $\pm$ 0.1 & 0.912 $\pm$ 0.174 &  \\ 
652 & 01:31:20.4 & 00:20:41.1 & 0.157 & 4.3 $\pm$ 0.5 & 1.38 & / & / & / & / &  \\ 
521 & 01:46:48.9 & -00:43:05.8 & 0.082 & 6.8 $\pm$ 0.7 & 1.07 & 18.4 $\pm$ 1.4 & 1.22 & 2.7 $\pm$ 0.3 & / &  \\ 
650 & 02:04:31.3 & 00:13:41.2 & 0.163 & 7.6 $\pm$ 0.8 & 1.08 & / & / & / & / &  \\ 
334 & 01:19:05.2 & -00:47:07.4 & 0.054 & 11.1 $\pm$ 0.9 & 1.12 & 26.8 $\pm$ 1.1 & 1.37 & 2.4 $\pm$ 0.2 & / &  \\ 
506 & 02:04:45.6 & -01:02:33.5 & 0.13 & 3.4 $\pm$ 0.4 & 1.09 & 10.4 $\pm$ 0.9 & 1.27 & 3.1 $\pm$ 0.5 & / &  \\ 
5 & 01:57:00.6 & -00:05:23.5 & 0.045 & 30.8 $\pm$ 1.2 & 1.02 & 79.9 $\pm$ 2.4 & 1.9 & 2.6 $\pm$ 0.1 & 1.94 $\pm$ 0.109 &  \\ 
364 & 01:09:04.5 & -01:12:13.1 & 0.063 & 5.5 $\pm$ 0.7 & 1.07 & 15.3 $\pm$ 1.3 & 1.41 & 2.8 $\pm$ 0.4 & / &  \\ 
277 & 01:28:46.5 & 00:16:59.9 & 0.105 & 9.5 $\pm$ 0.8 & 1.27 & / & / & / & / &  \\ 
202 & 01:02:13.6 & -00:26:13.8 & 0.067 & 12.0 $\pm$ 1.1 & 1.02 & 23.8 $\pm$ 1.4 & 1.66 & 2.0 $\pm$ 0.2 & / &  \\ 
734 & 02:04:04.0 & 00:25:24.8 & 0.175 & 3.4 $\pm$ 0.4 & 1.07 & / & / & / & / &  \\ 
753 & 01:05:23.5 & 00:56:08.6 & 0.151 & 0.9 $\pm$ 0.2 & 1.15 & / & / & / & / &  \\ 
151 & 01:42:33.1 & 00:13:10.8 & 0.08 & 5.5 $\pm$ 0.7 & 1.15 & 7.7 $\pm$ 1.6 & 1.33 & 1.4 $\pm$ 0.3 & 1.245 $\pm$ 0.244 &  \\ 
336 & 01:21:47.2 & -00:46:44.0 & 0.054 & 7.2 $\pm$ 0.9 & 1.1 & 21.3 $\pm$ 1.4 & 1.48 & 3.0 $\pm$ 0.4 & / &  \\ 
401 & 02:09:13.1 & 01:12:51.4 & 0.116 & 5.7 $\pm$ 0.5 & 1.07 & 12.1 $\pm$ 1.4 & 1.11 & 2.1 $\pm$ 0.3 & / &  \\ 
715 & 01:53:14.3 & -01:05:50.6 & 0.126 & 5.1 $\pm$ 0.6 & 1.11 & 11.3 $\pm$ 1.4 & 1.25 & 2.2 $\pm$ 0.4 & / &  \\ 
439 & 01:06:28.2 & -00:16:56.9 & 0.067 & 6.5 $\pm$ 0.6 & 1.11 & 15.1 $\pm$ 1.5 & 1.22 & 2.3 $\pm$ 0.3 & / &  \\ 
614 & 01:21:23.4 & 01:11:53.6 & 0.171 & 6.9 $\pm$ 0.6 & 1.23 & / & / & / & / &  \\ 
190 & 01:36:08.1 & 00:20:49.9 & 0.071 & 5.6 $\pm$ 0.5 & 1.2 & 14.7 $\pm$ 1.4 & 1.36 & 2.6 $\pm$ 0.3 & / &  \\ 
121 & 01:01:47.9 & -00:33:52.1 & 0.065 & 3.1 $\pm$ 0.6 & 1.05 & 9.5 $\pm$ 1.9 & 1.32 & 3.1 $\pm$ 0.8 & / &  \\ 
111 & 01:38:40.7 & -00:09:33.7 & 0.057 & 13.5 $\pm$ 1.2 & 1.02 & 31.7 $\pm$ 2.0 & 1.75 & 2.4 $\pm$ 0.3 & / &  \\ 
38 & 01:15:08.2 & 00:13:37.6 & 0.044 & 10.3 $\pm$ 1.0 & 1.06 & 22.0 $\pm$ 2.5 & 1.62 & 2.1 $\pm$ 0.3 & 0.632 $\pm$ 0.168 &  \\ 
267 & 01:24:48.9 & 00:54:36.1 & 0.033 & 67.5 $\pm$ 2.7 & 1.17 & 157.8 $\pm$ 4.3 & 2.18 & 2.3 $\pm$ 0.1 & 0.817 $\pm$ 0.074 &  \\ 
266 & 01:01:07.8 & 00:51:27.8 & 0.027 & 13.1 $\pm$ 1.6 & 1.03 & 32.3 $\pm$ 2.5 & 1.77 & 2.5 $\pm$ 0.4 & 1.028 $\pm$ 0.109 &  \\ 
264 & 01:28:30.3 & 00:00:14.5 & 0.026 & 20.7 $\pm$ 1.7 & 1.03 & 41.9 $\pm$ 2.2 & 1.97 & 2.0 $\pm$ 0.2 & 0.617 $\pm$ 0.085 &  \\ 
780 & 00:58:28.0 & -01:05:03.7 & 0.193 & 2.9 $\pm$ 0.6 & 1.1 & / & / & / & / &  \\ 
268 & 01:32:25.5 & 00:58:36.0 & 0.027 & 15.6 $\pm$ 1.6 & 1.09 & 47.4 $\pm$ 2.0 & 1.48 & 3.0 $\pm$ 0.3 & 0.518 $\pm$ 0.074 & \\
106 & 01:12:54.9 & 00:40:51.6 & 0.065 & 9.2 $\pm$ 0.8 & 1.02 & 21.3 $\pm$ 1.9 & 1.25 & 2.3 $\pm$ 0.3 & / &  \\ 
 \end{tabular} 
 } 
 \label{tab:RawData} 
 \end{table*}
  
 \begin{table*}   
 \ContinuedFloat 
 \caption{(cont.)} 
 { 
 \begin{tabular}{l c c c c c c c c c c c c c} 
 \\ 
\phantom{ID} & \phantom{RA} & \phantom{DEC} & \phantom{$z$} & \phantom{$\rm F_{CO(1-0)}$} & \phantom{$f_{\rm corr, \ CO(1-0)}$} & \phantom{$\rm F_{CO(2-1)}$} & \phantom{$f_{\rm corr, \ CO(2-1)}$} &  \phantom{$R_{21}$}&   \phantom{$\rm F_{\HI }$} &  \\
480 & 01:14:40.0 & 00:28:07.5 & 0.092 & 9.1 $\pm$ 0.8 & 1.21 & / & / & / & / &  \\ 
39 & 01:13:46.3 & 00:18:20.6 & 0.044 & 19.7 $\pm$ 1.6 & 1.1 & 74.4 $\pm$ 3.5 & 1.92 & 3.8 $\pm$ 0.3 & / &  \\ 
520 & 01:46:20.7 & -00:38:49.3 & 0.083 & 7.6 $\pm$ 0.8 & 1.02 & 15.1 $\pm$ 1.3 & 1.29 & 2.0 $\pm$ 0.3 & 1.891 $\pm$ 0.16 &  \\ 
59 & 01:54:54.2 & 00:48:40.4 & 0.027 & 54.3 $\pm$ 2.8 & 1.36 & 103.4 $\pm$ 3.9 & 2.25 & 1.9 $\pm$ 0.1 & 4.701 $\pm$ 0.092 &  \\ 
319 & 02:13:56.8 & 00:31:52.3 & 0.041 & 7.6 $\pm$ 1.3 & 1.14 & 20.2 $\pm$ 1.7 & 1.86 & 2.7 $\pm$ 0.5 & 2.424 $\pm$ 0.148 &  \\ 
55 & 01:24:28.7 & 01:08:19.2 & 0.031 & 14.8 $\pm$ 1.7 & 1.1 & 43.3 $\pm$ 3.1 & 1.76 & 2.9 $\pm$ 0.4 & / &  \\ 
18 & 02:11:14.5 & 01:05:15.4 & 0.04 & 8.0 $\pm$ 0.6 & 1.03 & 27.3 $\pm$ 1.3 & 1.6 & 3.4 $\pm$ 0.3 & 1.597 $\pm$ 0.13 &  \\ 
30 & 01:31:48.1 & 00:39:45.9 & 0.043 & 7.9 $\pm$ 0.7 & 1.14 & 18.6 $\pm$ 1.5 & 2.0 & 2.4 $\pm$ 0.3 & 0.883 $\pm$ 0.088 &  \\ 
37 & 01:15:08.8 & 00:15:56.9 & 0.043 & 19.0 $\pm$ 1.4 & 1.01 & 70.6 $\pm$ 2.4 & 1.41 & 3.7 $\pm$ 0.3 & 0.64 $\pm$ 0.216 &  \\ 
354 & 01:16:23.6 & 00:26:44.8 & 0.055 & 4.7 $\pm$ 0.5 & 1.09 & 15.7 $\pm$ 0.8 & 1.47 & 3.4 $\pm$ 0.4 & / &  \\ 
379 & 01:41:18.6 & -00:12:37.7 & 0.055 & 3.3 $\pm$ 0.6 & 1.1 & 6.9 $\pm$ 0.8 & 1.28 & 2.1 $\pm$ 0.4 & / &  \\ 
642 & 01:12:08.0 & -00:05:11.3 & 0.178 & 4.1 $\pm$ 0.6 & 1.1 & / & / & / & / &  \\ 
32 & 01:14:15.8 & 00:45:55.2 & 0.042 & 15.0 $\pm$ 0.9 & 1.24 & 41.8 $\pm$ 2.0 & 2.08 & 2.8 $\pm$ 0.2 & 0.65 $\pm$ 0.068 &  \\ 
649 & 02:05:01.8 & 00:16:13.5 & 0.161 & 6.5 $\pm$ 0.5 & 1.29 & / & / & / & / &  \\ 
437 & 00:58:14.9 & -00:13:37.0 & 0.071 & 11.7 $\pm$ 1.3 & 1.27 & 47.1 $\pm$ 2.0 & 1.5 & 4.0 $\pm$ 0.5 & / &  \\ 
204 & 01:06:11.5 & -00:53:59.6 & 0.067 & 12.8 $\pm$ 1.1 & 1.3 & 30.7 $\pm$ 1.5 & 1.46 & 2.4 $\pm$ 0.2 & / &  \\ 
292 & 02:01:43.4 & -01:04:11.4 & 0.043 & 5.2 $\pm$ 1.3 & 1.03 & 19.0 $\pm$ 3.6 & 1.88 & 3.6 $\pm$ 1.1 & / &  \\ 
514 & 01:57:18.3 & -01:10:11.9 & 0.081 & 5.2 $\pm$ 0.8 & 1.13 & 19.6 $\pm$ 2.2 & 1.35 & 3.8 $\pm$ 0.7 & / &  \\ 
65 & 01:24:38.1 & -00:03:46.4 & 0.028 & 10.2 $\pm$ 1.1 & 1.22 & 21.0 $\pm$ 1.8 & 1.95 & 2.1 $\pm$ 0.3 & 1.876 $\pm$ 0.117 &  \\
\hline
\end{tabular}
}
\end{table*}

\clearpage

\begin{table*} 
\caption{Derived physical quantities for our 78 Stripe82 galaxies: redshift $z$, stellar mass $\log \mstar$, star formation rate $\log {\rm SFR}$, metallicity $Z$, half-light radius $R_e$, dust temperature $T_{\rm dust}$, dust mass $\log M_{\rm dust}$, dust-based gas mass \lMgasdust , CO-based gas mass \lMgasCO, dynamical mass $\log M_{\rm dyn}$ and \HI\ mass $\log M_{\HI}$. For low-inclination systems (inc $< 25^{\circ}$), the values for $M_{\rm dyn}$ are marked with an asterisk to indicate that they are likely dominated by uncertainties in the inclination. The \HI\ mass $M_{\HI}$ is marked with a double asterisk when the value was obtained based on the \citet{Catinella2012} scaling relation rather than a direct \HI\ observation, as discussed in Section \ref{sec:HI}. {We stress that the uncertainties presented here are of statistical nature only and thus exclude systematics stemming from, e.g., the CO-to-\Hmol\ and dust-to-gas conversion factors.} }
{
\begin{tabular}{l l l l l l l l l l l l l}
\hline
ID & $z$ & $\log \mstar$ & $\log {\rm SFR}$ & $Z$ & $R_e$& $T_{\rm dust}$ &  $\log M_{\rm dust}$ & \lMgasdust  & \lMgasCO & $\log M_{\rm dyn}$  & $\log M_{\rm \HI }$ & \\
  &   & $ [ \Msun ]$ & $ [ \mpyr ] $ &  & $[{\rm kpc}]$& $[{\rm K}]$ &  $ [ \Msun ]$ &  [ \Msun ]  &  [ \Msun ] & $ [ \Msun ]$ & $ [ \Msun ]$ & \\
\hline
216 & 0.06 & 10.76 & 1.03 & 8.77 & 4.06 & 27.1 & 7.96 $\pm$0.01 & 9.87 $\pm$0.01 & 9.97 $\pm$0.05 & 10.43 $\pm$ 0.26* & 10.15 $\pm$0.29** &  \\ 
586 & 0.14 & 10.56 & 1.33 & 8.73 & 3.17 & 31.2 & 8.21 $\pm$0.03 & 10.16 $\pm$0.03 & 10.11 $\pm$0.07 & 10.56 $\pm$ 0.16 & 10.25 $\pm$0.29** &  \\ 
621 & 0.15 & 10.88 & 1.41 & 8.81 & 3.72 & 30.2 & 8.28 $\pm$0.04 & 10.16 $\pm$0.04 & 10.35 $\pm$0.06 & 10.79 $\pm$ 0.14 & 9.98 $\pm$0.29** &  \\ 
742 & 0.2 & 11.11 & 1.35 & 8.74 & 15.06 & 28.5 & 8.53 $\pm$0.03 & 10.47 $\pm$0.03 & 10.46 $\pm$0.06 & 11.56 $\pm$ 0.06 & 10.53 $\pm$0.29** &  \\ 
745 & 0.15 & 10.51 & 1.13 & 8.67 & 6.02 & 31.2 & 8.17 $\pm$0.03 & 10.18 $\pm$0.03 & 10.12 $\pm$0.06 & 10.34 $\pm$ 0.13 & 10.46 $\pm$0.29** &  \\ 
210 & 0.07 & 10.0 & 0.18 & 8.68 & 5.1 & 26.0 & 7.66 $\pm$0.05 & 9.65 $\pm$0.05 & 9.21 $\pm$0.09 & 10.2 $\pm$ 0.07 & 9.71 $\pm$0.29** &  \\ 
335 & 0.05 & 10.21 & 0.43 & 8.79 & 6.27 & 25.1 & 7.75 $\pm$0.03 & 9.65 $\pm$0.03 & 9.64 $\pm$0.07 & 10.58 $\pm$ 0.07 & 9.97 $\pm$0.29** &  \\ 
749 & 0.15 & 10.67 & 1.05 & 8.75 & 5.6 & 28.5 & 8.24 $\pm$0.03 & 10.17 $\pm$0.03 & 10.12 $\pm$0.06 & 10.84 $\pm$ 0.13 & 10.27 $\pm$0.29** &  \\ 
310 & 0.04 & 10.8 & 0.74 & 8.81 & 4.81 & 25.5 & 8.04 $\pm$0.01 & 9.92 $\pm$0.01 & 9.87 $\pm$0.06 & 10.78 $\pm$ 0.09 & 9.92 $\pm$0.29** &  \\ 
576 & 0.14 & 10.88 & 1.44 & 8.81 & 4.15 & 29.3 & 8.34 $\pm$0.03 & 10.22 $\pm$0.03 & 10.32 $\pm$0.06 & 10.79 $\pm$ 0.12 & 9.94 $\pm$0.29** &  \\ 
315 & 0.04 & 10.43 & 0.36 & 8.77 & 7.44 & 26.0 & 7.62 $\pm$0.03 & 9.53 $\pm$0.03 & 9.46 $\pm$0.07 & 10.64 $\pm$ 0.11 & 9.87 $\pm$0.05 &  \\ 
765 & 0.16 & 10.66 & 1.39 & 8.8 & 4.17 & 26.5 & 8.65 $\pm$0.02 & 10.54 $\pm$0.02 & 10.53 $\pm$0.05 & 10.94 $\pm$ 0.12* & 10.09 $\pm$0.29** &  \\ 
393 & 0.06 & 10.37 & 0.79 & 8.72 & 3.74 & 30.2 & 7.83 $\pm$0.01 & 9.79 $\pm$0.01 & 9.69 $\pm$0.06 & 10.29 $\pm$ 0.12 & 10.61 $\pm$0.02 &  \\ 
62 & 0.03 & 10.46 & 0.33 & 8.77 & 4.48 & 27.1 & 7.78 $\pm$0.01 & 9.7 $\pm$0.01 & 9.78 $\pm$0.06 & 10.73 $\pm$ 0.1 & 9.73 $\pm$0.04 &  \\ 
633 & 0.15 & 10.45 & 1.03 & 8.7 & 7.04 & 28.5 & 8.17 $\pm$0.05 & 10.14 $\pm$0.05 & 10.03 $\pm$0.07 & 10.12 $\pm$ 0.31 & 10.46 $\pm$0.29** &  \\ 
498 & 0.13 & 10.95 & 1.23 & 8.69 & 7.07 & 29.3 & 8.22 $\pm$0.03 & 10.21 $\pm$0.03 & 10.28 $\pm$0.06 & 11.0 $\pm$ 0.07 & 10.24 $\pm$0.29** &  \\ 
110 & 0.06 & 10.82 & 0.69 & 8.79 & 6.83 & 26.0 & 8.02 $\pm$0.01 & 9.92 $\pm$0.01 & 10.03 $\pm$0.05 & 10.5 $\pm$ 0.36* & 10.47 $\pm$0.29** &  \\ 
66 & 0.03 & 10.36 & 0.6 & 8.8 & 3.1 & 29.3 & 7.21 $\pm$0.02 & 9.11 $\pm$0.02 & 9.12 $\pm$0.07 & 10.22 $\pm$ 0.08 & 9.21 $\pm$0.29** &  \\ 
67 & 0.03 & 10.38 & -0.07 & 8.78 & 3.95 & 28.5 & 7.45 $\pm$0.01 & 9.36 $\pm$0.01 & 9.37 $\pm$0.06 & 10.51 $\pm$ 0.07 & 9.3 $\pm$0.06 &  \\ 
495 & 0.13 & 10.94 & 0.93 & 8.73 & 6.92 & 26.5 & 8.22 $\pm$0.05 & 10.17 $\pm$0.05 & 10.05 $\pm$0.06 & 10.92 $\pm$ 0.06 & 9.89 $\pm$0.29** &  \\ 
72 & 0.05 & 10.62 & 0.62 & 8.72 & 6.39 & 27.1 & 7.83 $\pm$0.01 & 9.8 $\pm$0.01 & 9.71 $\pm$0.06 & 10.76 $\pm$ 0.06 & 9.79 $\pm$0.29** &  \\ 
657 & 0.16 & 10.47 & 1.07 & 8.77 & 5.18 & 27.8 & 8.28 $\pm$0.04 & 10.19 $\pm$0.04 & 10.1 $\pm$0.06 & 10.45 $\pm$ 0.18 & 10.13 $\pm$0.29** &  \\ 
770 & 0.16 & 10.83 & 1.1 & 8.8 & 5.48 & 26.0 & 8.33 $\pm$0.05 & 10.22 $\pm$0.05 & 10.2 $\pm$0.06 & 10.53 $\pm$ 0.21 & 10.29 $\pm$0.29** &  \\ 
793 & 0.05 & 10.56 & 0.93 & 8.85 & 5.76 & 29.3 & 7.72 $\pm$0.01 & 9.57 $\pm$0.01 & 9.68 $\pm$0.06 & 10.56 $\pm$ 0.09 & 10.21 $\pm$0.07 &  \\ 
22 & 0.04 & 10.76 & 0.94 & 8.83 & 8.57 & 28.5 & 7.75 $\pm$0.01 & 9.61 $\pm$0.01 & 9.65 $\pm$0.06 & 10.93 $\pm$ 0.11 & 10.31 $\pm$0.04 &  \\ 
345 & 0.05 & 10.57 & 0.53 & 8.7 & 5.73 & 27.1 & 7.73 $\pm$0.02 & 9.7 $\pm$0.02 & 9.62 $\pm$0.06 & 10.74 $\pm$ 0.06 & 9.85 $\pm$0.07 &  \\ 
628 & 0.15 & 10.74 & 1.09 & 8.72 & 6.71 & 27.8 & 8.23 $\pm$0.05 & 10.19 $\pm$0.05 & 10.0 $\pm$0.08 & 10.82 $\pm$ 0.09 & 10.51 $\pm$0.29** &  \\ 
47 & 0.04 & 10.27 & 0.22 & 8.8 & 3.79 & 27.1 & 7.49 $\pm$0.02 & 9.38 $\pm$0.02 & 9.47 $\pm$0.07 & 10.22 $\pm$ 0.11 & 10.05 $\pm$0.06 &  \\ 
579 & 0.14 & 10.39 & 0.73 & 8.72 & 5.59 & 28.5 & 8.12 $\pm$0.05 & 10.07 $\pm$0.05 & 9.81 $\pm$0.08 & 10.59 $\pm$ 0.08 & 10.31 $\pm$0.29** &  \\ 
4 & 0.05 & 10.74 & 0.52 & 8.8 & 3.23 & 25.1 & 7.74 $\pm$0.05 & 9.63 $\pm$0.05 & 9.65 $\pm$0.06 & 10.65 $\pm$ 0.07 & 9.52 $\pm$0.29** &  \\ 
525 & 0.08 & 10.35 & 0.57 & 8.77 & 5.38 & 23.1 & 8.26 $\pm$0.09 & 10.17 $\pm$0.09 & 9.88 $\pm$0.05 & 10.34 $\pm$ 0.25* & 10.46 $\pm$0.1 &  \\ 
527 & 0.08 & 10.74 & 0.78 & 8.76 & 6.68 & 25.1 & 8.17 $\pm$0.03 & 10.09 $\pm$0.03 & 10.14 $\pm$0.05 & 11.4 $\pm$ 0.39* & 10.41 $\pm$0.08 &  \\ 
652 & 0.16 & 10.7 & 1.58 & 8.78 & 3.44 & 29.3 & 8.36 $\pm$0.03 & 10.27 $\pm$0.03 & 10.23 $\pm$0.06 & 10.36 $\pm$ 0.25 & 10.16 $\pm$0.29** &  \\ 
521 & 0.08 & 10.42 & 0.74 & 8.67 & 8.5 & 26.5 & 8.15 $\pm$0.03 & 10.15 $\pm$0.03 & 9.92 $\pm$0.06 & 10.69 $\pm$ 0.06 & 10.07 $\pm$0.29** &  \\ 
650 & 0.16 & 11.03 & 1.48 & 8.75 & 11.74 & 25.5 & 8.63 $\pm$0.03 & 10.57 $\pm$0.03 & 10.56 $\pm$0.06 & 11.16 $\pm$ 0.11 & 10.79 $\pm$0.29** &  \\ 
334 & 0.05 & 10.46 & 0.71 & 8.83 & 5.26 & 26.0 & 7.9 $\pm$0.03 & 9.77 $\pm$0.03 & 9.72 $\pm$0.05 & 10.56 $\pm$ 0.08 & 10.1 $\pm$0.29** &  \\ 
506 & 0.13 & 11.48 & 1.03 & 8.81 & 5.74 & 24.2 & 8.28 $\pm$0.05 & 10.16 $\pm$0.05 & 9.96 $\pm$0.06 & 11.32 $\pm$ 0.26* & 10.89 $\pm$0.29** &  \\ 
5 & 0.05 & 10.77 & 0.7 & 8.79 & 6.09 & 26.5 & 7.96 $\pm$0.01 & 9.86 $\pm$0.01 & 10.02 $\pm$0.07 & 10.78 $\pm$ 0.12 & 10.24 $\pm$0.02 &  \\ 
364 & 0.06 & 10.32 & 0.88 & 8.71 & 5.1 & 28.5 & 7.95 $\pm$0.02 & 9.92 $\pm$0.02 & 9.59 $\pm$0.07 & 10.5 $\pm$ 0.08 & 10.08 $\pm$0.29** &  \\ 
277 & 0.1 & 11.01 & 1.37 & 8.79 & 6.11 & 26.0 & 8.38 $\pm$0.02 & 10.28 $\pm$0.02 & 10.23 $\pm$0.05 & 10.98 $\pm$ 0.06 & 9.92 $\pm$0.29** &  \\ 
202 & 0.07 & 10.88 & 0.46 & 8.65 & 10.12 & 23.7 & 8.02 $\pm$0.08 & 10.04 $\pm$0.08 & 9.99 $\pm$0.07 & 11.21 $\pm$ 0.11 & 10.27 $\pm$0.29** &  \\ 
734 & 0.18 & 11.16 & 1.39 & 8.82 & 6.58 & 27.8 & 8.44 $\pm$0.03 & 10.31 $\pm$0.03 & 10.25 $\pm$0.06 & 11.6 $\pm$ 0.3* & 10.35 $\pm$0.29** &  \\ 
753 & 0.15 & 10.33 & 0.9 & 8.71 & 7.22 & 26.5 & 8.21 $\pm$0.14 & 10.18 $\pm$0.14 & 9.61 $\pm$0.08 & 10.26 $\pm$ 0.35 & 10.58 $\pm$0.29** &  \\ 
151 & 0.08 & 10.65 & 0.64 & 8.81 & 6.73 & 25.5 & 8.22 $\pm$0.11 & 10.1 $\pm$0.11 & 9.74 $\pm$0.07 & 10.85 $\pm$ 0.12 & 10.56 $\pm$0.08 &  \\ 
336 & 0.05 & 10.25 & 0.75 & 8.8 & 5.29 & 24.8 & 7.88 $\pm$0.03 & 9.77 $\pm$0.03 & 9.51 $\pm$0.07 & 10.89 $\pm$ 0.29* & 10.18 $\pm$0.29** &  \\ 
401 & 0.12 & 10.81 & 1.12 & 8.82 & 5.46 & 27.1 & 8.23 $\pm$0.03 & 10.11 $\pm$0.03 & 10.08 $\pm$0.06 & 10.85 $\pm$ 0.06 & 10.11 $\pm$0.29** &  \\ 
715 & 0.13 & 10.89 & 1.12 & 8.79 & 12.32 & 25.5 & 8.33 $\pm$0.05 & 10.22 $\pm$0.05 & 10.13 $\pm$0.07 & 11.2 $\pm$ 0.21 & 10.65 $\pm$0.29** &  \\ 
439 & 0.07 & 10.52 & 0.45 & 8.73 & 5.68 & 26.5 & 7.89 $\pm$0.02 & 9.84 $\pm$0.02 & 9.71 $\pm$0.06 & 10.59 $\pm$ 0.05 & 9.86 $\pm$0.29** &  \\ 
614 & 0.17 & 10.97 & 1.6 & 8.79 & 5.37 & 28.5 & 8.71 $\pm$0.02 & 10.61 $\pm$0.02 & 10.52 $\pm$0.05 & 10.54 $\pm$ 0.45 & 10.55 $\pm$0.29** &  \\ 
190 & 0.07 & 10.35 & 0.68 & 8.82 & 4.7 & 27.1 & 7.89 $\pm$0.03 & 9.77 $\pm$0.03 & 9.63 $\pm$0.06 & 10.37 $\pm$ 0.12 & 10.13 $\pm$0.29** &  \\ 
121 & 0.06 & 10.49 & 0.69 & 8.76 & 4.5 & 27.8 & 7.74 $\pm$0.03 & 9.67 $\pm$0.03 & 9.35 $\pm$0.08 & 10.44 $\pm$ 0.14 & 10.04 $\pm$0.29** &  \\ 
111 & 0.06 & 10.85 & 0.65 & 8.8 & 6.14 & 26.5 & 7.86 $\pm$0.02 & 9.75 $\pm$0.02 & 9.87 $\pm$0.07 & 10.9 $\pm$ 0.14 & 10.47 $\pm$0.29** &  \\ 
38 & 0.04 & 10.13 & 0.74 & 8.81 & 4.25 & 29.3 & 7.52 $\pm$0.02 & 9.39 $\pm$0.02 & 9.52 $\pm$0.07 & 10.51 $\pm$ 0.21 & 9.73 $\pm$0.1 &  \\ 
267 & 0.03 & 10.83 & 1.0 & 8.81 & 4.73 & 27.8 & 7.98 $\pm$0.0 & 9.86 $\pm$0.0 & 10.05 $\pm$0.06 & 10.79 $\pm$ 0.17 & 9.59 $\pm$0.04 &  \\ 
266 & 0.03 & 10.13 & 0.03 & 8.82 & 2.6 & 25.5 & 7.33 $\pm$0.03 & 9.2 $\pm$0.03 & 9.2 $\pm$0.08 & 10.24 $\pm$ 0.08 & 9.52 $\pm$0.04 &  \\ 
264 & 0.03 & 10.26 & -0.08 & 8.84 & 3.96 & 27.1 & 7.3 $\pm$0.01 & 9.16 $\pm$0.01 & 9.35 $\pm$0.07 & 10.37 $\pm$ 0.18 & 9.27 $\pm$0.06 &  \\ 
 \end{tabular} 
 } 
 \label{tab:DerivedData} 
 \end{table*}  
 \begin{table*}   
 \ContinuedFloat 
 \caption{(cont.)} 
 { 
 \begin{tabular}{l c c c c c c c c c c c c} 
 \\ 
\phantom{ ID} & \phantom{$z$} & \phantom{$\log \mstar$} & \phantom{$\log {\rm SFR}$} & \phantom{$Z$} & \phantom{$R_e$} & \phantom{$T_{\rm dust}$} &  \phantom{$\log M_{\rm dust}$} & \phantom{\lMgasdust}  & \phantom{\lMgasCO} & \phantom{$\log <M_{\rm dyn}>$}  & \phantom{$\log <M_{\rm \HI }>$} & \\ 	
780 & 0.19 & 11.17 & 1.34 & 8.72 & 4.11 & 26.5 & 8.49 $\pm$0.06 & 10.45 $\pm$0.06 & 10.32 $\pm$0.09 & 10.82 $\pm$ 0.22 & 9.92 $\pm$0.29** &  \\ 
268 & 0.03 & 10.11 & 0.15 & 8.76 & 2.71 & 27.8 & 7.4 $\pm$0.01 & 9.33 $\pm$0.01 & 9.25 $\pm$0.06 & 10.23 $\pm$ 0.08 & 9.2 $\pm$0.06 & \\
106 & 0.06 & 10.22 & 0.73 & 8.79 & 3.44 & 29.3 & 7.88 $\pm$0.02 & 9.77 $\pm$0.02 & 9.78 $\pm$0.06 & 10.24 $\pm$ 0.1 & 9.84 $\pm$0.29** &  \\
480 & 0.09 & 11.04 & 0.91 & 8.73 & 8.88 & 26.5 & 8.14 $\pm$0.03 & 10.09 $\pm$0.03 & 10.17 $\pm$0.06 & 11.3 $\pm$ 0.12 & 10.64 $\pm$0.29** &  \\ 
39 & 0.04 & 10.92 & 0.87 & 8.78 & 9.33 & 26.5 & 7.85 $\pm$0.01 & 9.76 $\pm$0.01 & 9.78 $\pm$0.07 & 11.0 $\pm$ 0.05 & 10.11 $\pm$0.29** &  \\ 
520 & 0.08 & 10.76 & 0.73 & 8.76 & 7.5 & 25.5 & 8.13 $\pm$0.03 & 10.05 $\pm$0.03 & 9.93 $\pm$0.06 & 10.76 $\pm$ 0.07 & 10.76 $\pm$0.04 &  \\ 
59 & 0.03 & 10.33 & 0.71 & 8.84 & 3.73 & 30.2 & 7.69 $\pm$0.0 & 9.55 $\pm$0.0 & 9.77 $\pm$0.07 & 10.35 $\pm$ 0.09 & 10.19 $\pm$0.01 &  \\ 
319 & 0.04 & 10.38 & 0.39 & 8.82 & 8.19 & 26.0 & 7.53 $\pm$0.02 & 9.4 $\pm$0.02 & 9.27 $\pm$0.09 & 10.63 $\pm$ 0.22 & 10.25 $\pm$0.03 &  \\ 
55 & 0.03 & 10.04 & 0.5 & 8.85 & 3.47 & 28.5 & 7.46 $\pm$0.01 & 9.31 $\pm$0.01 & 9.3 $\pm$0.07 & 10.44 $\pm$ 0.22 & 9.9 $\pm$0.29** &  \\ 
18 & 0.04 & 10.05 & 0.4 & 8.83 & 4.03 & 27.1 & 7.51 $\pm$0.02 & 9.37 $\pm$0.02 & 9.3 $\pm$0.06 & 10.09 $\pm$ 0.24* & 10.06 $\pm$0.03 &  \\ 
30 & 0.04 & 10.31 & -0.4 & 8.86 & 8.41 & 23.9 & 7.77 $\pm$0.08 & 9.61 $\pm$0.08 & 9.35 $\pm$0.07 & 10.74 $\pm$ 0.13 & 9.86 $\pm$0.04 &  \\ 
37 & 0.04 & 10.42 & 0.57 & 8.76 & 3.53 & 25.1 & 7.94 $\pm$0.01 & 9.86 $\pm$0.01 & 9.76 $\pm$0.06 & 10.43 $\pm$ 0.09 & 9.72 $\pm$0.13 &  \\ 
354 & 0.05 & 10.63 & 0.29 & 8.73 & 5.88 & 25.1 & 7.78 $\pm$0.12 & 9.73 $\pm$0.12 & 9.37 $\pm$0.06 & 10.76 $\pm$ 0.08 & 10.27 $\pm$0.29** &  \\ 
379 & 0.06 & 10.08 & 0.54 & 8.72 & 4.71 & 25.5 & 7.62 $\pm$0.04 & 9.58 $\pm$0.04 & 9.25 $\pm$0.08 & 10.47 $\pm$ 0.12 & 10.0 $\pm$0.29** &  \\ 
642 & 0.18 & 10.85 & 1.27 & 8.77 & 8.19 & 27.8 & 8.41 $\pm$0.04 & 10.33 $\pm$0.04 & 10.36 $\pm$0.07 & 11.0 $\pm$ 0.12 & 10.33 $\pm$0.29** &  \\ 
32 & 0.04 & 10.82 & 0.59 & 8.72 & 11.11 & 26.5 & 7.66 $\pm$0.02 & 9.62 $\pm$0.02 & 9.65 $\pm$0.06 & 10.83 $\pm$ 0.12 & 9.72 $\pm$0.04 &  \\ 
649 & 0.16 & 10.89 & 1.03 & 8.8 & 6.23 & 27.8 & 8.57 $\pm$0.02 & 10.46 $\pm$0.02 & 10.45 $\pm$0.05 & 12.11 $\pm$ 0.58* & 10.39 $\pm$0.29** &  \\ 
437 & 0.07 & 10.76 & 1.11 & 8.71 & 7.12 & 31.2 & 8.12 $\pm$0.01 & 10.08 $\pm$0.01 & 10.01 $\pm$0.06 & 10.74 $\pm$ 0.09 & 10.16 $\pm$0.29** &  \\ 
204 & 0.07 & 10.9 & 0.89 & 8.78 & 3.72 & 26.5 & 8.04 $\pm$0.02 & 9.95 $\pm$0.02 & 9.97 $\pm$0.06 & 10.75 $\pm$ 0.14 & 9.78 $\pm$0.29** &  \\ 
292 & 0.04 & 11.0 & 0.3 & 8.78 & 5.76 & 24.8 & 7.47 $\pm$0.05 & 9.38 $\pm$0.05 & 9.21 $\pm$0.11 & 11.2 $\pm$ 0.24 & 9.67 $\pm$0.29** &  \\ 
514 & 0.08 & 10.6 & 0.47 & 8.7 & 8.79 & 23.9 & 8.09 $\pm$0.08 & 10.07 $\pm$0.08 & 9.78 $\pm$0.07 & 10.79 $\pm$ 0.15 & 10.0 $\pm$0.29** &  \\ 
65 & 0.03 & 10.34 & -0.33 & 8.68 & 6.26 & 21.6 & 7.41 $\pm$0.03 & 9.4 $\pm$0.03 & 9.12 $\pm$0.07 & 10.7 $\pm$ 0.1 & 9.8 $\pm$0.03 &  \\
\hline \\
\end{tabular}
}
\end{table*}

\clearpage

\section{Postage stamps}
\label{app:postage_stamps}

In figure \ref{fig:postage_stamps} we present 3-colour $gri$ postage stamps for the $78$ galaxies targeted by our IRAM 30-m CO survey.

\begin{minipage}[t]{\textwidth}
\centering
\includegraphics[width=0.8\textwidth]{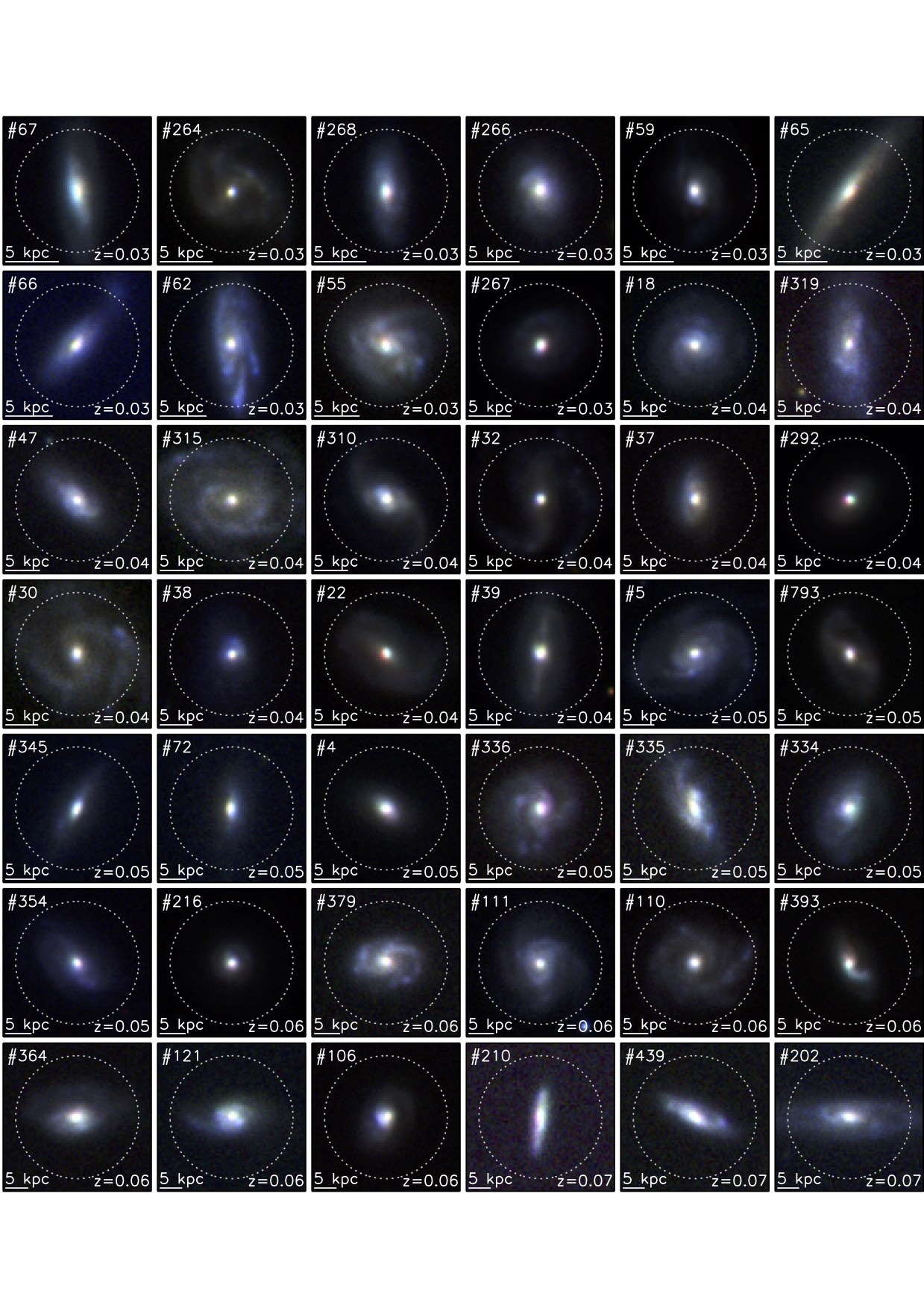} 
\captionof{figure}{SDSS 3-colour $gri$ postage stamps for the galaxies targeted by our IRAM 30-m CO(1-0) survey, ranked by redshift (indicated in the bottom right corner). For reference, we indicate the $22" $ beam of the IRAM 30-m telescope with a dotted circle.}
\label{fig:postage_stamps}
\end{minipage}

\begin{figure*}
\ContinuedFloat
\hspace*{-1cm}
\includegraphics[angle=90, width=1.1\textwidth]{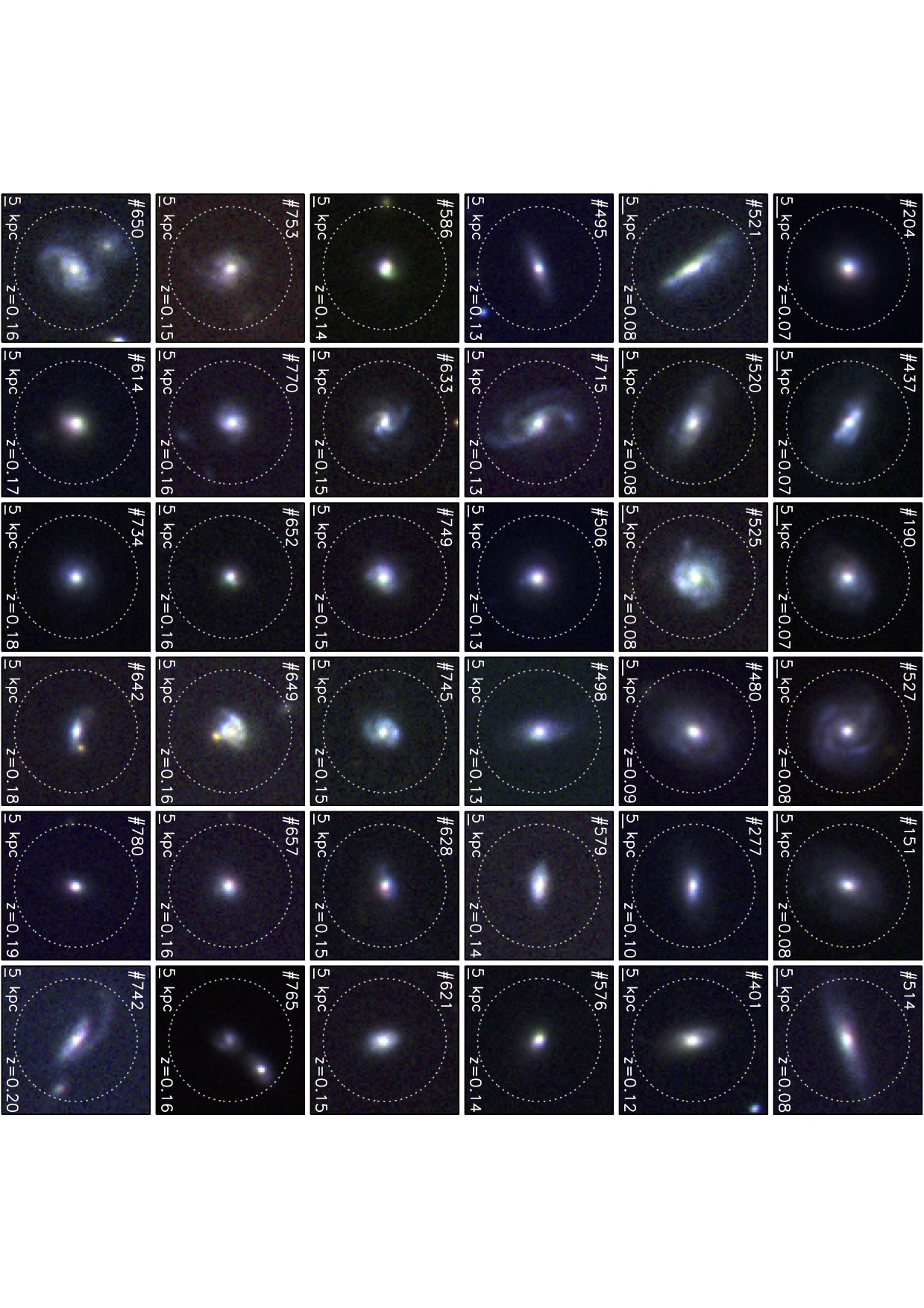}
\caption{(cont.)}
\end{figure*}
\clearpage

\section{Spectral gallery}
\label{app:spectra}

\begin{minipage}[t]{\textwidth}

In Figure \ref{fig:spectra}, we present the line profiles of all of our CO(1-0) observations, ranked by increasing redshift. Overplotted in red is the best-fitting rotating disk model. Figure \ref{fig:spectra21} shows a collection of the 56 CO(2-1) line profiles that were observed simultaneously. The \HI\  21 cm line spectra recorded for a subsample of our galaxies are shown in Figure \ref{fig:Ar_spectra}, sorted by increasing redshift as well. \\

\includegraphics[width=0.31\textwidth]{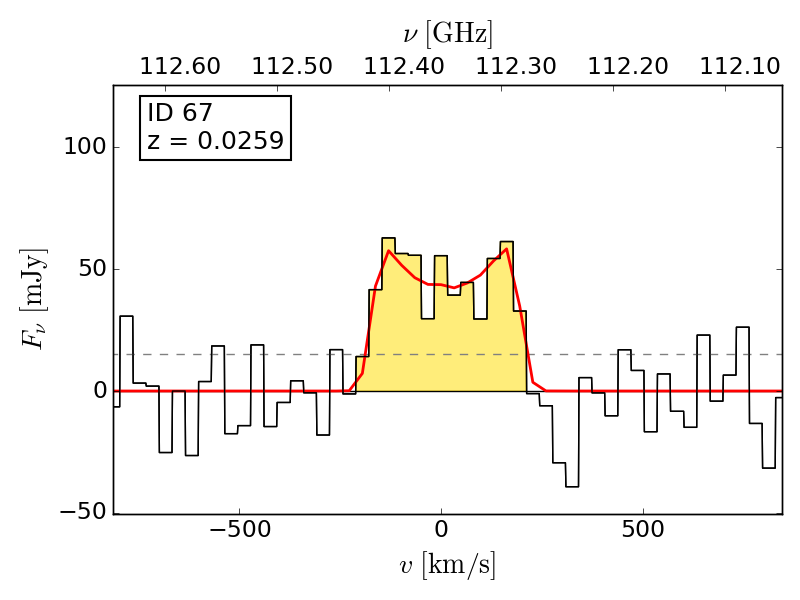}
\includegraphics[width=0.31\textwidth]{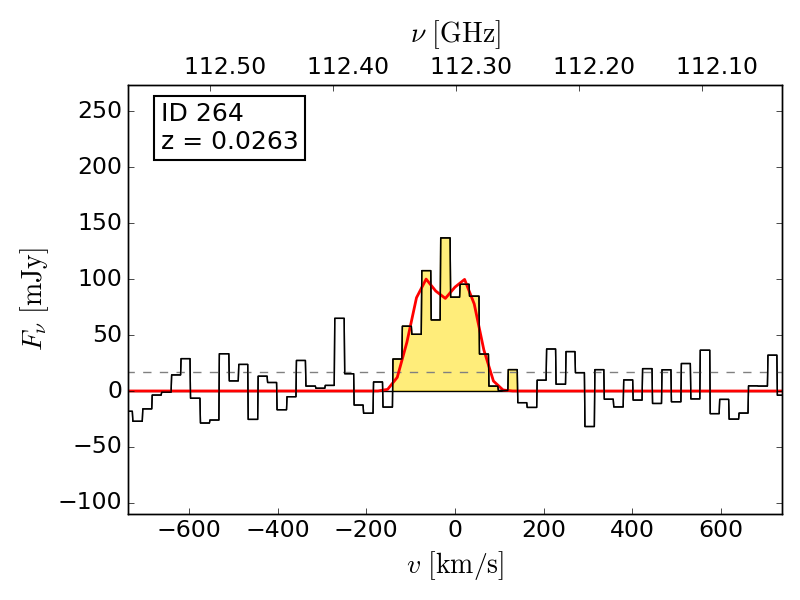}
\includegraphics[width=0.31\textwidth]{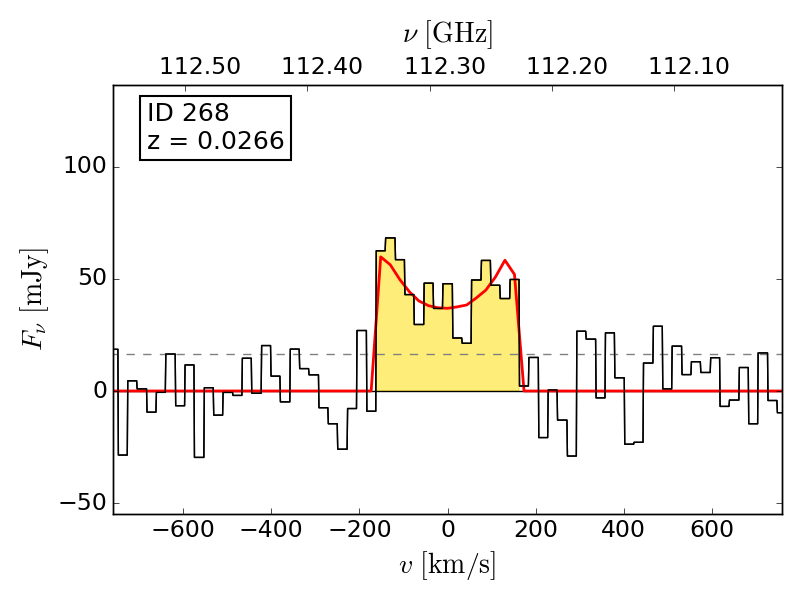}
\includegraphics[width=0.31\textwidth]{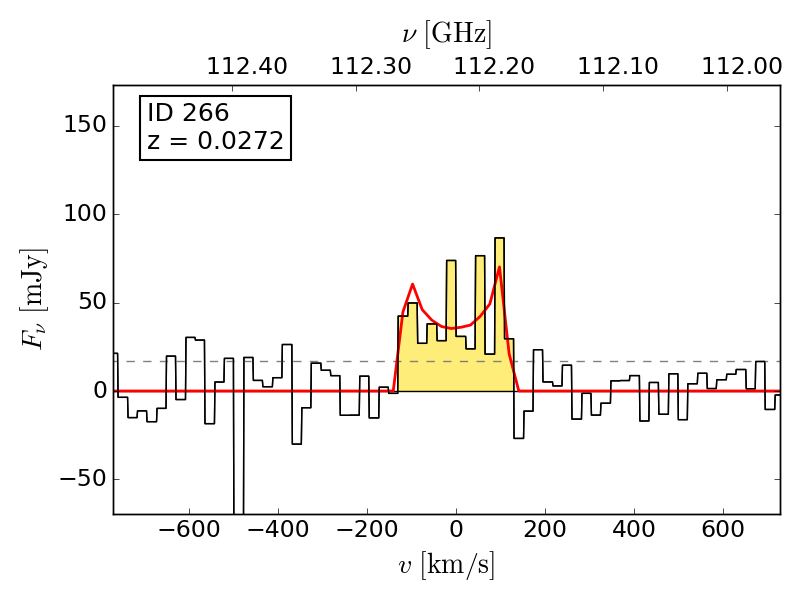}
\includegraphics[width=0.31\textwidth]{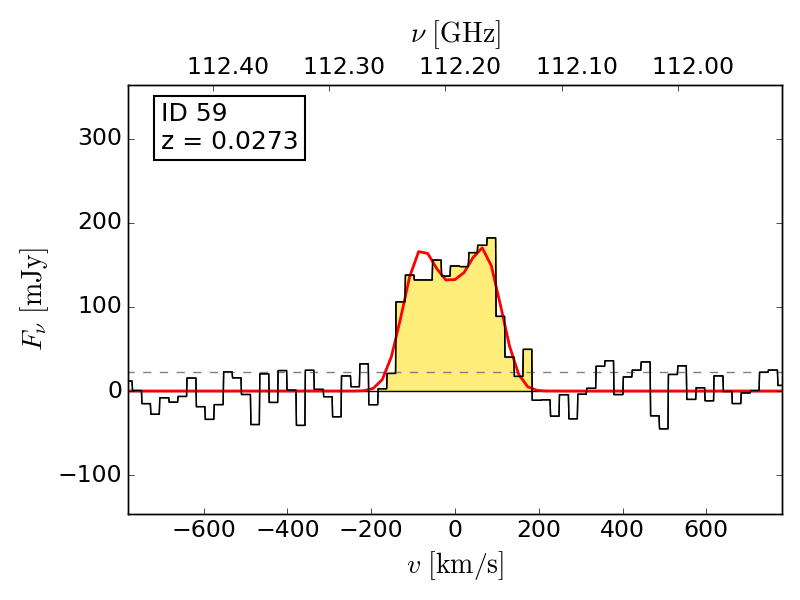}
\includegraphics[width=0.31\textwidth]{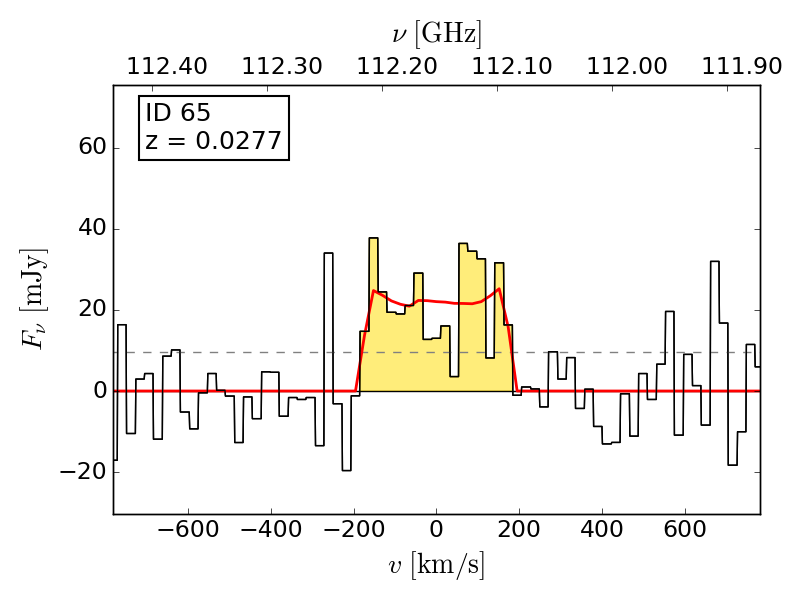}
\includegraphics[width=0.31\textwidth]{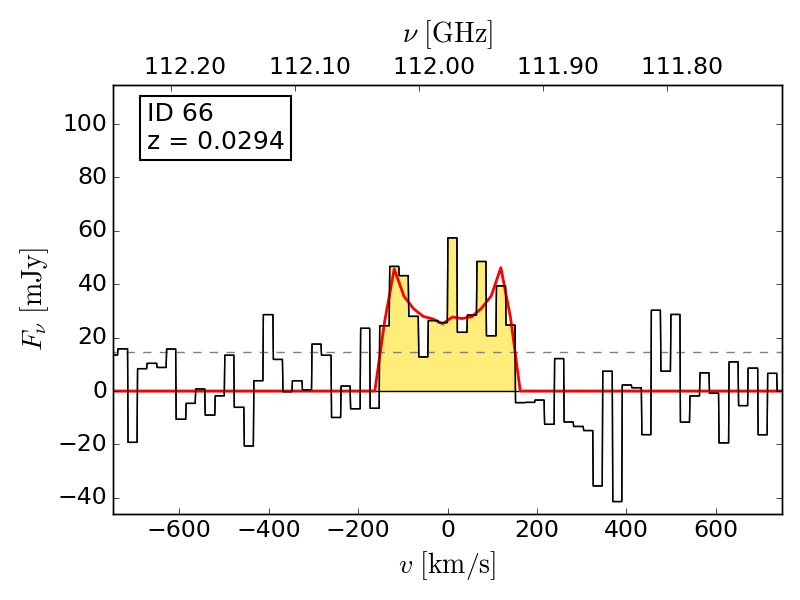}
\includegraphics[width=0.31\textwidth]{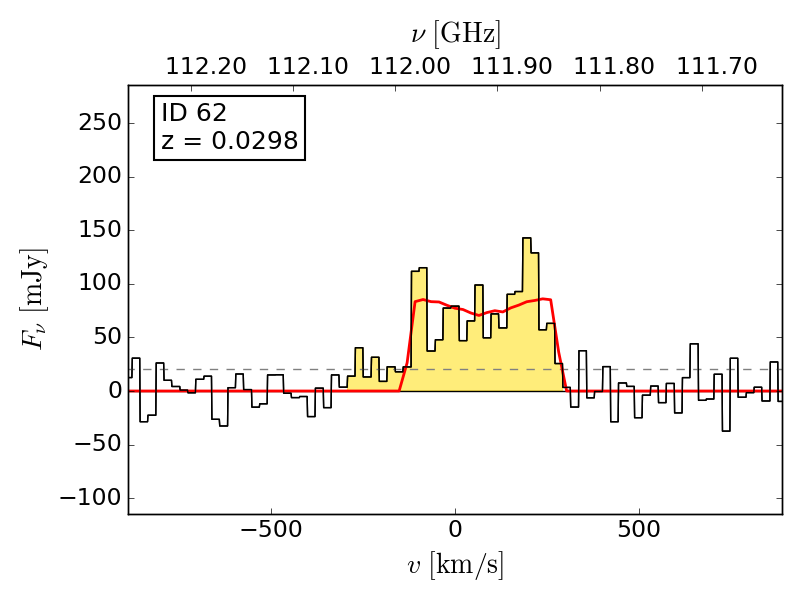}
\includegraphics[width=0.31\textwidth]{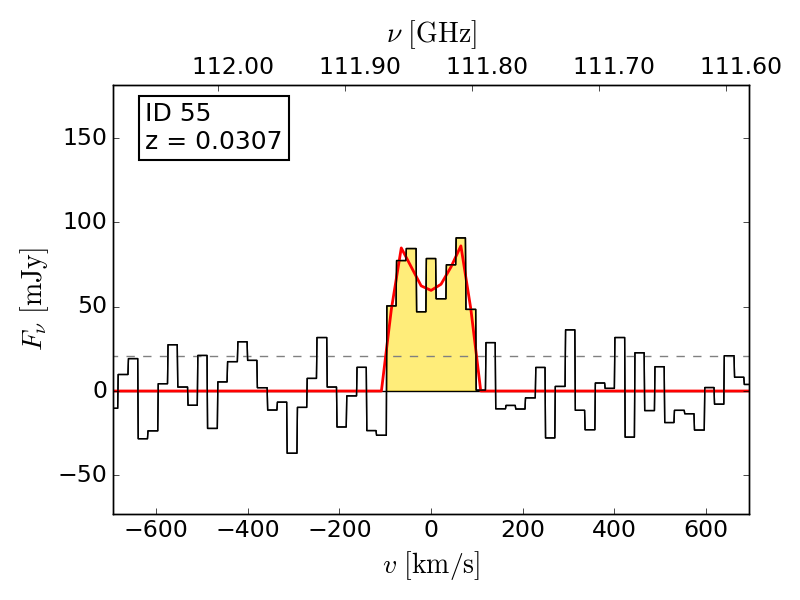}
\includegraphics[width=0.31\textwidth]{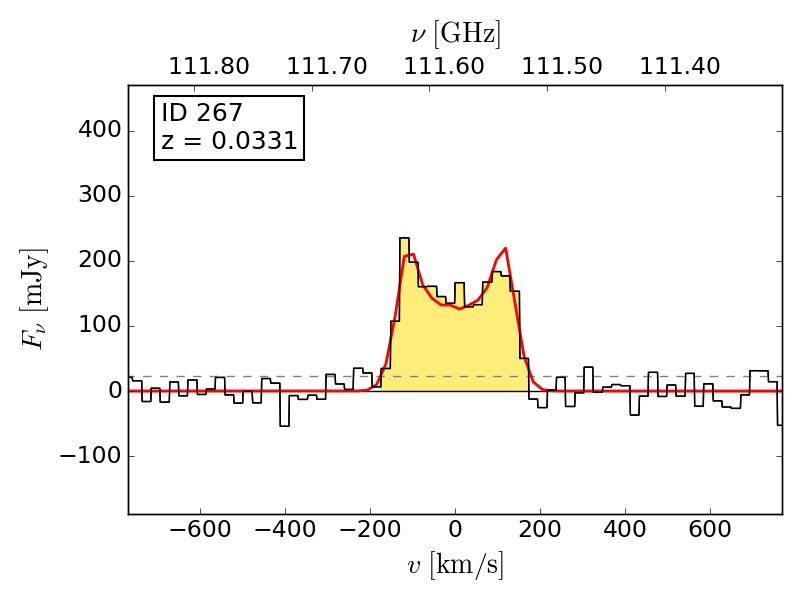}
\hspace*{0.5cm}
\includegraphics[width=0.31\textwidth]{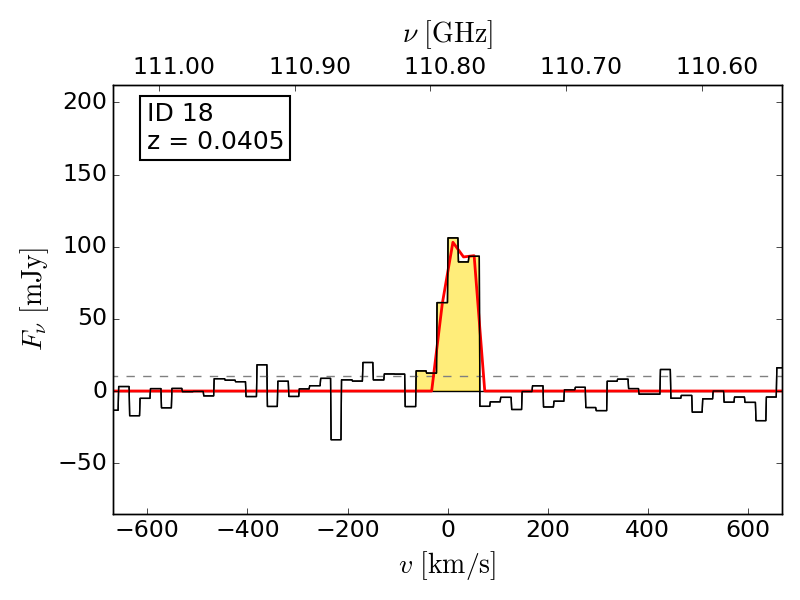}
\hspace*{0.5cm}
\includegraphics[width=0.31\textwidth]{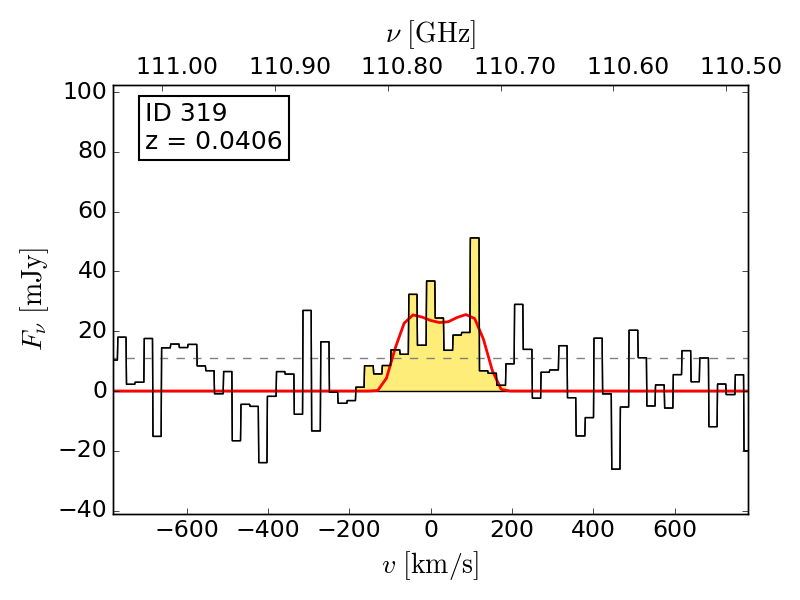}
\captionof{figure}{IRAM CO(1-0) line profiles for the 78 galaxies in our sample, sorted by increasing redshift. The red curve represents the line profile fit obtained by modelling the emission from a rotating disk (see Section \ref{sec:lpfit}). The yellow shaded region marks the line as adopted in the calculation of the integrated line flux, {and the dashed line denotes the $1 \sigma$ level. All flux densities are displayed as observed before the aperture correction was applied.}}
\label{fig:spectra}
\end{minipage}
\begin{figure*}
\ContinuedFloat
\includegraphics[width=0.31\textwidth]{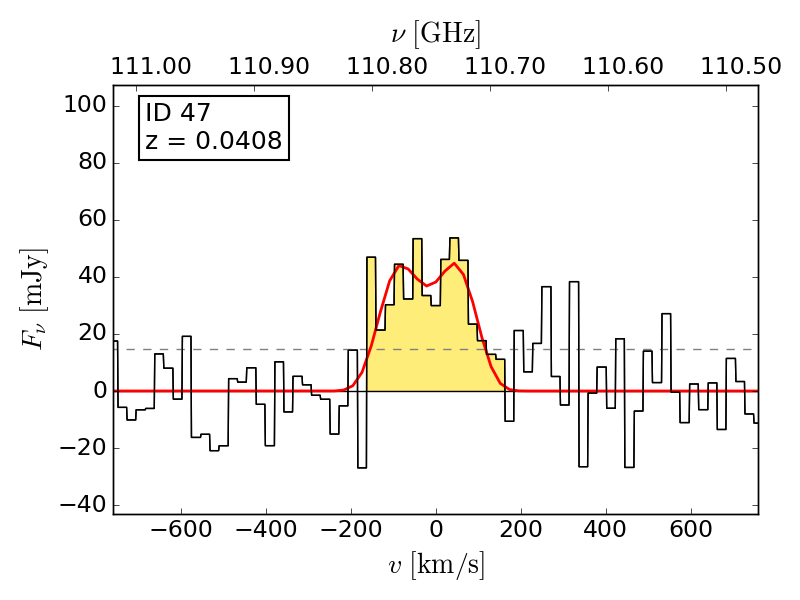}
\includegraphics[width=0.31\textwidth]{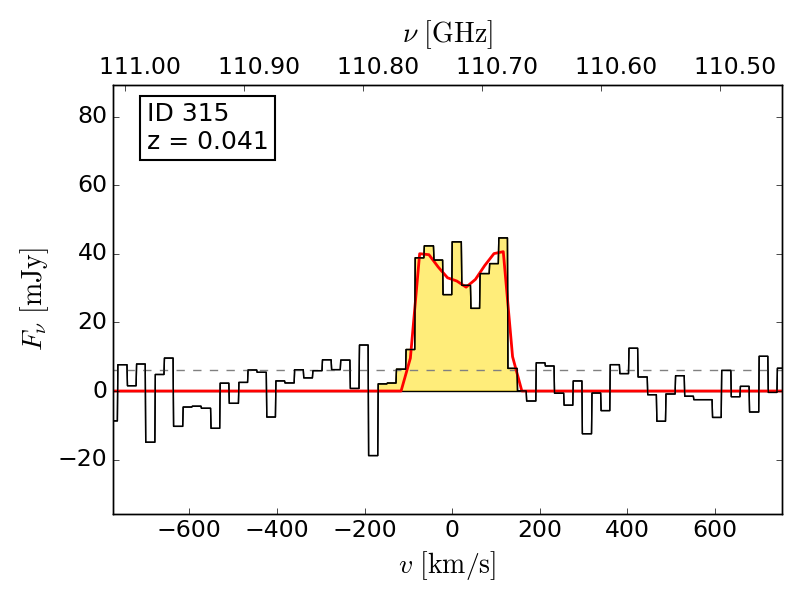}
\includegraphics[width=0.31\textwidth]{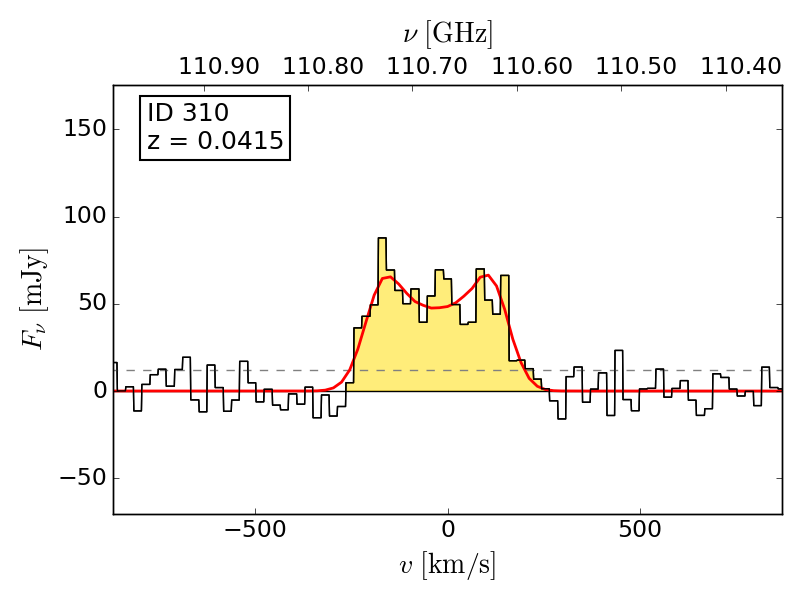}
\includegraphics[width=0.31\textwidth]{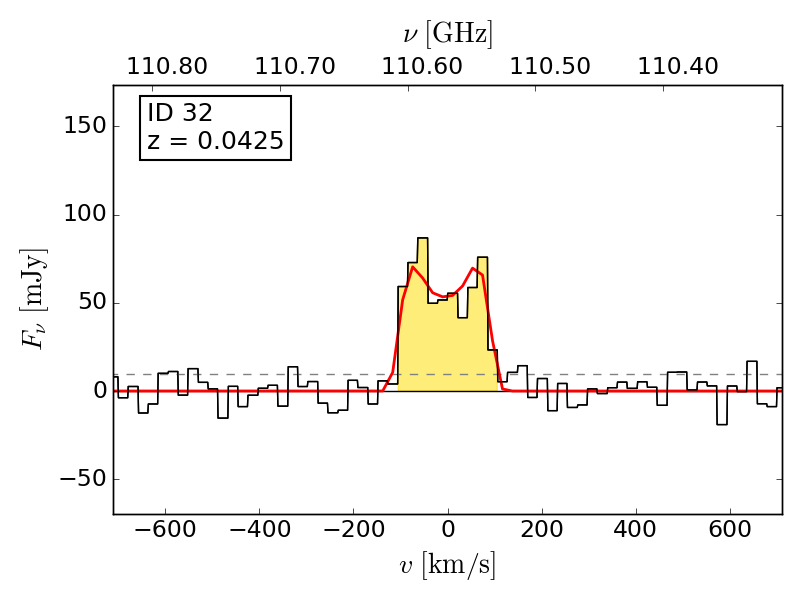}
\includegraphics[width=0.31\textwidth]{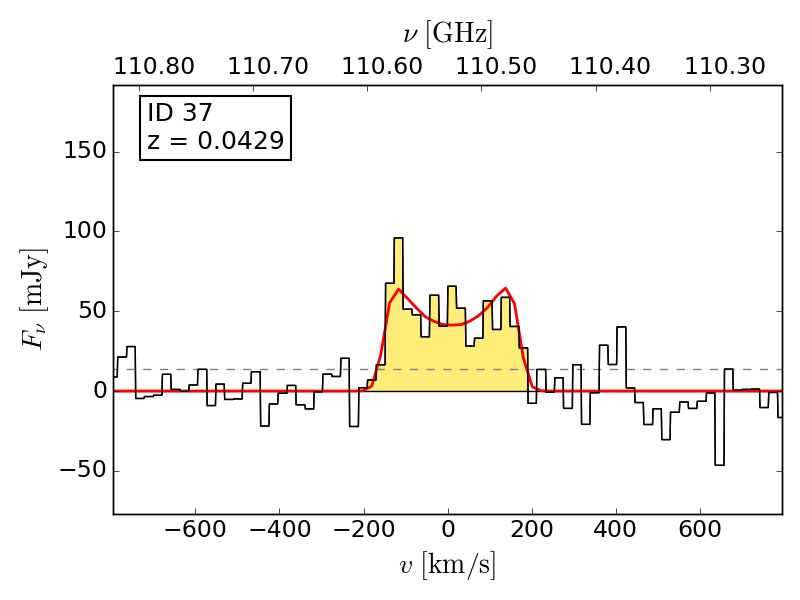}
\includegraphics[width=0.31\textwidth]{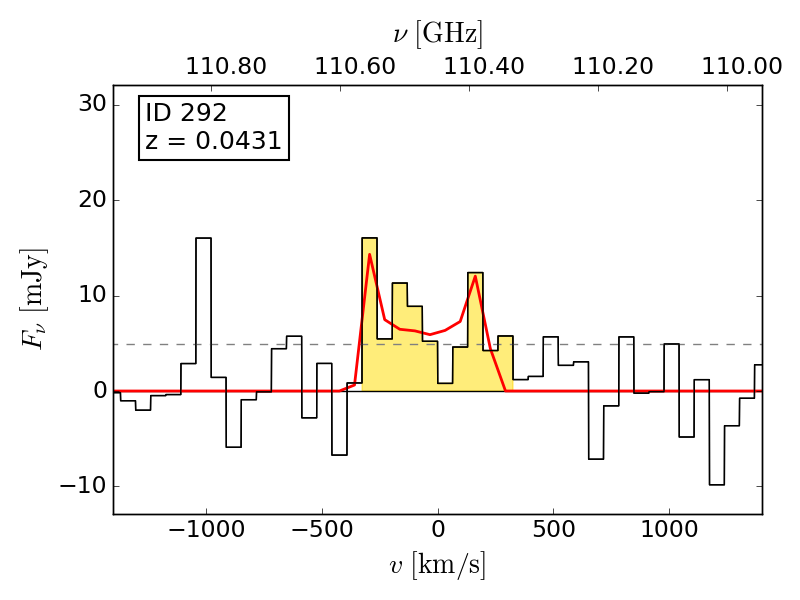}
\includegraphics[width=0.31\textwidth]{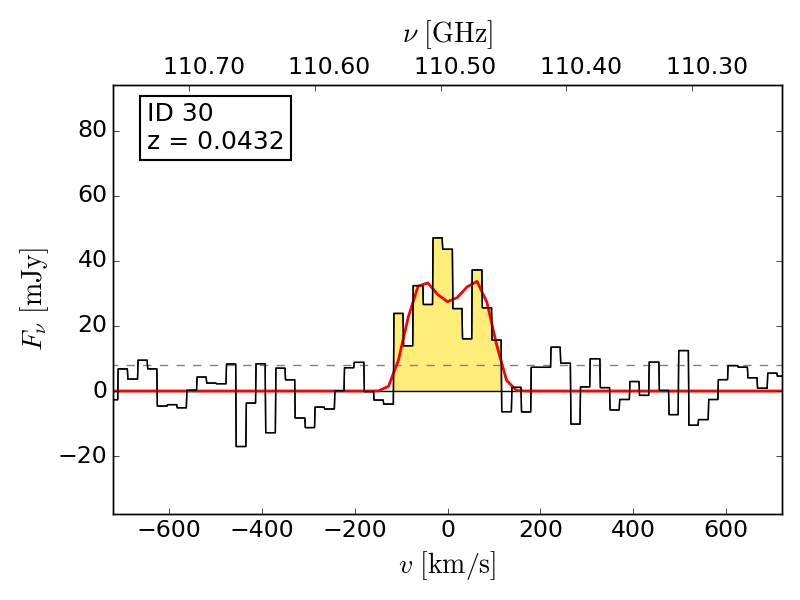}
\includegraphics[width=0.31\textwidth]{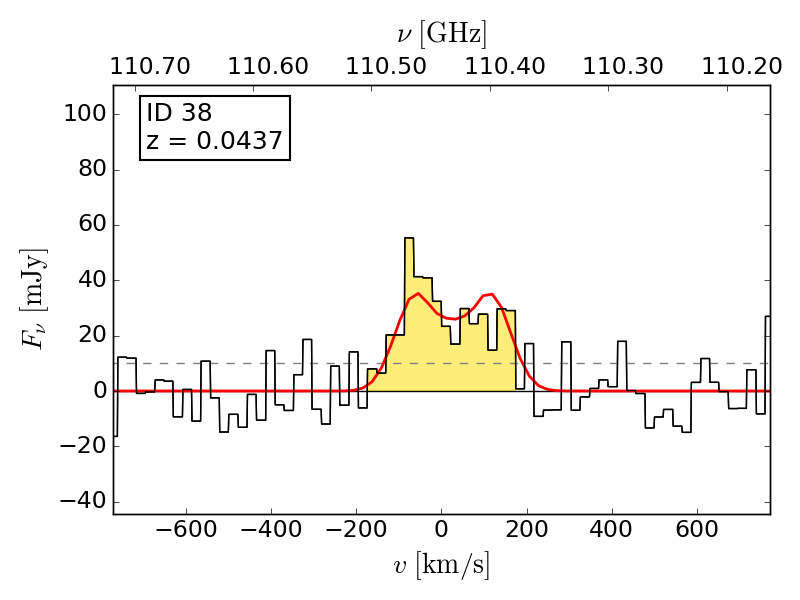}
\includegraphics[width=0.31\textwidth]{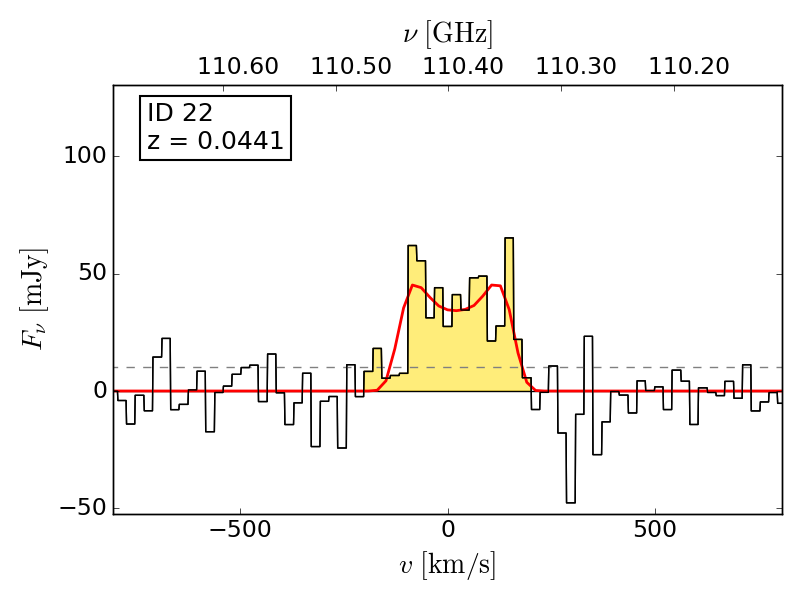}
\includegraphics[width=0.31\textwidth]{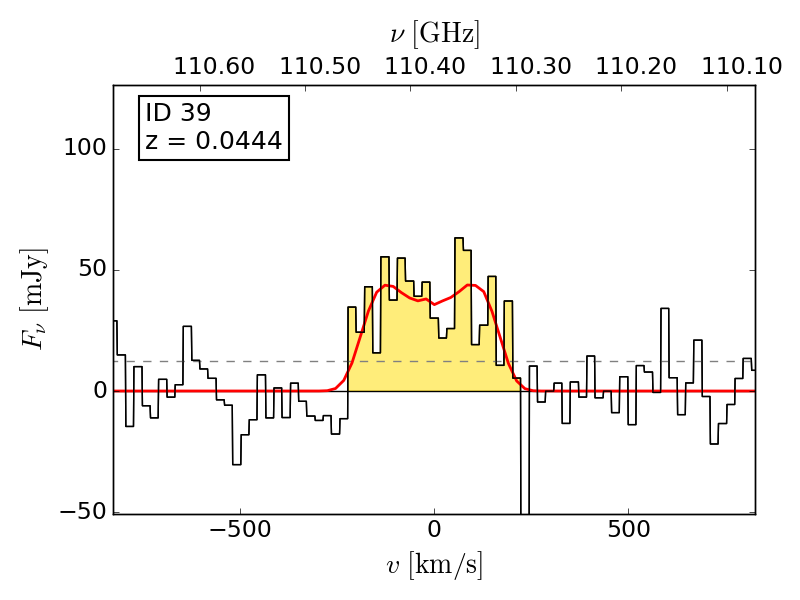}
\includegraphics[width=0.31\textwidth]{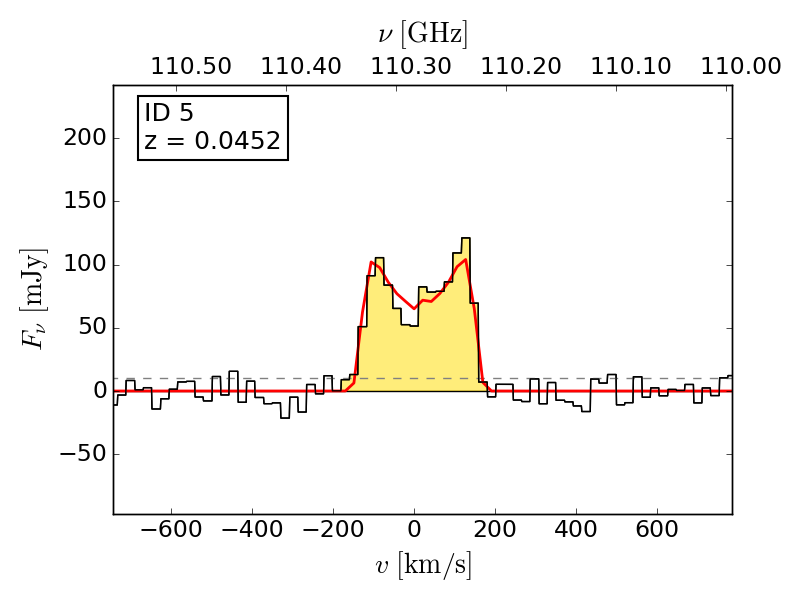}
\includegraphics[width=0.31\textwidth]{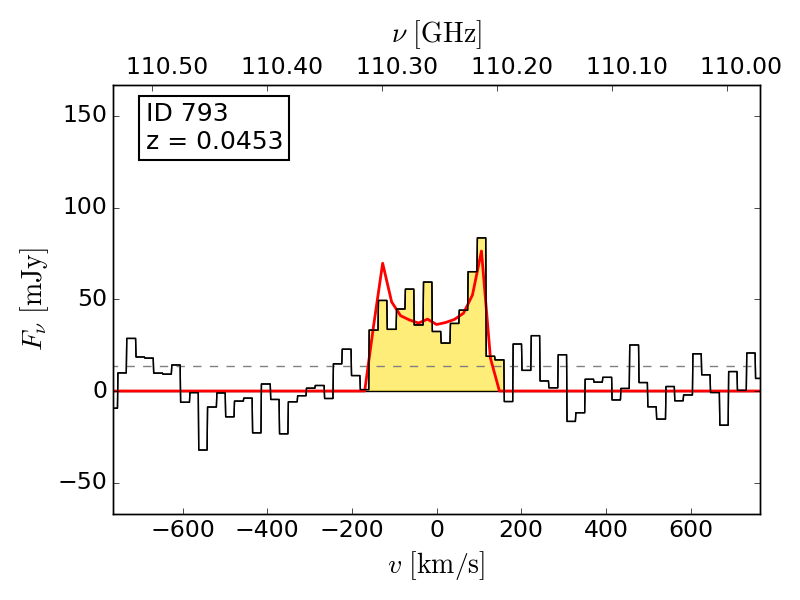}
\includegraphics[width=0.31\textwidth]{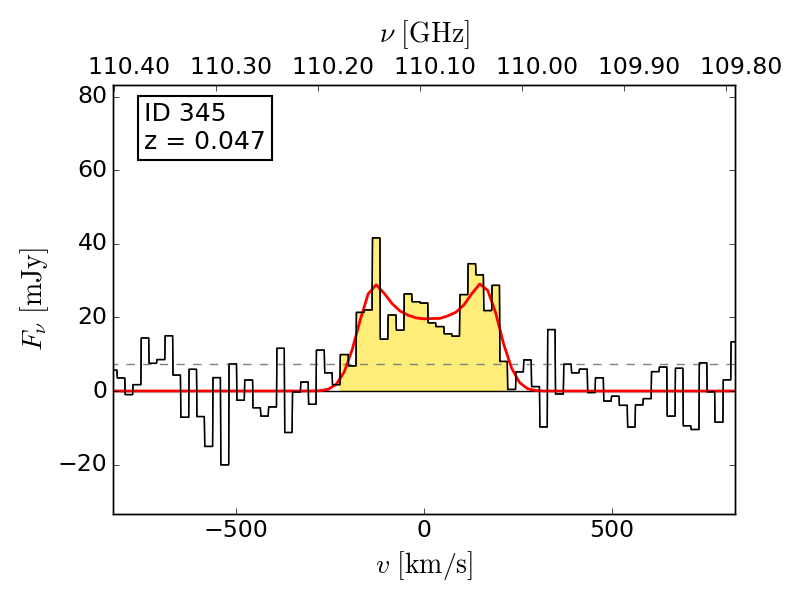}
\includegraphics[width=0.31\textwidth]{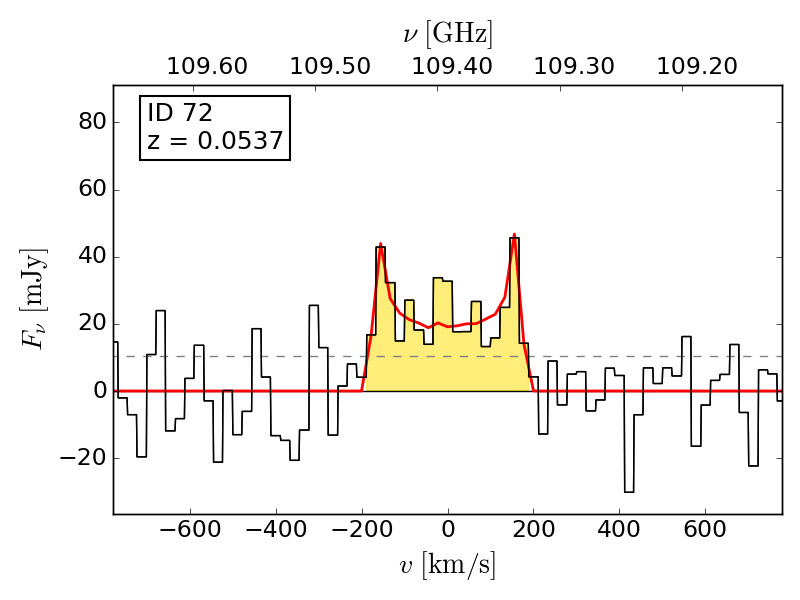}
\includegraphics[width=0.31\textwidth]{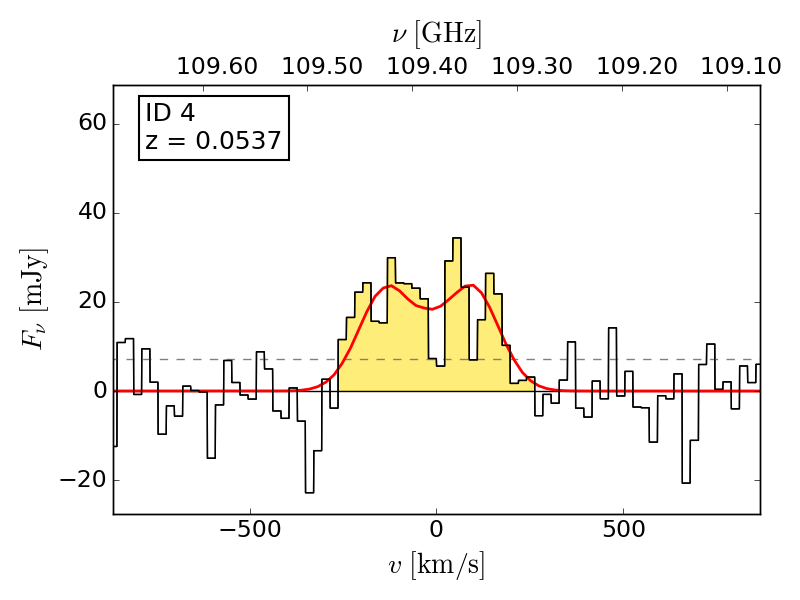}
\caption{(cont.)}
\end{figure*}
\begin{figure*}
\ContinuedFloat
\includegraphics[width=0.31\textwidth]{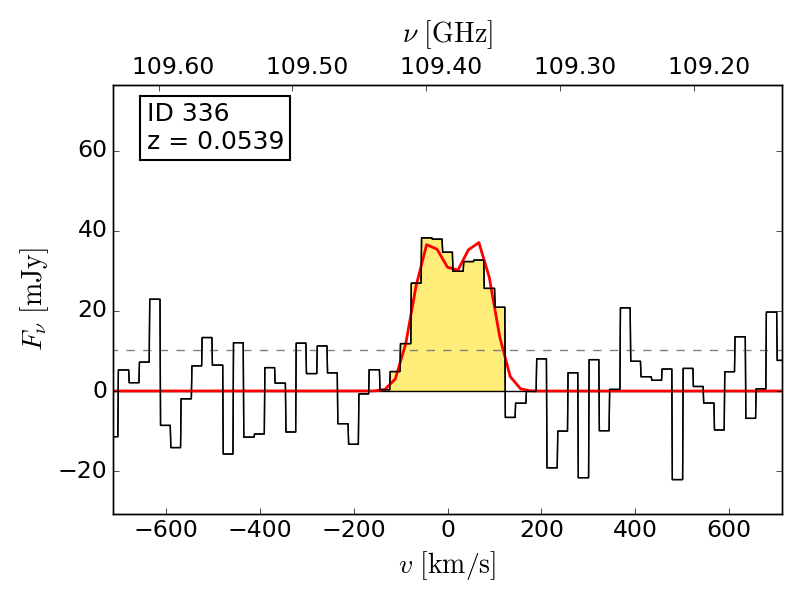}
\includegraphics[width=0.31\textwidth]{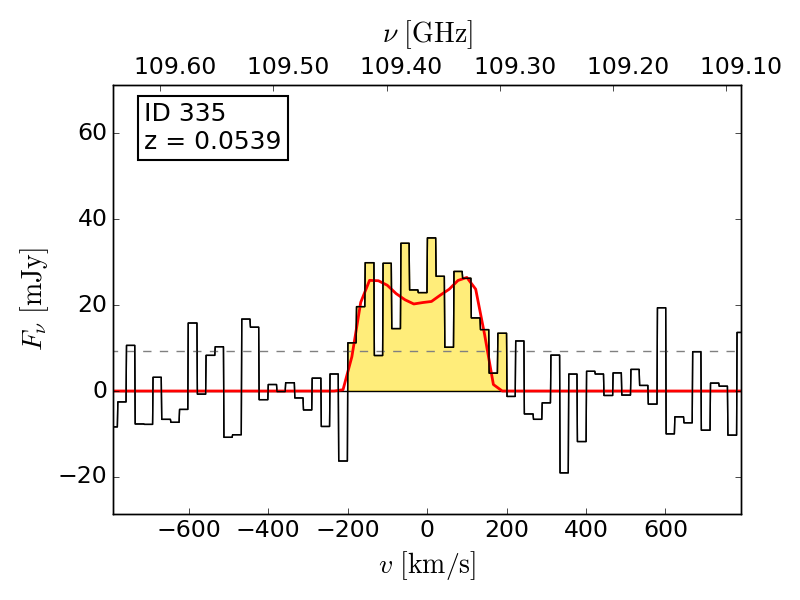}
\includegraphics[width=0.31\textwidth]{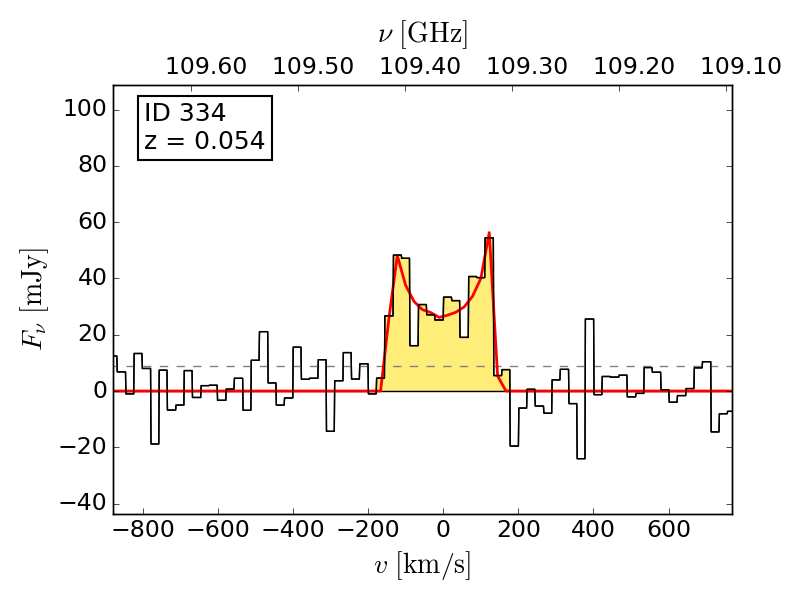}
\includegraphics[width=0.31\textwidth]{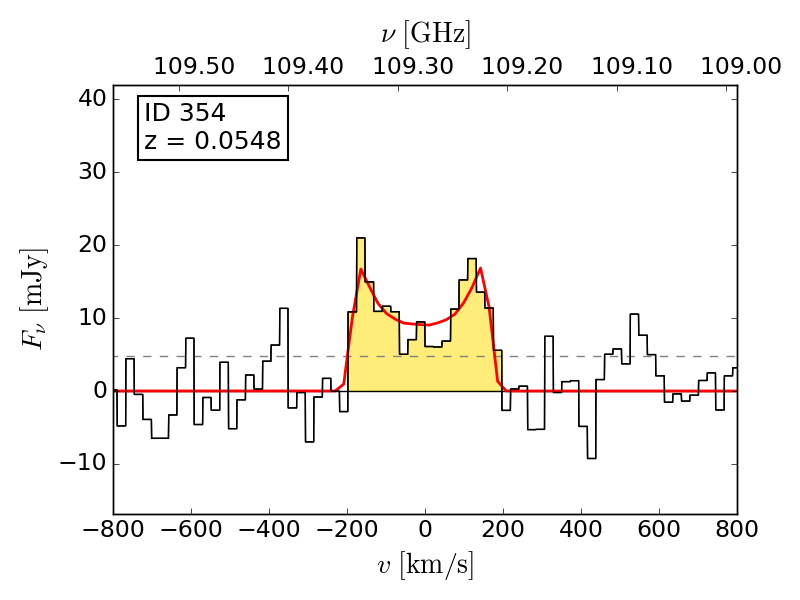}
\includegraphics[width=0.31\textwidth]{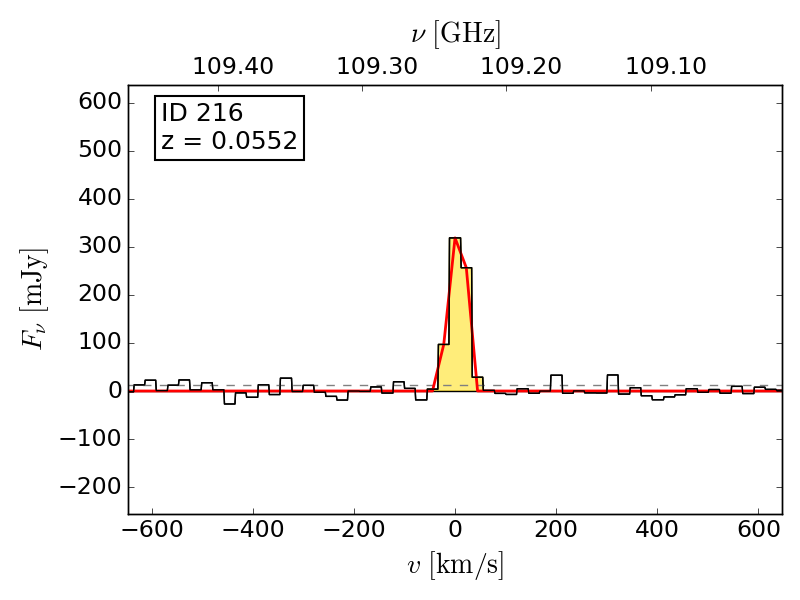}
\includegraphics[width=0.31\textwidth]{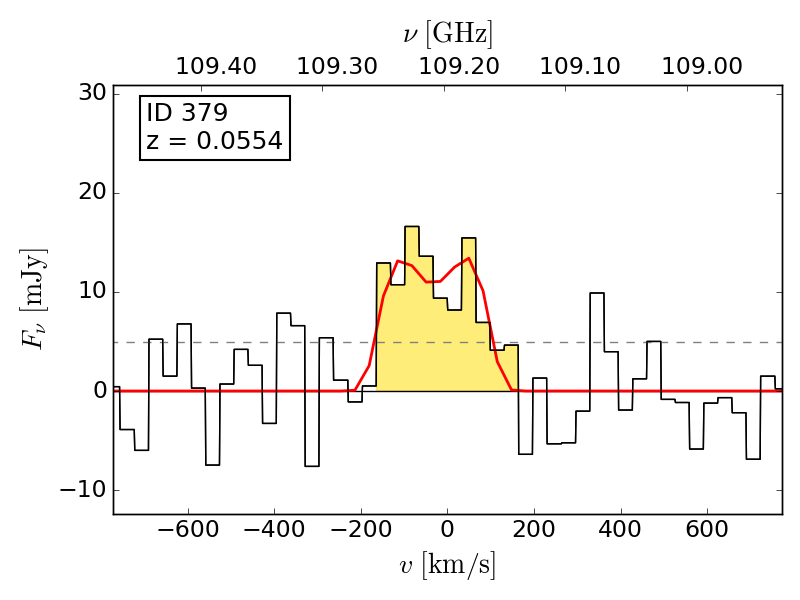}
\includegraphics[width=0.31\textwidth]{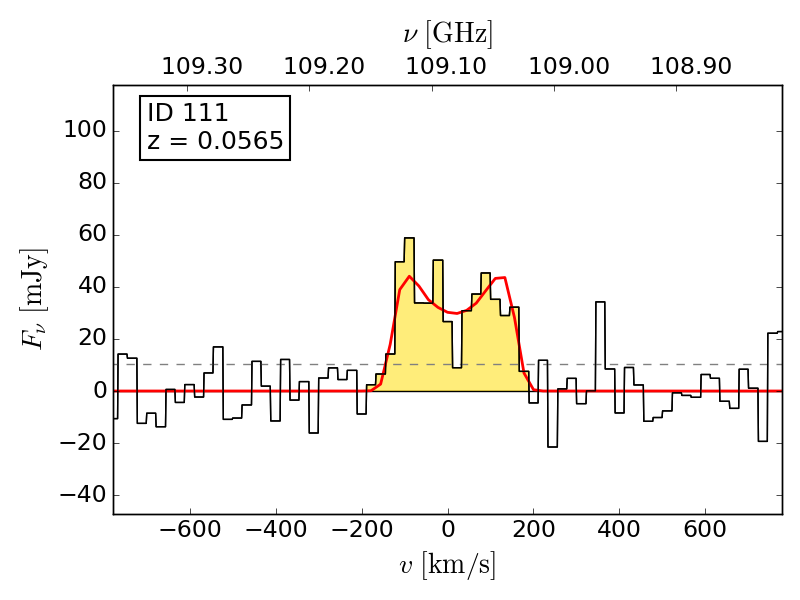}
\includegraphics[width=0.31\textwidth]{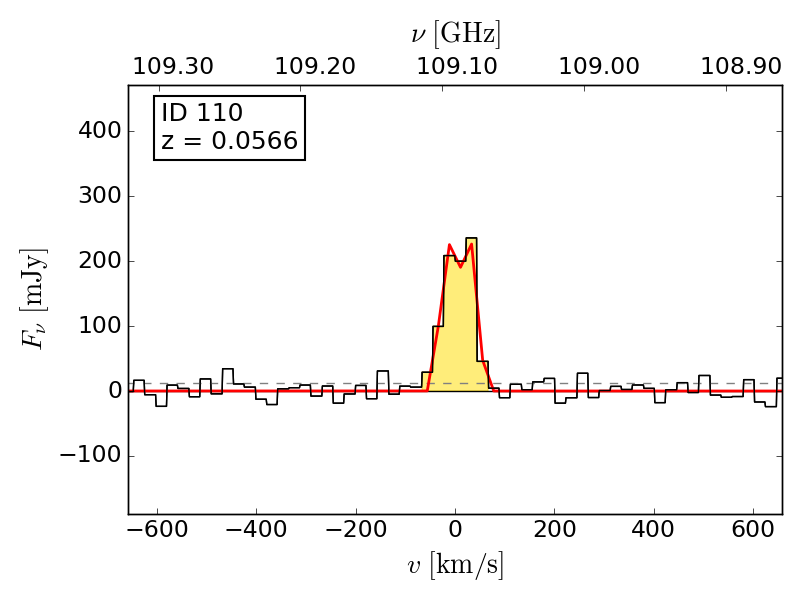}
\includegraphics[width=0.31\textwidth]{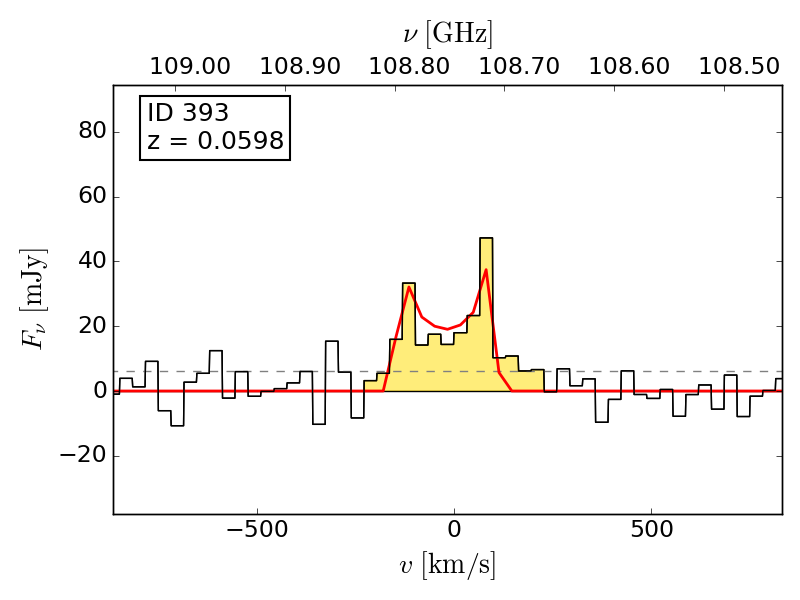}
\includegraphics[width=0.31\textwidth]{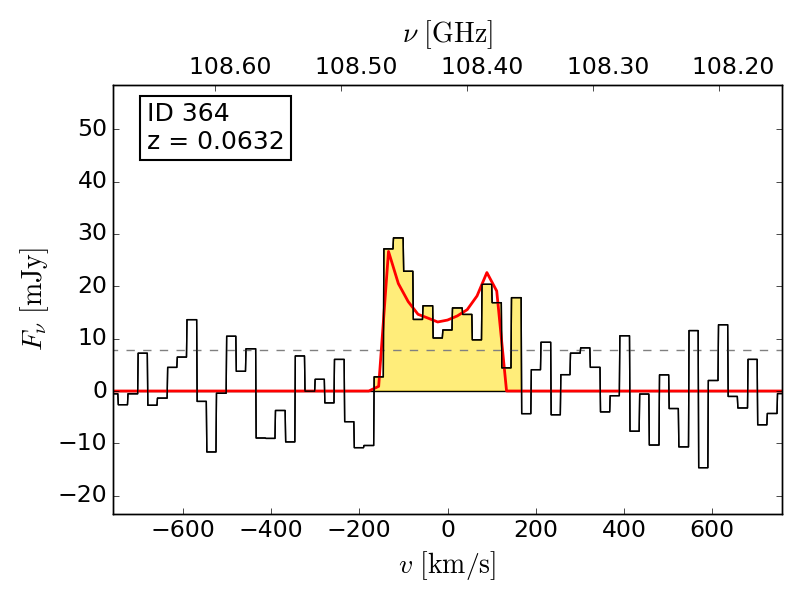}
\includegraphics[width=0.31\textwidth]{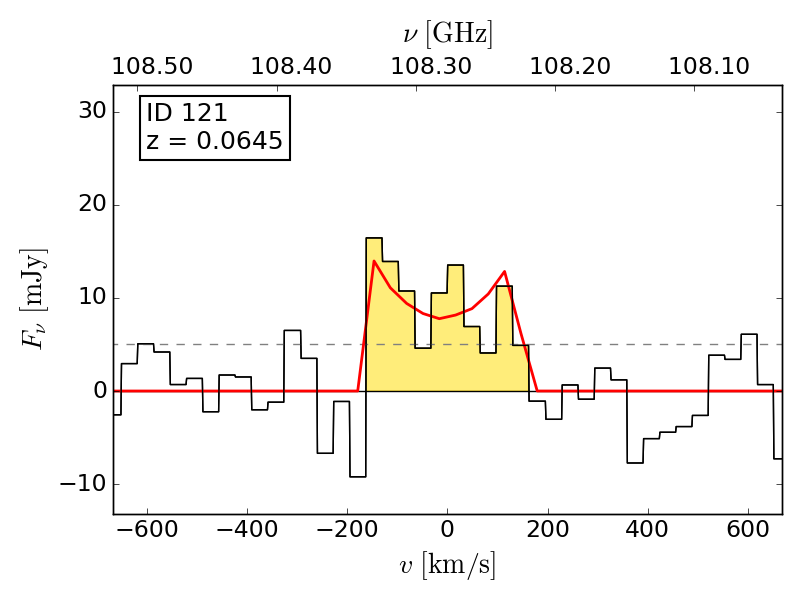}
\includegraphics[width=0.31\textwidth]{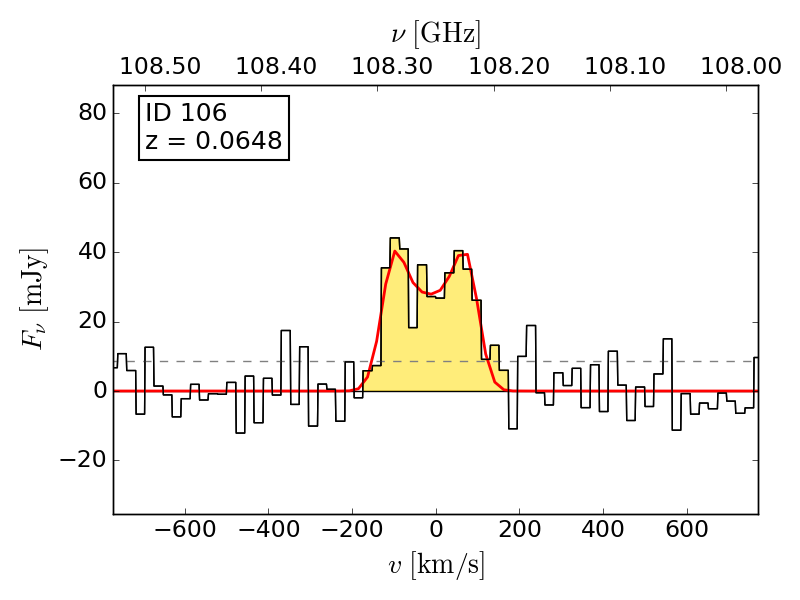}
\includegraphics[width=0.31\textwidth]{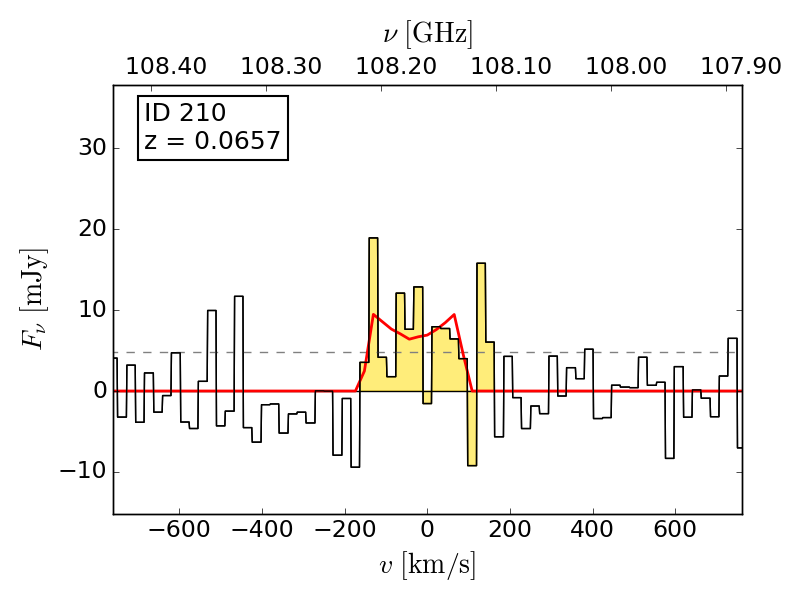}
\includegraphics[width=0.31\textwidth]{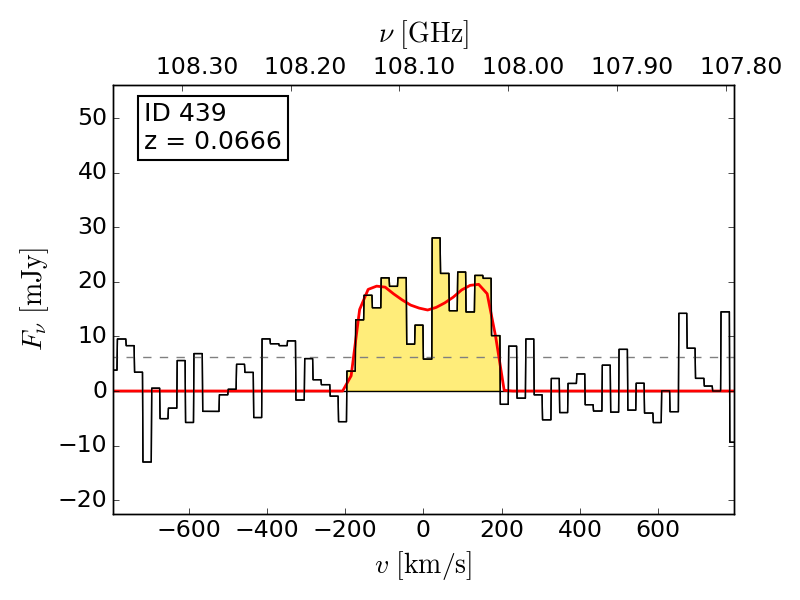}
\includegraphics[width=0.31\textwidth]{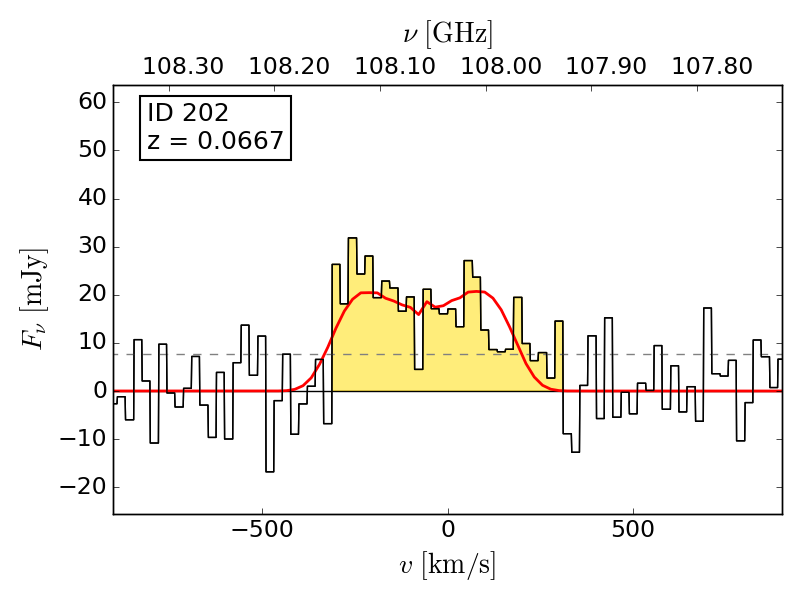}
\caption{(cont.)}
\end{figure*}
\begin{figure*}
\ContinuedFloat
\includegraphics[width=0.31\textwidth]{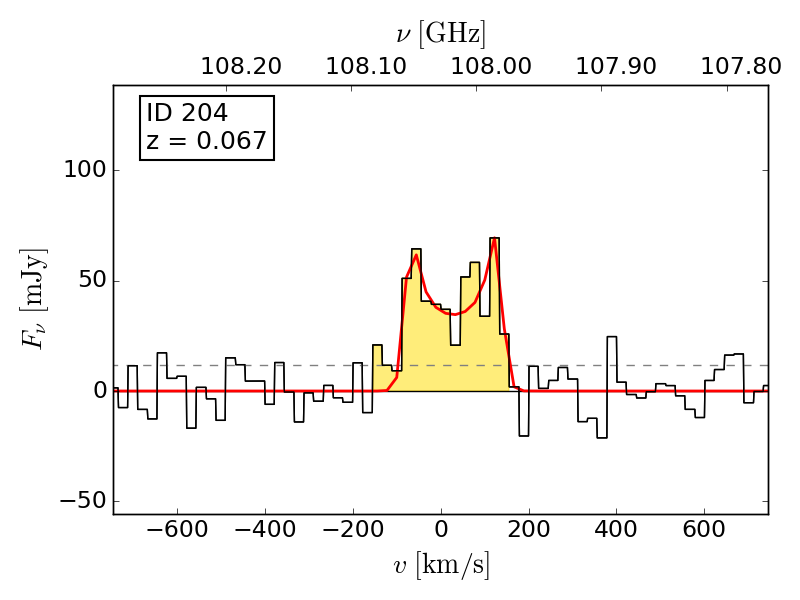}
\includegraphics[width=0.31\textwidth]{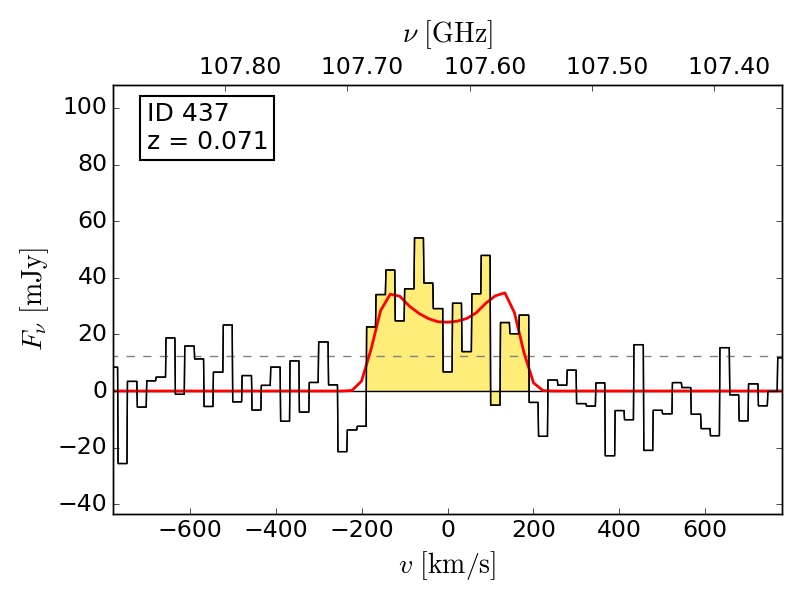}
\includegraphics[width=0.31\textwidth]{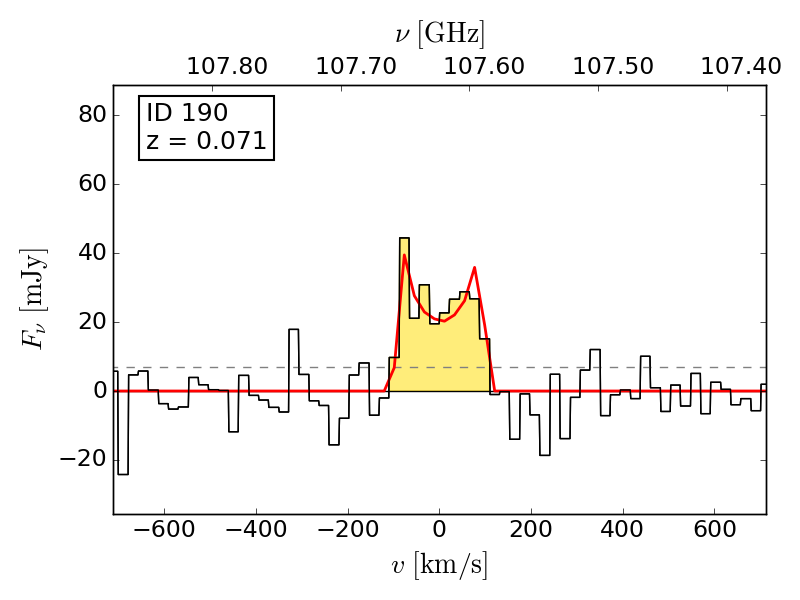}
\includegraphics[width=0.31\textwidth]{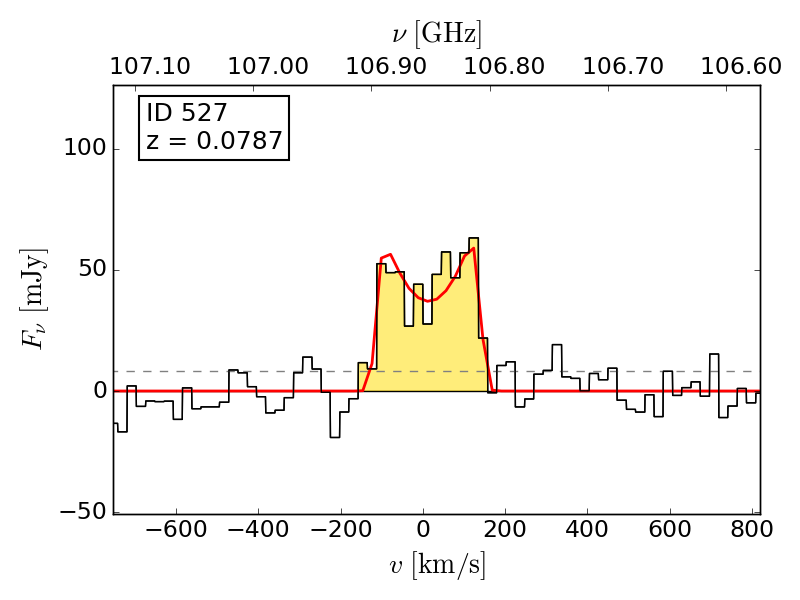}
\includegraphics[width=0.31\textwidth]{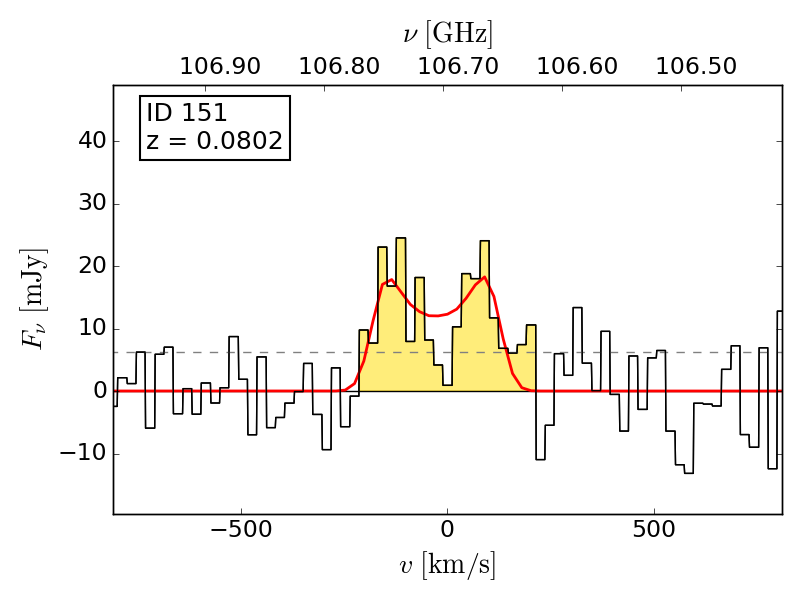}
\includegraphics[width=0.31\textwidth]{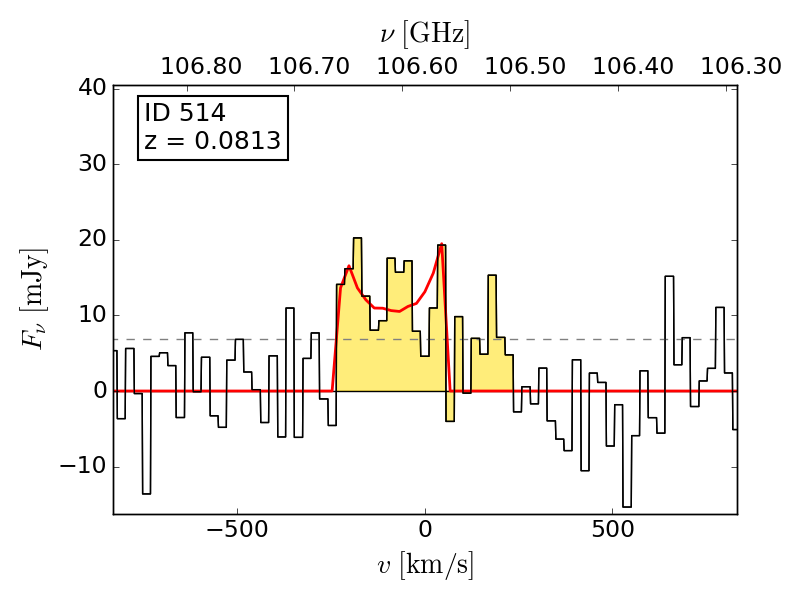}
\includegraphics[width=0.31\textwidth]{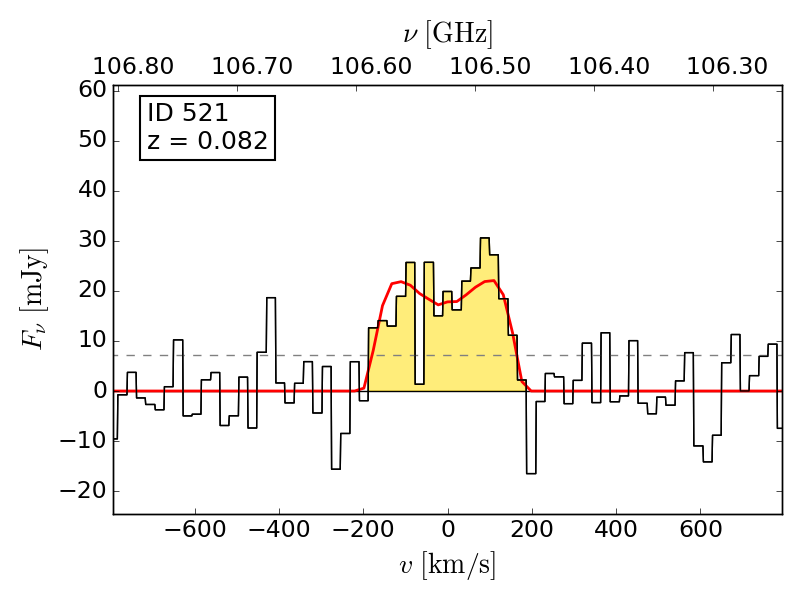}
\includegraphics[width=0.31\textwidth]{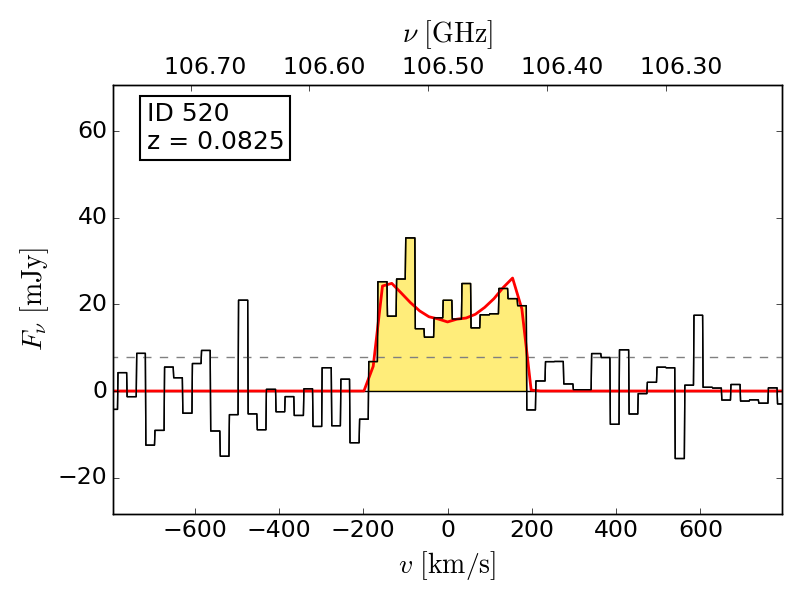}
\includegraphics[width=0.31\textwidth]{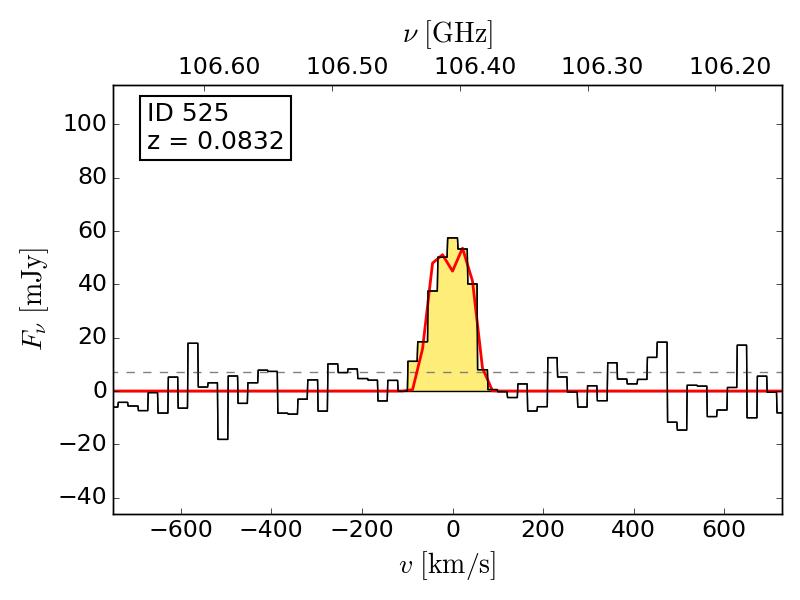}
\includegraphics[width=0.31\textwidth]{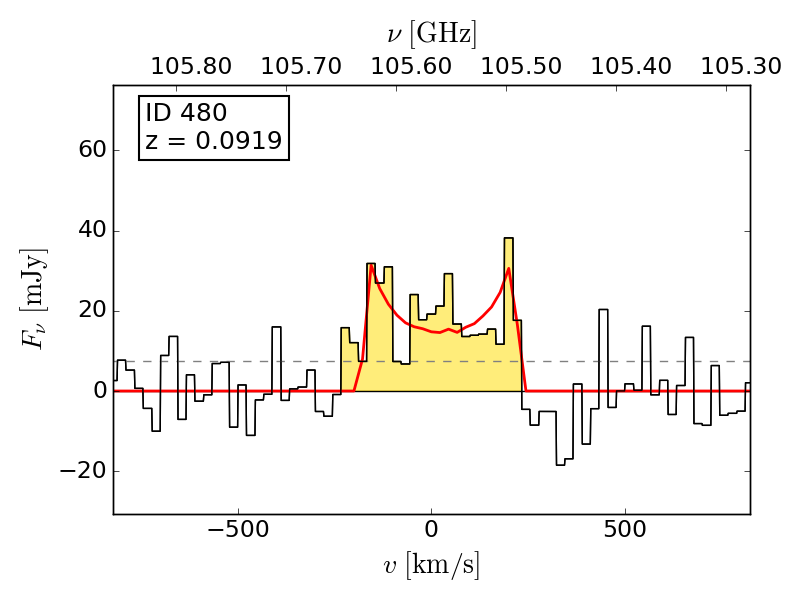}
\includegraphics[width=0.31\textwidth]{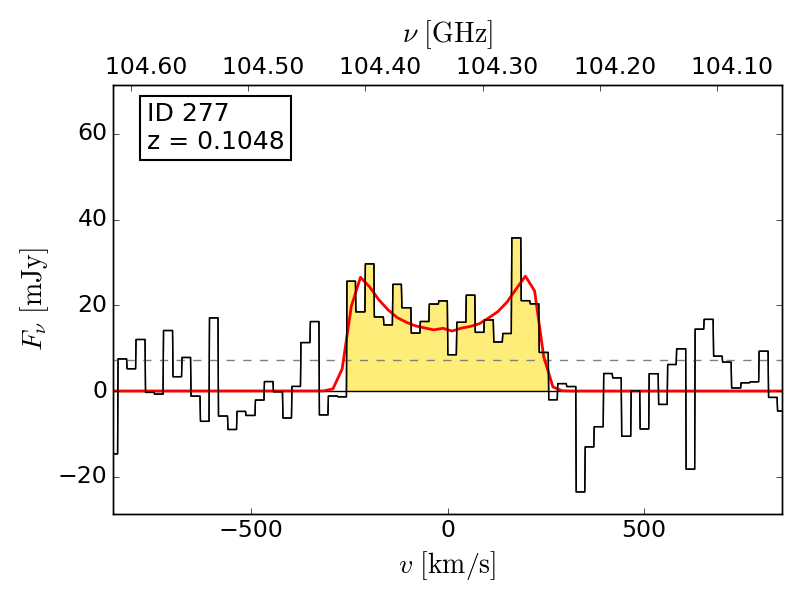}
\includegraphics[width=0.31\textwidth]{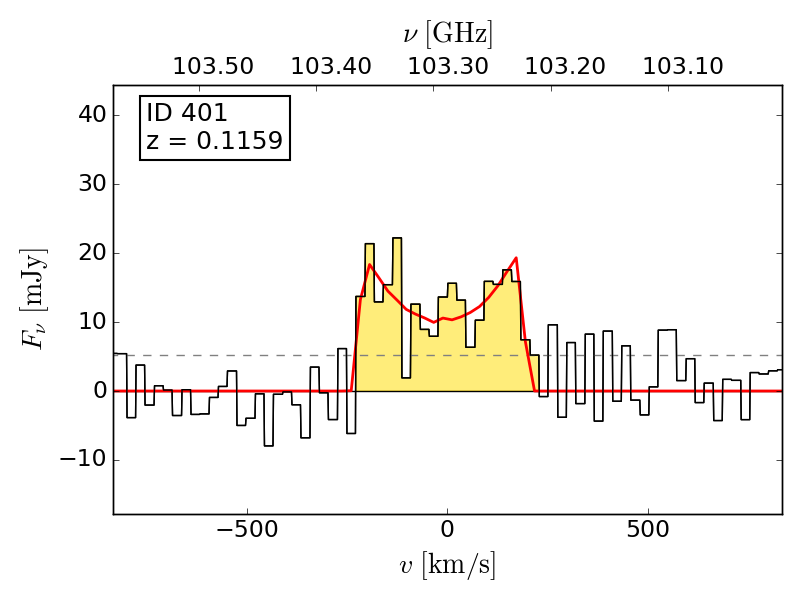}
\includegraphics[width=0.31\textwidth]{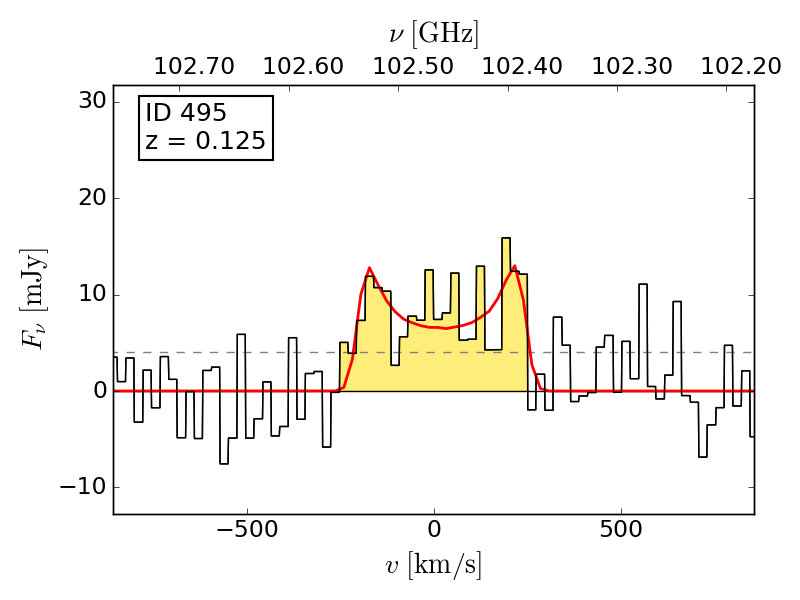}
\includegraphics[width=0.31\textwidth]{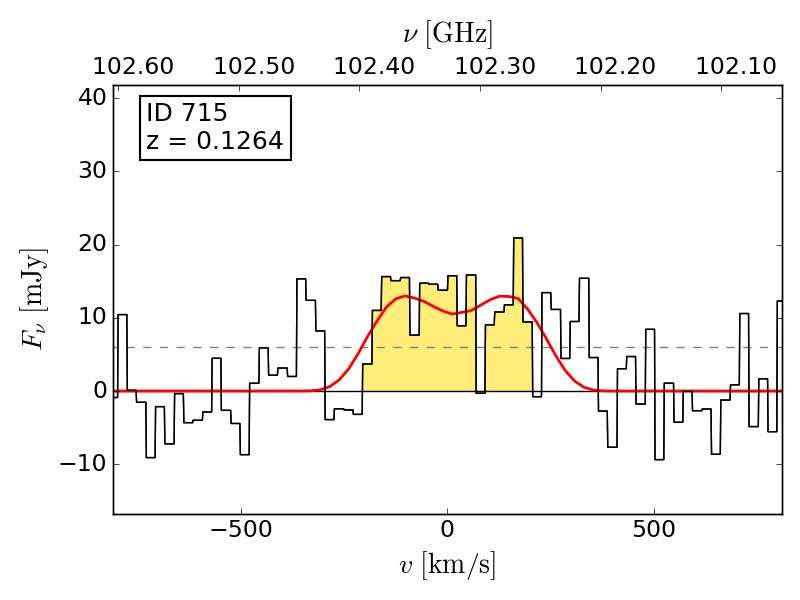}
\includegraphics[width=0.31\textwidth]{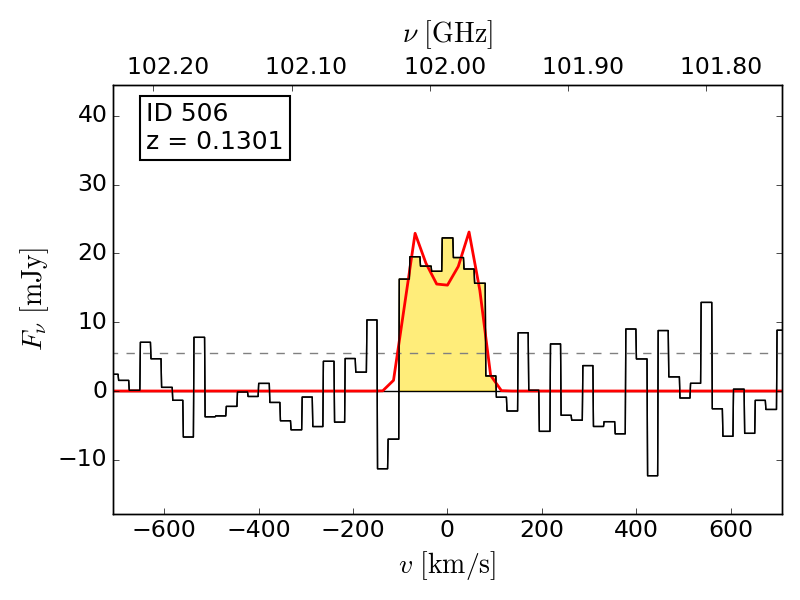}
\caption{(cont.)}
\end{figure*}
\begin{figure*}
\ContinuedFloat
\includegraphics[width=0.31\textwidth]{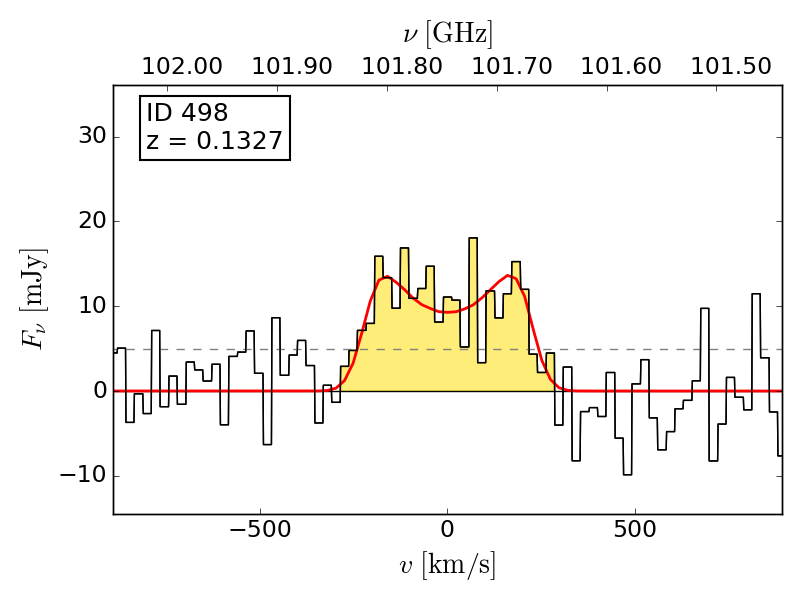}
\includegraphics[width=0.31\textwidth]{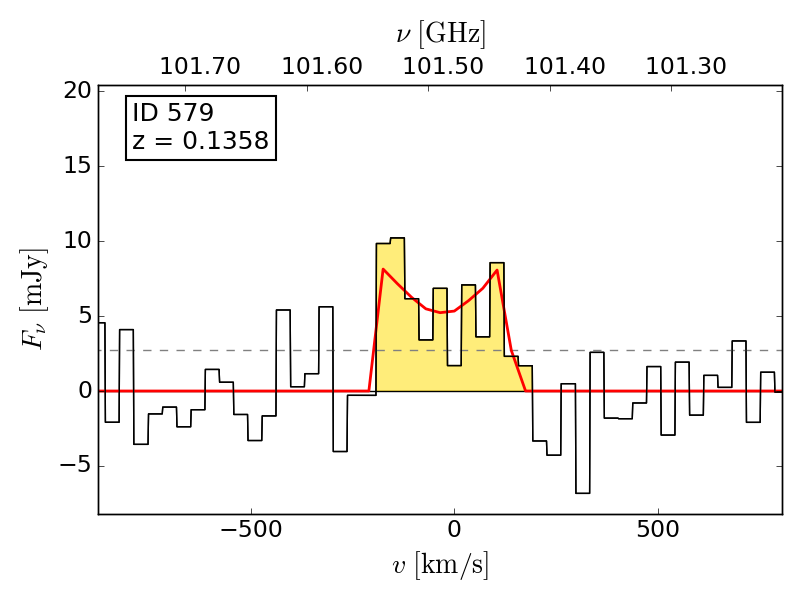}
\includegraphics[width=0.31\textwidth]{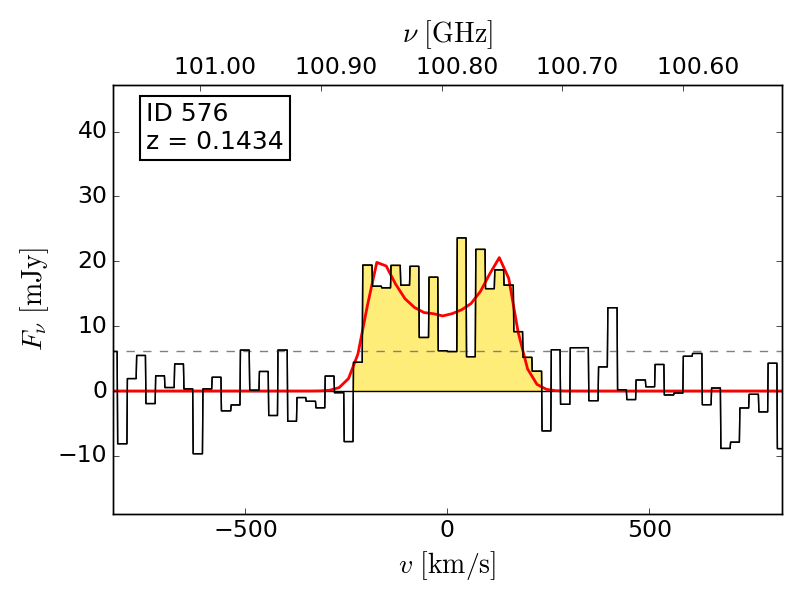}
\includegraphics[width=0.31\textwidth]{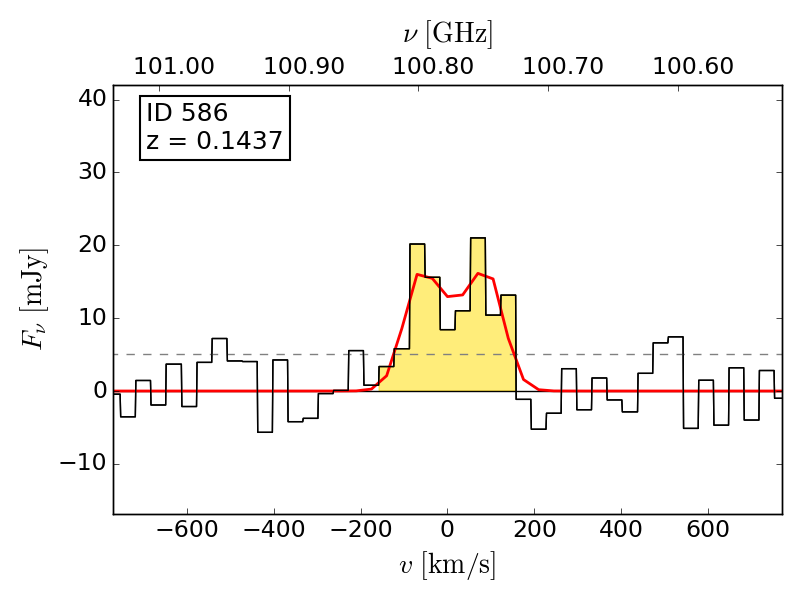}
\includegraphics[width=0.31\textwidth]{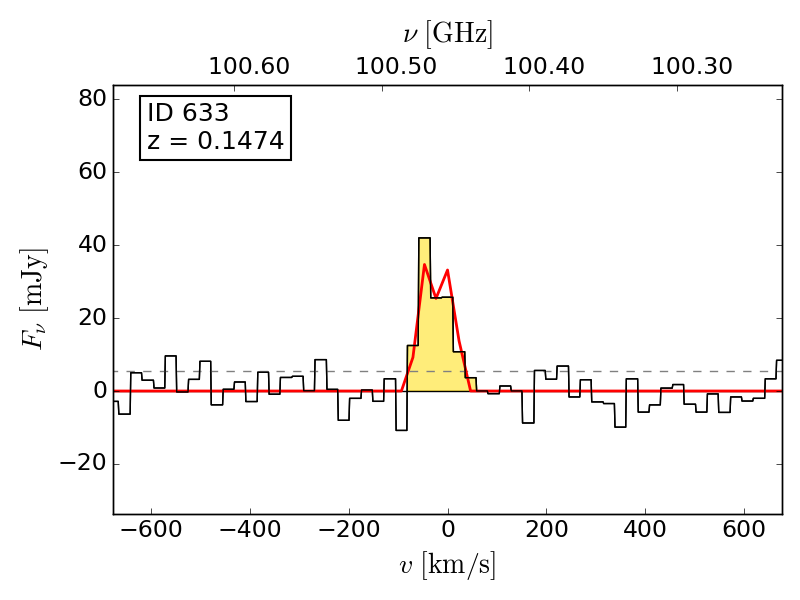}
\includegraphics[width=0.31\textwidth]{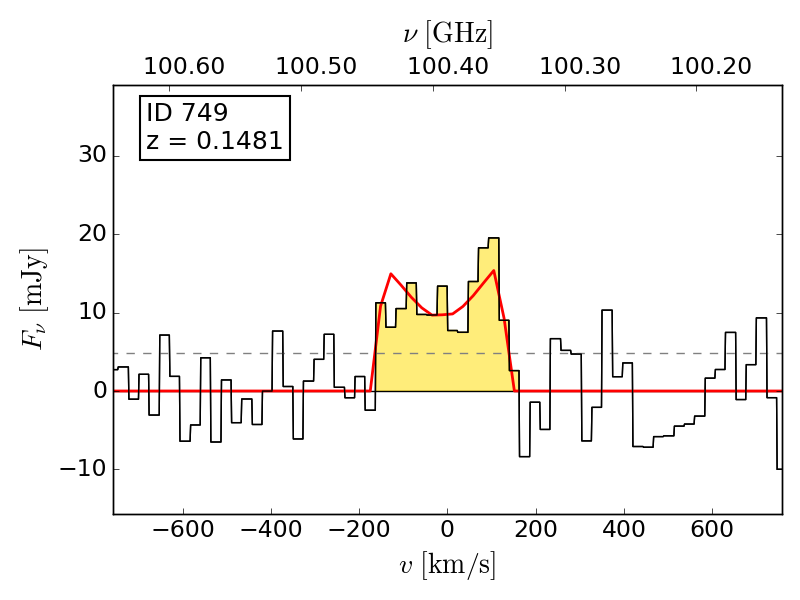}
\includegraphics[width=0.31\textwidth]{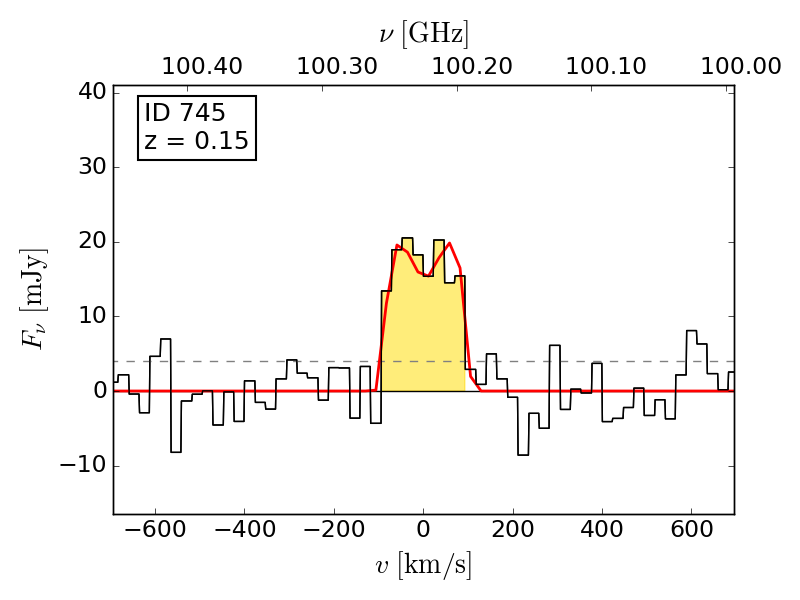}
\includegraphics[width=0.31\textwidth]{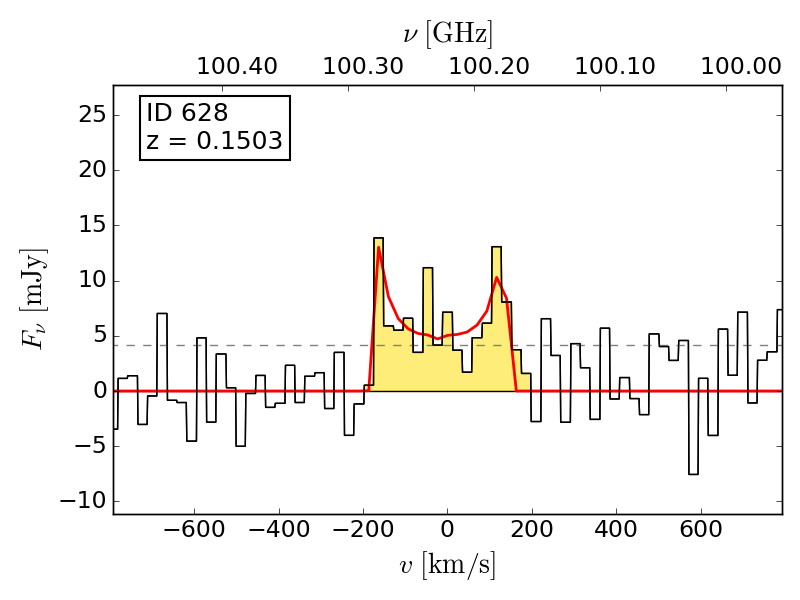}
\includegraphics[width=0.31\textwidth]{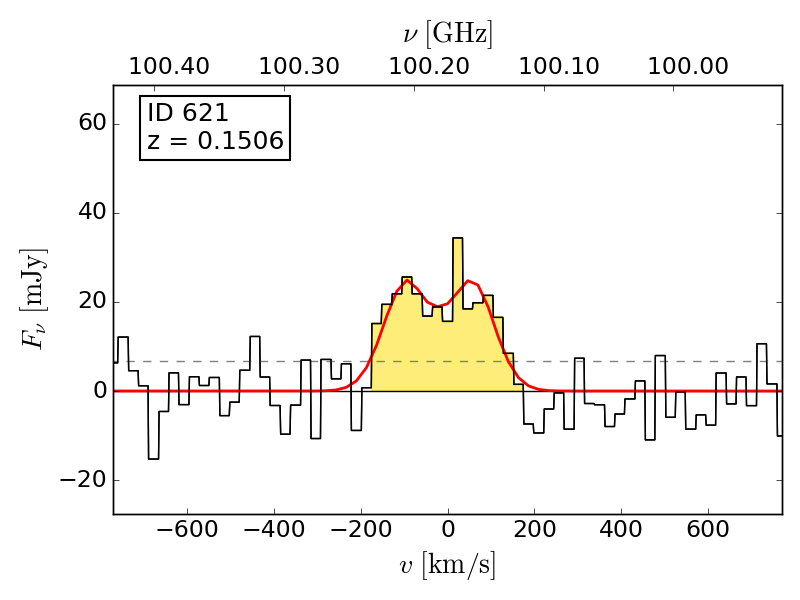}
\includegraphics[width=0.31\textwidth]{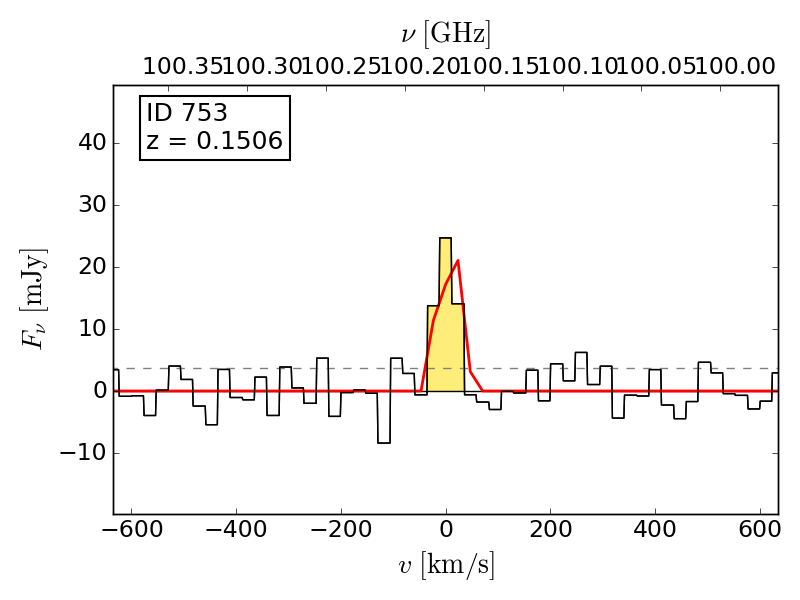}
\includegraphics[width=0.31\textwidth]{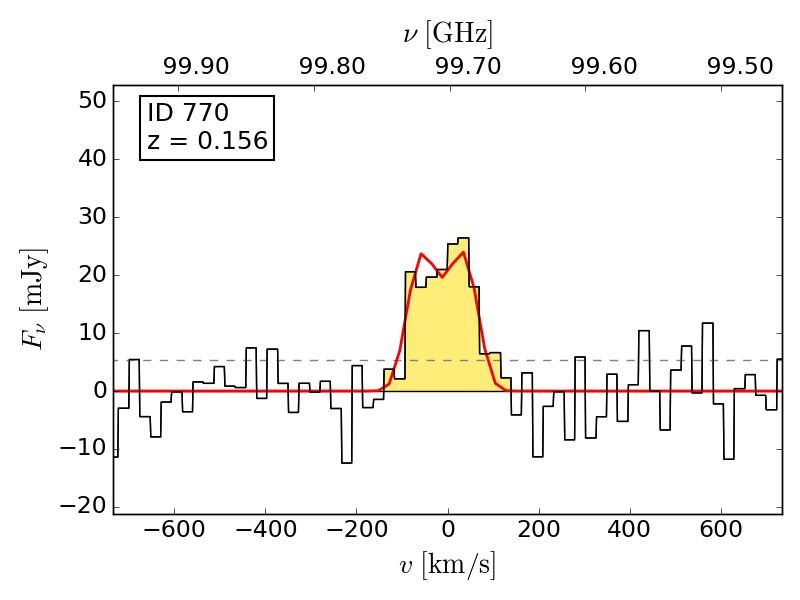}
\includegraphics[width=0.31\textwidth]{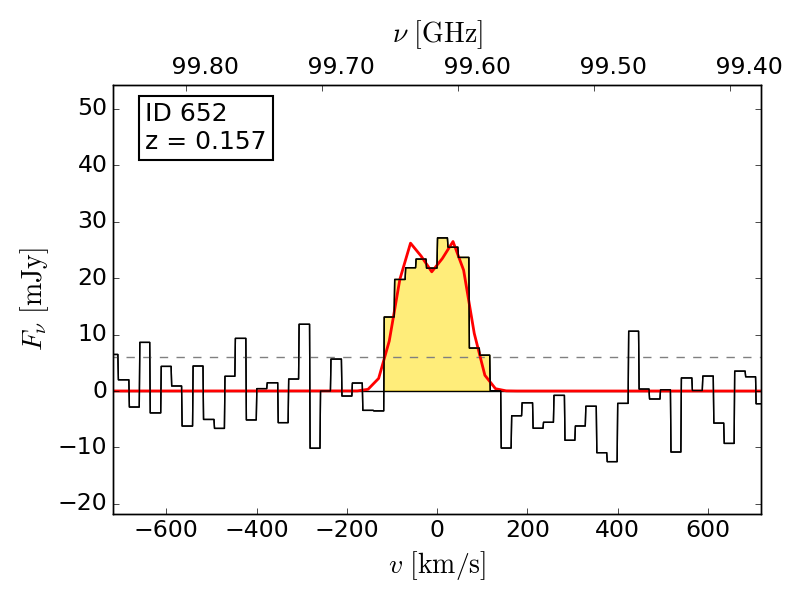}
\includegraphics[width=0.31\textwidth]{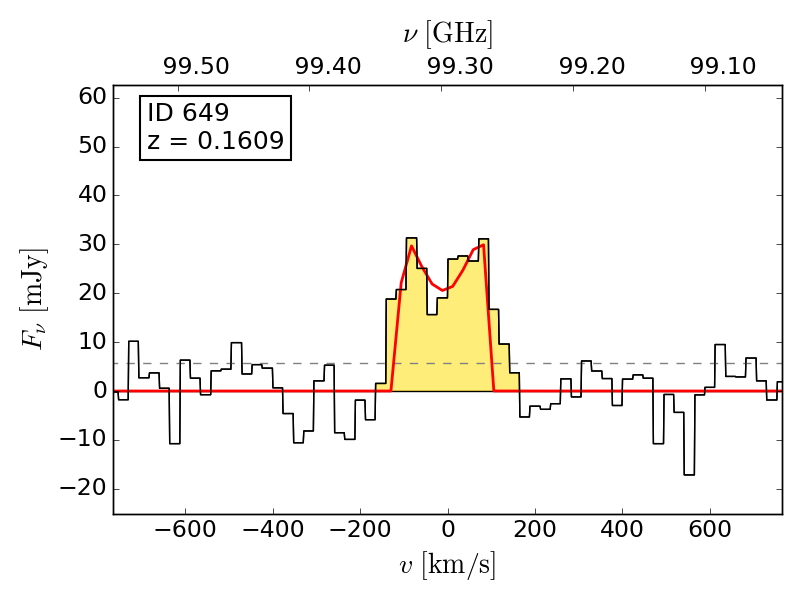}
\includegraphics[width=0.31\textwidth]{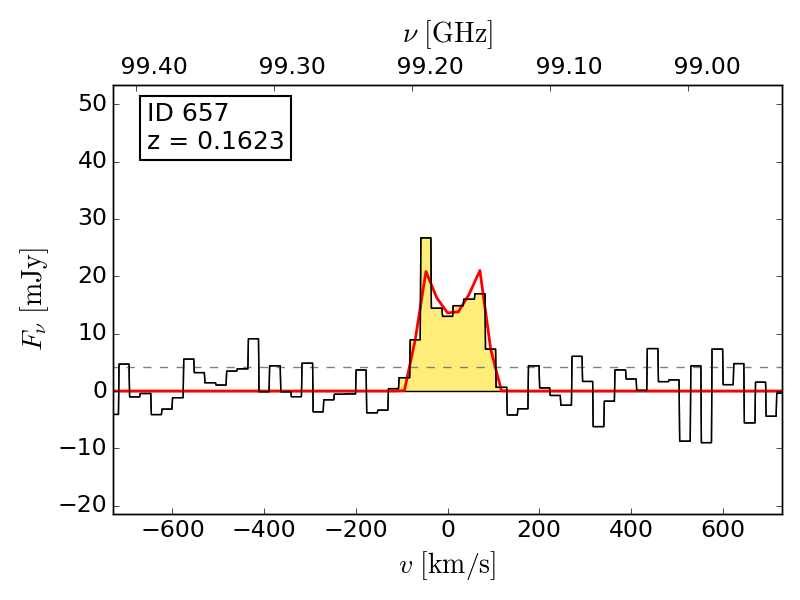}
\includegraphics[width=0.31\textwidth]{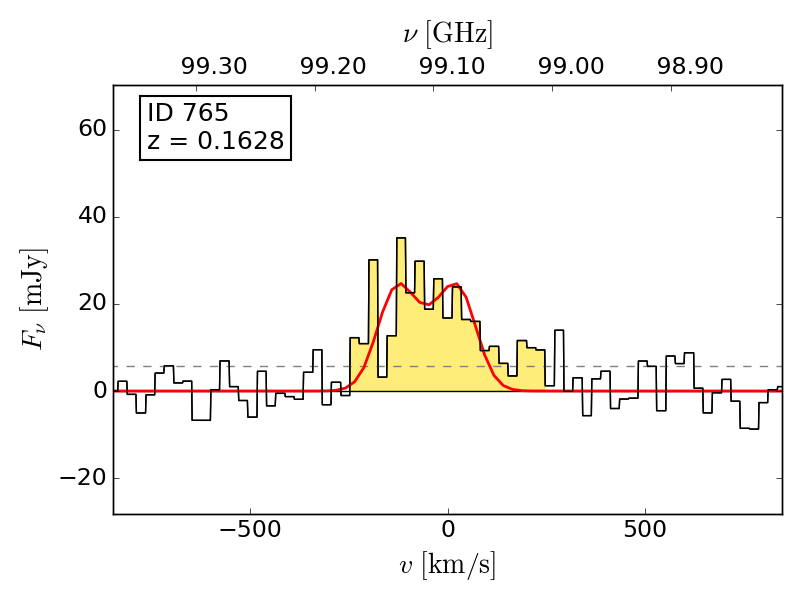}
\caption{(cont.)}
\end{figure*}
\begin{figure*}
\ContinuedFloat
\includegraphics[width=0.31\textwidth]{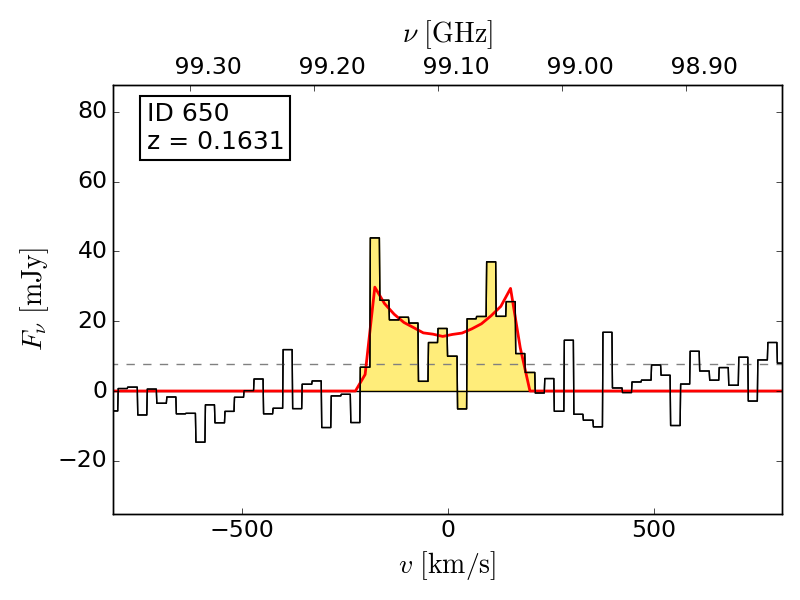}
\includegraphics[width=0.31\textwidth]{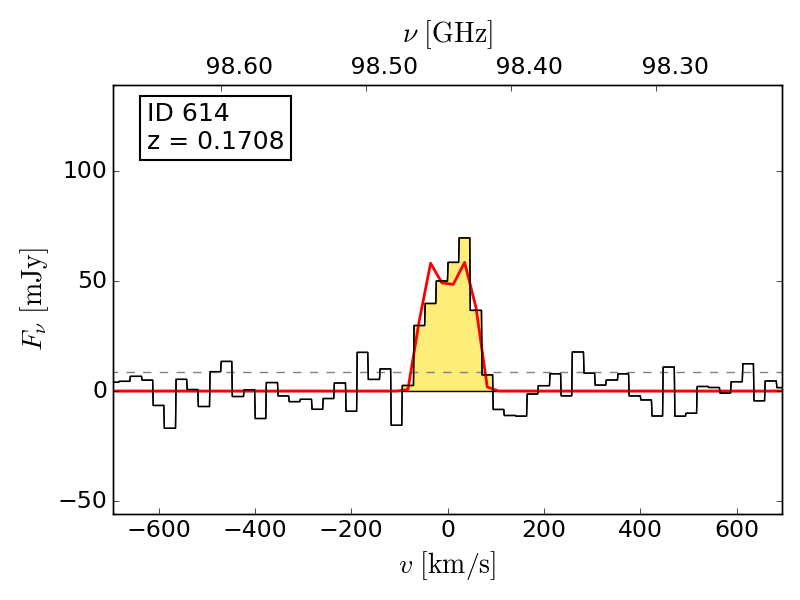}
\includegraphics[width=0.31\textwidth]{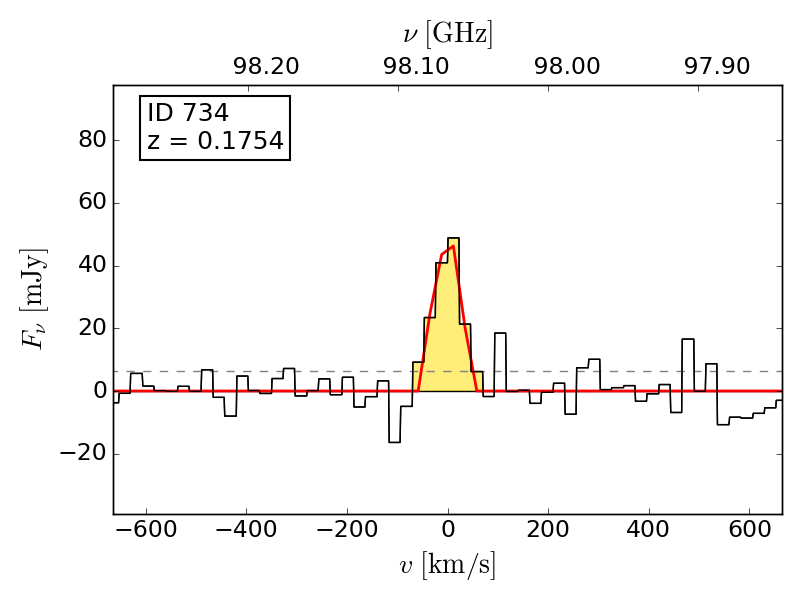}
\includegraphics[width=0.31\textwidth]{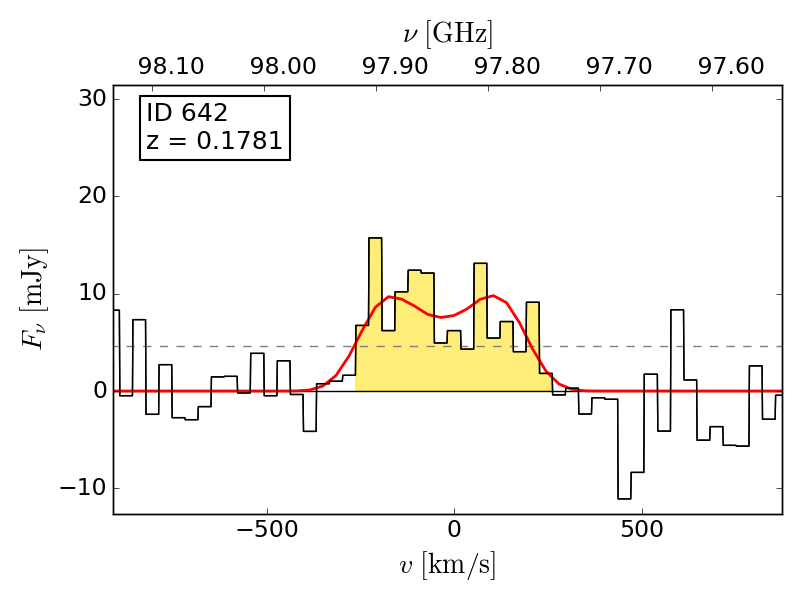}
\includegraphics[width=0.31\textwidth]{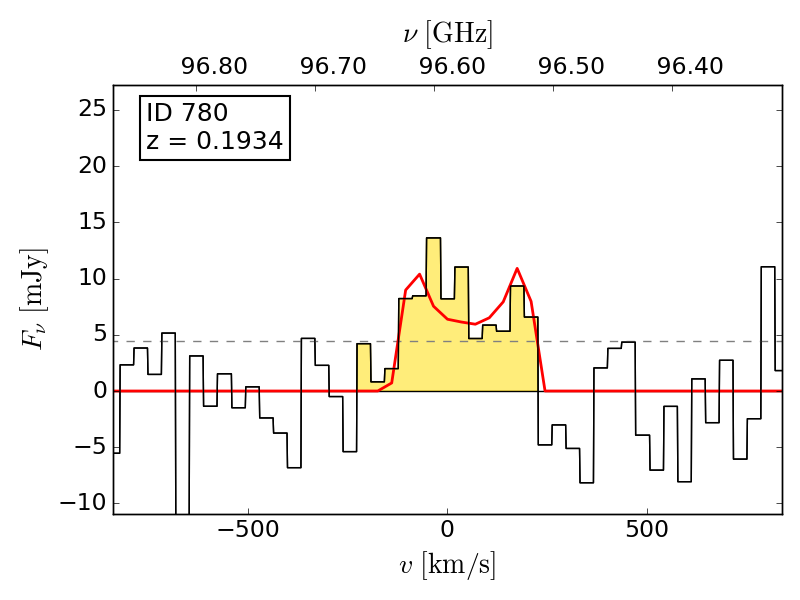}
\includegraphics[width=0.31\textwidth]{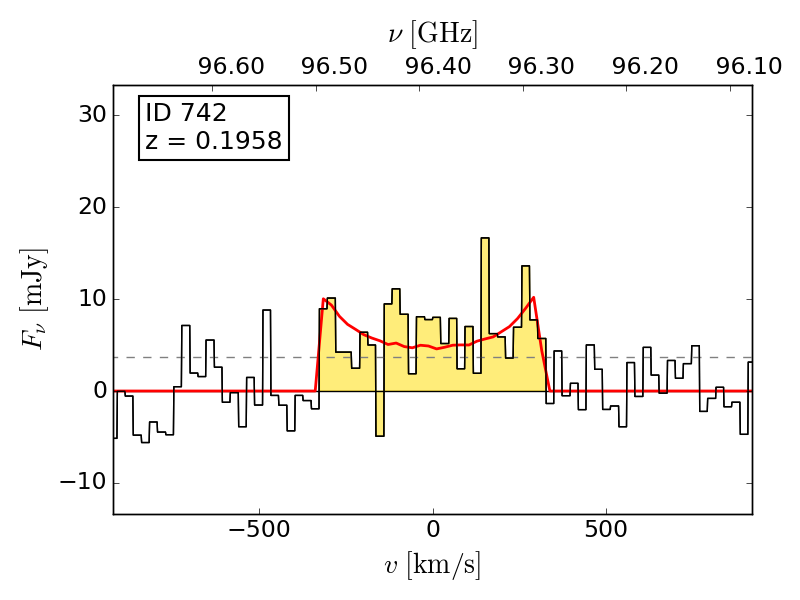}
\caption{(cont.)}
\end{figure*}

\clearpage

\begin{figure*}
\includegraphics[width=0.31\textwidth]{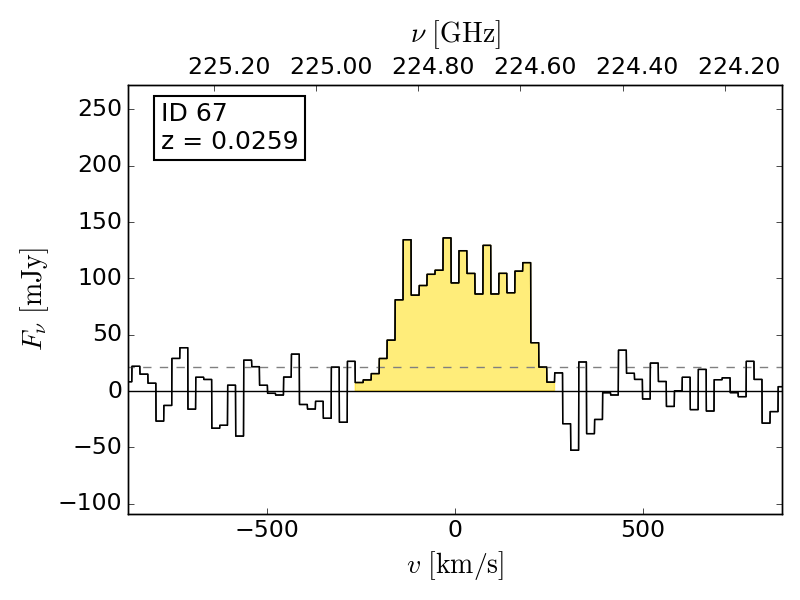}
\includegraphics[width=0.31\textwidth]{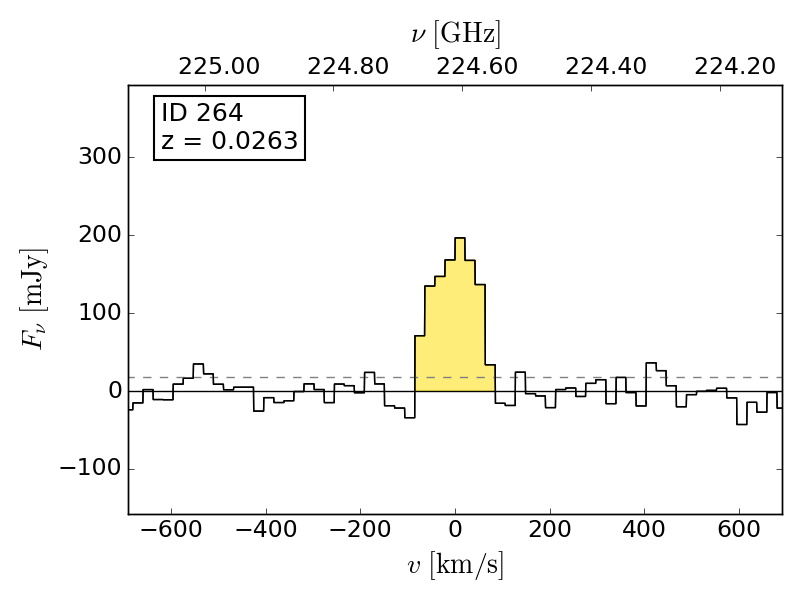}
\includegraphics[width=0.31\textwidth]{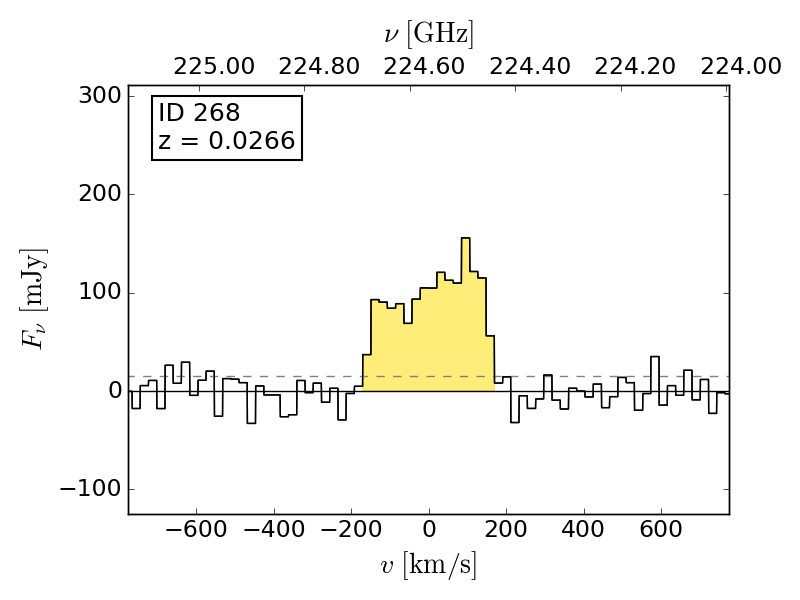}
\includegraphics[width=0.31\textwidth]{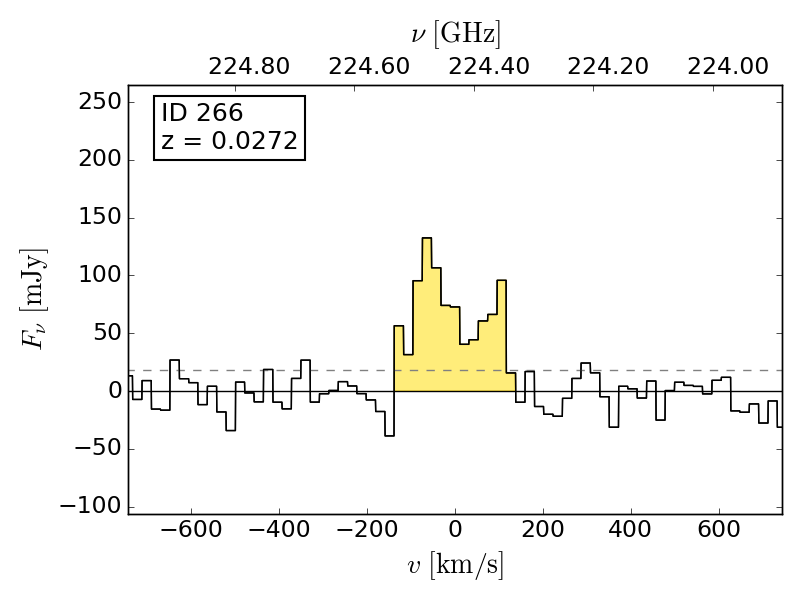}
\includegraphics[width=0.31\textwidth]{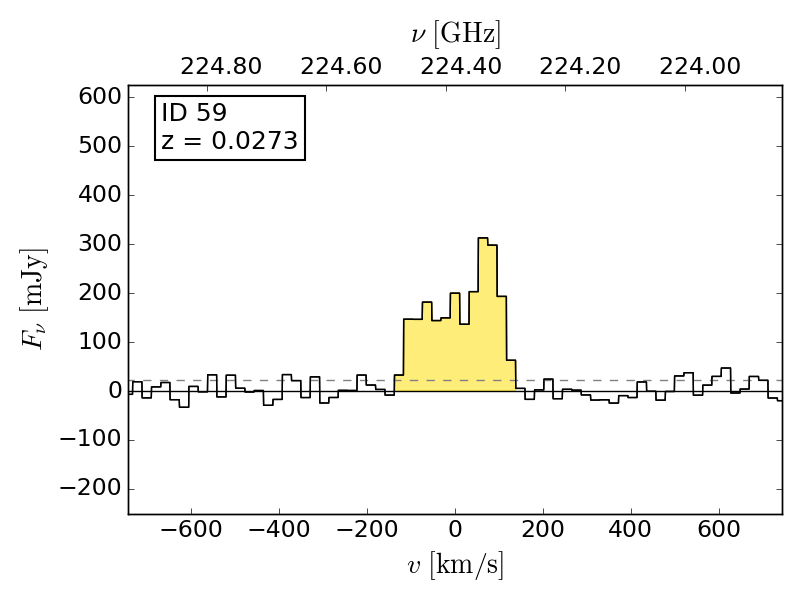}
\includegraphics[width=0.31\textwidth]{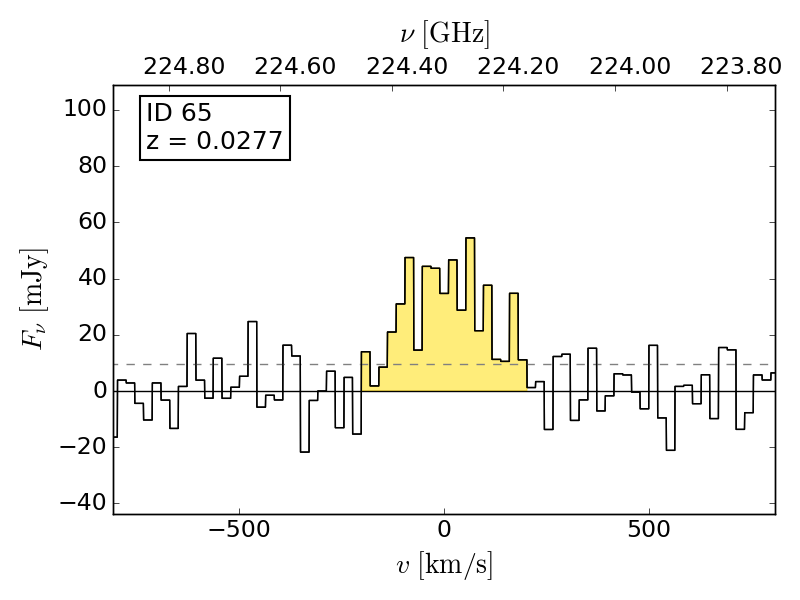}
\includegraphics[width=0.31\textwidth]{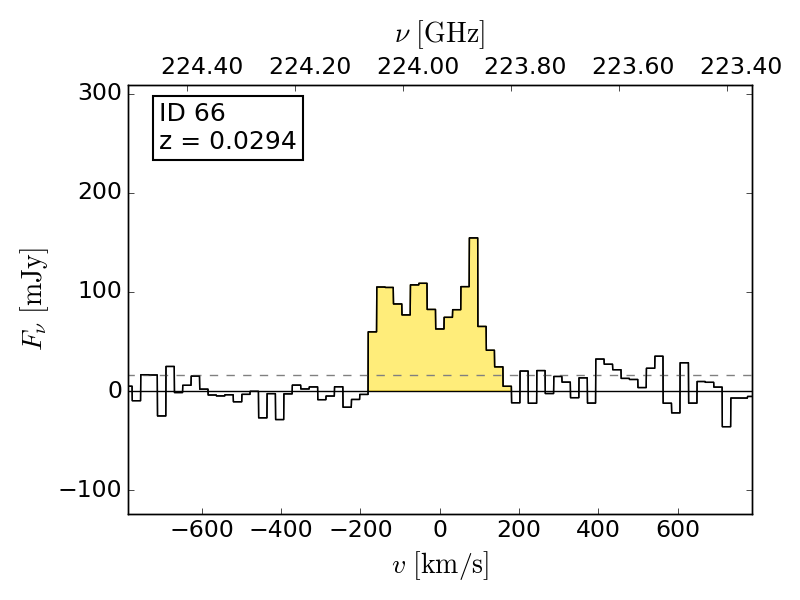}
\includegraphics[width=0.31\textwidth]{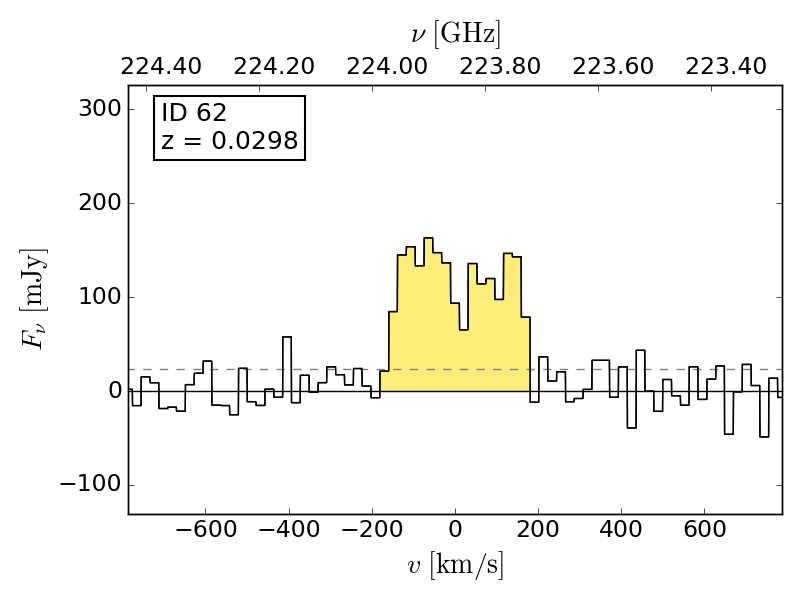}
\includegraphics[width=0.31\textwidth]{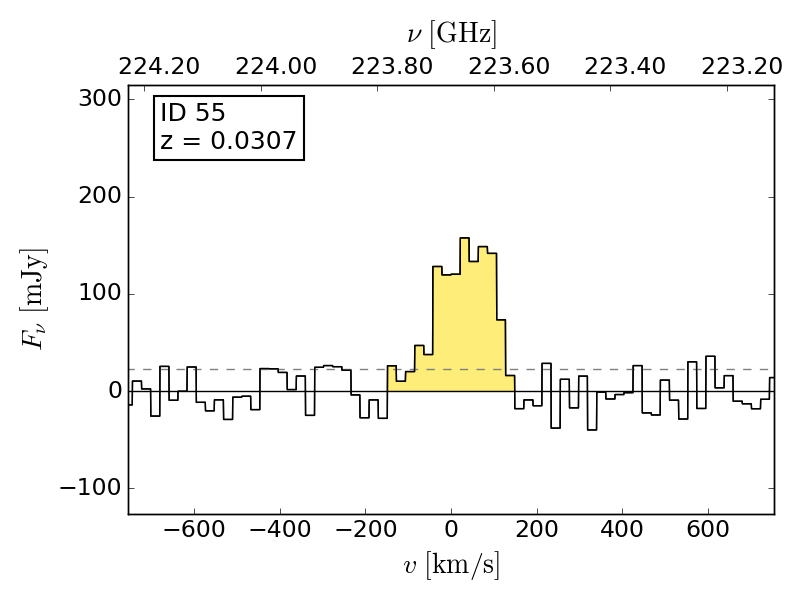}
\includegraphics[width=0.31\textwidth]{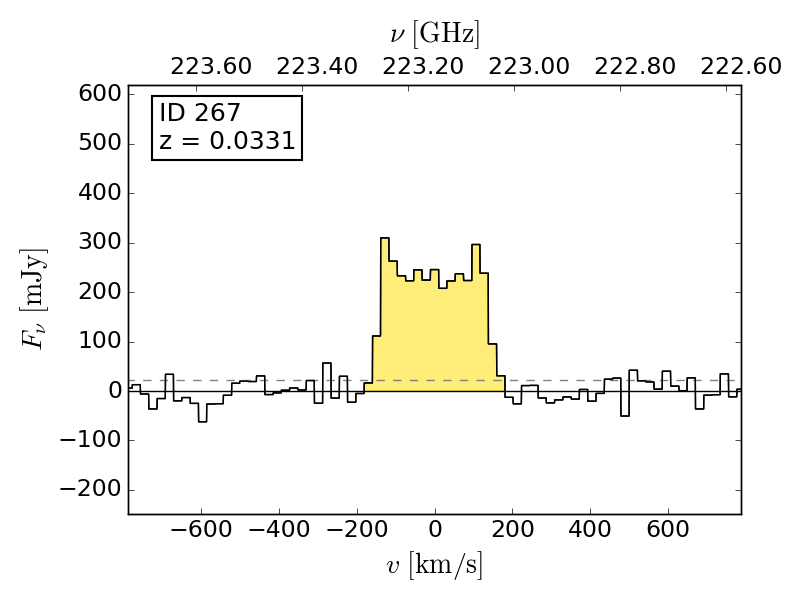}
\includegraphics[width=0.31\textwidth]{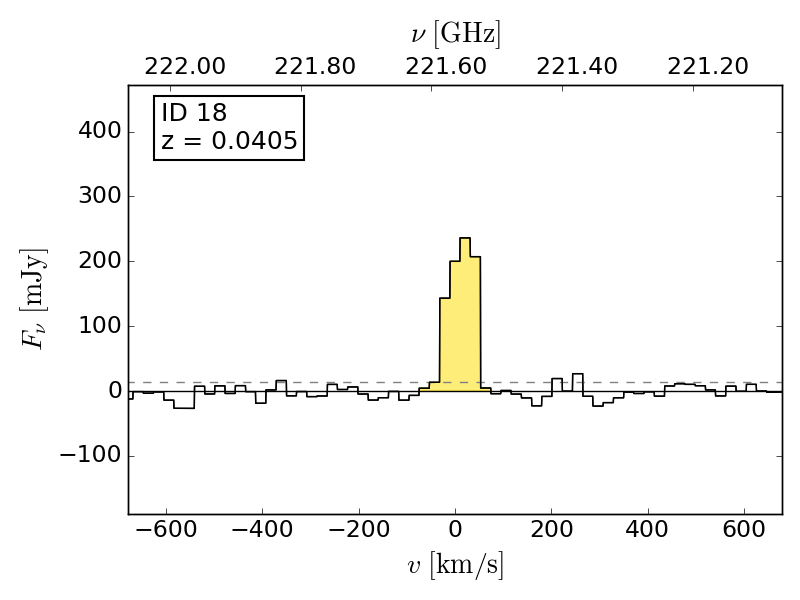}
\includegraphics[width=0.31\textwidth]{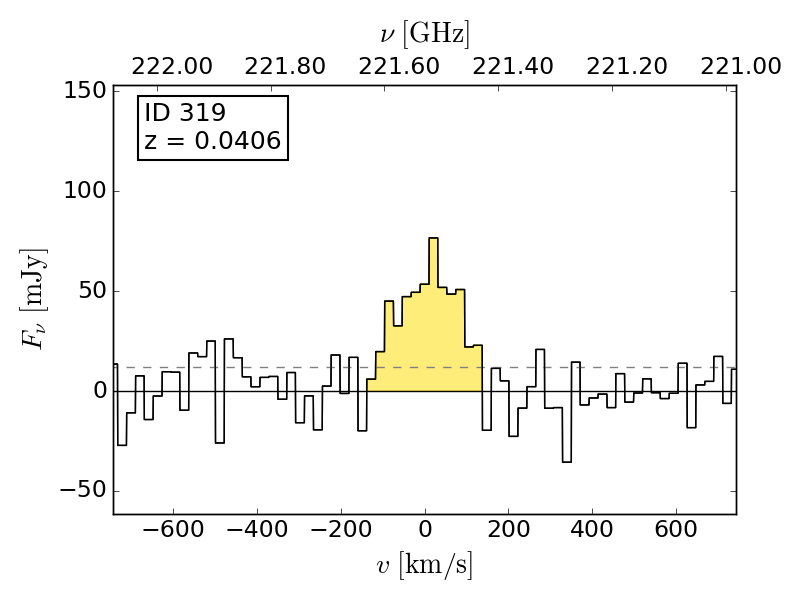}
\includegraphics[width=0.31\textwidth]{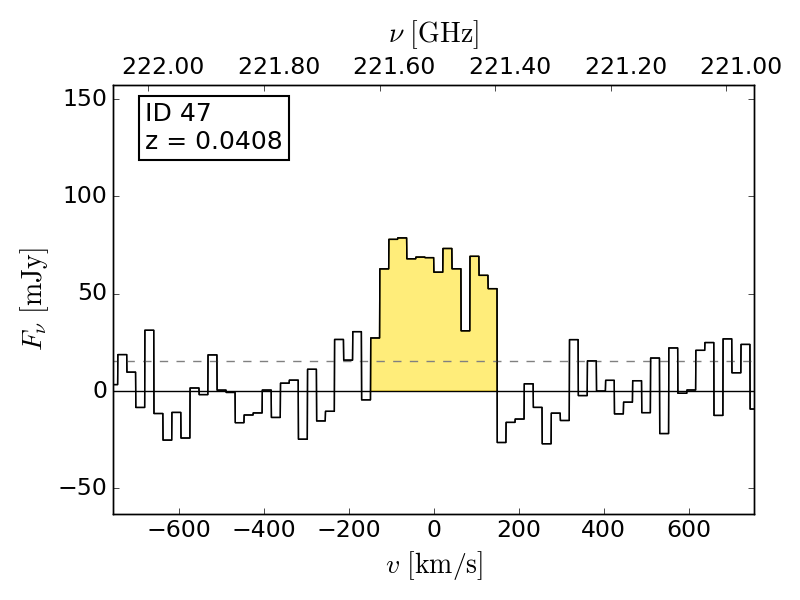}
\includegraphics[width=0.31\textwidth]{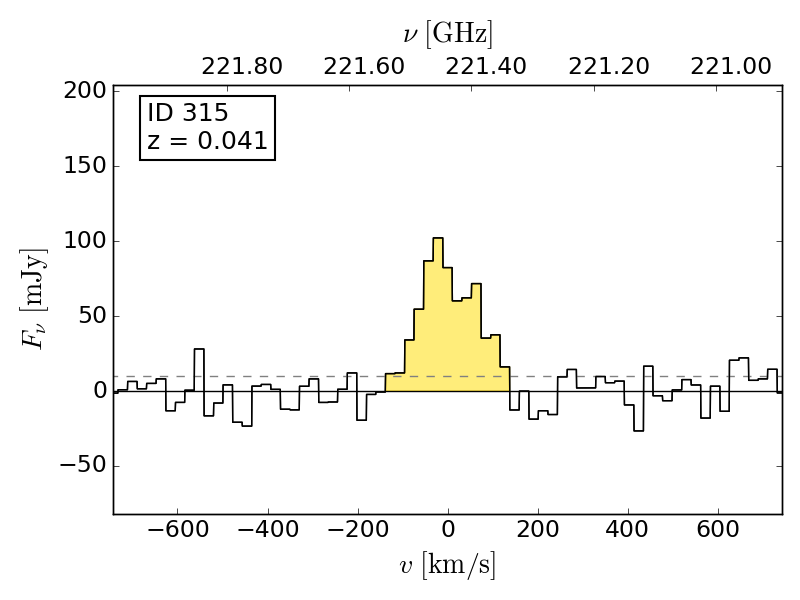}
\includegraphics[width=0.31\textwidth]{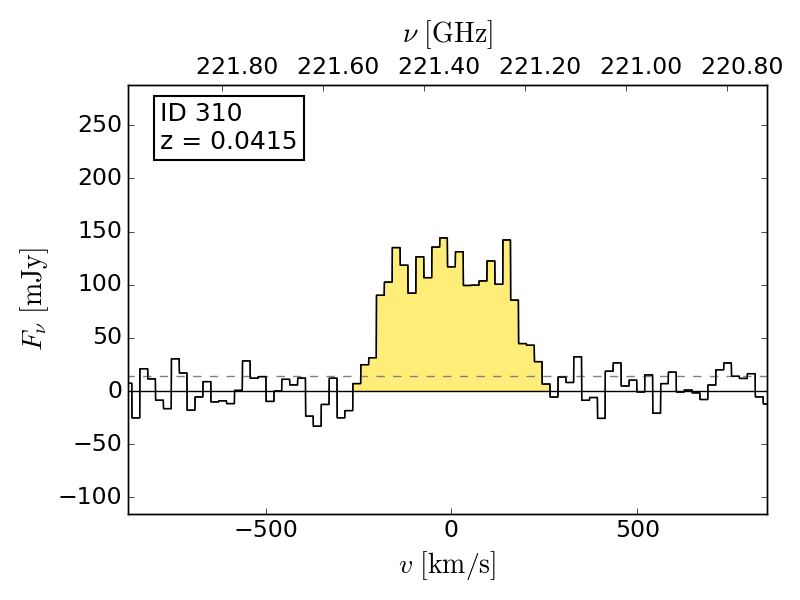}
\caption{IRAM CO(2-1) line profiles for 56 of our galaxies, sorted by increasing redshift. The yellow shaded region marks the line as adopted in the calculation of the integrated line flux, {and the dashed line denotes the $1 \sigma$ level. All flux densities are displayed as observed before the aperture correction was applied.}}
\label{fig:spectra21}
\end{figure*}
\begin{figure*}
\ContinuedFloat
\includegraphics[width=0.31\textwidth]{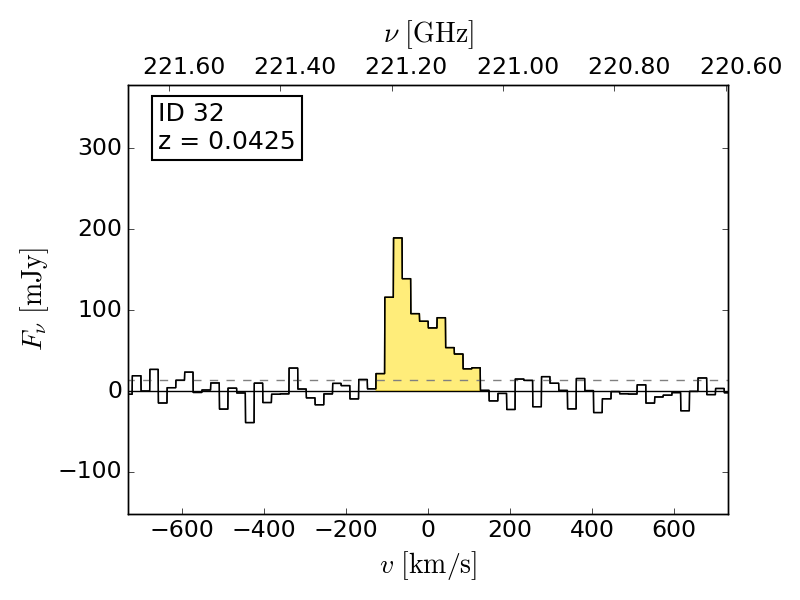}
\includegraphics[width=0.31\textwidth]{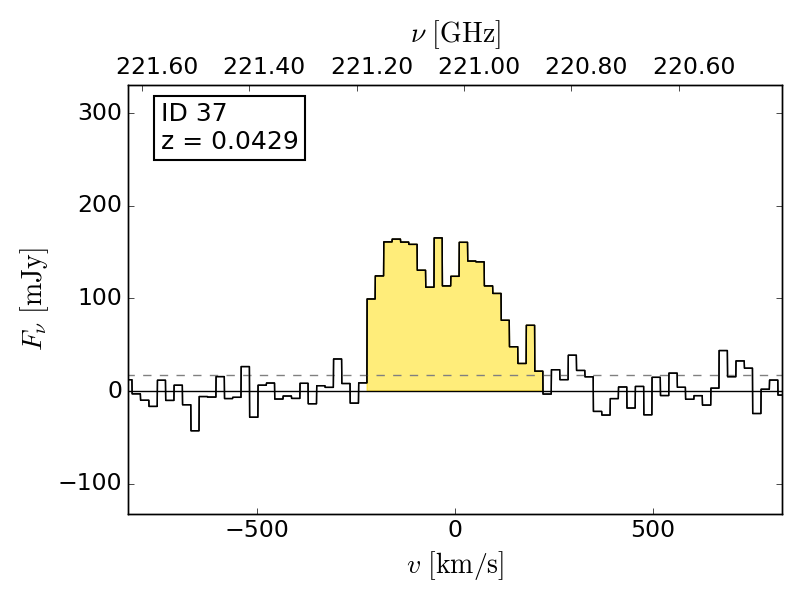}
\includegraphics[width=0.31\textwidth]{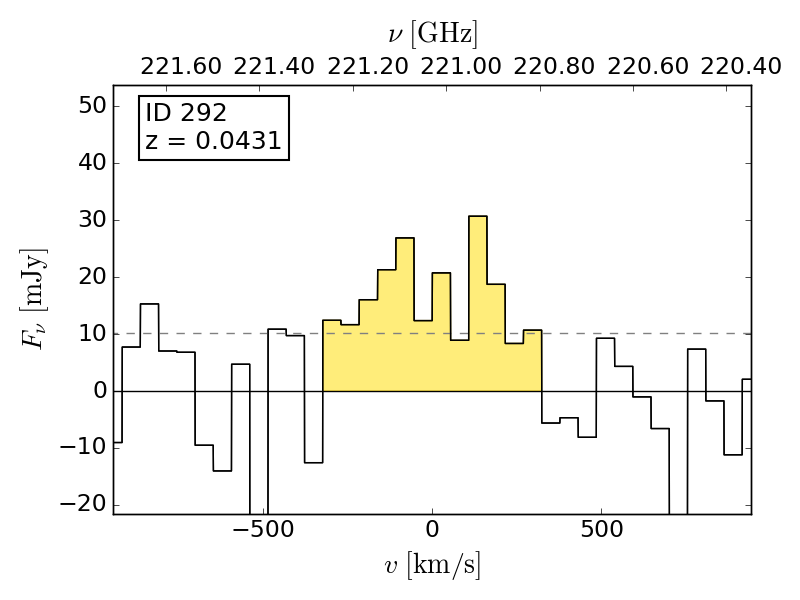}
\includegraphics[width=0.31\textwidth]{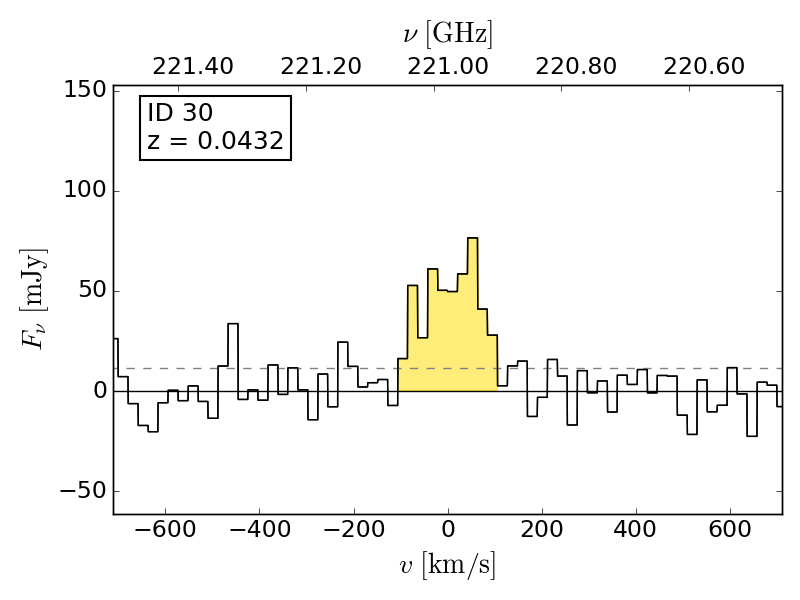}
\includegraphics[width=0.31\textwidth]{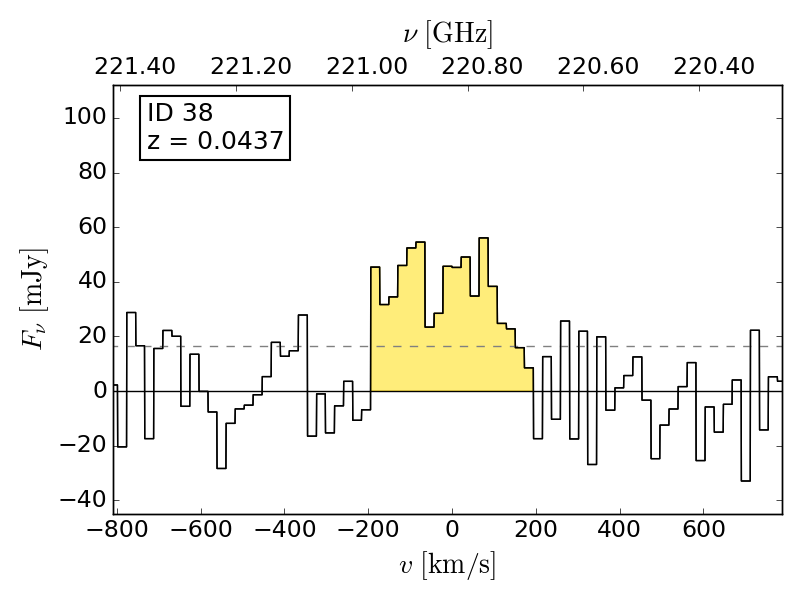}
\includegraphics[width=0.31\textwidth]{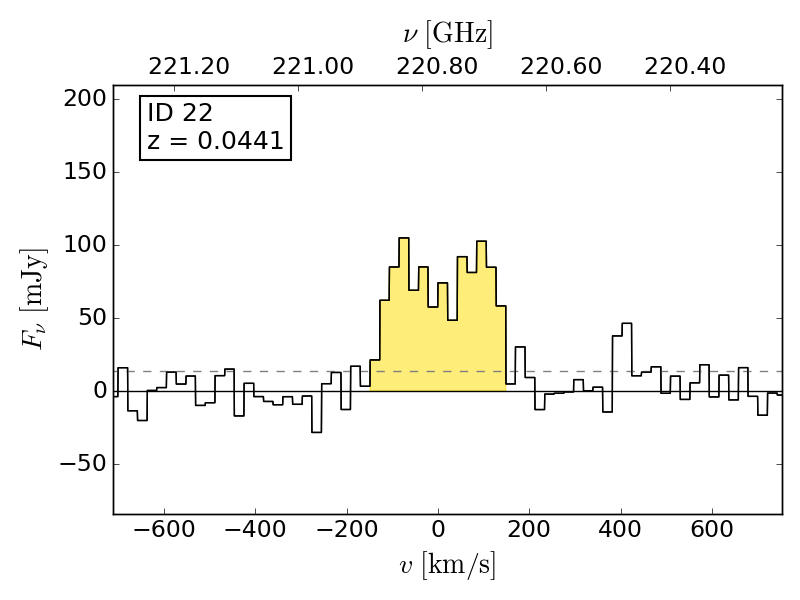}
\includegraphics[width=0.31\textwidth]{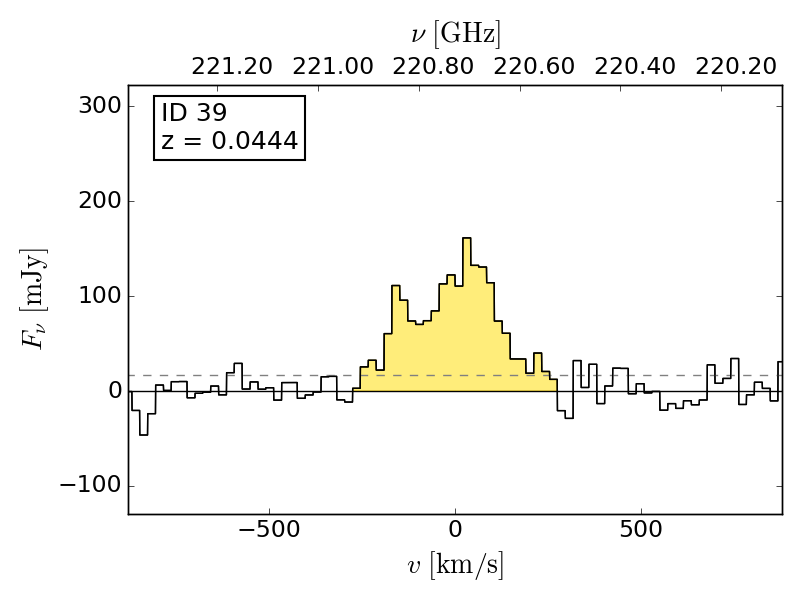}
\includegraphics[width=0.31\textwidth]{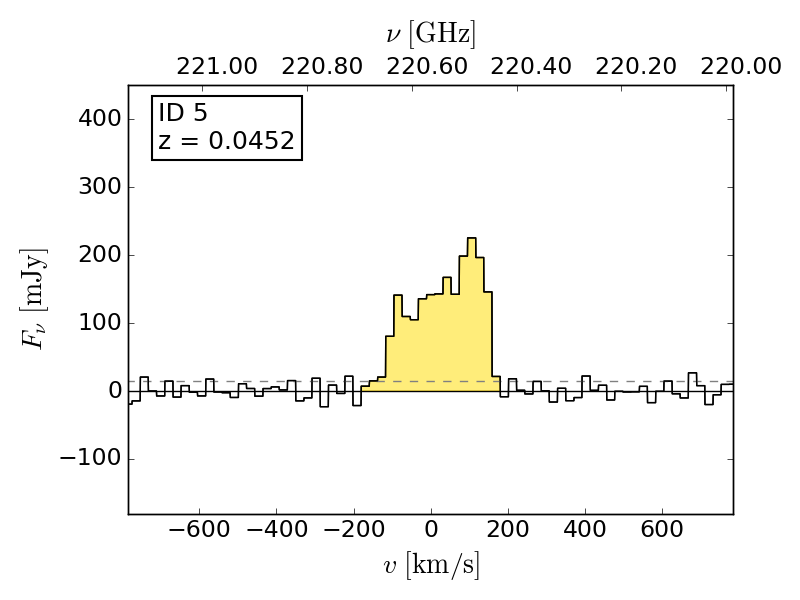}
\includegraphics[width=0.31\textwidth]{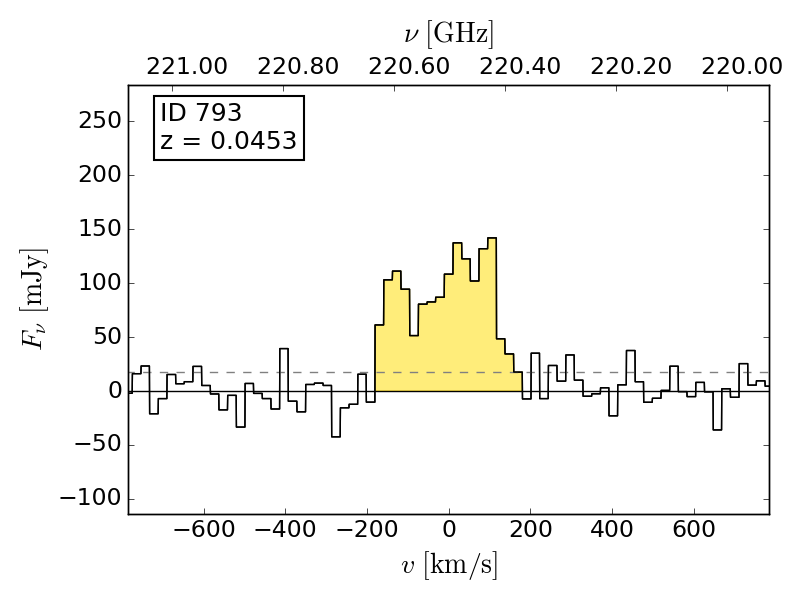}
\includegraphics[width=0.31\textwidth]{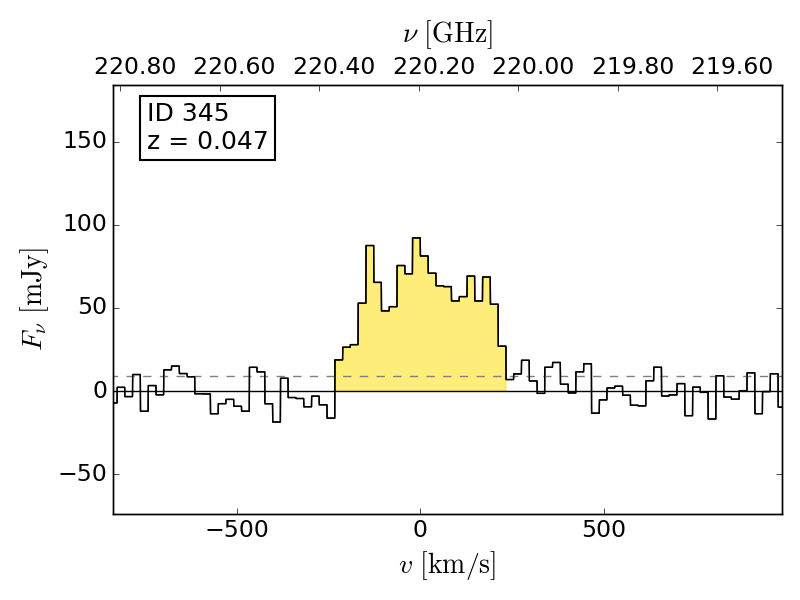}
\includegraphics[width=0.31\textwidth]{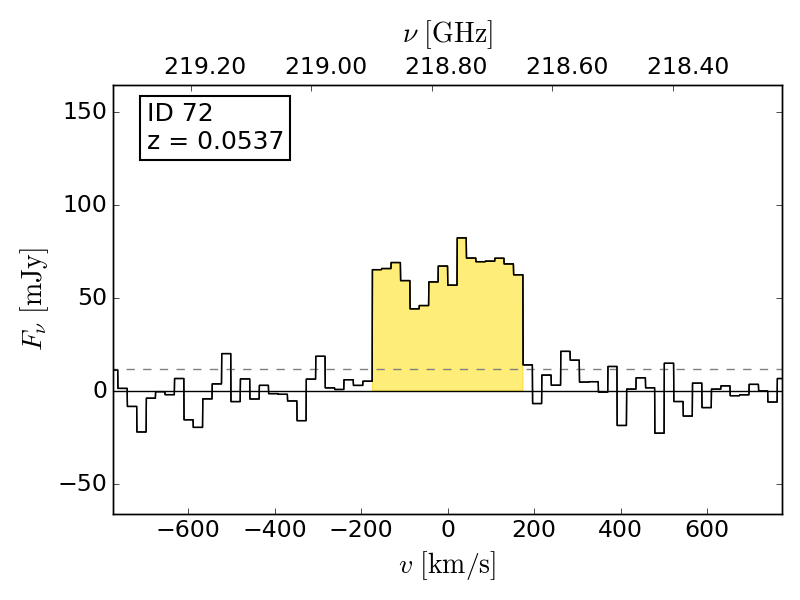}
\includegraphics[width=0.31\textwidth]{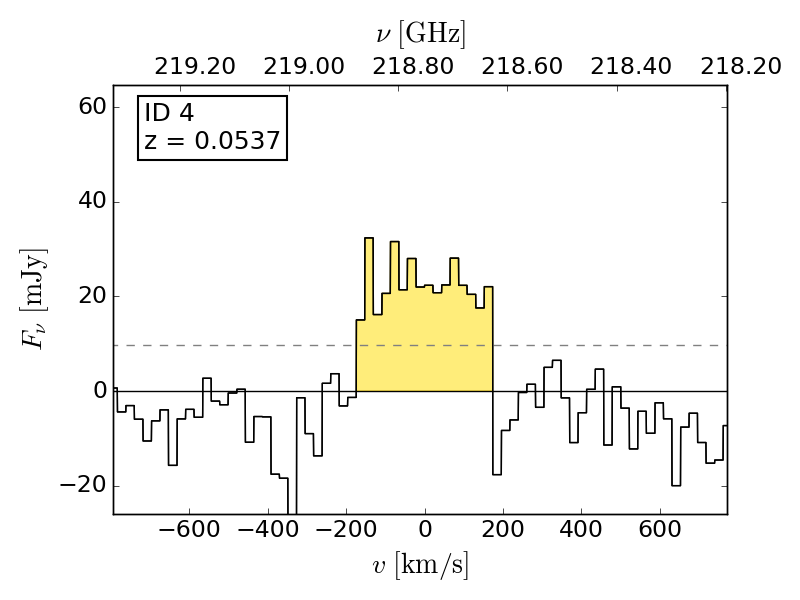}
\includegraphics[width=0.31\textwidth]{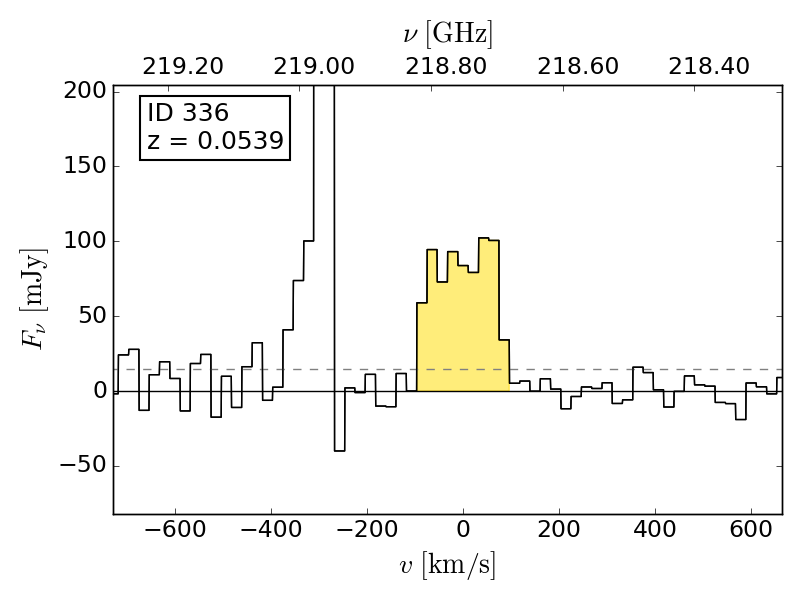}
\includegraphics[width=0.31\textwidth]{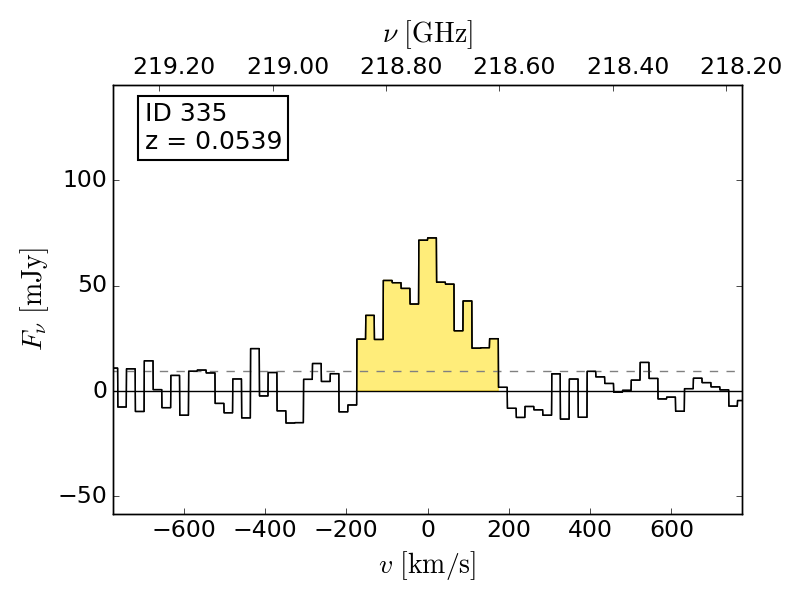}
\includegraphics[width=0.31\textwidth]{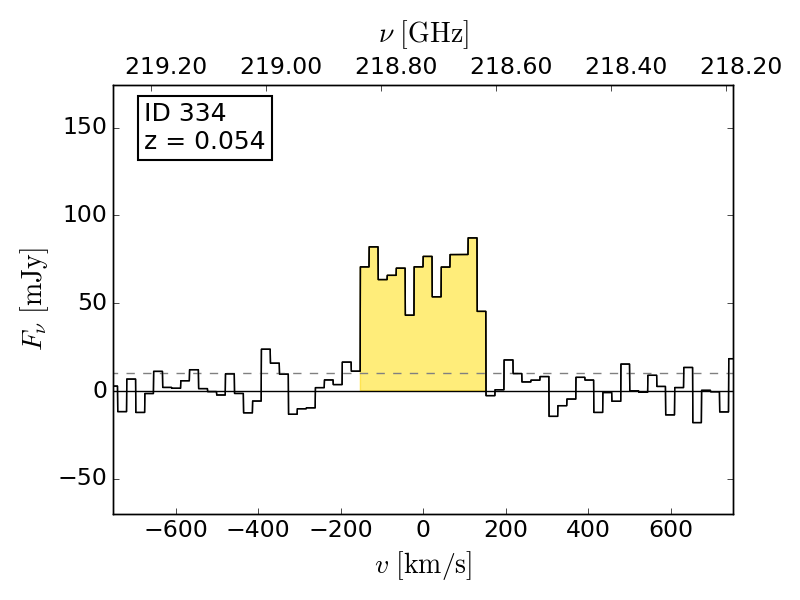}
\caption{(cont.)}
\end{figure*}
\begin{figure*}
\ContinuedFloat
\includegraphics[width=0.31\textwidth]{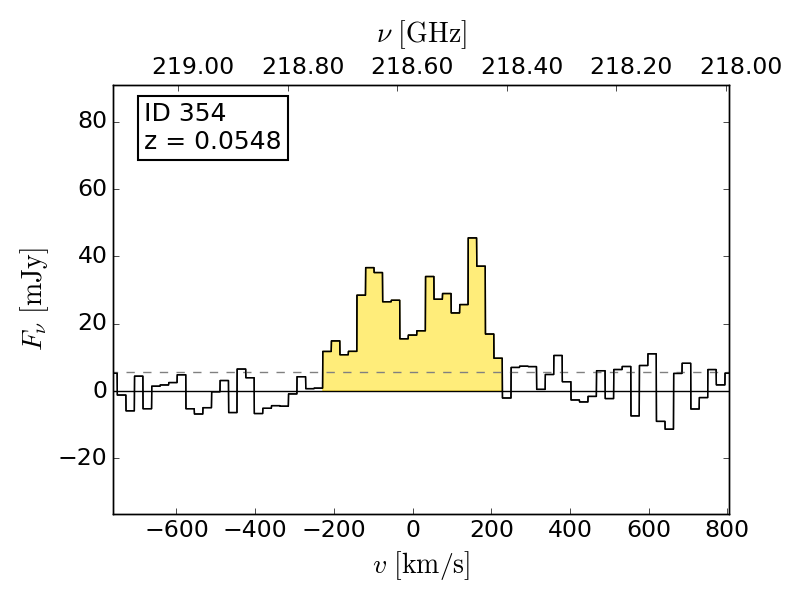}
\includegraphics[width=0.31\textwidth]{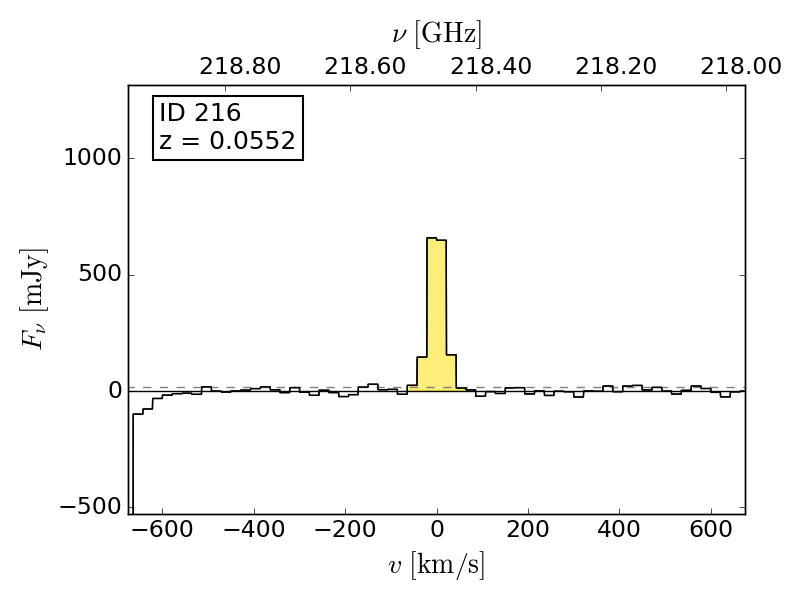}
\includegraphics[width=0.31\textwidth]{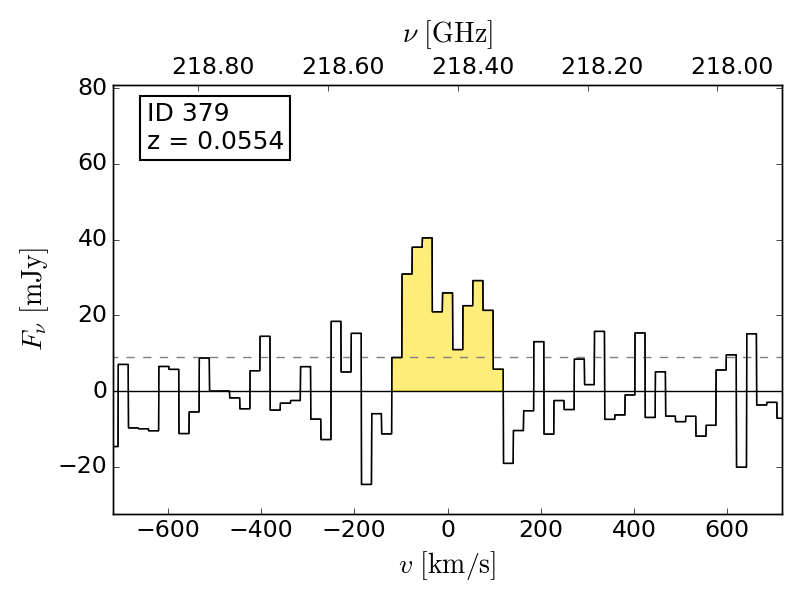}
\includegraphics[width=0.31\textwidth]{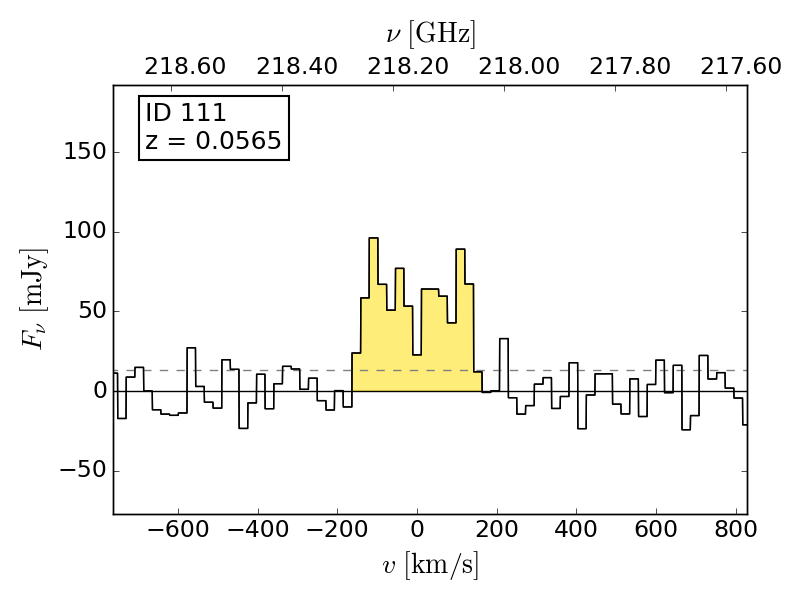}
\includegraphics[width=0.31\textwidth]{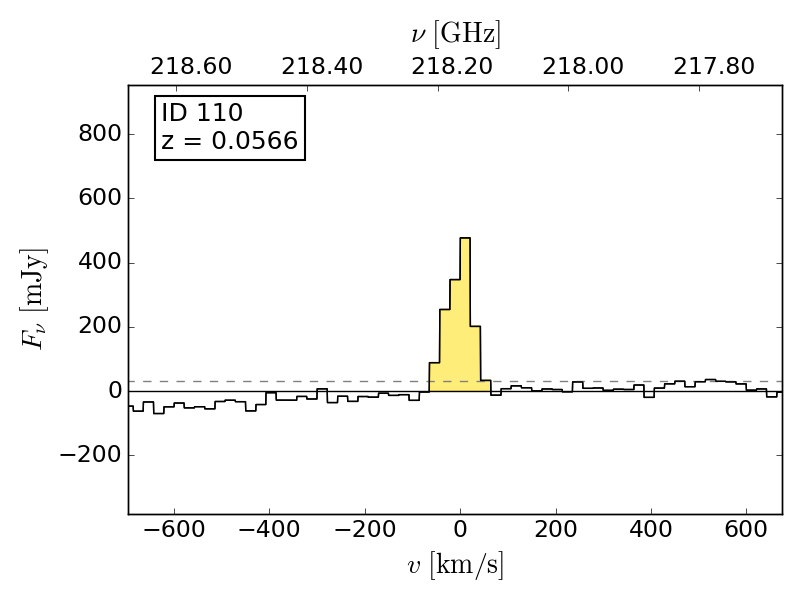}
\includegraphics[width=0.31\textwidth]{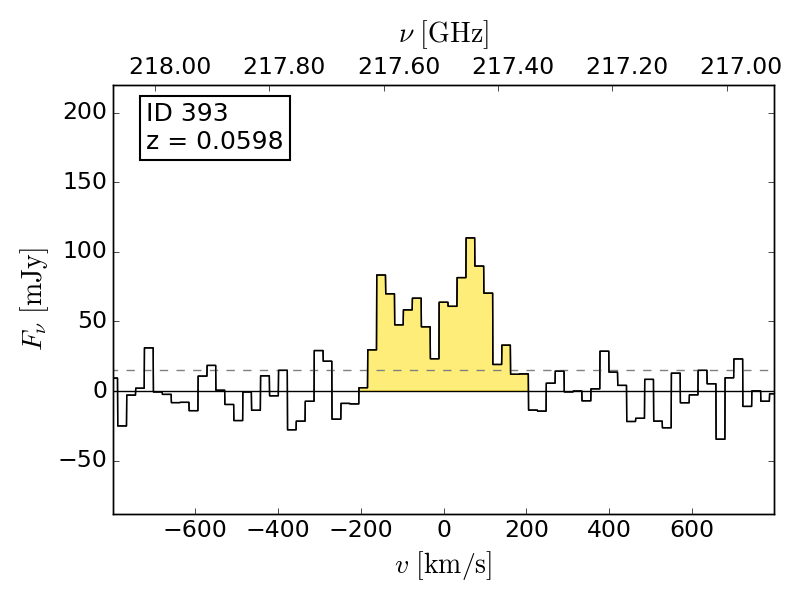}
\includegraphics[width=0.31\textwidth]{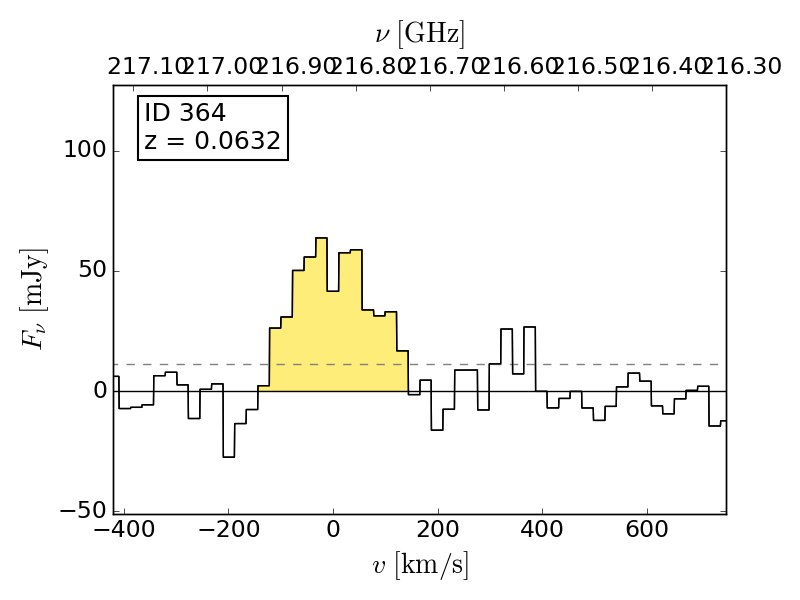}
\includegraphics[width=0.31\textwidth]{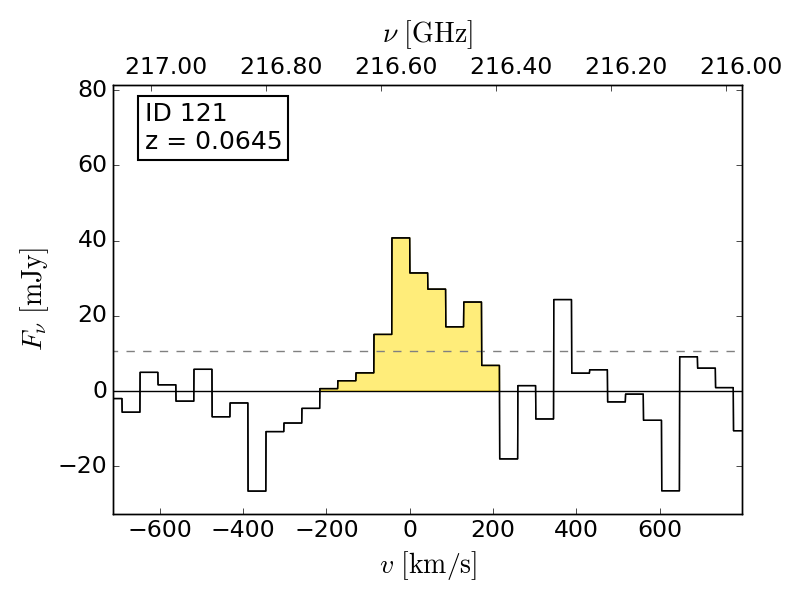}
\includegraphics[width=0.31\textwidth]{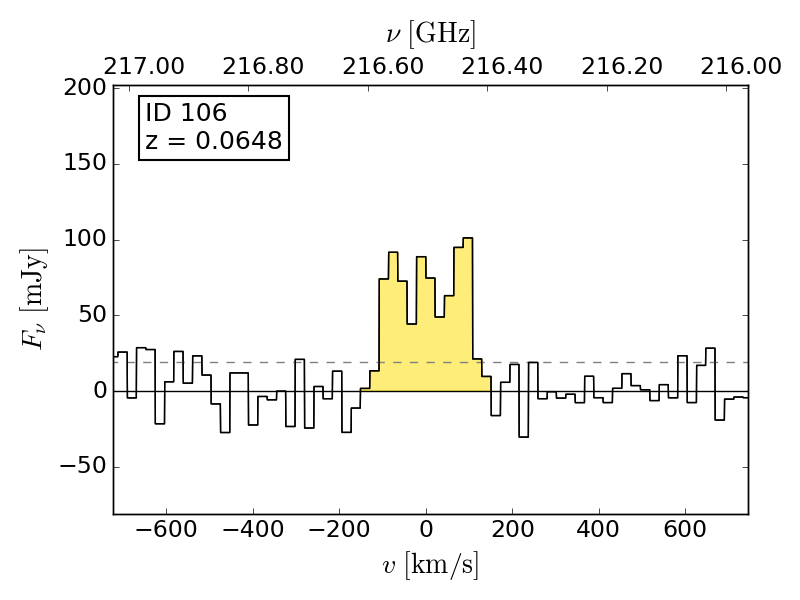}
\includegraphics[width=0.31\textwidth]{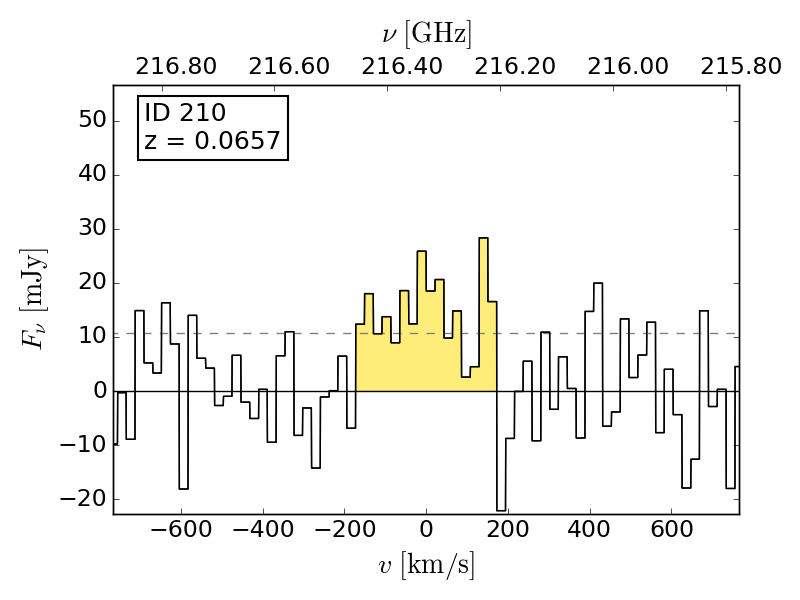}
\includegraphics[width=0.31\textwidth]{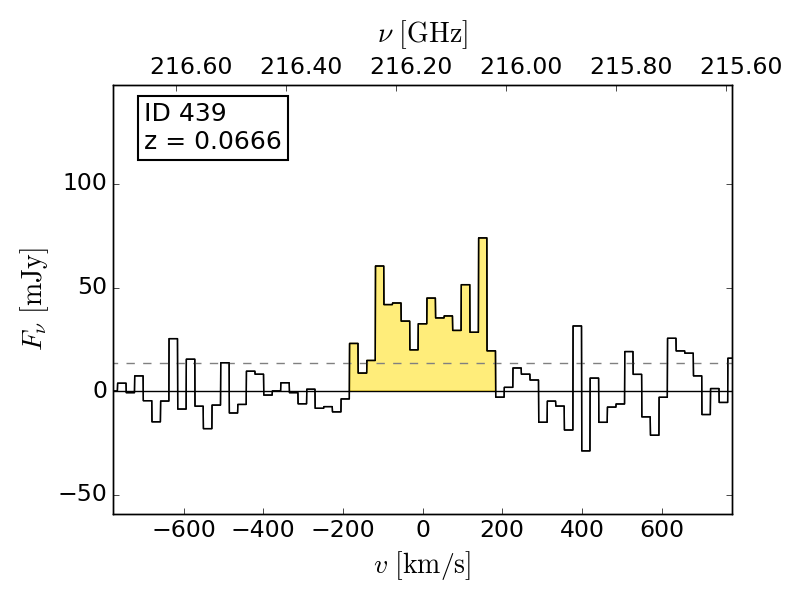}
\includegraphics[width=0.31\textwidth]{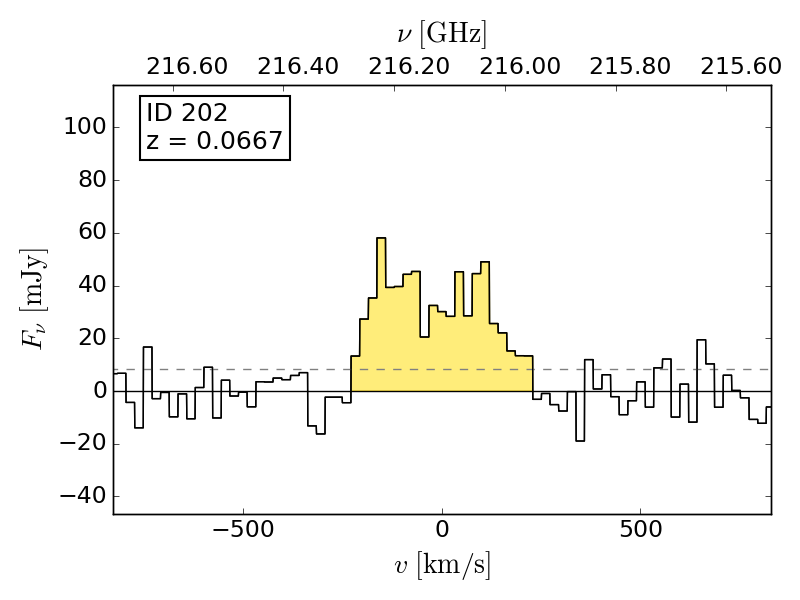}
\includegraphics[width=0.31\textwidth]{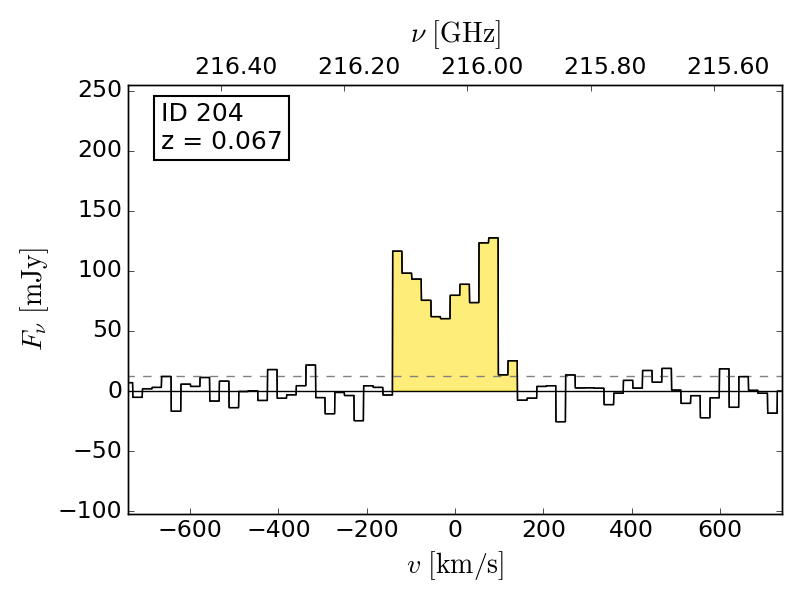}
\includegraphics[width=0.31\textwidth]{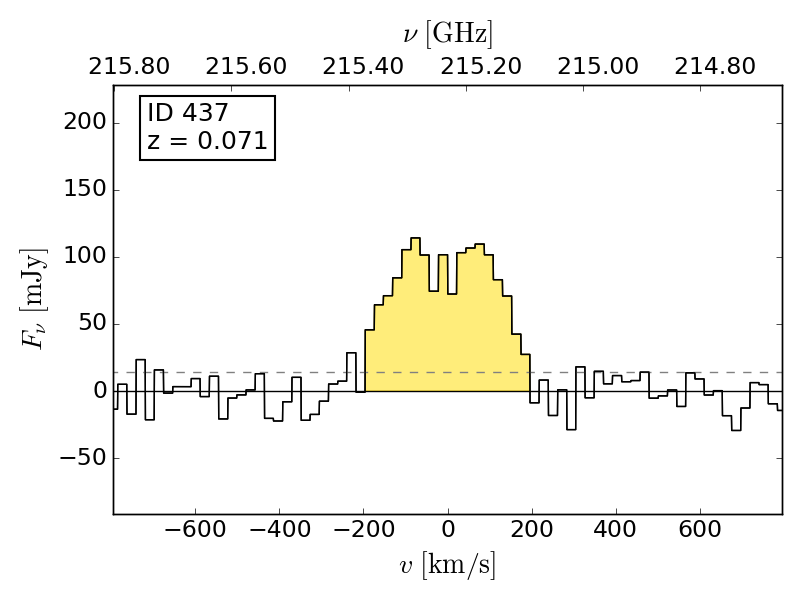}
\includegraphics[width=0.31\textwidth]{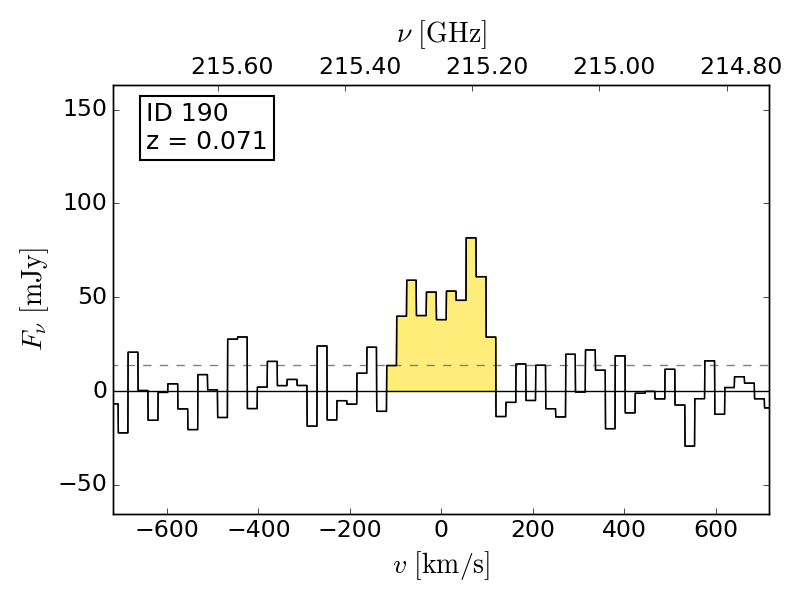}
\caption{(cont.)}
\end{figure*}
\begin{figure*}
\ContinuedFloat
\includegraphics[width=0.31\textwidth]{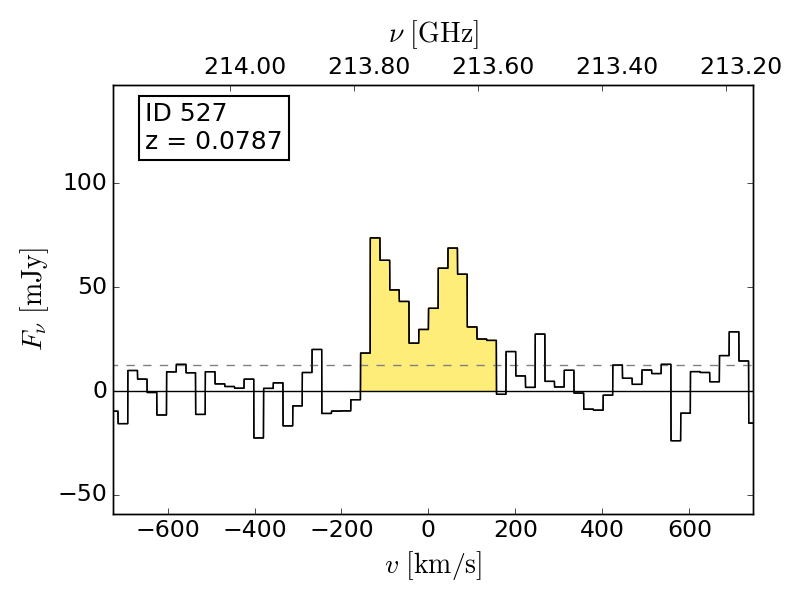}
\includegraphics[width=0.31\textwidth]{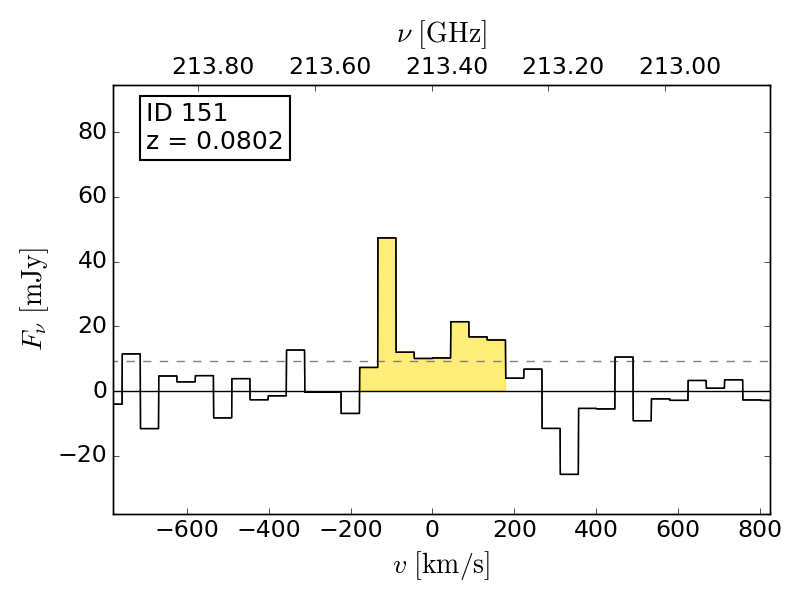}
\includegraphics[width=0.31\textwidth]{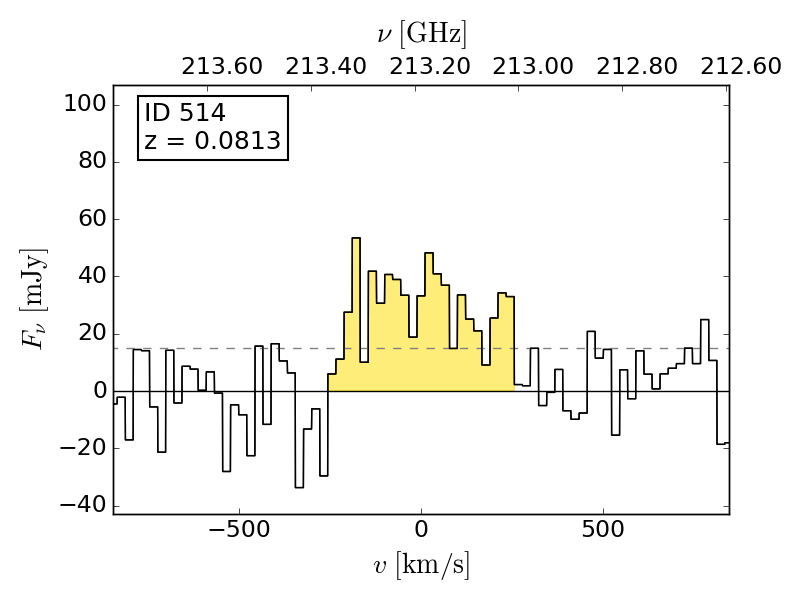}
\includegraphics[width=0.31\textwidth]{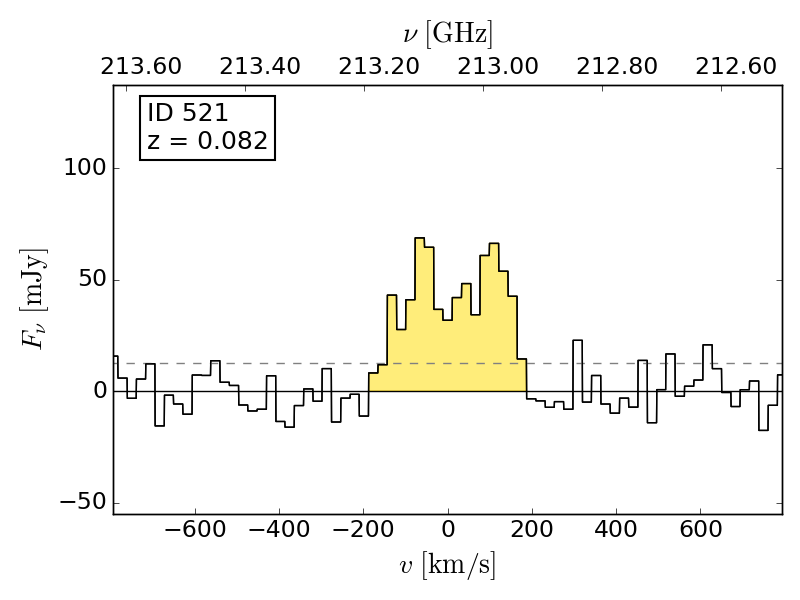}
\includegraphics[width=0.31\textwidth]{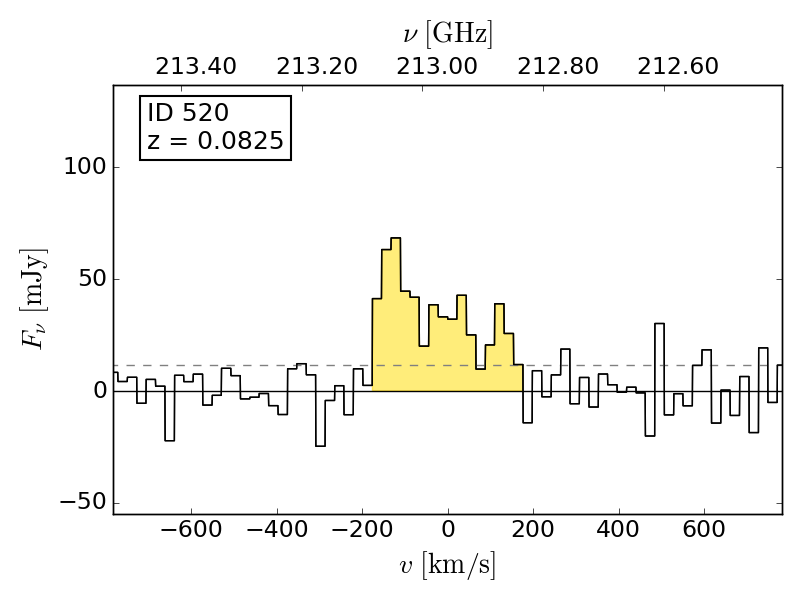}
\includegraphics[width=0.31\textwidth]{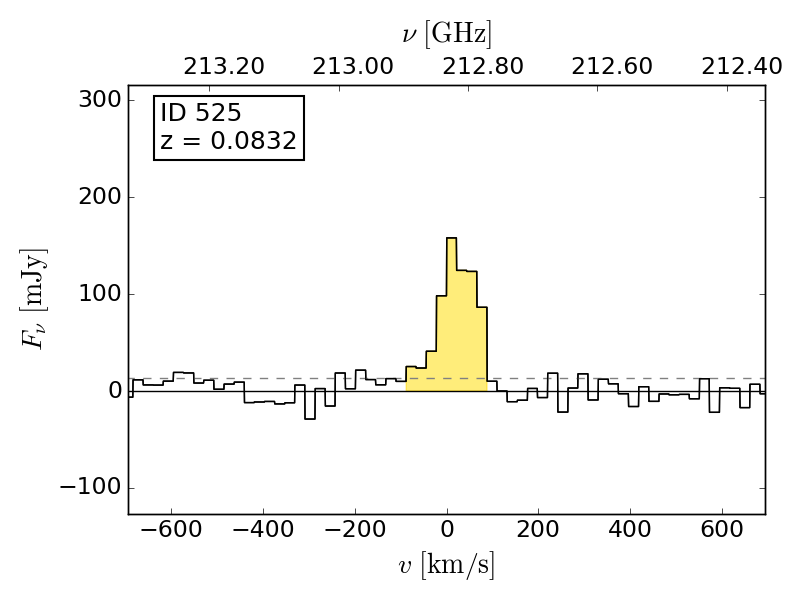}
\includegraphics[width=0.31\textwidth]{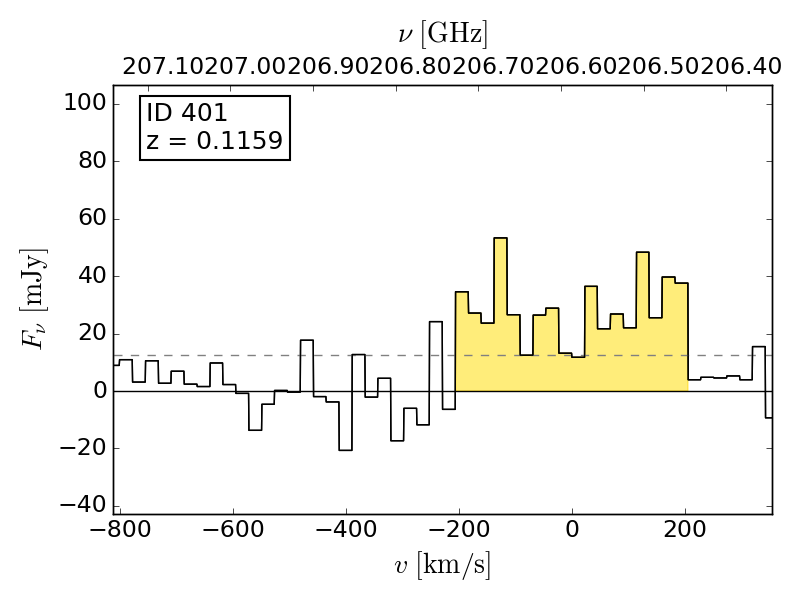}
\includegraphics[width=0.31\textwidth]{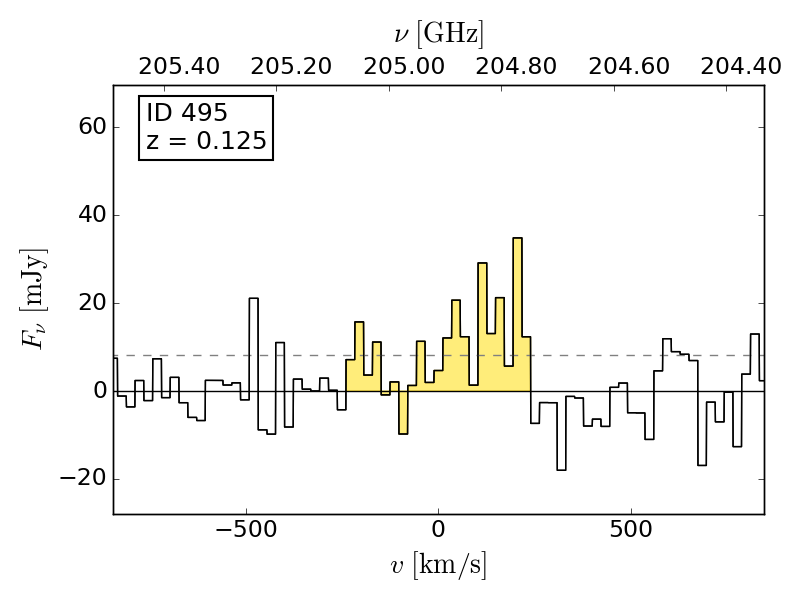}
\includegraphics[width=0.31\textwidth]{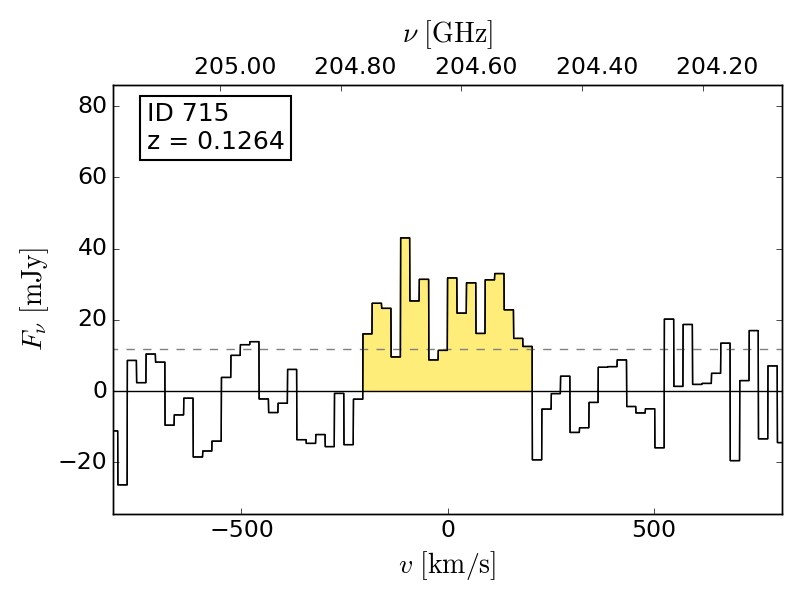}
\includegraphics[width=0.31\textwidth]{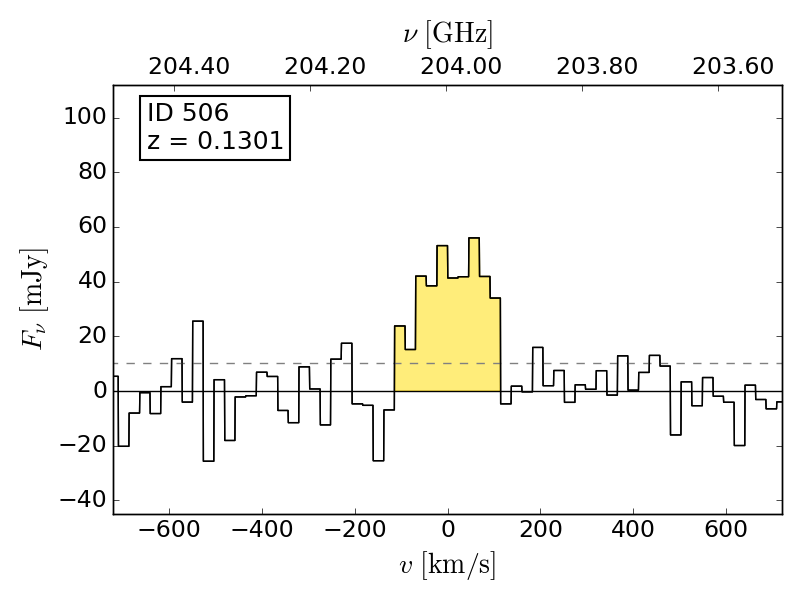}
\includegraphics[width=0.31\textwidth]{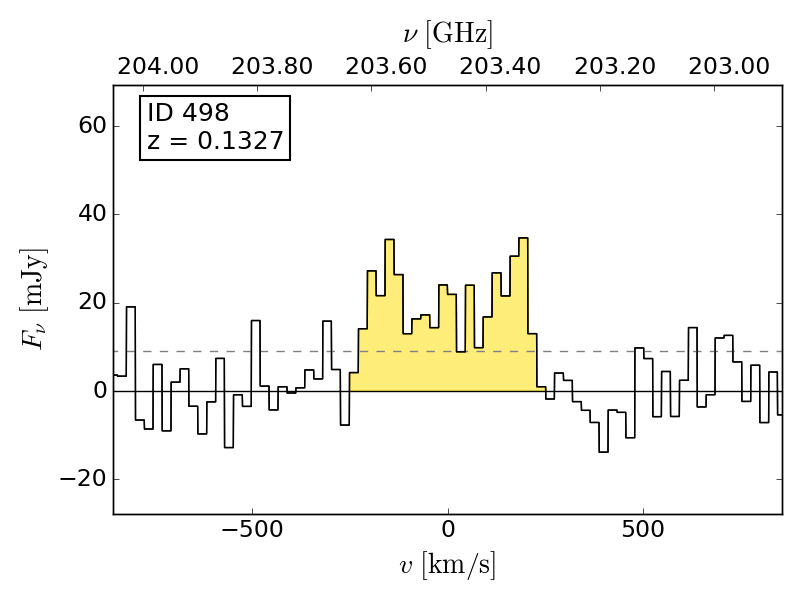}
\caption{(cont.)}
\end{figure*}


\begin{figure*}
\includegraphics[width=0.31\textwidth]{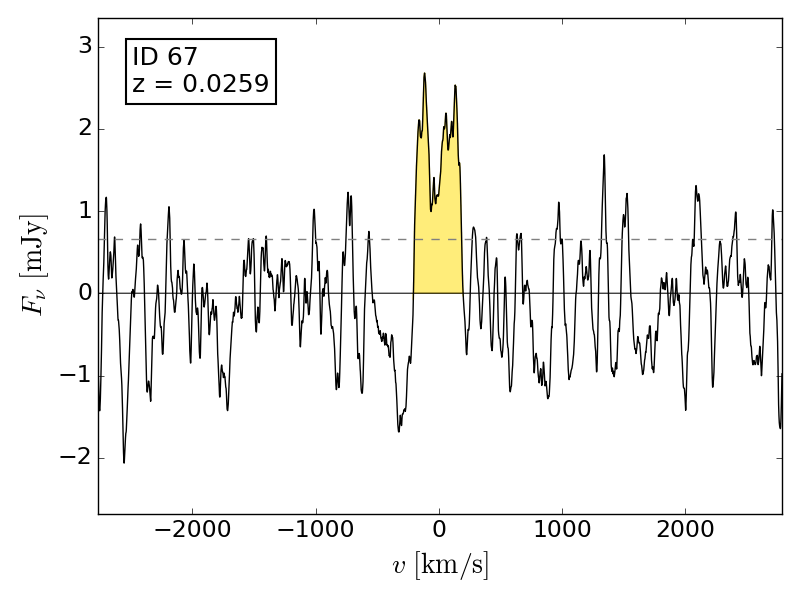}
\includegraphics[width=0.31\textwidth]{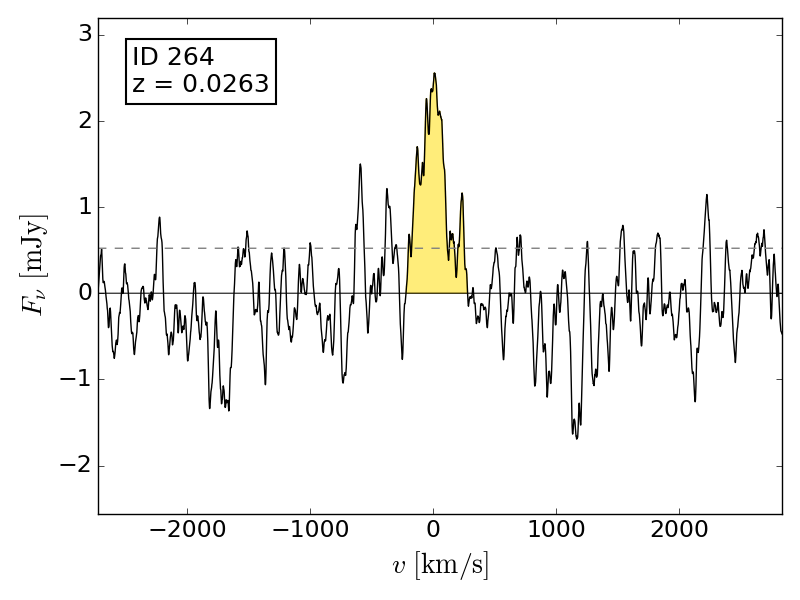}
\includegraphics[width=0.31\textwidth]{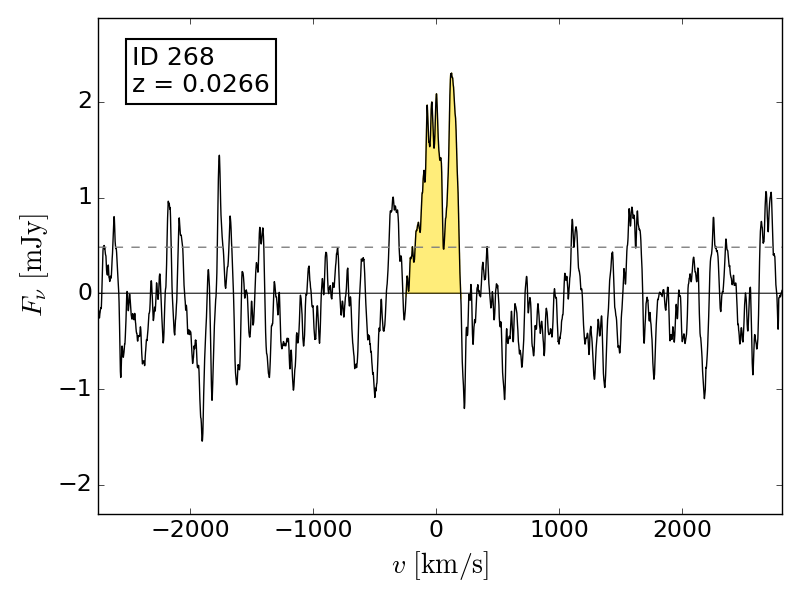}
\includegraphics[width=0.31\textwidth]{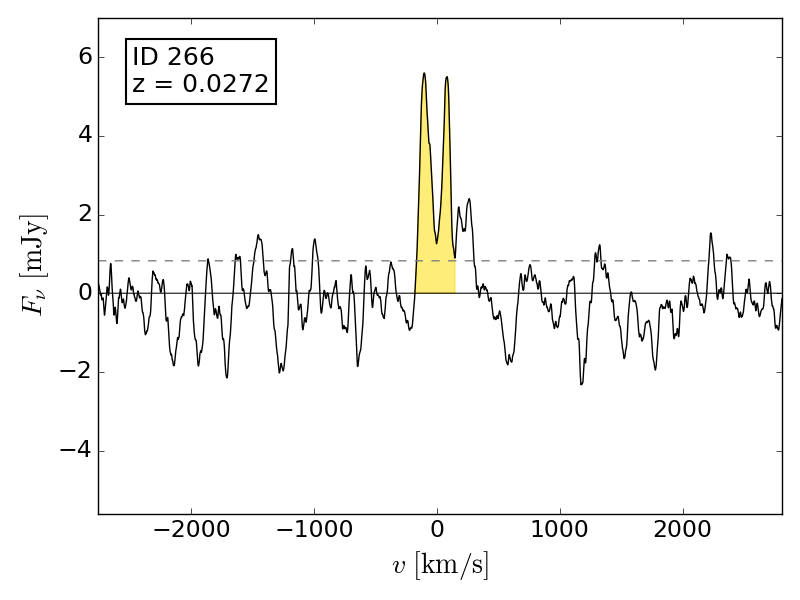}
\includegraphics[width=0.31\textwidth]{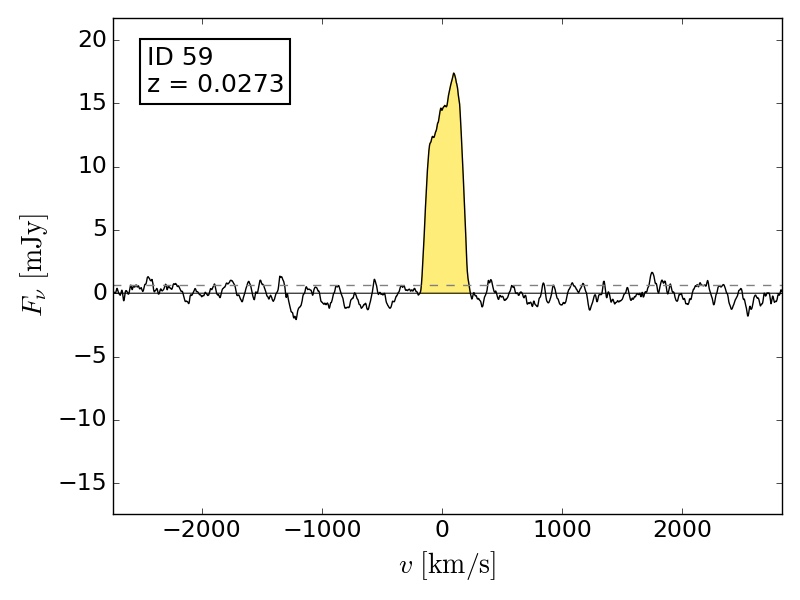}
\includegraphics[width=0.31\textwidth]{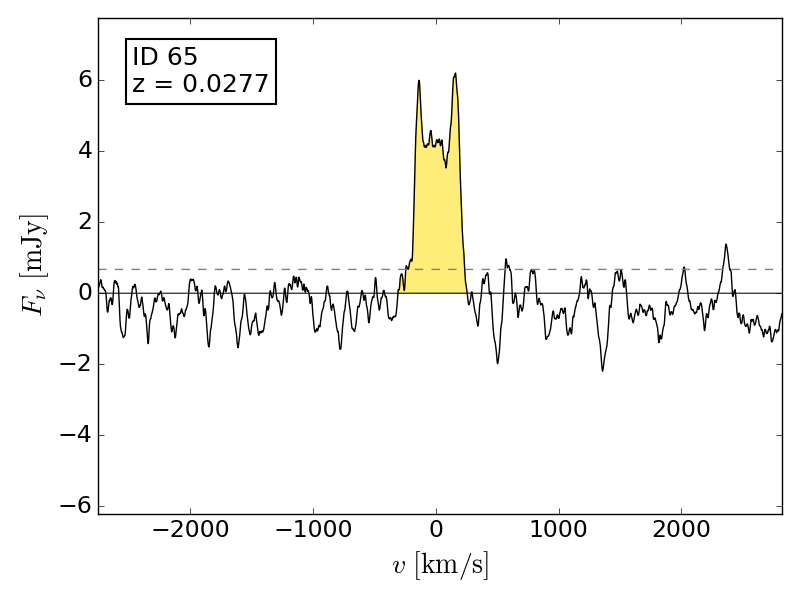}
\includegraphics[width=0.31\textwidth]{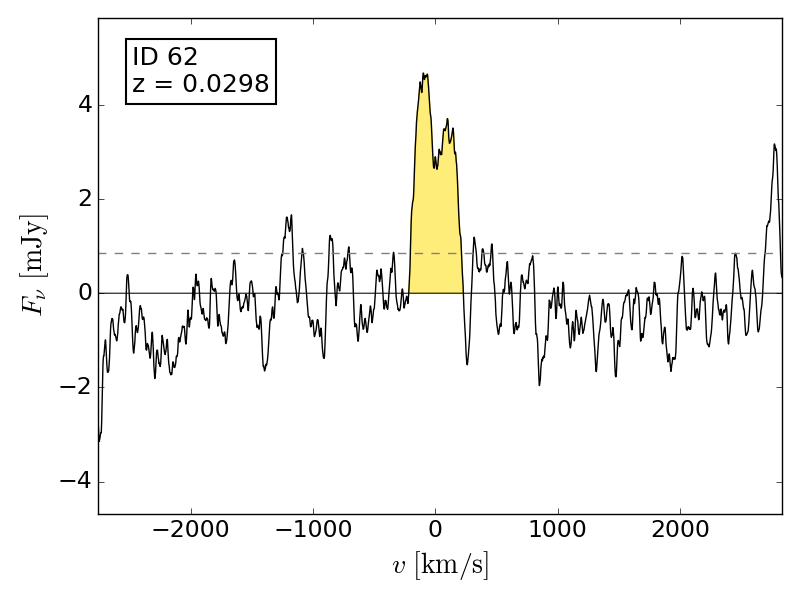}
\includegraphics[width=0.31\textwidth]{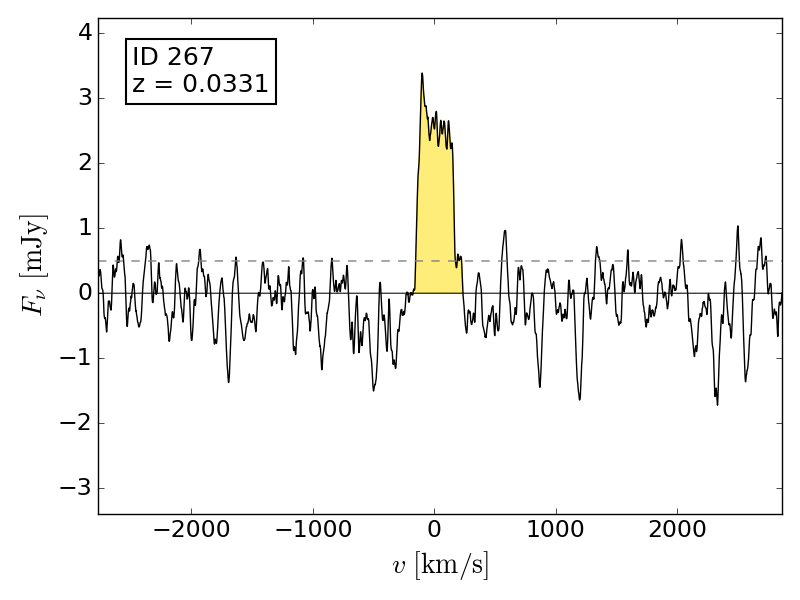}
\includegraphics[width=0.31\textwidth]{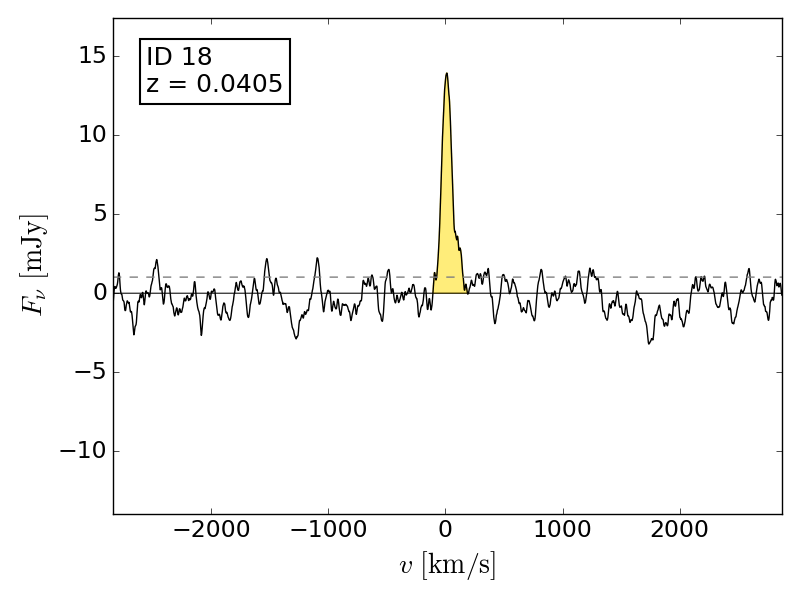}
\includegraphics[width=0.31\textwidth]{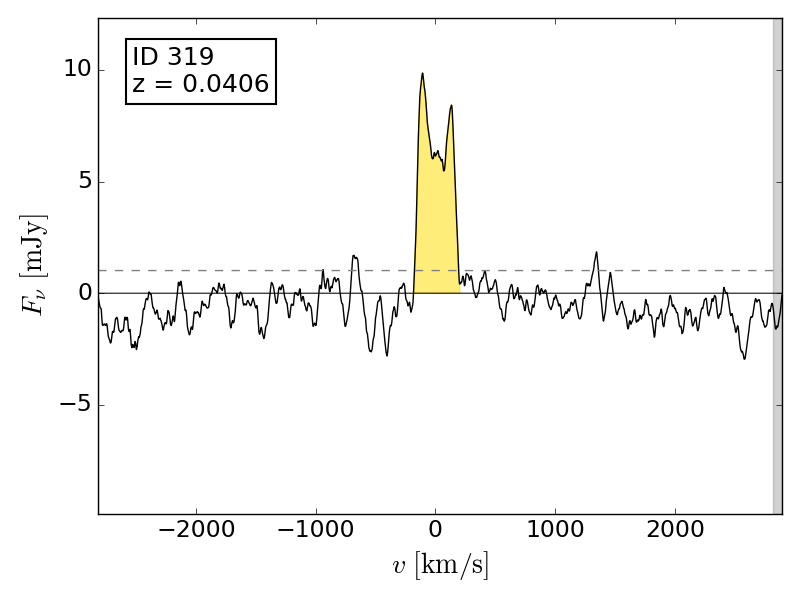}
\includegraphics[width=0.31\textwidth]{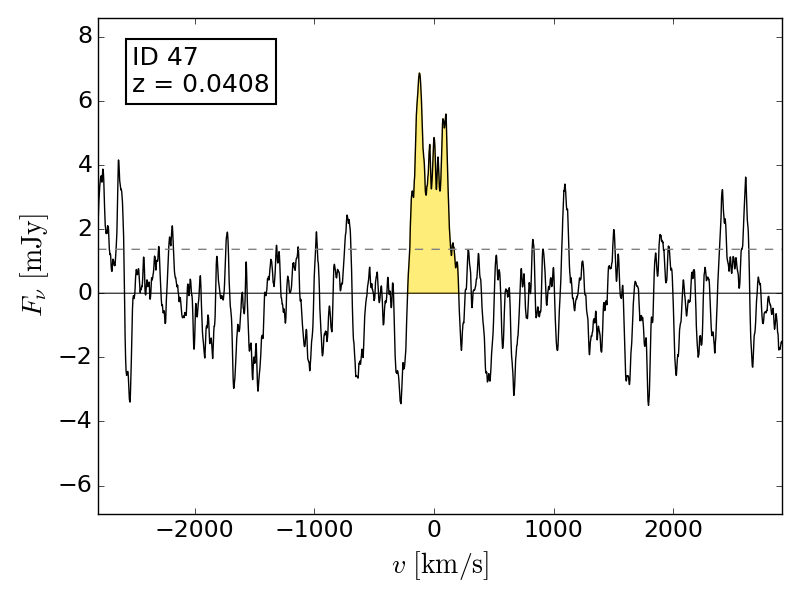}
\includegraphics[width=0.31\textwidth]{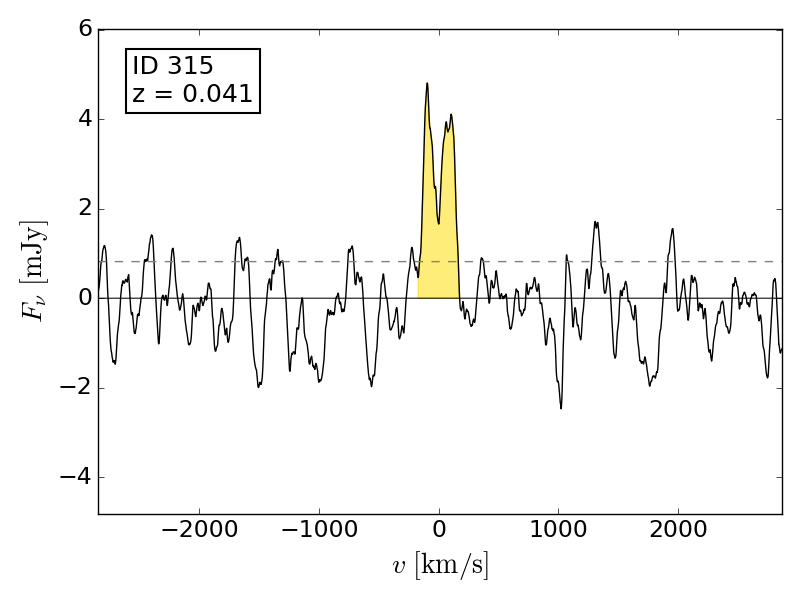}
\includegraphics[width=0.31\textwidth]{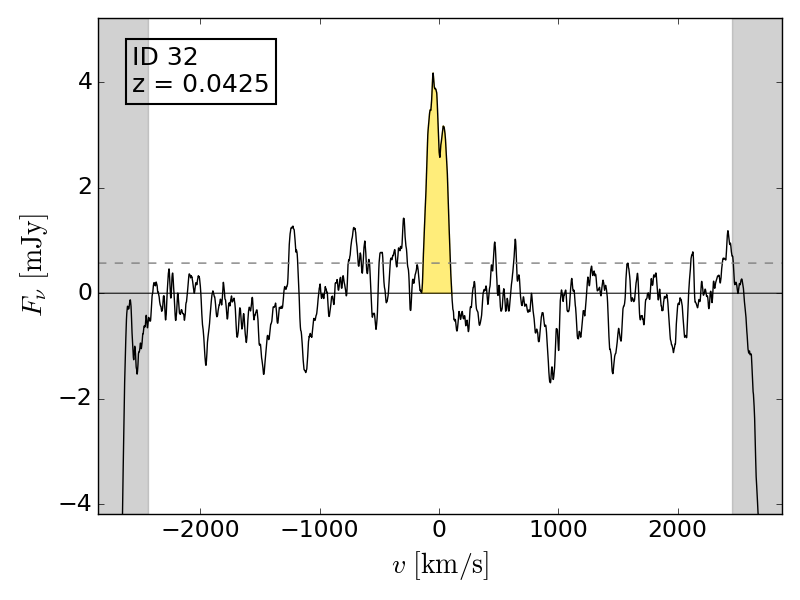}
\includegraphics[width=0.31\textwidth]{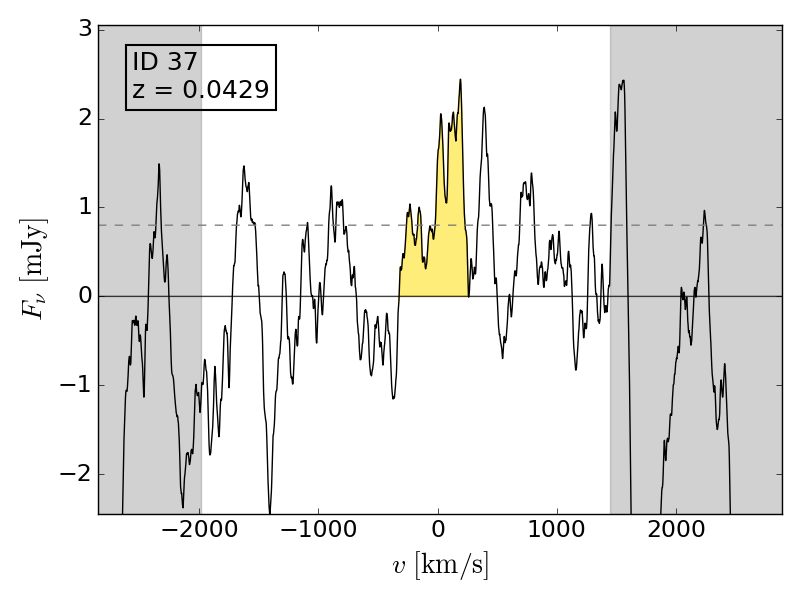}
\includegraphics[width=0.31\textwidth]{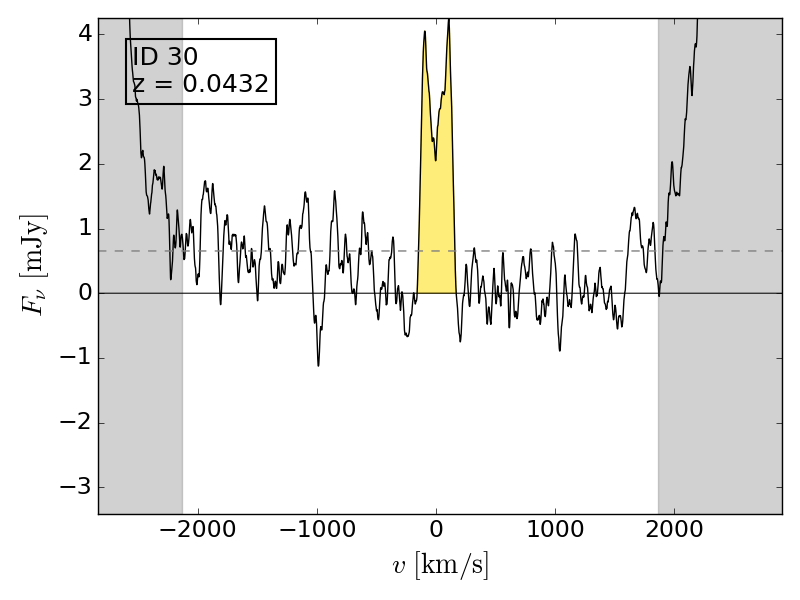}
\caption{\HI\ 21 cm line profiles for 24 galaxies in our sample, sorted by increasing redshift. The grey areas mark regions of interference from a different signal - these regions were excluded in the computation of the RMS error, which is used for calculating the uncertainties on integrated  \HI\ line fluxes, and \HI\ masses. The yellow shaded region marks the line as adopted in the calculation of the integrated line flux, {and the dashed line denotes the $1 \sigma$ level. The zero-point of the velocity axis is positioned at the predicted observed line frequency according to spectroscopic redshift from SDSS.}}
\label{fig:Ar_spectra}
\end{figure*}
\begin{figure*}
\ContinuedFloat
\includegraphics[width=0.31\textwidth]{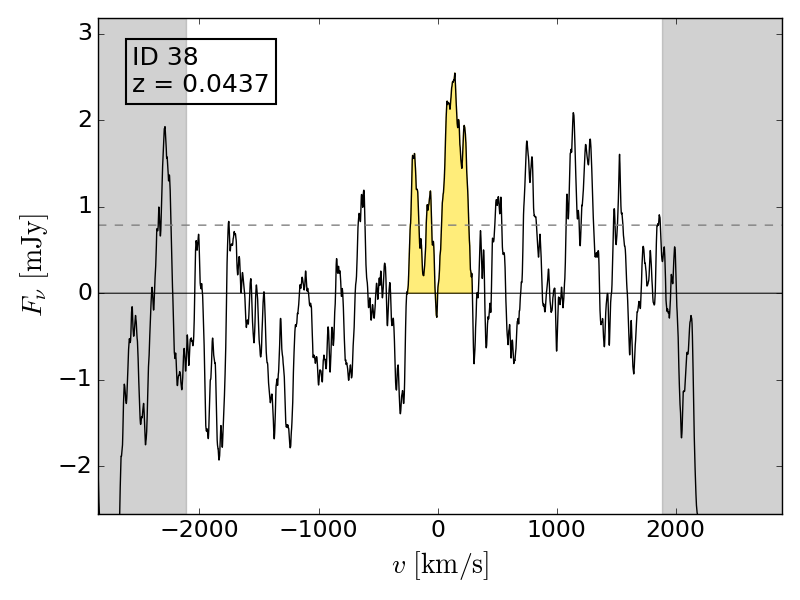}
\includegraphics[width=0.31\textwidth]{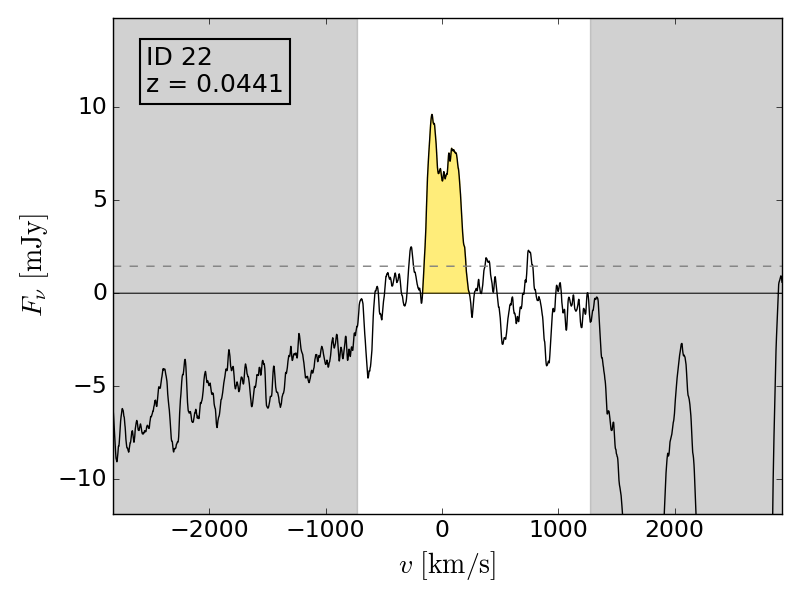}
\includegraphics[width=0.31\textwidth]{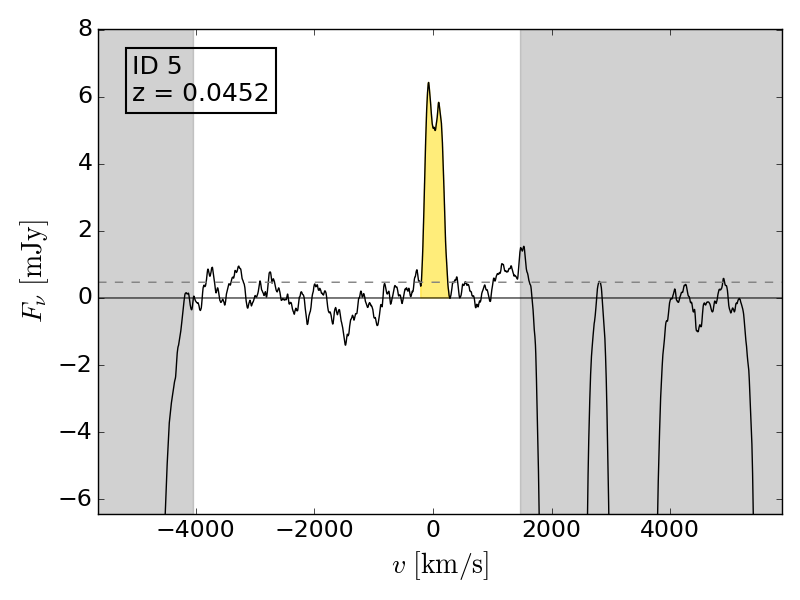}
\includegraphics[width=0.31\textwidth]{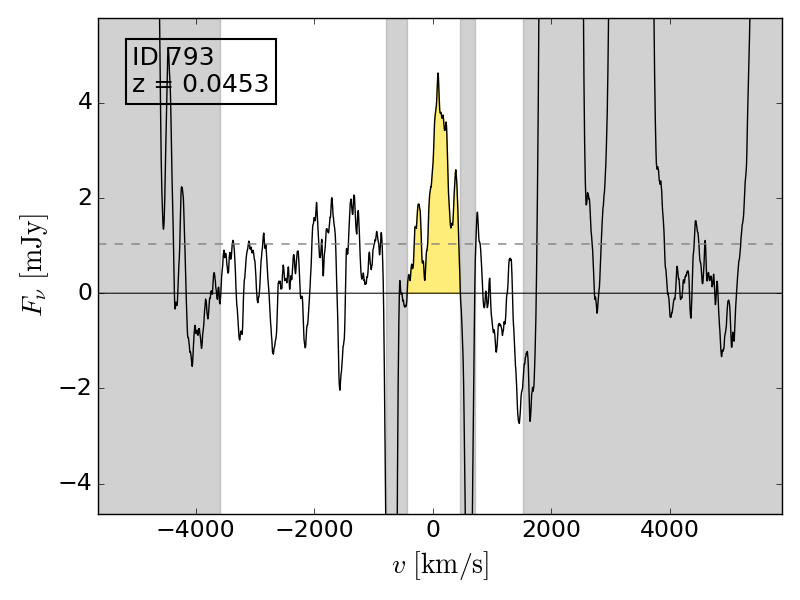}
\includegraphics[width=0.31\textwidth]{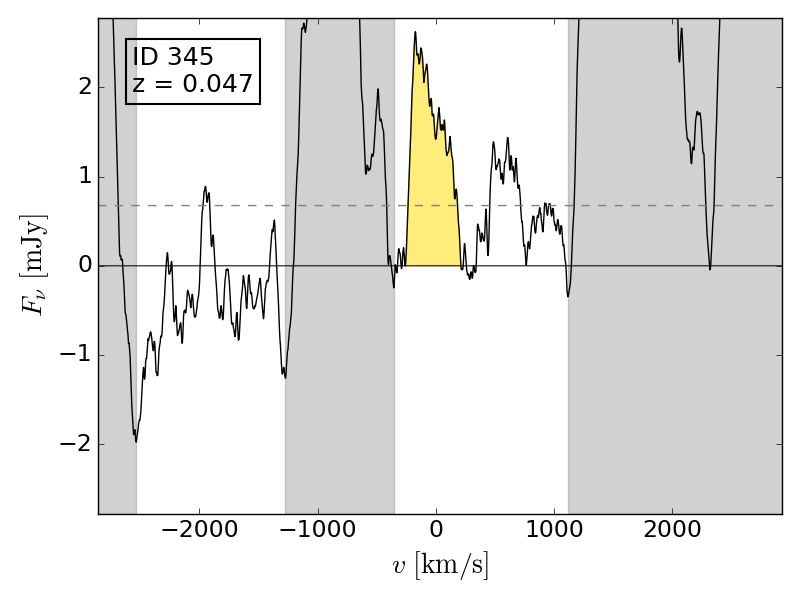}
\includegraphics[width=0.31\textwidth]{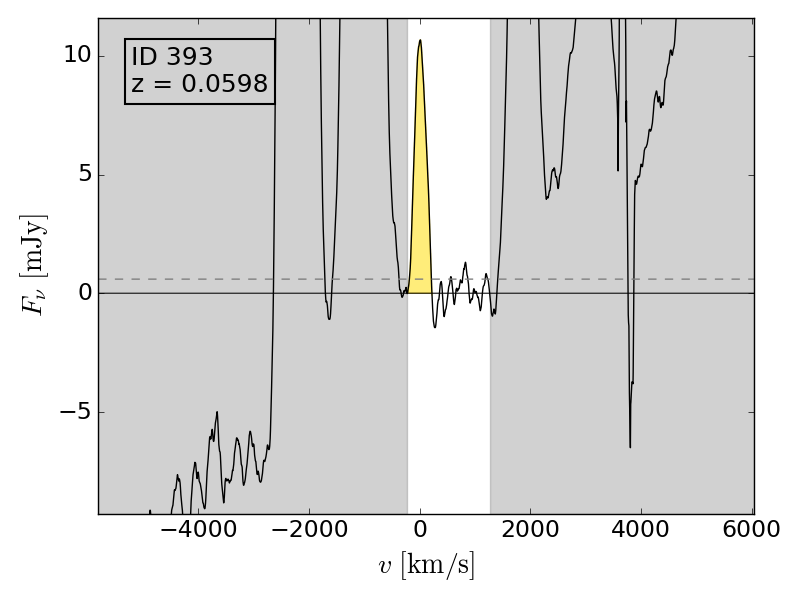}
\includegraphics[width=0.31\textwidth]{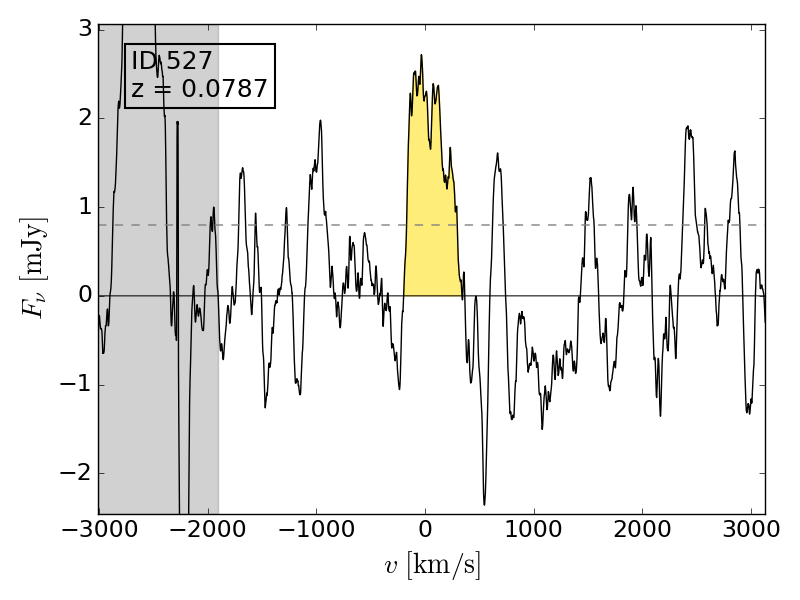}
\includegraphics[width=0.31\textwidth]{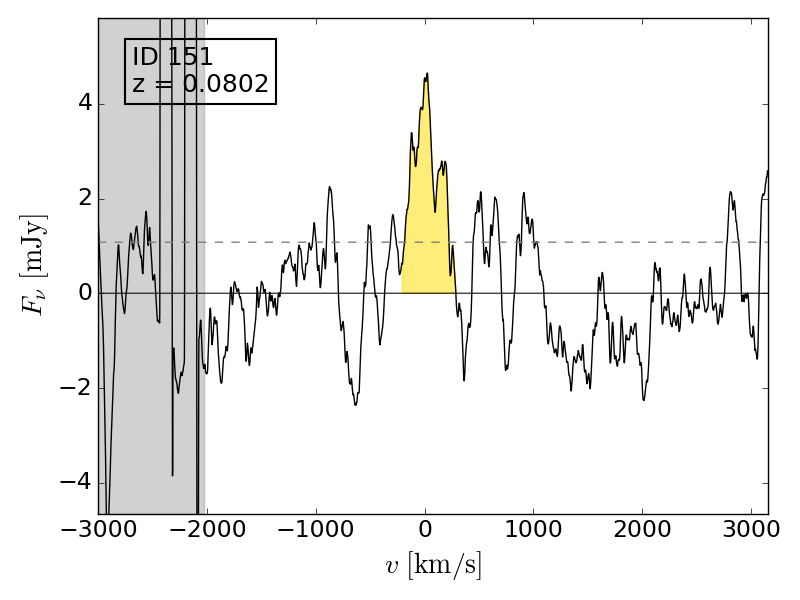}
\includegraphics[width=0.31\textwidth]{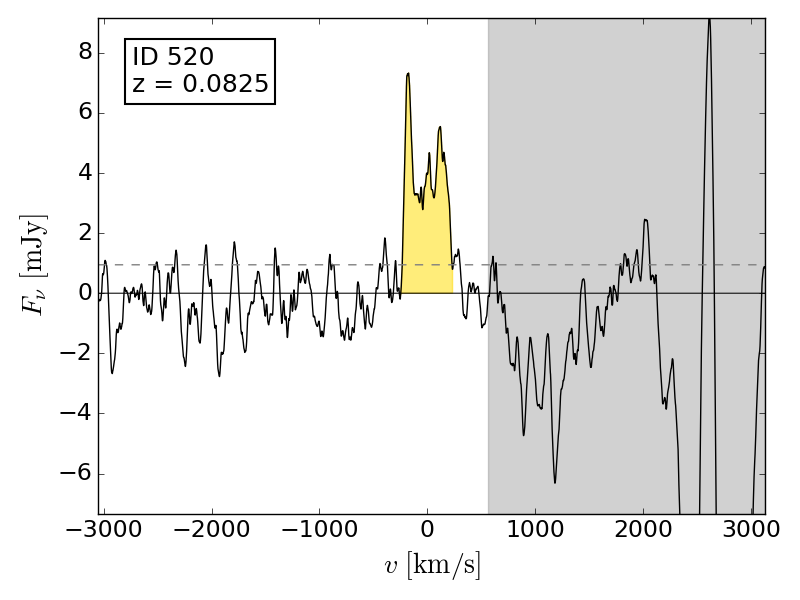}
\includegraphics[width=0.31\textwidth]{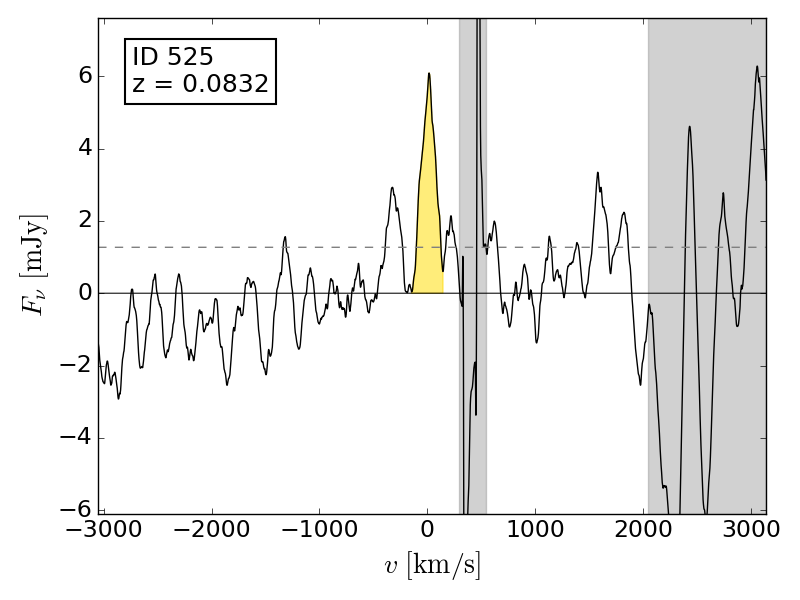}
\caption{(cont.)}
\end{figure*}


\end{document}

%% file: mycommands.tex
\newcommand{\todo}{\ifmmode {\color{red}\text{\Huge{\(\bullet\)}}} \else {\color{red} \Huge$\bullet$}\fi}
\newcommand{\tido}{\ifmmode {\bullet} \else $\bullet$\fi}

\newcommand{\E        }[1]{\ifmmode 10^{#1} \else $10^{#1}$\fi}
\newcommand{\tE        }[1]{\ifmmode \times10^{#1} \else $\times10^{#1}$\fi}
\newcommand{\til}{\ifmmode \sim \else $\sim$\fi}
\renewcommand{\~} {\ifmmode \sim \else $\sim$\fi}

\newcommand{\pc}	{\ifmmode {\rm pc} \else pc\fi}
\newcommand{\ld}	{\ifmmode {\rm l.d.} \else l.d.\fi}
\newcommand{\kms}	{\ifmmode {\rm km\,s}^{-1} \else km\,s$^{-1}$\fi}
\newcommand{\Jykms}	{\ifmmode {\rm Jy\,km\,s}^{-1} \else Jy\,km\,s$^{-1}$\fi}
\newcommand{\cc}	{\ifmmode {\rm cm}^{-3}    \else cm$^{-3}$\fi}
\newcommand{\cmii}	{\ifmmode {\rm cm}^{-2}    \else cm$^{-2}$\fi}
\newcommand{\ergs}	{\ifmmode {\rm erg\,s}^{-1} \else erg s$^{-1}$\fi}
\newcommand{\ergcms}	{\ifmmode {\rm erg\,cm}^{-2}\,{\rm s}^{-1} \else erg\,cm$^{-2}$\,s$^{-1}$\fi}
\newcommand{\ergcmsA}	{\ifmmode {\rm erg\,cm}^{-2}\,{\rm s}^{-1}\,{\rm\AA}^{-1}
\else erg\,cm$^{-2}$\,s$^{-1}$\,\AA$^{-1}$\fi}
\newcommand{  \ergcmsHz  }{\ifmmode{\rm erg\,cm}^{-2}\,{\rm s}^{-1}\,{\rm Hz}^{-1}
                       \else ergs\,cm$^{-2}$\,s$^{-1}$\,Hz$^{-1}$\fi}
\newcommand{\kev}	{\ifmmode {\rm keV} \else keV\fi}

\newcommand{\mic}	{\ifmmode {\rm \mu m} \else $\mu$m\fi}
\newcommand{\vFWHM}	{\ifmmode v_{\mbox{\tiny FWHM}} \else $v_{\mbox{\tiny FWHM}}$\fi}
\newcommand{\vBLR}	{\ifmmode v_{\mbox{\tiny BLR}} \else $v_{\mbox{\tiny BLR}}$\fi}
\newcommand{\sigBLR}	{\ifmmode \sigma_{\mbox{\tiny BLR}} \else $\sigma_{\mbox{\tiny BLR}}$\fi}
\newcommand{\vNLR}	{\ifmmode v_{\mbox{\tiny NLR}} \else $v_{\mbox{\tiny NLR}}$\fi}
\newcommand{\tauBLR}	{\ifmmode \tau_{\mbox{\tiny BLR}} \else $\tau_{\mbox{\tiny BLR}}$\fi}

\newcommand{\Hubble}	{\ifmmode {\rm km\,s}^{-1}\,{\rm Mpc}^{-1} \else km\,s$^{-1}$\,Mpc$^{-1}$\fi}
\newcommand{\NDunit}	{\ifmmode {\rm Mpc}^{-3} \else Mpc$^{-3}$\fi}
\newcommand{\LFunit}	{\ifmmode {\rm Mpc}^{-3}\,{\rm mag}^{-1} \else Mpc$^{-3}$\,mag$^{-1}$\fi}
\newcommand{\MFunit}	{\ifmmode {\rm Mpc}^{-3}\,{\rm dex}^{-1} \else Mpc$^{-3}$\,dex$^{-1}$\fi}

\newcommand{\Msun}{\ifmmode M_{\odot} \else $M_{\odot}$\fi}
\newcommand{\Lsun}{\ifmmode L_{\odot} \else $L_{\odot}$\fi}
\newcommand{\Zsun}{\ifmmode Z_{\odot} \else $Z_{\odot}$\fi}
\newcommand{\mpyr}{\ifmmode \Msun\,{\rm yr}^{-1} \else $\Msun\,{\rm yr}^{-1}$\fi}

\newcommand{\qnote}{\ifmmode q_{0} \else $q_{0}$\fi}
\newcommand{\Hnote}{\ifmmode H_{0} \else $H_{0}$\fi}
\newcommand{\hnote}{\ifmmode h_{0} \else $h_{0}$\fi}
\newcommand{\anote}{\ifmmode a_{0} \else $a_{0}$\fi}
\newcommand{\tnote}{\ifmmode t_{0} \else $t_{0}$\fi}

\def\gsim{\;\rlap{\lower 2.5pt \hbox{$\sim$}}\raise 1.5pt\hbox{$>$}\;}
\def\lsim{\;\rlap{\lower 2.5pt \hbox{$\sim$}}\raise 1.5pt\hbox{$<$}\;}

\newcommand{  \Halpha   }{\ifmmode {\rm H}\alpha \else H$\alpha$\fi}

\newcommand{  \ha       }{\Halpha}
\newcommand{  \Hbeta    }{\ifmmode {\rm H}\beta \else H$\beta$\fi}

\newcommand{  \hb       }{\Hbeta}
\newcommand{  \Hgamma   }{\ifmmode {\rm H}\gamma \else H$\gamma$\fi}
\newcommand{  \Hdelta   }{\ifmmode {\rm H}\delta \else H$\delta$\fi}
\newcommand{  \Lya      }{\ifmmode {\rm Ly}\alpha \else Ly$\alpha$\fi}
\newcommand{  \Lyb      }{\ifmmode {\rm Ly}\beta \else Ly$\beta$\fi}
\newcommand{  \Pa       }{\ifmmode {\rm P}\alpha \else P$\alpha$\fi}
\newcommand{  \Pb       }{\ifmmode {\rm P}\beta \else P$\beta$\fi}
\newcommand{  \Bra      }{\ifmmode {\rm Br}\alpha \else Br$\alpha$\fi}
\newcommand{  \Brg      }{\ifmmode {\rm Br}\gamma \else Br$\gamma$\fi}
\newcommand{  \hi       }{\ifmmode {\rm H}\,\textsc{i} \else H\,\textsc{i}\fi}
\newcommand{  \HI       }{\ifmmode {\rm H}\,\textsc{i} \else H\,\textsc{i}\fi}
\newcommand{  \hii      }{\ifmmode {\rm H}\,\textsc{ii} \else H\,\textsc{ii}\fi}
\newcommand{  \hei      }{\ifmmode {\rm He}\,\textsc{i} \else He\,\textsc{i}\fi}
\newcommand{  \heii     }{\ifmmode {\rm He}\,\textsc{ii} \else He\,\textsc{ii}\fi}
\newcommand{  \HeIIuv   }{\ifmmode {\rm He}\,\textsc{ii}\,\lambda1640 \else He\,\textsc{ii}\,$\lambda1640$\fi}
\newcommand{  \HeIIop   }{\ifmmode {\rm He}\,\textsc{ii}\,\lambda4686 \else He\,\textsc{ii}\,$\lambda4686$\fi}
\newcommand{  \CII	}{\ifmmode \left[{\rm C}\,\textsc{ii}\right]\,\lambda157.74\,\mu{\rm m} \else [C\,{\sc ii}]\ $\lambda157.74\,\mu{\rm m}$\fi}
\newcommand{  \cii	}{\ifmmode \left[{\rm C}\,\textsc{ii}\right] \else [C\,{\sc ii}]\fi}

\newcommand{  \ciii     }{\ifmmode {\rm C}\,\textsc{iii}\right] \else C\,\textsc{iii}]\fi}
\newcommand{  \CIII     }{\ifmmode {\rm C}\,\textsc{iii}\right]\,\lambda1909 \else C\,\textsc{iii}]\,$\lambda1909$\fi}
\newcommand{  \civ      }{\ifmmode {\rm C}\,\textsc{iv}  \else C\,\textsc{iv}\fi}
\newcommand{  \CIV      }{\ifmmode {\rm C}\,\textsc{iv}\,\lambda1549 \else C\,\textsc{iv}\,$\lambda1549$\fi}
\newcommand{  \nii      }{\ifmmode \left[{\rm N}\,\textsc{ii}\right]  \else [N\,\textsc{ii}]\fi}
\newcommand{\NII}{\ifmmode \left[{\rm N}\,\textsc{iii}\right]\,\lambda6583 \else [N\,{\sc iii}]\,$\lambda6583$\fi}
\newcommand{  \niii     }{\ifmmode {\rm N}\,\textsc{iii} \else N\,\textsc{iii}\fi}
\newcommand{  \niv      }{\ifmmode {\rm N}\,\textsc{iv}  \else N\,\textsc{iv}\fi}
\newcommand{  \NIVuv    }{\ifmmode {\rm N}\,\textsc{iv}\,\lambda1486 \else N\,\textsc{iv}\,$\lambda1486$\fi}
\newcommand{  \nv       }{\ifmmode {\rm N}\,\textsc{v}   \else N\,\textsc{v}\fi}
\newcommand{\oi}{\ifmmode \left[{\rm O}\,\textsc{i}\right] \else [O\,{\sc i}]\fi}
\newcommand{\OI}{\ifmmode \left[{\rm O}\,\textsc{i}\right]\,\lambda6300 \else [O\,{\sc i}]$\,\lambda6300$\fi}
\newcommand{\oii}{\ifmmode \left[{\rm O}\,\textsc{ii}\right] \else [O\,{\sc ii}]\fi}
\newcommand{\OII}{\ifmmode \left[{\rm O}\,\textsc{ii}\right]\,\lambda3727 \else [O\,{\sc ii}]\,$\lambda3727$\fi}
\newcommand{\oiii}{\ifmmode \left[{\rm O}\,\textsc{iii}\right] \else [O\,{\sc iii}]\fi}
\newcommand{\OIII}{\ifmmode \left[{\rm O}\,\textsc{iii}\right]\,\lambda5007 \else [O\,{\sc iii}]\,$\lambda5007$\fi}
\newcommand{  \OIIIuv   }{\ifmmode {\rm O}\,\textsc{iii}\,\lambda1663 \else O\,\textsc{iii}\,$\lambda1663$\fi}
\newcommand{  \oiv      }{\ifmmode {\rm O}\,\textsc{iv}  \else O\,\textsc{iv}\fi}
\newcommand{  \OIVuv    }{\ifmmode {\rm O}\,\textsc{iv}\,\lambda1402  \else O\,\textsc{iv}\,$\lambda1402$\fi}
\newcommand{  \OIVIR    }{\ifmmode {\rm O}\,\textsc{iv}\,25.9\,\mu {\rm m} \else O\,\textsc{iv}\,$25.9\,\mu$m\fi}
\newcommand{  \ovi      }{\ifmmode {\rm O}\,\textsc{vi}   \else O\,\textsc{vi}\fi}
\newcommand{  \Ovi      }{\ifmmode {\rm O}\,\textsc{vi}\,\lambda1035 \else O\,\textsc{vi}\,$\lambda1035$\fi}
\newcommand{  \nei      }{\ifmmode {\rm Ne}\,\textsc{i}   \else Ne\,\textsc{i}\fi}
\newcommand{  \neii     }{\ifmmode {\rm Ne}\,\textsc{ii}  \else Ne\,\textsc{ii}\fi}
\newcommand{  \NeiiIR   }{\ifmmode {\rm Ne}\,\textsc{ii}\,12.8\,\mu {\rm m} \else Ne\,\textsc{ii}\,$12.8\,\mu$m\fi}
\newcommand{  \neiii    }{\ifmmode {\rm Ne}\,\textsc{iii} \else Ne\,\textsc{iii}\fi}
\newcommand{  \neiv     }{\ifmmode {\rm Ne}\,\textsc{iv}  \else Ne\,\textsc{iv}\fi}
\newcommand{  \nev      }{\ifmmode {\rm Ne}\,\textsc{v}   \else Ne\,\textsc{v}\fi}
\newcommand{  \NevIR    }{\ifmmode {\rm Ne}\,\textsc{v}\,24.3\,\mu {\rm m} \else Ne\,\textsc{v}\,$24.3\,\mu$m\fi}
\newcommand{  \nevi     }{\ifmmode {\rm Ne}\,\textsc{vi}  \else Ne\,\textsc{vi}\fi}
\newcommand{  \mgi      }{\ifmmode {\rm Mg}\,\textsc{i} \else Mg\,\textsc{i}\fi}
\newcommand{  \mgii     }{\ifmmode {\rm Mg}\,\textsc{ii} \else Mg\,\textsc{ii}\fi}
\newcommand{  \MgII     }{\ifmmode {\rm Mg}\,\textsc{ii}\,\lambda2798 \else Mg\,\textsc{ii}\,$\lambda2798$\fi}
\newcommand{  \sii      }{\ifmmode {\rm S}\,\textsc{ii} \else S\,\textsc{ii}\fi}
\newcommand{  \siii     }{\ifmmode {\rm S}\,\textsc{iii} \else S\,\textsc{iii}\fi}
\newcommand{  \siv      }{\ifmmode {\rm S}\,\textsc{iv} \else S\,\textsc{iv}\fi}
\newcommand{  \sili     }{\ifmmode {\rm Si}\,\textsc{i}   \else Si\,\textsc{i}\fi}
\newcommand{  \silii    }{\ifmmode {\rm Si}\,\textsc{ii}  \else Si\,\textsc{ii}\fi}
\newcommand{  \Siliv    }{\ifmmode {\rm Si}\,\textsc{iv}  \else Si\,\textsc{iv}\fi}
\newcommand{  \SilIVuv  }{\ifmmode {\rm Si}\,\textsc{iv}\,\lambda1400  \else Si\,\textsc{iv}\,$\lambda1400$\fi}
\newcommand{  \AlIII   }{\ifmmode {\rm Al}\,\textsc{iii}\,\lambda1857 \else Al\,\textsc{iii}\,$\lambda1857$\fi}
\newcommand{  \Aliii   }{\ifmmode {\rm Al}\,\textsc{iii} \else Al\,\textsc{iii}\fi}
\newcommand{  \caii     }{\ifmmode {\rm Ca}\,\textsc{ii} \else Ca\,\textsc{ii}\fi}
\newcommand{  \feii     }{\ifmmode {\rm Fe}\,\textsc{ii} \else Fe\,\textsc{ii}\fi}
\newcommand{  \feiii    }{\ifmmode {\rm Fe}\,\textsc{iii} \else Fe\,\textsc{iii}\fi}
\newcommand{  \Kalpha   }{\ifmmode {\rm K}\alpha \else K$\alpha$\fi}

\newcommand{ \Lhb   }{\ifmmode L_{\hb} \else $L_{\hb}$\fi}
\newcommand{ \Lha   }{\ifmmode L_{\ha} \else $L_{\ha}$\fi}
\newcommand{ \fwhb  }{\ifmmode {\rm FWHM}\left(\hb\right) \else FWHM(\hb)\fi}
\newcommand{\sighb  }{\ifmmode \sigma\left(\hb\right) \else $\sigma\left(\hb\right)$\fi}
\newcommand{ \ewhb  }{\ifmmode {\rm EW}\left(\hb\right) \else EW(\hb)\fi}
\newcommand{ \fwha  }{\ifmmode {\rm FWHM}\left(\ha\right) \else FWHM(\ha)\fi}
\newcommand{ \ewha  }{\ifmmode {\rm EW}\left(\ha\right) \else EW(\ha)\fi}
\newcommand{ \Lmg   }{\ifmmode L\left(\mgii\right) \else $L\left(\mgii\right)$\fi}
\newcommand{ \fwmg  }{\ifmmode {\rm FWHM}\left(\mgii\right) \else FWHM(\mgii)\fi}
\newcommand{ \Lciv  }{\ifmmode L\left(\civ\right) \else $L\left(\civ\right)$\fi}
\newcommand{ \fwciv }{\ifmmode {\rm FWHM}\left(\civ\right) \else FWHM(\civ)\fi}
\newcommand{ \fwhm  }{\ifmmode {\rm FWHM} \else FWHM\fi} 
\newcommand{ \voff  }{\ifmmode v_{\rm off} \else $v_{\rm off}$\fi} 
\newcommand{ \vmax  }{\ifmmode v_{\rm max} \else $v_{\rm max}$\fi} 

\newcommand{ \mumg  }{\ifmmode \mu\left(\mgii\right) \else $\mu\left(\mgii\right)$\fi}
\newcommand{ \fmg   }{\ifmmode f\left(\mgii\right) \else $f\left(\mgii\right)$\fi}
\newcommand{ \muciv }{\ifmmode \mu\left(\civ\right) \else $\mu\left(\civ\right)$\fi}
\newcommand{ \fciv  }{\ifmmode f\left(\civ\right) \else $f\left(\civ\right)$\fi}

\newcommand{  \auvo     }{\ifmmode \alpha_{\nu,{\rm UVO}} \else $\alpha_{\nu,{\rm UVO}}$\fi}
\newcommand{  \Ledd     }{\ifmmode L_{\rm Edd} \else $L_{\rm Edd}$\fi}
\newcommand{  \lamLlam  }{\ifmmode \lambda L_{\lambda} \else $\lambda L_{\lambda}$\fi}
\newcommand{  \lLl      }{\ifmmode \lambda L_{\lambda} \else $\lambda L_{\lambda}$\fi}
\newcommand{  \nuLnu    }{\ifmmode \nu L_{\nu} \else $\nu L_{\nu}$\fi}
\newcommand{  \nLn      }{\ifmmode \nu L_{\nu} \else $\nu L_{\nu}$\fi}
\newcommand{  \Luv      }{\ifmmode L_{1450} \else $L_{1450}$\fi}
\newcommand{  \Lop      }{\ifmmode L_{5100} \else $L_{5100}$\fi}
\newcommand{  \lLop     }{\ifmmode \log\left(\Lop/\ergs\right) \else $\log\left(\Lop/\ergs\right)$\fi}
\newcommand{  \Lthree   }{\ifmmode L_{3000} \else $L_{3000}$\fi}
\newcommand{  \lLthree  }{\ifmmode \log\left(\Lthree/\ergs\right) \else $\log\left(\Lthree/\ergs\right)$\fi}
\newcommand{  \Lsix      }{\ifmmode L_{6200} \else $L_{6200}$\fi}
\newcommand{  \lLisx     }{\ifmmode \log\left(\Lop/\ergs\right) \else $\log\left(\Lop/\ergs\right)$\fi}
\newcommand{  \Lxray    }{\ifmmode L_{\rm X} \else $L_{\rm X}$\fi}
\newcommand{  \Lhard    }{\ifmmode L_{\rm 2-10} \else $L_{\rm 2-10}$\fi}
\newcommand{  \Lsoft    }{\ifmmode L_{\rm 0.5-2} \else $L_{\rm 0.5-2}$\fi}

\newcommand{\Fthree}{\ifmmode F_{3000} \else $F_{3000}$\fi}
\newcommand{\fuv}{\ifmmode f_{\lambda}\left(1450{\rm \AA}\right) \else $f_{\lambda}\left(1450 {\rm \AA}\right)$\fi}
\newcommand{\fthree}{\ifmmode f_{\lambda}\left(3000{\rm \AA}\right) \else $f_{\lambda}\left(3000{\rm \AA}\right)$\fi}
\newcommand{\fH}{\ifmmode f_{\lambda}\left(1.65\micron\right) \else
$f_{\lambda}\left(1.65\micron\right)$\fi}

\newcommand{\fbol}{\ifmmode f_{\rm bol} \else $f_{\rm bol}$\fi}
\newcommand{\fbolwv}{\ifmmode f_{\rm bol}\left(\lambda\right) \else $f_{\rm bol}\left(\lambda\right)$\fi}
\newcommand{\fbolopt}{\ifmmode f_{\rm bol}\left(5100{\rm \AA}\right) \else $f_{\rm bol}\left(5100{\rm \AA}\right)$\fi}
\newcommand{\fbolthree}{\ifmmode f_{\rm bol}\left(3000{\rm \AA}\right) \else $f_{\rm bol}\left(3000{\rm \AA}\right)$\fi}
\newcommand{\fboluv}{\ifmmode f_{\rm bol}\left(1450{\rm \AA}\right) \else $f_{\rm bol}\left(1450{\rm \AA}\right)$\fi}

\newcommand{\fobs}{\ifmmode f_{\rm obs} \else $f_{\rm obs}$\fi}

\newcommand{  \mbh      }{\ifmmode M_{\rm BH} \else $M_{\rm BH}$\fi}
\newcommand{  \lmbh     }{\ifmmode \log\left(\mbh/\Msun\right) \else $\log\left(\mbh/\Msun\right)$\fi} 
\newcommand{  \lledd    }{\ifmmode L/L_{\rm Edd} \else $L/L_{\rm Edd}$\fi}
\newcommand{  \mmedd    }{\ifmmode \dot{m}/\dot{m}_{\rm \,Edd} \else $\dot{m}/\dot{m}_{\rm \,Edd}$\fi}
\newcommand{  \Lbol     }{\ifmmode L_{\rm bol} \else $L_{\rm bol}$\fi}
\newcommand{  \lbol     }{\ifmmode L_{\rm bol} \else $L_{\rm bol}$\fi}
\newcommand{  \lLbol    }{\ifmmode \log\left(\Lbol/\ergs\right) \else $\log\left(\Lbol/\ergs\right)$\fi} 
\newcommand{  \Lagn     }{\ifmmode L_{\rm AGN} \else $L_{\rm AGN}$\fi}
\newcommand{  \lagn     }{\ifmmode L_{\rm AGN} \else $L_{\rm AGN}$\fi}

\newcommand{  \tgrow     }{\ifmmode t_{\rm growth} \else $t_{\rm growth}$\fi}
\newcommand{  \tUni      }{\ifmmode t_{\rm Universe} \else $t_{\rm Universe}$\fi}

\newcommand{  \Mindot	}{\ifmmode \dot{M}_{\rm infall} \else $\dot{M}_{\rm infall}$\fi}
\newcommand{  \Mbhdot	}{\ifmmode \dot{M}_{\rm BH} \else $\dot{M}_{\rm BH}$\fi}

\newcommand{  \Maddot	}{\ifmmode \dot{M}_{\rm AD} \else $\dot{M}_{\rm AD}$\fi}

\newcommand{  \Mdot	}{\ifmmode \dot{M} \else $\dot{M}$\fi}

\newcommand{  \as	}{\ifmmode a_{\rm *} \else $a_{\rm *}$\fi}
\newcommand{  \avec	}{\ifmmode \vec{a}_{\rm *} \else $\vec{a}_{\rm *}$\fi}
\newcommand{  \re	}{\ifmmode \eta      	 \else $\eta$\fi}
\newcommand{  \RISCO	}{\ifmmode R_{\rm ISCO}  \else $R_{\rm ISCO}$\fi}
\newcommand{  \rg	}{\ifmmode r_{\rm g}  \else $r_{\rm g}$\fi}
\newcommand{  \rS	}{\ifmmode r_{\rm S}  \else $r_{\rm S}$\fi}

\newcommand{  \mseed    }{\ifmmode M_{\rm seed} \else $M_{\rm seed}$\fi}
\newcommand{  \mbul     }{\ifmmode M_{\rm Bulge} \else $M_{\rm Bulge}$\fi} 
\newcommand{  \mstar    }{\ifmmode M_{\star} \else $M_{\star}$\fi} 
\newcommand{  \mgal     }{\ifmmode M_{*} \else $M_{*}$\fi} 
\newcommand{  \mhost    }{\ifmmode M_{\rm Host} \else $M_{\rm Host}$\fi}
\newcommand{  \mmsmall  }{\ifmmode M_{\rm BH}/M_{*} \else $M_{\rm BH}/M_{*}$\fi}
\newcommand{  \mmlarge  }{\ifmmode M_{*}/M_{\rm BH} \else $M_{*}/M_{\rm BH}$\fi}

\newcommand{  \mmwp     }{\ifmmode \left(M_{*}/M_{\rm BH}\right) \else $\left(M_{*}/M_{\rm BH}\right)$\fi}
\newcommand{  \ml       }{\ifmmode M_{*}/L_{*} \else $M_{*}/L_{*}$\fi}
\newcommand{  \mlwp     }{\ifmmode \left(M_{*}/L\right) \else $\left(M_{*}/L\right)$\fi}
\newcommand{  \mlk      }{\ifmmode \left(M_{*}/L_{K}\right) \else $\left(M_{*}/L_{K}\right)$\fi}
\newcommand{  \sigs     }{\ifmmode \sigma_{*} \else $\sigma_{*}$\fi}
\newcommand{  \Reff     }{\ifmmode R_{\rm e} \else $R_{\rm e}$\fi}
\newcommand{  \nser     }{\ifmmode n_{\rm s} \else $n_{\rm s}$\fi}
\newcommand{  \LFIR     }{\ifmmode L_{\rm FIR} \else $L_{\rm FIR}$\fi}
\newcommand{  \Lfir     }{\ifmmode L_{\rm FIR} \else $L_{\rm FIR}$\fi}

\newcommand{  \mdyn     }{\ifmmode M_{\rm dyn} \else $M_{\rm dyn}$\fi} 
\newcommand{  \mgas     }{\ifmmode M_{\rm gas} \else $M_{\rm gas}$\fi} 
\newcommand{  \mh       }{\ifmmode M_{\rm h} \else $M_{\rm h}$\fi}
\newcommand{  \mhalo    }{\ifmmode M_{\rm halo} \else $M_{\rm halo}$\fi}

\newcommand{  \Lcii     }{\ifmmode L_{\cii} \else $L_{\cii}$\fi}

\newcommand{\bj}{\ifmmode b_{\rm J} \else $b_{\rm J}$\fi}

\newcommand{\iab}{\ifmmode i_{\rm AB} \else $i_{\rm AB}$\fi}

\newcommand{\jab}{\ifmmode J_{\rm AB} \else $J_{\rm AB}$\fi}
\newcommand{\hab}{\ifmmode H_{\rm AB} \else $H_{\rm AB}$\fi}
\newcommand{\kab}{\ifmmode K_{\rm AB} \else $K_{\rm AB}$\fi}

\newcommand{\jveg}{\ifmmode J_{\rm Vega} \else $J_{\rm Vega}$\fi}
\newcommand{\hveg}{\ifmmode H_{\rm Vega} \else $H_{\rm Vega}$\fi}
\newcommand{\kveg}{\ifmmode K_{\rm Vega} \else $K_{\rm Vega}$\fi}

\newcommand{  \Chisq    }{\ifmmode \chi^{2} \else $\chi^{2}$}
\newcommand{  \nelec    }{\ifmmode n_{e} \else $n_{e}$\fi}     
\newcommand{  \nh       }{\ifmmode n_{H} \else $n_{H}$\fi}     
\newcommand{  \Ncol     }{\ifmmode N_{col} \else $N_{col}$\fi} 
\newcommand{  \NH       }{\ifmmode N_{H} \else $N_{H}$\fi}     

\newcommand{\MgasCO}{\ifmmode M_{\rm gas, \, CO} \else $M_{\rm gas, \, CO}$\fi}
\newcommand{\Mgasdust}{\ifmmode M_{\rm gas, \, dust} \else $M_{\rm gas, \, dust}$\fi}
\newcommand{\lMgasCO}{\relax\ifmmode \log \left( M_{\rm gas, \, CO} \right) \else $ \log \left( M_{\rm gas, \, CO}  \right)$\fi}
\newcommand{\lMgasdust}{\ifmmode  \log \left( M_{\rm gas, \, dust} \right) \else $ \log \left( M_{\rm gas, \, dust} \right)$\fi}
\newcommand{\Mdust}{\ifmmode M_{\rm dust} \else $M_{\rm dust}$\fi}
\newcommand{\lMdust}{\ifmmode  \log \left( M_{\rm dust} \right) \else $ \log \left( M_{\rm dust} \right)$\fi}

\newcommand{\Hmol }{\ifmmode {\rm H_2} \else ${\rm H_2}$\fi}

\def\deg{\hbox{$^\circ$}}

\def\micron{\hbox{$\mu$m}}
\def\ion#1#2{#1$\;${\small\rm\@Roman{#2}}\relax}

